\RequirePackage{fix-cm}
\documentclass[superscriptaddress,secnumarabic,amssymb,amsmath,nobibnotes,aps,prd,showkeys,showpacs,nofootinbib,twocolumn,a4paper,epjc3]{svjour3}
\smartqed  
\RequirePackage{graphicx}
 \RequirePackage{mathptmx}      
\RequirePackage{latexsym}
\RequirePackage[numbers,sort&compress]{natbib}
\RequirePackage[colorlinks,citecolor=blue,urlcolor=blue,linkcolor=blue]{hyperref}
\journalname{Eur. Phys. J. C}
\usepackage{amsmath}
\usepackage{amsfonts}
\usepackage{amssymb}
\usepackage{widetext}
\usepackage{subfigure}
\usepackage{epstopdf}
\usepackage{epsf}
\def\e{\mathbf{e}}

\def\udot{\dot{u}}
\def\ex{e_1{}^1}
\def\ey{e_2{}^2}
\def\ez{e_3{}^3}

\def\y{\vartheta}
\def\z{\varphi}

\begin{document}

\title{Einstein-æther models III: conformally static metrics, perfect fluid and scalar fields}

\author{Genly Leon\thanksref{e1,addr1} \and  Alfredo D. Millano \thanksref{e2,addr1} \and  Joey Latta \thanksref{e3,addr2}.
}
\thankstext{e1}{e-mail: genly.leon@ucn.cl}
\thankstext{e2}{alfredo.millano@alumnos.ucn.cl}
\thankstext{e3}{lattaj@mathstat.dal.ca} 
\institute{Departamento  de  Matem\'aticas,  Universidad  Cat\'olica  del  Norte, Avda. Angamos  0610,  Casilla  1280  Antofagasta,  Chile\label{addr1} \and Department of Mathematics and Statistics,
 Dalhousie University, Halifax, Nova Scotia, Canada  B3H 3J5 \label{addr2}
}
\date{Received: 06 October 2020/ Accepted: 03 December 2020}

\maketitle

\begin{abstract}
The asymptotic properties of conformally static metrics in Einstein-æther theory with a perfect fluid source and a scalar field are analyzed. In case of perfect fluid, some relativistic solutions  are recovered such as: Minkowski spacetime, the Kasner solution,  a flat FLRW space and static orbits depending on the barotropic parameter $\gamma$. To analyze locally the behavior of the solutions near a sonic line $v^2=\gamma-1$, where $v$ is the tilt, a new ``shock" variable is used. Two new equilibrium point on this line are found.  These points do not exist in General Relativity when $1 <\gamma<2 $. In the limiting case of General Relativity these points represent stiff solutions with extreme tilt. Lines of equilibrium points associated with a change of causality of the homothetic vector field are found in the limit of General Relativity. For non-homogeneous scalar field $\phi(t,x)$ with potential  $V(\phi(t,x))$ the symmetry of the conformally static metric restrict the scalar fields to be considered to  $ \phi(t,x)=\psi (x)-\lambda t,   V(\phi(t,x))= e^{-2 t} U(\psi(x))$,  $U(\psi)=U_0 e^{-\frac{2 \psi}{\lambda}}$. An exhaustive analysis (analytical or numerical) of the stability conditions is provided for some particular cases.
\end{abstract}

\maketitle
\date{\today }

\section{Introduction}

According to the measurements from  type Ia supernovae \cite{Riess:1998cb} the Universe  is experiencing an accelerated expansion due to an unknown 
``Dark Energy'' source, that was introduced in the standard cosmological model to account for $68\%$ of the energy content of the universe \cite{Ade:2013sjv}. Measurements of anisotropies of the cosmic microwave background (CMB) from experiments including the WMAP
\cite{Bennett:2012zja} and Planck \cite{Ade:2013zuv} satellites, have provided strong support for the standard $\Lambda$CDM model of cosmology
where $\Lambda$ is a cosmological constant. However, there are some tensions with local measurements of the Hubble expansion rate from supernovae Ia \cite{Riess:2011yx} and other cosmological data \cite{Ade:2014xna}, that settled this model under question mark. A very well-known issue of  $\Lambda$CDM model is that the energy density comprised in a cosmological constant
$\Lambda$ has to be fine-tuned by $\sim ~55$ orders of magnitude to account for the present acceleration \cite{Martin:2012bt}. Therefore, various attempts to explain the cosmic acceleration within General Relativity (GR) were proposed as alternatives to $\Lambda$CDM,  as well as several alternatives that abandon GR and modify the Einstein-Hilbert action.  Within the last group, a very interesting alternative is the so-called  Einstein-{\ae}ther theory, which is an effective field theory in which the Hilbert action is modified by the introduction of a dynamical timelike unit vector field, $u^{a}$, the æther, which is covariantly coupled, at Lagrangian level, up to the second order derivatives of the spacetime metric $g_{ab}$, excluding total derivatives. The unitarity is imposed by introducing a Lagrangian multiplier in action. This theory has some features that make it of interest to mathematicians, physics, and to cosmologists.  These are: a) it violates the Lorentz invariance, but preserves locality and covariance; b) it has some imprints on the inflationary scenery; c) it satisfies conditions for linearized stability, positive energy, and vanishing of preferred-frame post-Newtonian parameters;   d) for generic values of the coupling constants, the æther and the metric isotropizes (although for large angles or large angle derivatives of the tilt angle there is a runaway behavior in which the anisotropies increases with time, and some singularities may appear); and e) every hypersurface-orthogonal Einstein æther solution is a Ho\v{r}ava solution, etc., see, e.g.,  \cite{Gasperini:1986ym,Kostelecky:1989jp,Jacobson:2000xp,Carroll:2004ai,Eling:2004dk,Lim:2004js,Kanno:2006ty,Zlosnik:2006zu,Donnelly:2010cr,Carruthers:2010ii,Barrow:2012qy,Sandin:2012gq,Alhulaimi:2013sha,Jacobson:2013xta,Blas:2014aca,Coley:2015qqa,Latta:2016jix,Alhulaimi:2017ocb,VanDenHoogen:2018anx,Coley:2019tyx,Leon:2019jnu,Roumeliotis:2018ook,Roumeliotis:2019tvu,Paliathanasis:2020bgs,Paliathanasis:2019pcl,Barausse:2011pu,Donnelly:2011df,Eling:2006df,Eling:2006ec,Eling:2006xg,Eling:2007xh,Elliott:2005va,Foster:2005dk,Garfinkle:2007bk,Garfinkle:2011iw,Heinicke:2005bp,Hikin:2010ry,Jacobson:2011cc,Jacobson:2010mx,Jacobson:2007fh,Jacobson:2008aj,Jacobson:2004ts,Jacobson:2015mra,Jacobson:2014mda,Nakashima:2010nq,Pujolas:2011sk,Seifert:2007fr,Seifert:2006kv}.

Einstein-æther theory has  applications in various anisotropic and inhomogeneous contexts.  In \cite{Coley:2015qqa} it was implemented the 1+3 orthonormal frame formalism, adopting the comoving æther gauge, to obtain evolution equations in normalized variables, which are suitable for numerical calculations and for phase space analyzes. 
Spatially homogeneous Kantowski-Sachs models were studied. e.g., in \cite{Coley:2015qqa,Latta:2016jix,Alhulaimi:2017ocb,VanDenHoogen:2018anx,Paliathanasis:2019hvm}. In \cite{Alhulaimi:2017ocb} the scalar field interacts to both the æther field expansion and shear scalars through the potential. The stability against spatially curvature and anisotropic perturbations was studied. The late-time attractor of the theory is the vacuum de-Sitter expansionary phase.
Static metrics for non-tilted a perfect fluid with  linear and polytropic equations of state, and with a scalar field with exponential or monomial potentials, were studied in \cite{Coley:2015qqa,Coley:2019tyx,Leon:2019jnu}.
Other solutions were examined elsewhere: vacuum Bianchi Type V  \cite{Roumeliotis:2019tvu}; Friedmann--Lema\^{\i}tre--Robertson--Walker metric (FLRW) \cite{Paliathanasis:2019pcl,Roumeliotis:2018ook}; a Locally Rotationally Symmetric (LRS) Bianchi Type III  \cite{Roumeliotis:2018ook}; modified scalar field cosmology with interactions between the scalar field and the æther  \cite{Paliathanasis:2020bgs} based on Einstein--æther theories by \cite{Kanno:2006ty} and  \cite{Donnelly:2010cr}. An emphasis was set on the issue of the existence of solutions of the reduced equations, the classification of the singularities, and the stability analysis. 

These theories are different from Scalar-tensor theories, and they are similar to the particle creation, bulk viscosity, and varying vacuum
theories, or varying-mass dark matter particles theories \cite{Pan:2016jli,Li:2009mf,Pan:2016bug,Basilakos:2009wi,Basilakos:2009ms,Pan:2017ios,Oikonomou:2016pnq,Leon:2009dt}. In \cite{Paliathanasis:2020axi} the Einstein-æther theory which incorporates a scalar field nonminimally coupled to the æther through an effective coupling $B\left(\phi\right)= 6 B_0 \phi^2$  \cite{Kanno:2006ty} was studied.  It was found there are five families of scalar field potentials on the form $V_{A}\left(  \phi\right)  =V_{0}\phi^{p}+V_{1}\phi^{r}$, where $p,r$ are specific constants, which lead to Liouville--integrable systems, and which admit conservation laws  quadratic  in the momenta. Following an analogous strategy in \cite{Paliathanasis:2020pax} were determined  exact and analytic solutions of the gravitational field equations in Einstein-æther scalar model field with a Bianchi I background space with nonlinear interactions of the scalar field with the æther field.  Conservation laws for the field equations for specific forms of the unknown functions such that the field equations are Liouville integrable were derived. Furthermore, the evolution of the anisotropies was studied by determining the equilibrium points and analyzing their stability.

This paper is the third of a series of works devoted to Einstein-æther theory with perfect fluids and scalar fields. In paper I \cite{Coley:2019tyx}, the field equations in the Einstein-æther theory for static spherically symmetric spacetimes and a perfect fluid source, and subsequently with the addition of a scalar field (with an exponential self-interacting potential) were investigated. Appropriate dynamical variables which facilitate the study of the equilibrium points of the resulting dynamical system were introduced. In addition, the dynamics at infinity was discussed. The qualitative properties of the solutions are of particular interest, as well as their asymptotic behavior and whether they admit singularities. A number of new solutions were presented. 
Continuing this line, in paper II \cite{Leon:2019jnu} the existence of analytic solutions for the field equations in the Einstein-æther theory for a static spherically symmetric spacetime was investigated. A detailed dynamical system analysis of the field equations was provided.

This paper is  focused on the study of timelike self-similar (TSS) spherically symmetric models with perfect fluid and/or scalar fields, using the covariant decomposition $1+3$  \cite{vanElst:1996dr,Goliath:1998mx,Coley:2015qqa,wainwrightellis1997}. This formalism  is well-suited for performing qualitative  and numerical analysis.
TSS spherically symmetric models are characterized by a 4-dimensional symmetric homothetic group $H_4$ acting multiply transitively on 3-dimensional timelike surfaces.

This paper is organized as follows. 
In section \ref{section2.2} the $1+3$ orthonormal frame formalism  is summarized. 
In section \ref{aetheory}  the action for the Einstein-æther Theory is presented, following \cite{Jacobson:2008aj,Garfinkle:2011iw,Carroll:2004ai}. 
In section \ref{model3}, TSS spherically symmetric models with perfect fluid are studied using the covariant decomposition $1+3$   \cite{Coley:2015qqa,wainwrightellis1997}, following the approach given in \cite{Goliath:1998mx}. 
In section \ref{thetaperfectfluid}   different stability conditions of the equilibrium solutions of dynamical systems will be established.  Numerical methods will be used to support and validate the analytical results.
In section \ref{SECT:4.1}  a non-homogeneous scalar field $\phi(t,x)$ with potencial  $V(\phi(t,x))$ which satisfies the symmetry of the conformally static metric \cite{Coley:2002je} is studied. 
The $\theta$-normalization procedure will be implemented in section \ref{Sect:4.2}. An exhaustive analysis (analytical or numerical) of the stability conditions is provided for some particular cases.
Section \ref{ch_5} is devoted to conclusions.
   
\section{The $1+3$ orthonormal frame formalism}
\label{section2.2}
In the $1+3$ orthonormal frame formalism \cite{vanElst:1996dr,Coley:2015qqa,wainwrightellis1997} the metric can be expressed as:
\begin{equation}
ds^2=-N^2dt^2+(\ex)^{-2}dx^2+(\ey)^{-2}(d\y^2+\sin^2 \y d\z^2),
\end{equation}
where $N, \ex$ and $\ey$ are functions of $t$ and $x$. \newline
The Killing Vector Fields are given by \cite{Stephani:2003tm}:
\begin{equation}
\partial_{\z},\quad \cos{\z} \partial_{\y}-\sin{\z} \cot{\y} \partial_{\z},\quad \sin{\z} \partial_{\y}+\cos{\z} \cot{\y} \partial_{\z}.
\end{equation}
The frame vectors written in coordinate form are:
\begin{equation}
\e_0=N^{-1}\partial_t,\quad \e_1=\ex \partial_x,\quad \e_2=\ey \partial_{\y},\quad \e_3=\ez \partial_{\z}.
\end{equation}
where $\ez =\ey / \sin{\y}.$ 

Therefore, the kinematic variables are restricted by:
\begin{equation}
\sigma_{\alpha \beta}=\text{diag}(-2\sigma_+,\sigma_+,\sigma_+),\quad \omega_{\alpha \beta}=0,\quad \udot_\alpha=(\udot_1,0,0),
\end{equation}
where 
\begin{equation}
\udot_1=\e_1 \ln{N}.
\end{equation}
The spatial commutators are given by:
\begin{equation}
a_{\alpha}=(a_1,a_2,0),\quad 
n_{\alpha \beta=}\begin{pmatrix}
0 & 0 & n_{13}\\
0 & 0 & 0 \\
n_{13} & 0 & 0
\end{pmatrix},
\end{equation}
where
\begin{equation}
a_1=\e_{1} \ln{\ey},\quad a_2=n_{13}=-\frac{1}{2}\ey\cot{\y}.
\end{equation}
There are restrictions over the matter components:
\begin{equation}
q_{\alpha}=(q_1,0,0),\quad \pi_{\alpha \beta}=diag(-2\pi_+,\pi_+,\pi_+).
\end{equation}
The frame rotation $\Omega_{\alpha \beta}$ is zero.
The cosmological constant is chosen to be 0 for simplicity. 
\newline 
The quatity $n_{13}$ only appears in the equations together with $\e_2 n_{13}$ through the Gauss spatial curvature of the $2$-spheres:
\begin{equation}
{}^2K:=2(\e_2-2n_{13})n_{13},
\end{equation}
which is simplified to: 
\begin{equation}
{}^2K=(\ey)^2.
\end{equation}
Hence, the dependence on $\y$ does not appears explicitly in the equations. The quantity ${}^2K$ is used instead of $\ey$ to write the field equations. The spatial curvatures simplify to:
\begin{subequations}
\begin{equation}
{}^3S_{\alpha \beta}=\text{diag}(-2\; {}^3S_+,{}^3S_+,{}^3S_+),
\end{equation}
with ${}^3R$ y ${}^3S_+$ defined by:
\begin{equation}
{}^3R=4\e_1a_1-6a_{1}^2+2\; {}^2K, 
{}^3S_+=-\frac{1}{3}\e_1a_1+\frac{1}{3}{}^2K.
\end{equation}
\end{subequations}
The components of the  Weyl curvature are simplified to:
\begin{equation}
E_{\alpha \beta}=\text{diag}(-2E_+,E_+,E_+), \quad H_{\alpha \beta}=0,
\end{equation}
with $E_+$  given by:
\begin{equation}
E_+=H\sigma_+ + \sigma_{+}^2+{}^3S_+ -\frac{1}{2}\pi_+.
\end{equation}
Using the following simplifications,
\begin{equation*}
{}^2K=K,\quad \udot_1=\udot,\quad a_1=a,
\end{equation*}
the essential variables are:
\begin{equation}
N,\quad \ex,\quad K,\quad H,\quad \sigma_+,\quad a,\quad \mu,\quad q_1,\quad p,\quad \pi_+,
\end{equation}
and the auxiliary variables are 
\begin{equation}
{}^3 K,\quad {}^3 S_+,\quad \udot.
\end{equation}
The field equations are written as:
\begin{subequations}
\begin{align}
&\e_0 \ex = (-H+2\sigma_+)\ex \\
&\e_0 K = -2(H+\sigma_+)K \\
&\e_0 H = -H^2-2\sigma_{+}^2 +
\frac{1}{3}(\e_1+\udot-2a)\udot-\frac{1}{6}(\mu+3p) \\
&\e_0 \sigma_+  = -3H\sigma_+-\frac{1}{3}(\e_1+\udot+a)\udot-{}^3S_+ +\pi_+ \\
&\e_0 a = (-H+2\sigma_+)a-(\e_1+\udot)(H+\sigma_+)\\
&\e_0 \mu = -3H(\mu+p)-(\e_1+2\udot-2a)q_1-6\sigma_+\pi_+ \\
&\e_0 q_1 = (-4H+2\sigma_+)q_1-\e_1p-(\mu+p)\udot \nonumber \\ & +2(\e_1+\udot-3a)\pi_+
\end{align}
\end{subequations}
The restrictions are the Gauss and Codazzi equations together with the definition of $a$ are the following:
\begin{subequations}
\begin{align}
0 & = 3H^2+\frac{1}{2}{}^3R-3\sigma_{+}^2-\mu,\\
0 & =  -2\e_1(H+\sigma_+)+6a\sigma_+ +q_1,\\
0 & =  (\e_1-2a)K,
\end{align}
\end{subequations}
where the spatial curvatures are given by:
\begin{subequations}
\begin{equation}
{}^3R=4\e_1a-6a^2+2K, \quad 
{}^3S_+=-\frac{1}{3}\e_1 a+\frac{1}{3}K.
\end{equation}
\end{subequations}
Afterwards, the lapse function $N$ is provided specifying the time gauge and, since there are not evolution equations for $p$ and $\pi_+$, they should be specified by the fluid model through equations of state for $p$ and the transport equation  for $\pi_+$.

\section{Einstein-æther Gravity}
\label{aetheory}

The action for the Einstein-æther Theory is the most general covariant functional involving partial derivatives of order at most two (not including total derivatives) of the space-time metric $g_{ab}$ and a vector field $u^a$, called æther  \cite{Jacobson:2008aj,Garfinkle:2011iw,Carroll:2004ai} given by: 
\begin{align}
& S=\int d^{4}x\sqrt{-g}\left[  \frac{1}{2}R +\mathcal{L}_{\text{æ}} + M \left(  u^{c}u_{c
} + 1\right) + \mathcal{{L}}_m  \right]  ,\label{action}
\end{align}
where:
\begin{equation}\label{aeLagrangian}
\mathcal{L}_{\text{æ}} \equiv -K^{a b}{}_{c d}\nabla_{a}u^{c
}\nabla_{b}u^{d},
\end{equation} is the Einstein-æther lagrangian \cite{Jacobson:2008aj} with:
\begin{equation}
K^{a b}{}_{c d}\equiv c_{1}g^{a b}g_{c d} + c_{2}
\delta_{c}^{a}\delta_{d}^{b} + c_{3}\delta_{d}^{a}
\delta_{c}^{b} + c_{4}u^{a}u^{b}g_{c d}.
\end{equation}
That action contains the Einstein-Hilbert term $ \frac{1}{2}R$, wherein: $R$ denotes the Ricci scalar, $g_{a b}$ denotes the metric tensor, and $K^{a b}{}_{c d}$ is a tensor of four indices corresponding to the kinetic terms of the æther. It contains four dimensionless constants  $c_{i}$ and $M$ is the Lagrange multiplier that forces unitarity of the æther vector,  $u^{c}u_{c
}=-1$. That is, $u_{c}$ is a timelike vector \cite{Garfinkle:2011iw}.
The signature of the metric $g_{ab}$  is $({-}{+}{+}{+})$.  Physical units are such that  $c=1, \kappa^2\equiv 8\pi G=1,$ where $c$ is the speed of light. 

The field equations of the Einstein--æther theory accounts for \cite{Donnelly:2010cr,Jacobson:2000xp}:
\begin{itemize}
\item The effects of anisotropy and inhomogeneities (e.g., curvature) on the geometry of the spherically symmetric models under consideration.

\item The contribution from the energy-momentum tensor $T _{ab}^{\text{æ}}$ of the æther field, which depends on the dimensionless parameters $ c_i, i=1, \ldots 4$. In General Relativity all $ c_i = 0 $, hence  the Einstein's field equations are generalized. 
To study the effects of matter, the values corresponding to General Relativity, or values close to them, can be substituted.

\item When studying the phenomenology of theories within a preferred framework, and particularly,  in the isotropic and spatially homogeneous universe, it is generally assumed  the æther field will be aligned with the cosmic frame
(natural resting frame preferred by the CMB) and therefore is related to the expansion rate of the universe.

\item 
In principle, in spherically symmetric models the preferred frame determined by the æther can be different (that is, tilted) to the CMB rest frame. This adds additional terms to the energy-momentum tensor of the æther $T _{ab}^{\text{æ}}$, for example, an hyperbolic angle of tilt, $v$, which measures the æther boost with respect the CMB rest frame
\cite{Kanno:2006ty,Carruthers:2010ii}. 
In homogeneous but spatially anisotropic models, it is expected that the hyperbolic inclination angle $v$ will decay along with its derivative as $t\rightarrow +\infty$ \cite{Coley:2004jm,Coley:2006nw}.  
\end{itemize}
All spherically symmetric æther fields are  hypersurface-- orthogonal. Therefore, all spherically symmetric solutions of the æther theory will also be solutions in the infrared limit of gravity of Ho\v{r}ava. The opposite is not true in general, but it is true for solutions with spherical symmetry with regular center \cite{Jacobson:2008aj}.
When spherical symmetry is imposed, the æther is hypersurface-- orthogonal, and it has zero twist. Therefore, without loss of
generality it is possible to make $ c_4 $ zero \cite{Jacobson:2008aj}.
After redefining parameters to remove $ c_4 $, the parameter space is 3-dimensional. The $ c_i $ contributes to the effective Newtonian gravitational constant $ G $. Then a parameter $ c_i $ can be specified to make $ 8 \pi G = 1 $. The remaining two parameters characterize two non-trivial physical quantities, for example, the Schwarzschild mass and radius of a matter distribution. The other restrictions imposed on the $ c_i $ are summarized in \cite{Jacobson:2008aj} and in equations 43-46 in \cite{Barausse:2011pu}.

The field equations obtained by variation of (\ref{action}) with respect to 
$g^{ab}$, $u^a$, and $M$ are respectively given by  \cite{Garfinkle:2007bk}:
\begin{subequations}
\begin{eqnarray}
{G_{ab}} &=& {T^{TOT}_{ab}}
\label{EFE2}
\\
M {u_b} &=& {\nabla _a} {{J^a}_b}+ c_4 \udot_a \nabla_b u^a
\label{evolveu}
\\
{u^a}{u_a} &=& -1,\label{unit}
 \end{eqnarray}
 \end{subequations}
where $G_{ab}$ is the Einstein tensor of the metric $g_{ab}$,  ${T^{TOT}_{ab}}$
is the total energy-momentum tensor, ${T^{TOT}_{ab}}=T _{ab}^{\text{æ}}+T^{mat}_{ab}$,
where $T^{mat}_{ab}$ is the total contribution of matter. 
$T^{mat}_{ab}$ will be omitted for the moment (and will be added later for models with perfect fluid and with scalar field), starting with vacuum case ($\mathcal{{L}}_m=0$). 
The quantities ${J^a}_b,\ {\udot_a}$
and the æther energy-momentum tensor $T _{ab}^{\text{æ}}$ are given by:
\begin{subequations}
\begin{align} 
{{J^a}_m} & =
-{{K^{ab}}_{mn}}{\nabla_b}{u^n}
\label{J} \\
{\dot u_a} &= {u^b}{\nabla _b}{u_a},
\label{a} \\
{T _{ab}^{\text{æ}}} &= 2c_{1}(\nabla_{a}u^{c}\nabla_{b}u_{c}-
\nabla^{c}u_{a}\nabla_{c}u_{b})  \nonumber\\
&  - 2[\nabla_{c}(u_{(a} J^{c}{}_{b)}) + \nabla_{c}(u^{c
}J_{(a b)}) - \nabla_{c}(u_{(a}J_{b)}{}^{c})] \nonumber\\
&  -2 c_4 \udot_a \udot_b + 2 M u_a u_b + g_{a b}\mathcal{L}_{\text{æ}}. \label{aestress}
\end{align}
\end{subequations}
Taking the contraction of \eqref{evolveu} with $u^b$ and with the induced metric  $h^{b c}:= g^{b c}+u^b u^c$ the following equations are obtained: 
\begin{subequations}
\label{aether_eqs}
\begin{align}
\label{definition:lambda}
&M = - u^b \nabla_a J^a_b-c_4 \udot_a \udot^a,\\
\label{restriction_aether}
& 0 = h^{b c}\nabla_a J^a_b + c_4 h^{b c} \udot_a \nabla_b u^a.
\end{align}
\end{subequations}
Equation \eqref{definition:lambda} is used as the definition of the Lagrange multiplier. The second system of equations give compatibility conditions that the æther vector must satisfy. 
\section{Timelike self-similar spherically symmetric  perfect fluid models}
\label{model3}

In the diagonal homothetic formulation, the line element can be written in diagonal form, where one of the coordinates adapts to the homothetic symmetry \cite{Bogoyavlensky}:
\begin{align}
\label{metricTSS}
& d\tilde{s}^2 = e^{2 t} ds^2= \nonumber \\
& e^{2 t}\Big[- b_1^{-2}(x)  dt^2 + dx^2   +b_2^{-2}(x) (d\y^2 + \sin^2 (\y)  d\z^2)\Big].
\end{align}
For the conformally static metric with line element given by \eqref{metricTSS}, the following scalars can be defined:
\begin{small}
\begin{align}
& \theta=\frac{\sqrt{3}}{3}\left(2\alpha -\beta\right), \quad \sigma= \frac{\sqrt{3}}{3}\left(-\alpha + 2 \beta\right), \ \nonumber \\
& \alpha=3\widehat{\beta^0}, \quad  \beta= 3\widehat{\beta^+}, \quad  b_1^{-1}=e^{\beta^0-2\beta^+},  \quad  b_2^{-1}=e^{\beta^0+\beta^+},    
\end{align}
\end{small}
where  $\widehat{...}$ denotes the derivative with respect to the spatial variable $x$. The quantities $\alpha$ and $\beta$ are respectively the expansion scalar and the shear scalar of the normal congruence to the symmetry surface of the static universe $\left(\mathcal{M}, d{s}^2\right)$, conformally related to the physical spacetime $\left(\mathcal{M}, d {\tilde{s}}^2\right)$ through the homothetic factor $e^{2 t}$. A non-tilted æther vector  $\mathbf{u}=e^{-t }b_1 \partial_t$ is considered. 
\newline 
Assuming that the matter content of the physical universe $\left(\mathcal{M}, d\tilde{s}^2\right)$ is a perfect fluid, that is specified by a 4-velocity vector field  $\mathbf{v}$ given by
$\mathbf{v} = \Gamma e^{-t }(-b_1 \partial_t + v \partial_x),\quad   \Gamma = (1-v^2)^{-\frac12}$, where $v$ is the tilt parameter, which is a funtion of $x$, with $-1\leq v \leq 1$. The equation of state parameter $p = (\gamma-1) \mu, \quad 1 \leq \gamma < 2$ is chosen for the perfect fluid (unless otherwise stated). By convenience a function that depends only on  $x$:   $\mu_t=\frac{e^{-2 t } \left((\gamma -1) v^2+1\right) \mu}{1-v^2}$ is defined, which represents the energy density of the fluid  measured by an observer associated with the homothetic symmetry.
On the other hand, if the content of matter is that of a non-homogeneous scalar field, $ \phi(t, x) $, with its self-interaction potential $V(\phi (t, x)) $, these must respect the homothecy of the conformally static symmetry associated with the line element \eqref{metricTSS}, so they have to be of the form \cite{Coley:2002je}: $ \phi(t,x)=\psi (x)-\lambda t, \quad  V(\phi(t,x))= e^{-2 t} U(\psi(x))$,  $U(\psi)=U_0 e^{-\frac{2 \psi}{\lambda}}$,
where by convenience it is assumed $\lambda>0$, such that for $\psi>0$, $U\rightarrow 0$ as $\lambda \rightarrow 0$, which restrict the kind of scalar field potentials to be considered.  

Using the metric  \eqref{metricTSS}, the lagrangian \eqref{aeLagrangian} becomes:
\begin{align}
&\mathcal{L}_{u} = \frac{1}{3} e^{-2 t} \left(( c_{1}- c_{4}) {\sigma}^2-9  b_{1}^2 ( c_{1}+3 c_{2}+ c_{3})\right).
\end{align}
The Lagrange multiplier  \eqref{definition:lambda} is calculated as:
\begin{align}
& M=-3 e^{-2 t}b_1^2 (c_{1}+c_{2}+ c_{3}) \nonumber \\
& -\frac{1}{3} e^{-2 t} {\sigma}({\sigma}(- c_{1}+ c_{3}+2  c_{4})+2 c_{3} {\theta}) -\frac{ c_{3} e^{-2 t} \widehat{{\sigma}}}{\sqrt{3}}.
\end{align}
The æther  equation \eqref{restriction_aether} is reduced to: 
\begin{equation}
{e^{-t}  b_{1} {\sigma} (2  c_{1}+3  c_{2}+ c_{3}- c_{4})}=0.
\end{equation}
The trace of the intrinsic Ricci 3-curvature of the spatial 3-surfaces orthogonal to $\mathbf{u}$ is given by
\begin{align}
& {}^{*} R =-\frac{2}{3} e^{-2 t} \left(-3  b_{2}^2+2 \sqrt{3} \left(\widehat{ {\theta}}+\widehat{ {\sigma}}\right)+3 ( {\theta}+ {\sigma})^2\right).
\end{align}
Now the æther parameters are re-defined as \cite{Jacobson:2013xta}:
$c_\theta = c_2 + (c_1 + c_3)/3,\ c_\sigma = c_1 + c_3,\ c_\omega = c_1 - c_3,\ c_a = c_4 - c_1$, corresponding to terms in the Lagrangian relative to expansion, shear scalar, acceleration and  twist of the æther. To impose the condition of the æther \eqref{restriction_aether} is taken $c_a = 3 c_{\theta} $, $ (2 c_ {1} +3 c_ {2} + c_ {3} - c_ {4}) = 0$.
Therefore, the parameter space is reduced to a constant, $ c_\theta $.
\newline
The æther energy-moment tensor can be expressed by:
\begin{small}
\begin{equation}
\label{Taether}
{T^{\text{æ}}}_{a}^{b}=e^{-2 t}\left(
\begin{array}{cccc}
  \mu & -  q & 0 & 0 \\
-  q &   p-2 \pi & 0 & 0 \\
 0 & 0 &   p+\pi & 0 \\
 0 & 0 & 0 &   p+\pi \\
\end{array}
\right);
\end{equation}
\end{small}
\noindent $\mu = c_{\theta } \left(9 b_1^2-4  {\theta}  {\sigma}-2 \sqrt{3} \widehat{ {\sigma}}-3  {\sigma}^2\right),
  p =  \frac{1}{3} c_{\theta } \left(9 b_1^2- {\sigma}^2\right)$, \newline $q =   -2 \sqrt{3} b_1 c_{\theta }  {\sigma},   \pi =   -\frac{2}{3} c_{\theta } {\sigma}^2$ are the effective energy density, isotropic pressure, energy flux, and anisotropic pressure of the æther, measured by an observer associated with the homothetic symmetry in the static universe $\left(\mathcal{M}, d{s}^2\right)$; therefore,  depending only on $x$. 
     
The matter energy-momentum tensor is given by: 
 \begin{small}
\begin{align}
\label{Tm}
& {T^{m}}_{a}^{b}=\nonumber \\
& e^{-2 t} \left(
\begin{array}{cccc}
  \mu_{t}  & -\frac{\gamma  v  \mu_{t}}{(\gamma -1) v ^2+1} & 0 & 0 \\
 -\frac{  \gamma  v  \mu_{t}}{(\gamma -1) v ^2+1} & \frac{  \left(v ^2+\gamma -1\right) \mu_{t}}{(\gamma -1) v ^2+1} & 0 & 0 \\
 0 & 0 & -\frac{  (\gamma -1) \left(v ^2-1\right) \mu_{t}}{(\gamma -1) v ^2+1} & 0 \\
 0 & 0 & 0 & -\frac{  (\gamma -1) \left(v ^2-1\right) \mu_{t}}{(\gamma -1) v ^2+1} \\
\end{array}
\right).
\end{align}
\end{small}
Using the Einstein equations, the Jacobi identities and the contracted Bianchi identities, a system of ordinary differential equations for the frame vectors and the comutation functions, and an extra equation for the æther are obtained. The comoving gauge is chosen for æther; leaving as a degree of freedom a reparametrization of the spatial variables and the temporal variable. 

The final equations are:\\ 
\noindent{\bf Propagation Equations:}  
\begin{subequations}
\label{modelouno}
\begin{align}
&\widehat{ {\theta}}=-\sqrt{3} b_{2}^2-\frac{ {\sigma}\left(2 C_2  {\sigma}+ {\theta}\right)}{\sqrt{3}}-\frac{\sqrt{3} \gamma \mu_{t} v^2}{(\gamma -1) v^2+1}, \label{EQ48 }\\ 
&\widehat{ {\sigma}}=-\frac{ {\sigma} (2  {\theta}+ {\sigma})}{\sqrt{3}}+\frac{\sqrt{3} \mu_{t} \left(2-3 \gamma +(\gamma -2)
   v^2\right)}{2 C_2 \left((\gamma -1) v^2+1\right)},\\ 
&\widehat{b_1}= \frac{b_1  {\sigma}}{\sqrt{3}},\\
&\widehat{b_2}= -\frac{b_2 ( {\theta}+ {\sigma})}{\sqrt{3}},\\
&\widehat{v}=  \frac{\left(1-v^2\right)}{\sqrt{3} \gamma  \left(1-\gamma +v^2\right)} \Big\{ \gamma  v (2 (\gamma -1) \theta +\gamma  \sigma) \nonumber \\
& +\sqrt{3} b_1\left((\gamma -1) (3\gamma -2)+(\gamma -2) v^2\right)\Big\}.
\end{align}
\end{subequations}
\noindent{\bf Equation for  $\mu_t$:}
	\begin{equation}
	\mu_t= \frac{\left((\gamma -1) v^2+1\right) }{3 \left(1-\gamma -v^2\right)}\left(C_2 \left(3 b_1^2+\sigma ^2\right)+3 b_2^2-\theta^2\right),
	\end{equation}
\noindent{\bf	Auxiliary equation:}
\begin{align}
& \widehat{\mu _{t}}=\frac{\mu_{t}}{\sqrt{3} \left(\gamma -v^2-1\right) \left((\gamma -1) v^2+1\right)} \times \nonumber \\
 &  \Big\{ \gamma  \left(\sigma+(\gamma -1) v^4 (2 \theta +\sigma )-v ^2 ((4 \gamma -6) \theta+\gamma  \sigma )\right)\nonumber \\
& +2 \sqrt{3} b_1 v \left((7-3 \gamma ) \gamma +(\gamma  (2 \gamma -5)+4) v^2-4\right)\Big\}.
\end{align}
\noindent{\bf Restriction:}
\begin{equation}
\gamma  \mu _{t} v-\frac{2 C_2 b_1 \sigma \left((\gamma -1) v^2+1\right)}{\sqrt{3}}=0.\end{equation}
To simplify the notation the following parameters are introduced $C_1=1-2c_\sigma, C_2=1+3c_\theta,C_3=1+c_a$, such that by choosing $C_1=1, C_2=1, C_3=1,$  General Relativity is recovered.
The condition  $c_a = 3 c_{\theta }$, implies $C_2=C_3$, and the parameter  $C_1$ does not appears explicitly in the equations, only on the definition of the Lagrange multiplier. 

For an ideal gas with $\gamma=1$  the matter energy-momentum tensor is simplified to: 
\begin{equation}
\label{Tmgas-ideal}
{T^{m}}_{a}^{b}=  e^{-2 t}  \left(
\begin{array}{cccc}
  \mu _{t} & -  \mu _{t} v & 0 & 0 \\
 -  \mu _{t} v  &   \mu _{t} v^2 & 0 & 0 \\
 0 & 0 & 0 & 0 \\
 0 & 0 & 0 & 0 \\
\end{array}
\right),
\end{equation}
where $v$ measures the inclination of the fluid relative to the 4-velocity of the æther. The equations reduce to
\begin{subequations}
\label{YYYYY}
\begin{align}
& \widehat{\theta}= -\sqrt{3} b_2^2-\frac{\sigma\left(2 C_2 \sigma+\theta \right)}{\sqrt{3}}-\sqrt{3} \mu_{t} v^2,\\
& \widehat{\sigma}= -\frac{\sqrt{3}\mu_{t}( \left(v^2+1\right)}{2
   C_2}-\frac{\sigma (2 \theta+\sigma)}{\sqrt{3}},\\
& \widehat{b_1}= \frac{b_1 \sigma}{\sqrt{3}},\\
& \widehat{b_2}=  -\frac{b_2 (\theta+\sigma)}{\sqrt{3}},\\
& \widehat{\mu_t}= \mu_{t}\left(\frac{v^2 (\sigma-2 \theta)-\sigma}{\sqrt{3} v^2}-2 b_1 v\right),\\
& \widehat{v}= b_1 \left(v^2-1\right)-\frac{\sigma \left(v^2-1\right)}{\sqrt{3} v},
\end{align}
\end{subequations}
with restrictions
\begin{subequations}
\label{YYYYY-rest}
\begin{align}
& \mu_{t} v-\frac{2 C_2 b_1 \sigma}{\sqrt{3}}=0,\\
& -C_2 \left(3 b_1^2+\sigma^2\right)-3 b_2^2+\theta^2-3 \mu_{t} v^2=0.
\end{align}
\end{subequations}
\newline $C_2 = 1$ needs to be set to recover General Relativity . For this reason is natural to consider $C_1 =\mathcal{O}(1)$, meanwhile it can be assumed  $C_2>0$. The restriction
$\frac{\left(\gamma  v^2-v^2+1\right)}{\gamma +v^2-1}\geq 0$ is imposed. That is, $0<\gamma \leq 1, 1-\gamma \leq v^2\leq 1$, or  $1<\gamma <2, -1\leq v\leq 1$.
Due to the usual energy condition for the fluid, which is expressed as $ 1 <\gamma <2 $, the second condition is satisfied. Together with the energy condition $ \mu_t \geq 0 $, lead to
\begin{equation}
C_2 \left(3 b_1^2+\sigma ^2\right)+3 b_2^2 \leq \theta^2.
\end{equation}
By hypothesis $C_2>0$,   $\theta^2$ 
is the dominant quantity, and the terms on the left hand side of the above inequality are both non-negative. This suggests to considering $\theta$-normalized equations. 

\section{$\theta$-normalized equations}
\label{thetaperfectfluid}

In this section four specific models will be studied by using  the following normalized variables,
\begin{align}
{\Sigma} =\frac{\sigma}{\theta},\quad  A=\frac{\sqrt{3} b_1}{\theta},\quad K=\frac{3 b_2^2}{\theta^2},\quad {\Omega}_t =\frac{3\mu_t}{\theta^2},
\end{align}
and the radial coordinate 
$$ f'=\frac{df}{d \eta} := \frac{\sqrt{3}\widehat{f}}{\theta}.$$
A parameter  ${r}$ is defined in analogous way to the ``Hubble gradient parameter'' ${r}$, by 
\begin{equation}
   \widehat{\theta}=-r {\theta}^2,
\end{equation}
\begin{equation}
\label{defnr}
{r}=\frac{{\Sigma} \left(2 C_2 {\Sigma} +1\right)}{\sqrt{3}}+\frac{\gamma  {\Omega}_t v^2}{\sqrt{3} \left((\gamma -1) v^2+1\right)}+\frac{K}{\sqrt{3}},
\end{equation}
The normalized equations of interest are: 
\newline 
\noindent{\bf Propagation equations:} 
\begin{subequations}
\label{system29}
\begin{align}
&{\Sigma}'=-{{\Sigma}} \left({{\Sigma}}-\sqrt{3} {r}+2\right)+\frac{{{\Omega}_t} \left(-3 \gamma +(\gamma -2) v^2+2\right)}{2 C_2 \left((\gamma -1) v^2+1\right)},\\
& A'=A \left({{\Sigma}}+\sqrt{3} {r}\right),\\
& K'=	2 K \left(-{{\Sigma}}+\sqrt{3} {r}-1\right),\\
& v'=\frac{\left(v^2-1\right)}{\gamma  \left(\gamma -v^2-1\right)} \Big\{\gamma  v \left(\gamma  {{\Sigma}}+2 \gamma -2\right) \nonumber\\
   &  +A \Big[(\gamma -1) (3 \gamma -2)+ (\gamma -2)
   v^2\Big]\Big\}. \label{system29v}
	\end{align}    
\end{subequations}
\newline
\noindent{\bf Equation for ${\Omega}_t$:}
\begin{equation}
\label{defnOmegat}
{\Omega}_{t}=\frac{\left(\gamma  v^2-v^2+1\right) \left(1-C_2{{\Sigma}}^2- C_2 A^2 -K\right)}{\gamma +v^2-1}.
\end{equation}
\newline
\noindent{\bf Auxiliary equation:}
\begin{align}
& {\Omega}_{t}'= 2 \sqrt{3} {r} {\Omega}_{t}+\frac{{\Omega}_{t}}{\left(\gamma -v^2-1\right) \left(\gamma  v^2-v^2+1\right)} \times \nonumber \\
& 
\Big\{\gamma  {{\Sigma}}+(\gamma -1) \gamma  \left({{\Sigma}}+2\right) v^4 -\gamma  v^2 \left(\gamma  {{\Sigma}}+4 \gamma -6\right) \nonumber \\
& +2 A v \Big[(\gamma  (2 \gamma -5)+4) v^2-  (\gamma -1) (3 \gamma -4) \Big]\Big\},
	\end{align}
	\newline
\noindent{\bf Restriction:}
\begin{equation}
\label{constraintmod1}
\gamma  {\Omega}_t v -2 A C_2 {\Sigma} \left((\gamma -1) v^2+1\right)=0.
	\end{equation}
It is easily verified that the previous equations are invariant under the discrete transformation 
\begin{equation}
({\Sigma},A,K,v)\rightarrow({\Sigma},-A,K,-v). \end{equation}
Therefore, it is enough to analyze the region $A\geq 0$. 

Assuming $v^2\neq \gamma-1$, and substituting the expressions \eqref{defnr} and \eqref{defnOmegat} the system  \eqref{system29} is reduced to: 
\begin{widetext}
\begin{subequations}
\label{Eq:22}
\begin{align}
   & \Sigma'= -\frac{1}{2 C_2 \left(\gamma +v^2-1\right)} \Bigg\{A^2 C_2 \left(-3 \gamma +v^2 (\gamma +2 \gamma  C_2 \Sigma -2)+2\right)   +K \left(-3 \gamma -2 (\gamma -1) C_2 \Sigma +v^2 (\gamma +2 (\gamma -1) C_2 \Sigma -2)+2\right) \nonumber \\
   & +\left(C_2 \Sigma
   ^2-1\right) \left(-3 \gamma -4 (\gamma -1) C_2 \Sigma +(\gamma -2) v^2 (2 C_2 \Sigma +1)+2\right)\Bigg\},
   \\
   & A'= \frac{A }{\gamma +v^2-1} \Bigg\{v^2 \left(2 \Sigma  (C_2 \Sigma +1)-\gamma 
   \left(C_2 \left(A^2+\Sigma ^2\right)-1\right)\right)   +2 (\gamma -1) \Sigma  (C_2 \Sigma +1)-(\gamma -1) K \left(v^2-1\right)\Bigg\}, \\
   & K'= \frac{2 K }{\gamma +v^2-1} \Bigg\{v^2 \left(-C-2 \gamma A^2 +\gamma -(\gamma
   -2) C_2 \Sigma ^2-1\right) +(\gamma -1) \left(2 C_2 \Sigma ^2-1\right)-(\gamma -1) K \left(v^2-1\right)\Bigg\},\\ 
   & v'=\frac{\left(v^2-1\right) \left(A \left(3 \gamma ^2-5 \gamma +(\gamma -2)
   v^2+2\right)+\gamma  v (\gamma  (\Sigma +2)-2)\right)}{\gamma  \left(\gamma -v^2-1\right)}.
\end{align}
\end{subequations}
\end{widetext}
In particular, four specific models will be studied, these are: models with extreme tilt \eqref{extreme_tilt_0}; presureless perfect fluid \eqref{3.41-3.43}; the reduced system \eqref{eq:3.57} in the invariant set $ A = v = 0 $; and the general system \eqref{reducedsyst}. The case $v^2= \gamma-1$ will be studied in section \ref{SL}. The invariant sets  $v=\pm 1$ will be analyzed in section \ref{tiltfluidoperfecto}. In section \ref{invP6} are calculated invariant manifolds of $P_6$ using analytical tools. 
Section \ref{fluidosinpresion} is devoted to the study of the ideal gas ($\gamma=1$). The general case $v\neq 0$ corresponding to tilted fluid will be studied in section \ref{general}. The equilibrium points with $v=0, A=0$ will be studied in section \ref{vcero}. 
In section \ref{summary6}  the sinks and sources for the model with perfect fluid are summarized.  Finally, in section  \ref{DiscusionC1} results will be summarized and the relation with previous results in the literature  will be discussed. 

Noticing that the gradient of the restriction \eqref{constraintmod1} is zero at the equilibrium points: $(\Sigma,A,K,v)=(0,0,1,0),(0,0,1,\pm 1)$,  the stability analysis of these points will be performed conserving the four eigenvalues due to the restriction being degenerated at these equilibrium points. 
\begin{figure*}[t!]
    \centering
    \subfigure[$\widetilde{M}^{+}$]{\includegraphics[scale=0.45]{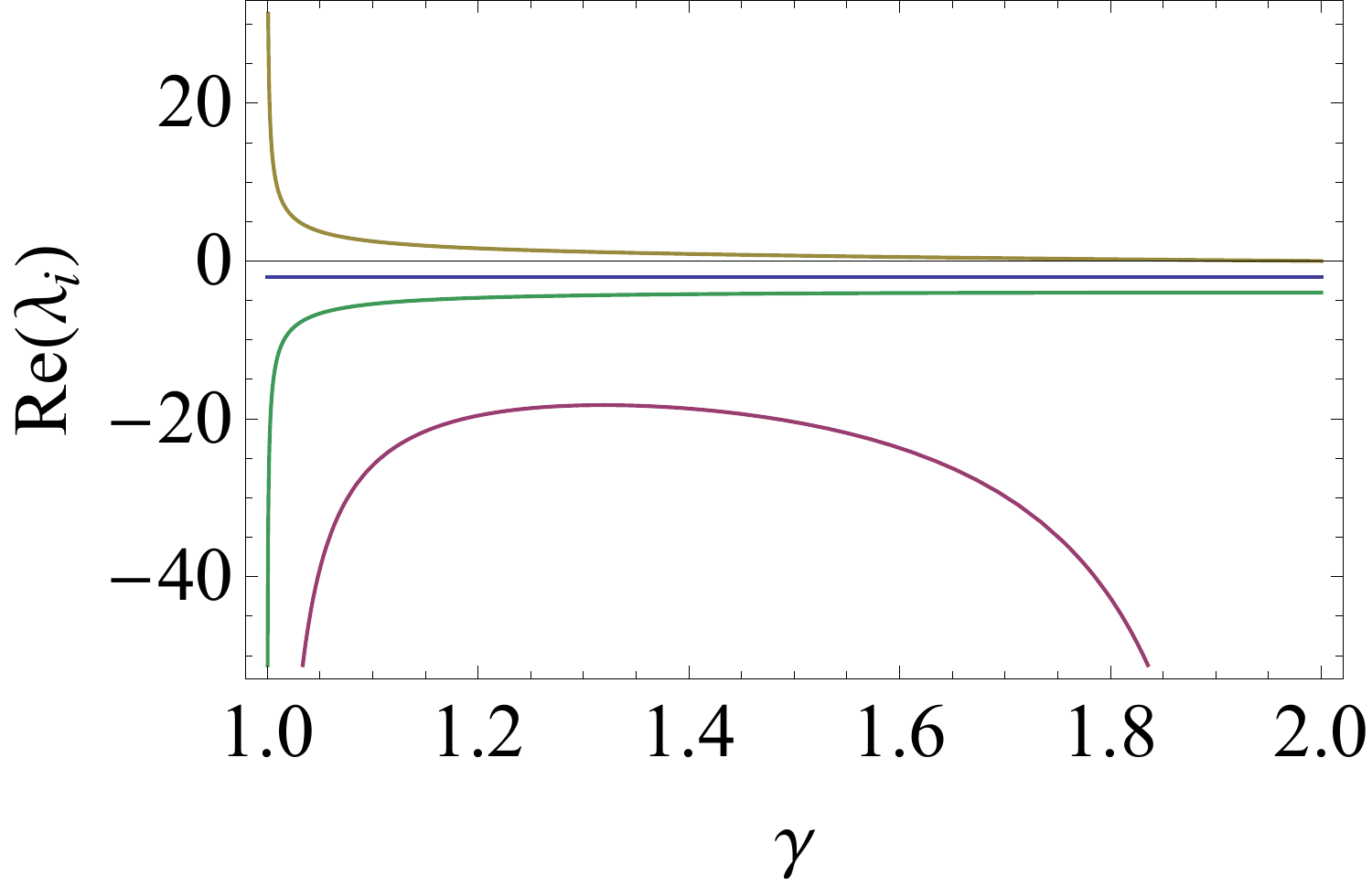}} \hspace{2cm}
    \subfigure[$\widetilde{M}^{-}$]{\includegraphics[scale=0.45]{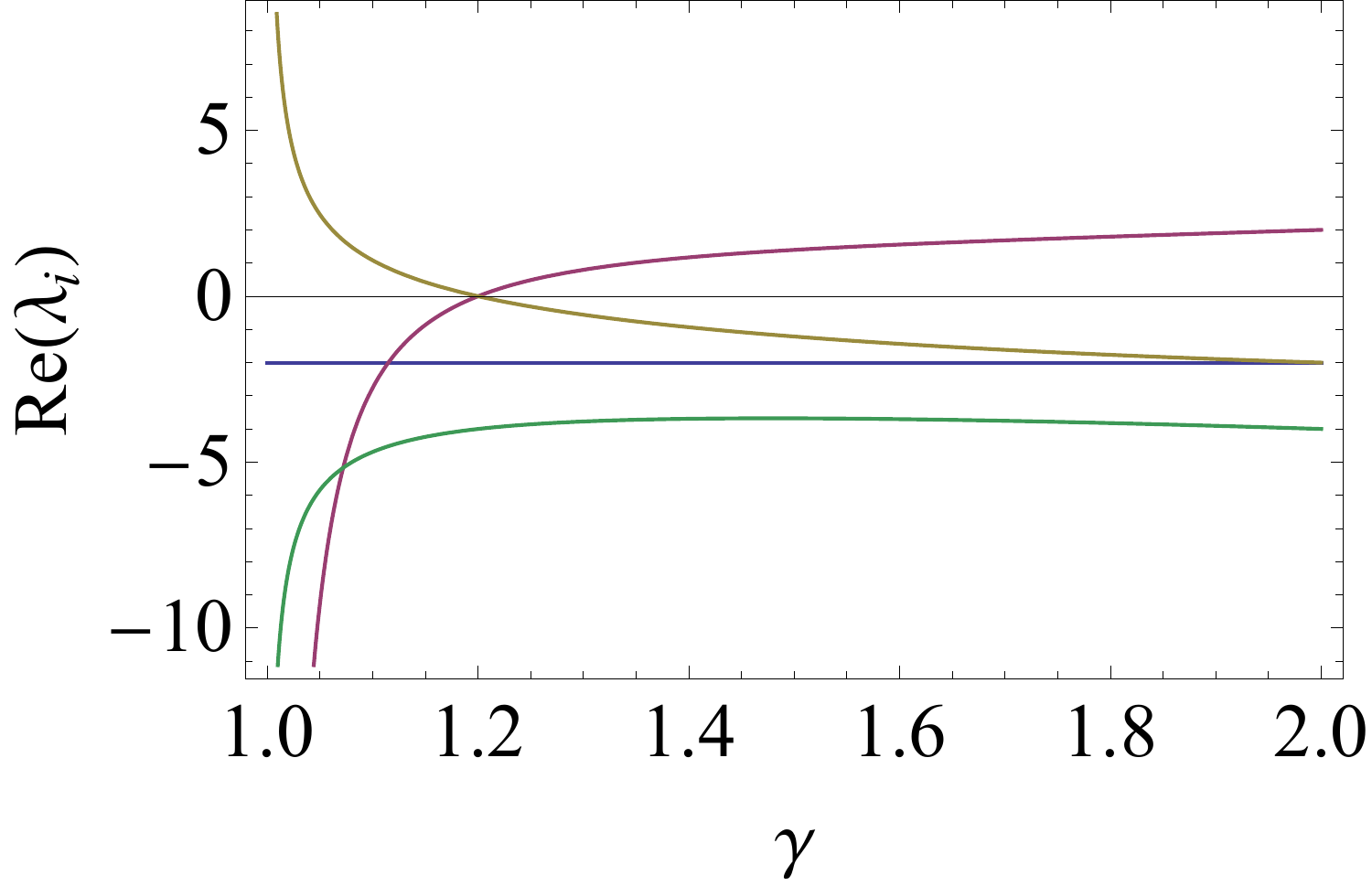}}
    \caption[{Real parts of the eigenvalues of the equilibrium point $\widetilde{M}^{\pm}$ for $C_2=1$.}]{\label{eigenM} Real part of the eigenvalues of the equilibrium point $\widetilde{M}^{\pm}$ for $C_2=1$.}
   \end{figure*}
The equilibrium point $N_1:(\Sigma,A,K,v)=(0,0,1,0)$ has eigenvalues $\{-2,-1,1,2\}$, so, it is a hyperbolic saddle. The equilibrium points $N_{2,3}:(\Sigma,A,K,v)=(0,0,1,\pm 1)$ have eigenvalues \newline $\left\{0,-1,1,\frac{4}{\gamma -2}+4\right\}$, so, they are non-hyperbolic saddles.

Additionally,  \eqref{reducedsyst} admits the equilibrium points \newline  $\widetilde{M}^{\pm}:({\Sigma}, {A}, K, v)=\left(0,1,0, f_\pm(\gamma)\right),\quad \Omega_t=0$, \newline $f_\pm(\gamma)=\frac{(\gamma -1) \gamma  \pm \left(\sqrt{(\gamma -1) \left((\gamma
   -1) \gamma ^2+(2-\gamma ) (3 \gamma -2)\right)}\right)}{2-\gamma }$, \newline  existing for $C_2=1$. The eigenvalues of  $\widetilde {M}^{+}$ are:  \newline 
\noindent\( {\left\{-2,\frac{1}{(-2+\gamma ) (-1+\gamma )}\left(\gamma  \left(8+(-4+\gamma ) \gamma -\Delta(\gamma)\right)+\right.\right.}\\
 {\left.4 \left(-1+\Delta(\gamma)\right)\right),} 
 {\frac{1}{2 (-1+\gamma ) \gamma }\left(-2-\Delta(\gamma)+\right.}\\
 {\gamma  \left(6-\gamma  (3+\gamma )+\Delta(\gamma)\right)+}\\
 {\surd \left((-1+\gamma ) \left(-4 \left(2+\Delta(\gamma)\right)+\right.\right.}
 {\gamma  \left(4 \left(5+\Delta(\gamma)\right)+\right.}\\
 {\gamma  \left(-28+6 \Delta(\gamma)+\gamma  \right.}\\
 {\left.\left.\left.\left.\left.\left(29+\gamma  (-11+2 \gamma )-2 \Delta(\gamma)\right)\right)\right)\right)\right)\right),}\\
 {-\frac{1}{2 (-1+\gamma ) \gamma }\left(2+\Delta(\gamma)-\right.}
 {\gamma  \left(6-\gamma  (3+\gamma )+\Delta(\gamma)\right)+}\\
 {\surd \left((-1+\gamma ) \left(-4 \left(2+\Delta(\gamma)\right)+\right.\right.}
 {\gamma  \left(4 \left(5+\Delta(\gamma)\right)+\right.}\\
 {\gamma  \left(-28+6 \Delta(\gamma)+\right.}\\
 {\left.\left.\left.\left.\left.\left.\gamma  \left(29+\gamma  (-11+2 \gamma )-2 \Delta(\gamma)\right)\right)\right)\right)\right)\right)\right\}}\). \newline
The eigenvalues of $\widetilde {M}^{-}$ are: \newline
\noindent\( {\left\{-2,\frac{1}{(-2+\gamma ) (-1+\gamma )}\left(-4 \left(1+\Delta(\gamma)\right)+\right.\right.}\\
 {\left.\gamma  \left(8+(-4+\gamma ) \gamma +\Delta(\gamma)\right)\right),} 
 {\frac{1}{2 (-1+\gamma ) \gamma }\left(-2+\Delta(\gamma)-\right.}\\
 {\gamma  \left(-6+\gamma  (3+\gamma )+\Delta(\gamma)\right)+}\\
 {\surd \left((-1+\gamma ) \left(4 \left(-2+\Delta(\gamma)\right)+\right.\right.} 
 {\gamma  \left(-4 \left(-5+\Delta(\gamma)\right)+\right.}\\
 {\gamma  \left(-28-6 \Delta(\gamma)+\gamma  \right.} 
 {\left.\left.\left.\left.\left.\left(29+\gamma  (-11+2 \gamma )+2 \Delta(\gamma)\right)\right)\right)\right)\right)\right),}\\
 {-\frac{1}{2 (-1+\gamma ) \gamma }\left(2-\Delta(\gamma)+\right.}
 {\gamma  \left(-6+\gamma  (3+\gamma )+\Delta(\gamma)\right)+}\\
 {\surd \left((-1+\gamma ) \left(4 \left(-2+\Delta(\gamma)\right)+\right.\right.}
 {\gamma  \left(-4 \left(-5+\Delta(\gamma)\right)+\right.}\\
 {\gamma  \left(-28-6 \Delta(\gamma)+\right.}\\
 {\left.\left.\left.\left.\left.\left.\gamma  \left(29+\gamma  (-11+2 \gamma )+2 \Delta(\gamma)\right)\right)\right)\right)\right)\right)\right\}}\), \newline
where $\Delta(\gamma)=\sqrt{(-1+\gamma ) (-4+\gamma  (8+(-4+\gamma ) \gamma ))}$. 

In Figure \ref{eigenM}  the real parts of the eigenvalues of the equilibrium point $\widetilde{M}^{\pm}$ for $C_2=1$ are depicted, showing in general that it is a hyperbolic saddle. For $\gamma=\frac{6}{5}$, the eigenvalues of $\widetilde{M}^{-}$ are $\{-2,0,0,-4\}$, so, it is non-hyperbolic. 

\subsection{Surface of non-extendibility of solutions}
\label{SL}
For $\gamma>1$,  $v^2= \gamma-1$  represents a surface of non-extendibility of the solutions, and it is called sonic surface. 
The curve parametrized by  $\Sigma$, 
\begin{equation}
SL_{\pm}: A=-\frac{\gamma 
   \varepsilon  (\gamma  (\Sigma
   +2)-2)}{4 (\gamma
   -1)^{3/2}},  v=\varepsilon\sqrt{\gamma -1},   \varepsilon=\pm 1,
\end{equation} is called sonic line. The solutions diverge in a finite time when the solutions approach the sonic surface $v^2 = \gamma-1 $. The only way it can be passed through the sonic surface is when the numerator of the equation \eqref{system29v} also vanishes, this is through the sonic line $SL_{\pm} $. In $SL_{\pm}$ both the denominator and the numerator of \eqref{system29v} are zero. This indicates the presence of a singularity of the system \eqref{system29}. As a difference with General Relativity, for $1<\gamma<2$ and $C_2= \frac{\gamma
   ^2}{4 (\gamma -1)^2}$ the system \eqref{reducedsyst} admits the following equilibrium points:  
\begin{enumerate}
    \item[$SL_1$:] $\Sigma =
   \frac{2 (\gamma -1)}{\gamma
   },v=\sqrt{\gamma -1}, A=
  - \frac{\gamma  (\gamma  (\Sigma
   +2)-2)}{4 (\gamma
   -1)^{3/2}}$, 
    \item[$SL_2$:] $\Sigma =
   -\frac{2 (\gamma -1)}{\gamma
   },v= -\sqrt{\gamma -1}, A=
   \frac{\gamma  (\gamma  (\Sigma
   +2)-2)}{4 (\gamma
   -1)^{3/2}}$,
\end{enumerate}
which lie on the sonic line. These points do not exist in General Relativity when $1 <\gamma<2 $. When $\gamma = 2, C_2 =1$ these points exist, and since $\gamma = 2$ the fluid behaves like stiff matter. Additionally, if $ \gamma = 2, C_2 = 1$, these points correspond to models with extreme tilt ($v = \varepsilon$), $SL_1: \Sigma = 1, A = -2, v = 1$, and $SL_2: \Sigma = -1, A = 0, v = -1$.
On the sonic surface the inequality $ C_2 (A ^ 2 + \Sigma ^ 2) \leq 1 $ must be satisfied, which corresponds to $ K \geq 0$, which imposes additional conditions on the parameters. To analyze locally the behavior of the solutions near this sonic line, the new ``shock" variable $\xi$ is introduced: \begin{equation}
    \frac{d \xi}{d \eta}= \frac{1}{(\gamma-1)-v^2},
\end{equation} \newline that leads to the system 
\begin{subequations}
\begin{widetext}
\begin{align}
  & \frac{d \Sigma}{d \xi}=  \frac{\Sigma \left(1-\gamma
   +v^2\right) }{\gamma 
   v} \Big\{A^2 \gamma 
   C_2 v+A \left(3 \gamma +2
   (\gamma -1) C_2 \Sigma
   +v^2 (-(\gamma +2 (\gamma -1)
   C_2 \Sigma
   -2))-2\right)  +\gamma  v
   \left(1-C_2 \Sigma
   ^2\right)\Big\},
  \\
  & \frac{d A}{d \xi}= -\frac{A \left(-\gamma
   +v^2+1\right) \left(\gamma  v
   \left(1-A^2 C_2\right)+2
   \Sigma  \left(A (\gamma -1)
   C_2
   \left(v^2-1\right)+\gamma 
   v\right)+\gamma  C_2
   \Sigma ^2 v\right)}{\gamma 
   v},\\
 & \frac{d v}{d \xi}= \frac{\left(v^2-1\right)
   \left(A \left(3 \gamma ^2-5
   \gamma +(\gamma -2)
   v^2+2\right)+\gamma  v (\gamma 
   (\Sigma +2)-2)\right)}{\gamma
   },
\end{align}
\end{widetext}
\end{subequations}

\begin{figure*}[t!]
    \centering
    \subfigure[$C_2=0.5$]{\includegraphics[scale=0.28]{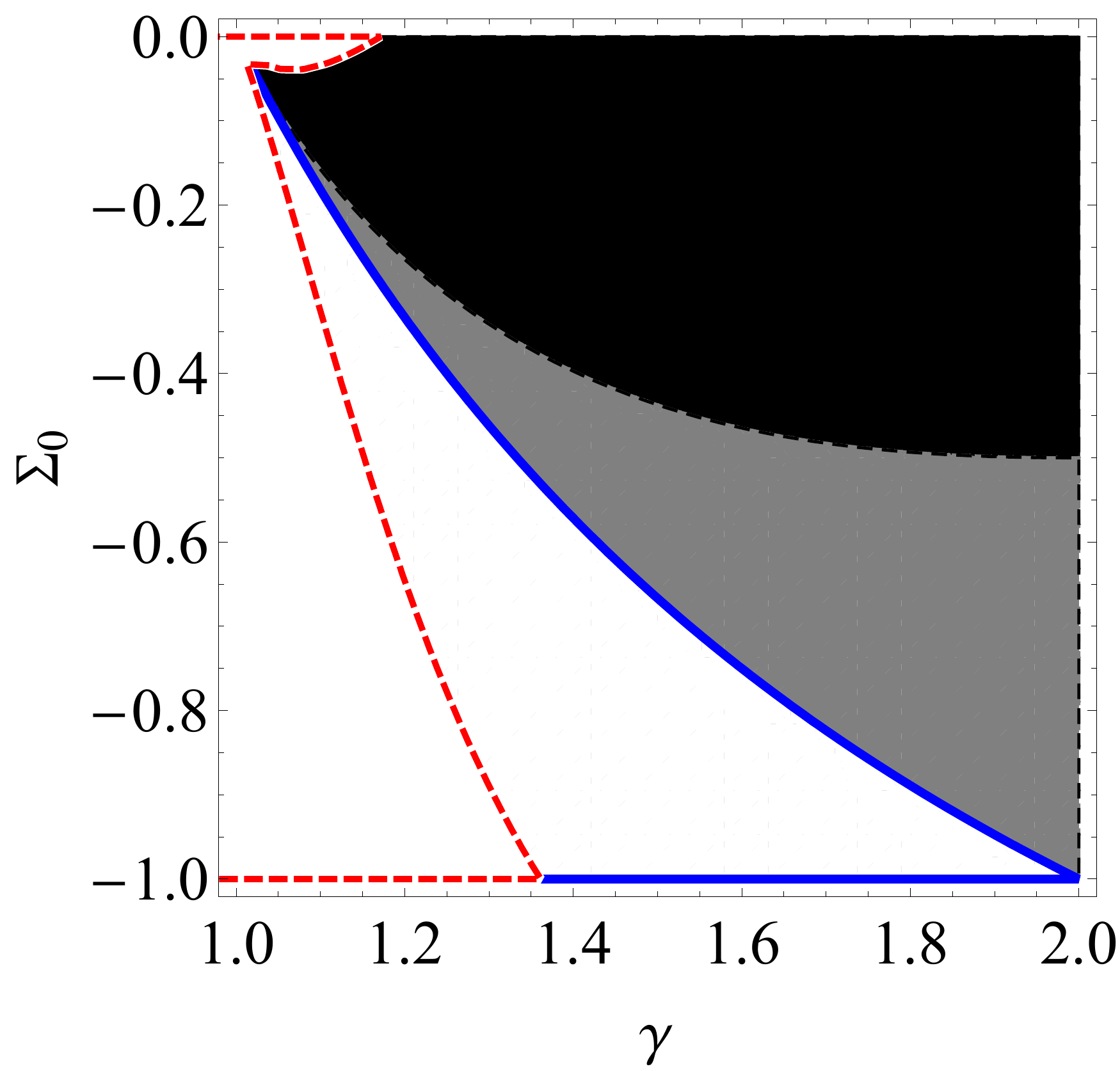}}\hspace{1cm}
	\subfigure[\label{SL-stabilitya} $C_2=1$]{\includegraphics[scale=0.28]{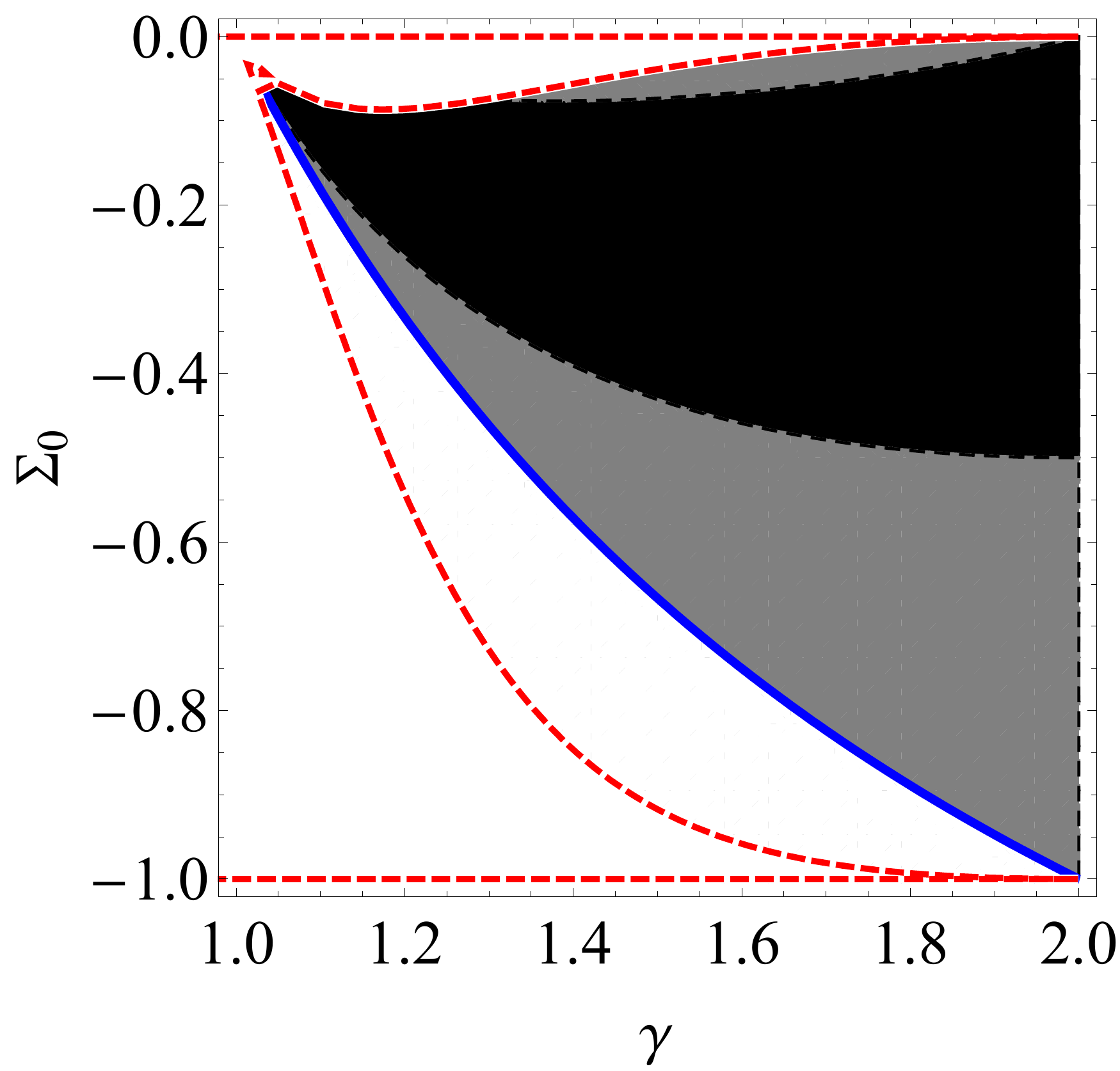}}\hspace{1cm}
	\subfigure[$C_2=2$]{\includegraphics[scale=0.28]{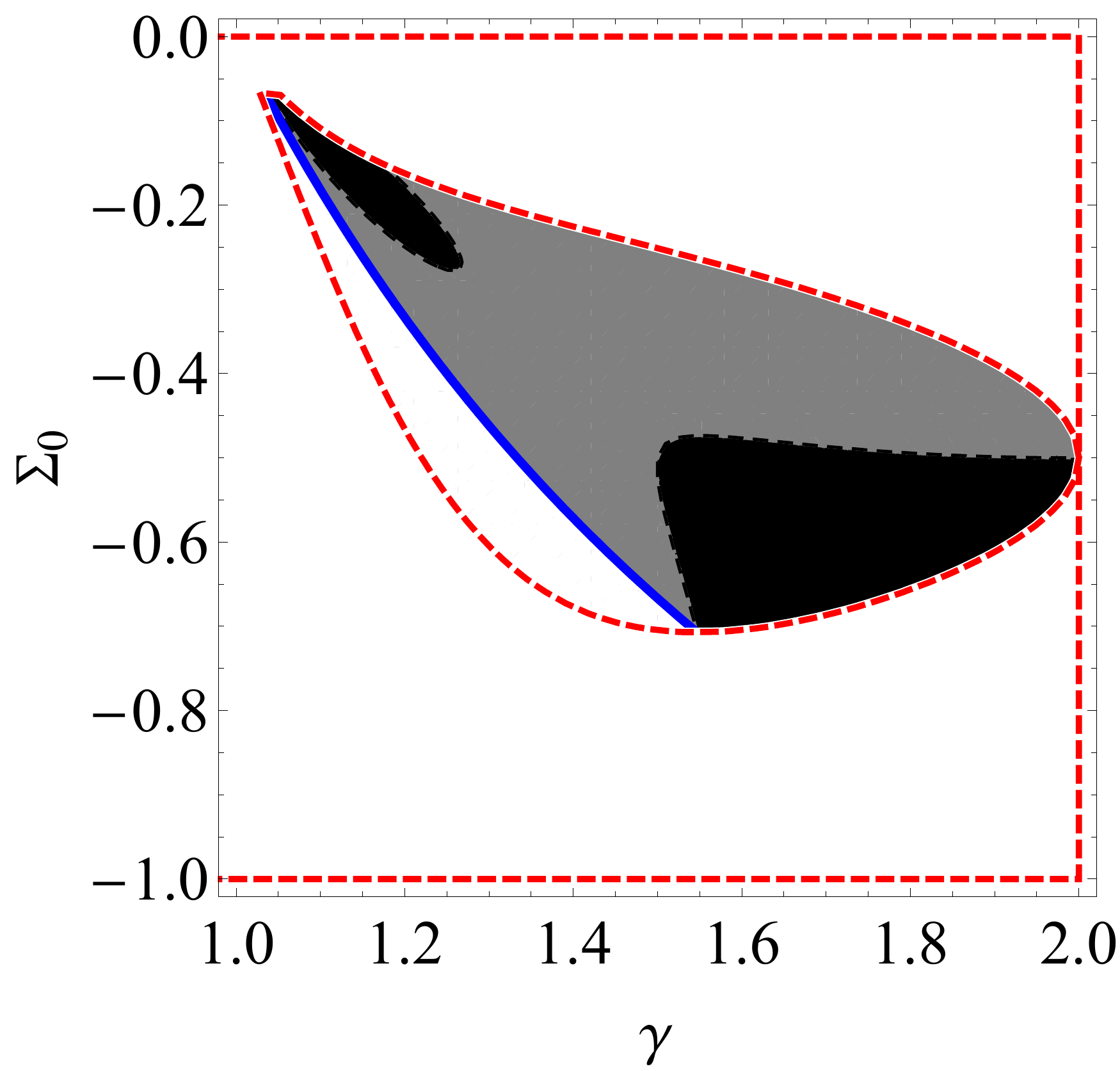}}\hspace{1cm}
    \caption[{\label{SL-stability} Stability regions for $SL_{-}: v_0=-\sqrt{\gamma-1}, A=\frac{\gamma 
   (\gamma  (\Sigma
   +2)-2)}{4 (\gamma
   -1)^{3/2}}$, for $C_2=0.5, 1, 2$  y $\gamma\in[1, 2]$.}]{\label{SL-stability} Stability regions for $SL_{-}: v=-\sqrt{\gamma-1}, A=\frac{\gamma 
   (\gamma  (\Sigma
   +2)-2)}{4 (\gamma
   -1)^{3/2}}$, for $C_2=0.5, 1, 2$ y $\gamma\in[1, 2]$, where $v_0=-\sqrt{\gamma-1}, \Sigma_0$ denote the values of $v, \Sigma$ in an arbitrary fixed point on the sonic curve.}
   \end{figure*}
The new variable is not monotonic since $(\gamma-1) -v^ 2$ can change the sign, so the system is not suitable to do qualitative/asymptotic analysis of the system since it does not represent a dynamical system. However, the system is suitable for numerical integration in a neighborhood of the sonic line $SL_{\pm}$, and for the local stability analysis of $SL_{\pm}$ at the perturbative level. \newline  
$SL_{\pm}$ can be parametrized by:
\begin{small}
\begin{align}
    & \gamma = v_0^2+1,\;  A= -\frac{\left(v_0^2+1\right) \left(\Sigma_0 +(\Sigma_0 +2)
   v_0^2\right)}{4 v_0^3},\; \Sigma=\Sigma_0. 
\end{align}
\end{small}
\newline 
Defining the equation of state parameter  $\omega=\gamma-1$, we deduce that $\omega=v_0^2$. \newline 
Defining the following linear perturbations:
\begin{small}
\begin{align}
& \delta_{v}=v-v_0, \; \delta_{A}=A+\frac{\left(v_0^2+1\right) \left(\Sigma_{0}+(\Sigma_{0}+2)
   v_0^2\right)}{4 v_0^3}, \; \delta_{\Sigma}= \Sigma-\Sigma_0,
\end{align}
\end{small}
the evolution equations of the perturbations are given by: 
\begin{widetext}
\begin{subequations}
\begin{align}
  & \frac{d \delta_\Sigma }{d \xi}=  \delta_v \Bigg\{\frac{\Sigma_0 \left((C_2+2) v_0^4+2 (C_2-2)
   v_0^2+C_2-2\right)}{2 v_0}   +\frac{\Sigma_0^2 \left((5 C_2+1) v_0^6-(C_2+3)
   v_0^4+(3 C_2-5) v_0^2+C_2-1\right)}{2 v_0^3} \nonumber \\
   & +\frac{C_2 \Sigma_0^3 \left(-3 v_0^4+2
   v_0^2+1\right)^2}{8 v_0^5}\Bigg\},
   \end{align}
  \begin{align}
  & \frac{d \delta_A }{d \xi}= \delta_v
   \Bigg\{-\frac{\left(v_0^2+1\right) \left(C_2 \left(v_0^2+1\right)^2-4 v_0^2\right)}{4
   v_0^2} -\frac{\Sigma_0 \left(v_0^2+1\right) \left(11 C_2 v_0^6+(C_2-20)
   v_0^4+(9 C_2-4) v_0^2+3 C_2\right)}{8 v_0^4}  \nonumber \\
   & -\frac{\Sigma_0^2
   \left(v_0^2+1\right) \left(19 C_2 v_0^8-4 (C_2+4) v_0^6+2 (C_2-8)
   v_0^4+12 C_2 v_0^2+3 C_2\right)}{16 v_0^6}  -\frac{C_2 \Sigma_0^3
   \left(3 v_0^6+v_0^4-3 v_0^2-1\right)^2}{32 v_0^8}\Bigg\},
   \end{align}
   \begin{align}
  & \frac{d \delta_v}{d \xi}= \delta_v
   \left(v_0^4+\frac{\Sigma_0 \left(v_0^2-1\right) \left(v_0^2+1\right)^2}{2
   v_0^2}-1\right) +\frac{v_0 \left(v_0^2-1\right) \left(4 \delta_A v_0^3+\delta_\Sigma  \left(v_0^2+1\right)^2\right)}{v_0^2+1}.
\end{align}
\end{subequations}
The eigenvalues of the equilibrium point
$(\delta_\Sigma, \delta_A, \delta_v)=(0,0,0)$ are $\lambda_1=0$ and the roots  $\lambda_2$ y $\lambda_3$ of the polynomial: 
\begin{align}
  & P(\lambda):=-\lambda ^2+\frac{\lambda  \left(v_0^4-1\right)
   \left(\Sigma_0+(\Sigma_0+2) v_0^2\right)}{2 v_0^2}   -\left(v_0^2-1\right) v_0^2 \left(C_2 \left(v_0^2+1\right)^2-4
   v_0^2\right)  \nonumber \\
   & -\Sigma_0 \left(v_0^2-1\right) \left((5 C_2-1) v_0^6-(C_2+9)
   v_0^4+(3 C_2+1) v_0^2+C_2+1\right) \nonumber \\
   & +\frac{\Sigma_0^2 \left(v_0^2-1\right) \left(-C_2 \left(-3 v_0^4+2
   v_0^2+1\right)^2-12 v_0^2+2 \left(v_0^2+6\right) v_0^6-2\right)}{4
   v_0^2},  
\end{align}
say, 
\noindent\( {\left\{\lambda_{2}=\frac{\left(-1+ v_0^2\right) \left(1+ v_0^2\right)^2 \Sigma_0}{4  v_0^2}-\right.} 
 {\frac{1}{4  v_0^{13} \left(1+ v_0^2\right)}\left(2  v_0^{13}+2  v_0^{15}-2  v_0^{17}-2  v_0^{19}+\surd \left(- v_0^{22}
\left(-1+ v_0^2\right) \left(1+ v_0^2\right)^2 \right.\right.}\\
 {\left(\Sigma_0^2+(-1+4 C_2)  v_0^{10} (2+\Sigma_0) (2+9 \Sigma_0)+ v_0^2 \Sigma_0 (4+(11+4
C_2) \Sigma_0)-2  v_0^6 \right.} 
 {(-2+(-8+\Sigma_0) \Sigma_0+4 C_2 (-2+(-6+\Sigma_0) \Sigma_0))} \\
 {- v_0^8 (68+\Sigma_0 (152+51
\Sigma_0)+} 
 {\left.\left.\left.\left.16 C_2 \left(-2+\Sigma_0+3 \Sigma_0^2\right)\right)+2  v_0^4 (2+\Sigma_0 (12+25
\Sigma_0+8 C_2 (1+\Sigma_0)))\right)\right)\right),}\\\\
 {\lambda_{3}=\frac{\left(-1+ v_0^2\right) \left(1+ v_0^2\right)^2 \Sigma_0}{4  v_0^2}+\frac{1}{4  v_0^{13}
\left(1+ v_0^2\right)}\left(-2  v_0^{13}-2  v_0^{15}+2  v_0^{17}+2  v_0^{19}+\right.}\\
 {\surd \left(- v_0^{22} \left(-1+ v_0^2\right) \left(1+ v_0^2\right)^2 \left(\Sigma_0^2+(-1+4 C_2)  v_0^{10}
(2+\Sigma_0) (2+9 \Sigma_0)+ v_0^2 \Sigma_0 \right.\right.}\\
 {(4+(11+4 C_2) \Sigma_0)-2  v_0^6 (-2+(-8+\Sigma_0) \Sigma_0+4 C_2 (-2+(-6+\Sigma_0) \Sigma_{0}))-} 
 { v_0^8 \left(68+\Sigma_0 (152+51 \Sigma_0)+16 C_2 \left(-2+\Sigma_0+3 \Sigma_0^2\right)\right)+} \\
 {\left.\left.\left.\left.2  v_0^4 (2+\Sigma_0 (12+25 \Sigma_0+8 C_2 (1+\Sigma_0)))\right)\right)\right)\right\}}\).
\end{widetext}
A zero eigenvalue appears because it is a curve of equilibrium points. Also, the curve:
\begin{align*}
&\delta_{\Sigma}(\Sigma_0)= \Sigma-\Sigma_0,  \delta_{A}(\Sigma_0)=A+\frac{\left(v_0^2+1\right) \left(\Sigma_{0}+(\Sigma_{0}+2)
   v_0^2\right)}{4 v_0^3}, \\
   & \delta_{v}(\Sigma_0)=v-v_0, 
\end{align*}
has the tangent vector at the point $(\delta_\Sigma, \delta_A, \delta_v)=(0,0,0)$:  
$$\frac{d}{d \Sigma_0}\left(\delta_{\Sigma}(\Sigma_0),
\delta_{A}(\Sigma_0),
\delta_{v}(\Sigma_0)\right)=\left(1,-\frac{\left(v_0^2+1\right)^2}{4 v_0^3},0\right).$$ 
The eigenvector associated with the zero eigenvalue:
$$\mathbf{v}=\left(-\frac{4 v_0^3}{\left(v_0^2+1\right)^2},1,0\right),$$ 
is parallel to the vector tangent to the curve at the point. Then the curve is normally--hyperbolic, so the stability can be studied considering only the signs of the nonzero eigenvalues.

For the stability analysis of the sonic line  $SL_{\pm}$,  the following invariant sets are identified: 
\begin{enumerate}
    \item $A_0=0$:  $\frac{\left(v_0^2+1\right) \left(\Sigma_0+(\Sigma_0+2) v_0^2\right)}{4
   v_0^3}=0$, 
    \item $K_0=0$: $1-C_2 \left(\Sigma_0^2+\frac{\left(v_0^2+1\right)^2 \left(\Sigma_0+(\Sigma_0+2) v_0^2\right)^2}{16 v_0^6}\right)=0$,
\end{enumerate}
both invariant sets determine curves in the space of parameters  $v_0,\Sigma_0$, which, due to the fact that they are invariant cannot be passed by orbits. For the discussion the following existence conditions will be imposed:
\begin{small}
\begin{align*}
   &\frac{\left(v_0^2+1\right) \left(\Sigma_0+(\Sigma_0+2) v_0^2\right)}{4
   v_0^3}\geq 0, \\
   &  1-C_2 \left(\Sigma_0^2+\frac{\left(v_0^2+1\right)^2 \left(\Sigma_0+(\Sigma_0+2) v_0^2\right)^2}{16 v_0^6}\right)\geq 0.
\end{align*}
\end{small}
In Figure \ref{SL-stability}  stability regions for $SL_{-}: v_0=-\sqrt{\gamma-1}, A=\frac{\gamma 
(\gamma  (\Sigma_0   +2)-2)}{4 (\gamma   -1)^{3/2}}$ are depicted in the space of parameters, for (a) $C_2=0.5$, (b) $C_2=1$ and (c) $C_2=2$ and $\gamma\in[1, 2]$, where $v_0=-\sqrt{\gamma-1}, \Sigma_0$ denote the values of   $v, \Sigma$ in a fixed point at the sonic curve. The unshaded region represents the region where $ A_0 <0 $ or $ K_0 <0 $, which is the non-physical region. The dotted red line corresponds to $ K = 0$  and the thick blue line corresponds to $ A = 0$. The region represented in gray color corresponds to the region where $ (\delta_\Sigma, \delta_A, \delta_v) = (0,0,0)$ is a hyperbolic saddle. The region represented in black color corresponds to the region where $ (\delta_\Sigma, \delta_A, \delta_v) =(0,0,0) $ is stable.
Figure \ref{SL-stabilitya} reproduces the stability results shown in Figure 1 of \cite{Goliath:1998mx} (with the exception of a strip in the parameter space, represented in gray, where the point equilibrium $(\delta_ \Sigma, \delta_A, \delta_v) = (0,0,0) $ is a hyperbolic saddle, see upper right corner in Figure \ref{SL-stabilitya}, whose analysis was omitted in \cite{ Goliath:1998mx}). For $SL_{+} $ the existence condition $\frac{\left(v_0 ^ 2 + 1 \right) \left (\Sigma_0 + (\Sigma_0 + 2) v_0 ^ 2 \right)} {4 v_0 ^ 3} \geq 0, $
it is not verified, so they are outside the physical region and their analysis is omitted.

\subsection{Invariant sets $v=\pm 1$}
\label{tiltfluidoperfecto}
In this section, the following invariant sets are studied $ v = \pm 1 $, which corresponds to  extreme tilt. Then, from the equations \eqref{constraintmod1}, \eqref{defnOmegat}, and \eqref{defnr} it follow:
\begin{subequations}
\begin{align}
& {\Omega}_t = \pm 2  A C_2 {\Sigma},\\
& K=1- C_2 A^2  -C_2 \Sigma ^2 \mp  2  A C_2 {\Sigma},\\
& \sqrt{3} r=  1+\Sigma+C_2 {\Sigma}^2- C_2 A^2.   
\end{align}
\end{subequations}
Afterwards,  the following reduced  2-dimensional system is obtained:  
\begin{subequations}
\label{extreme_tilt_0}
\begin{align}
&{\Sigma}'=-{{\Sigma}} \left(1 \pm 2  A-C_2( {\Sigma}^2- A^2)\right), \\
 & A'=A \left( 1+ 2{\Sigma}+C_2( {\Sigma}^2- A^2)\right),
	\end{align}
\end{subequations}	
where $\pm 1$ denotes the sign of $v$. 

\begin{figure*}
\centering
	\subfigure[]{\includegraphics[scale=0.6]{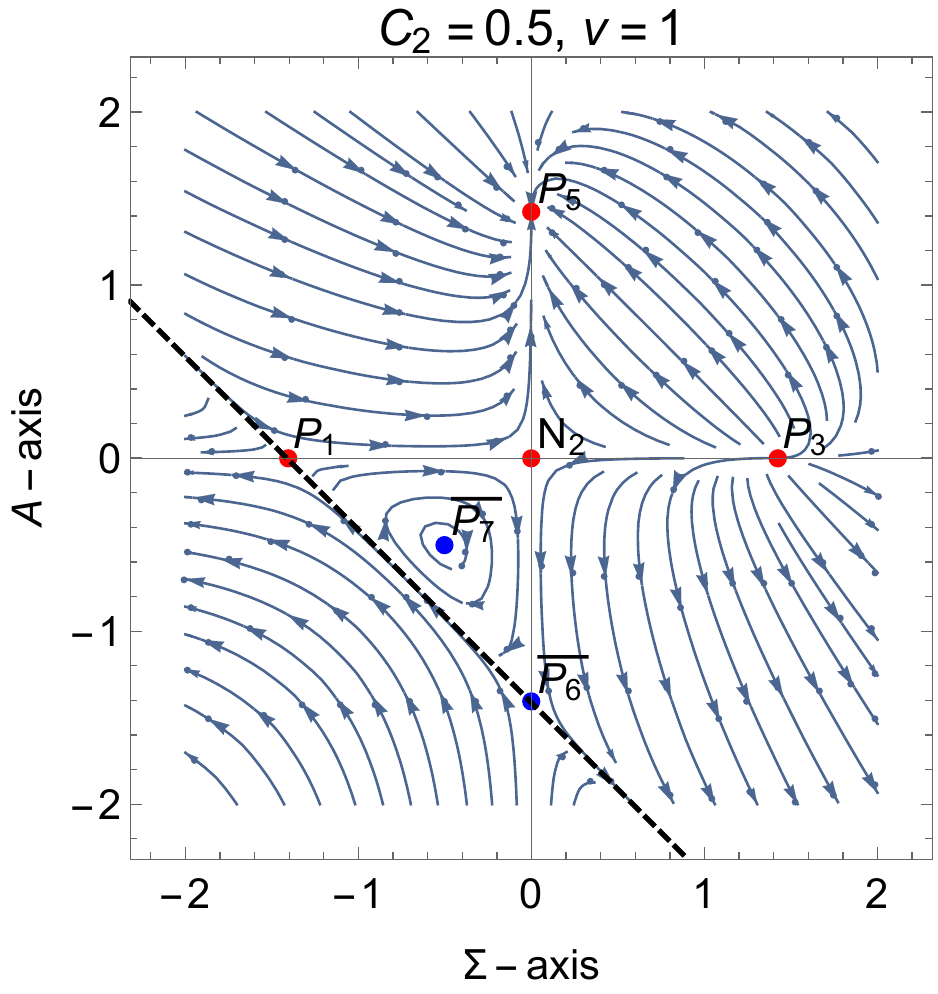}}
	\subfigure[]{\includegraphics[scale=0.6]{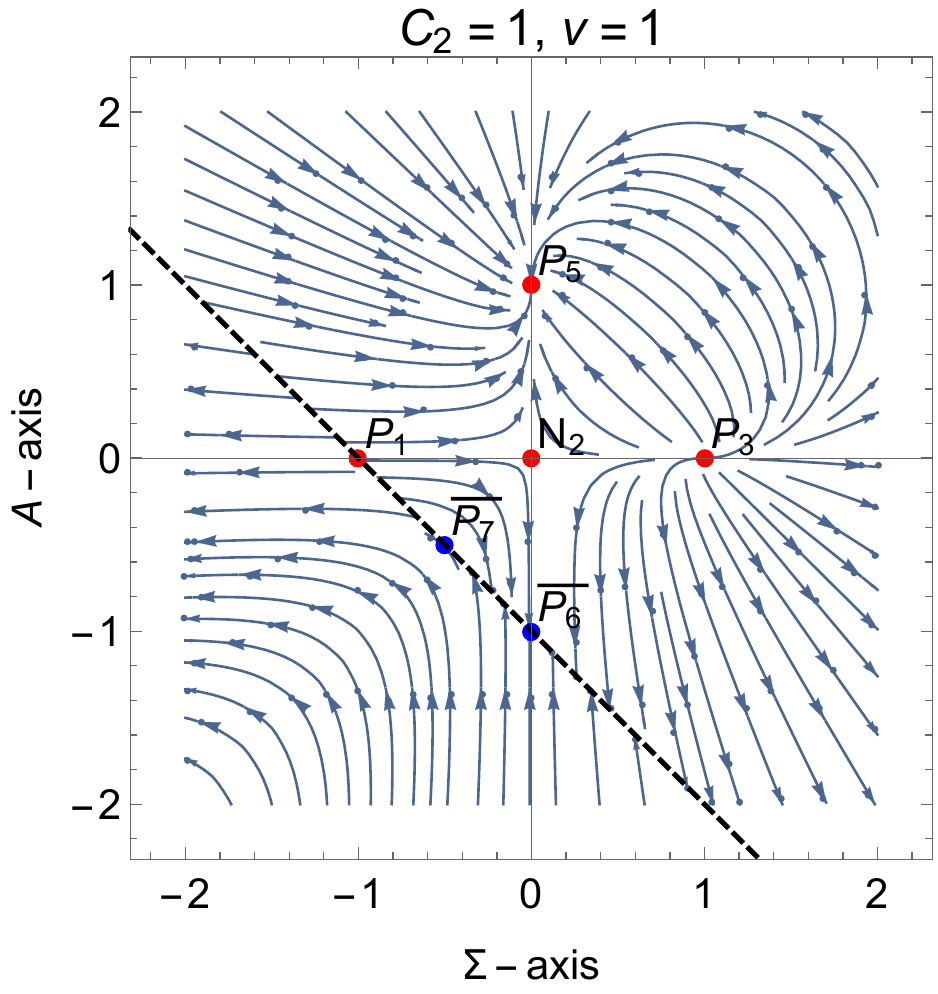}}
	\subfigure[]{\includegraphics[scale=0.6]{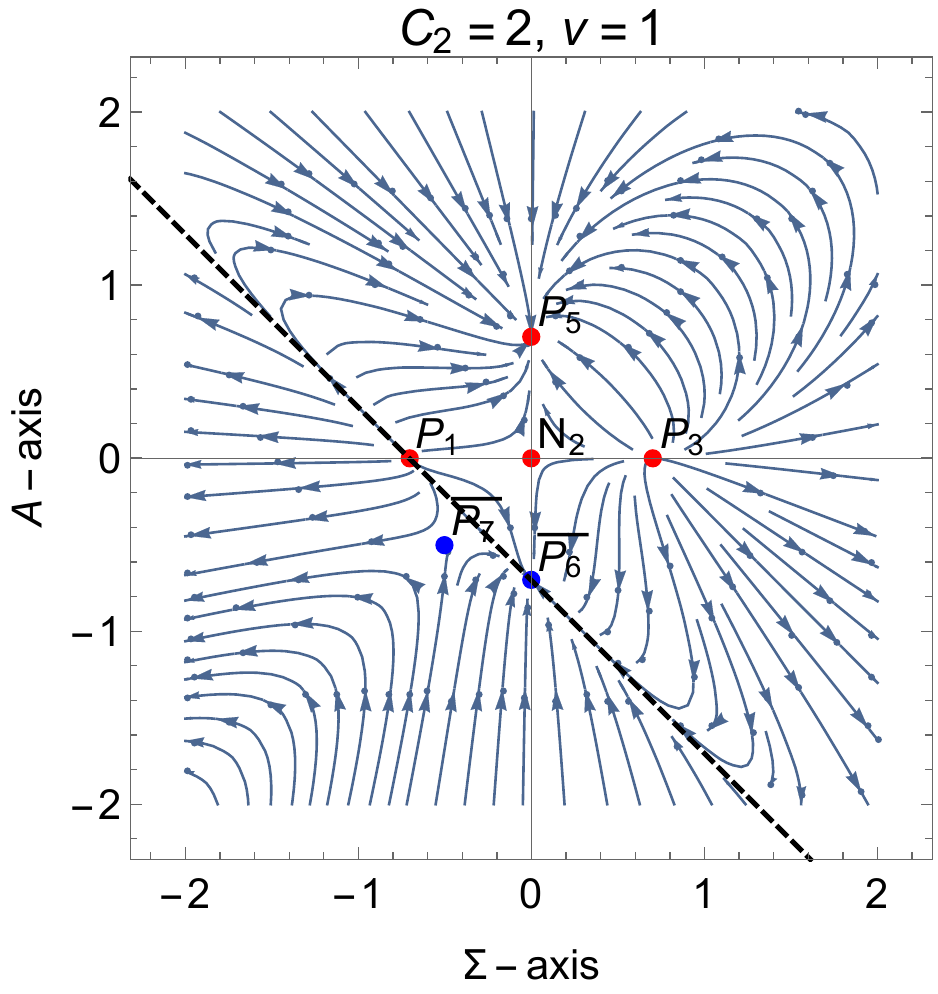}}
	\subfigure[]{\includegraphics[scale=0.6]{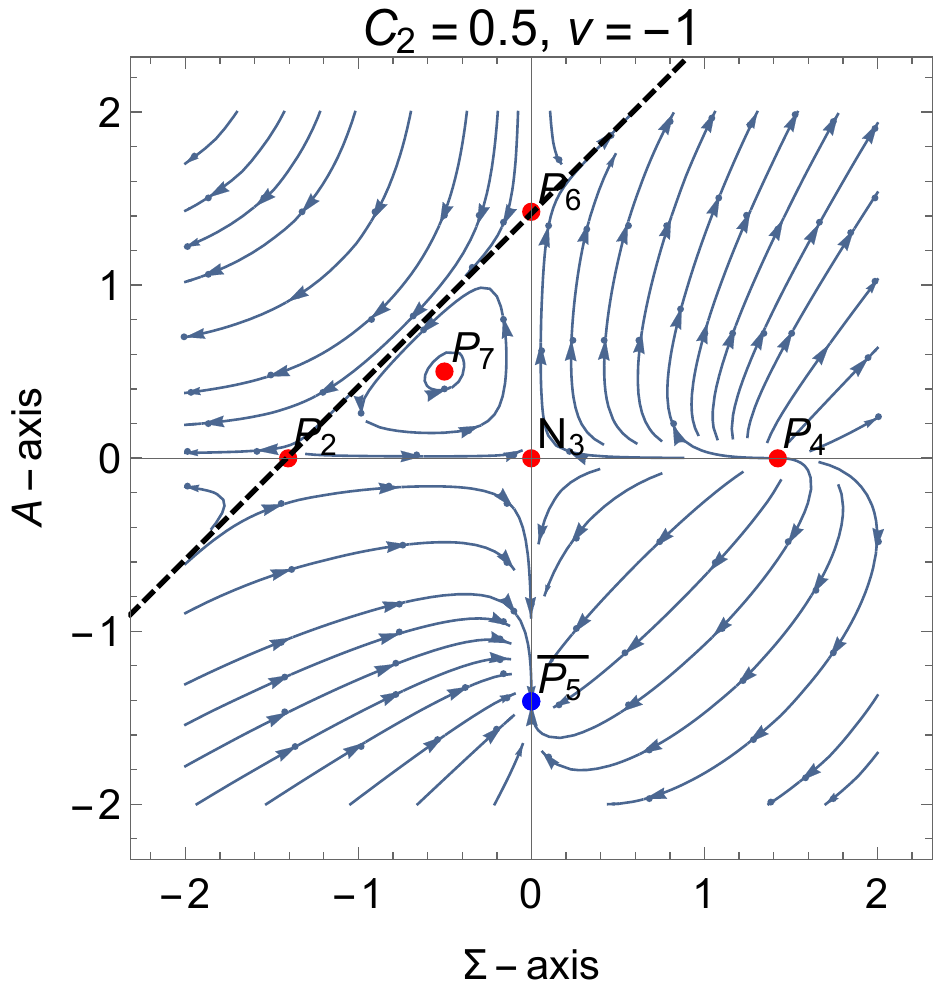}}
	\subfigure[]{\includegraphics[scale=0.6]{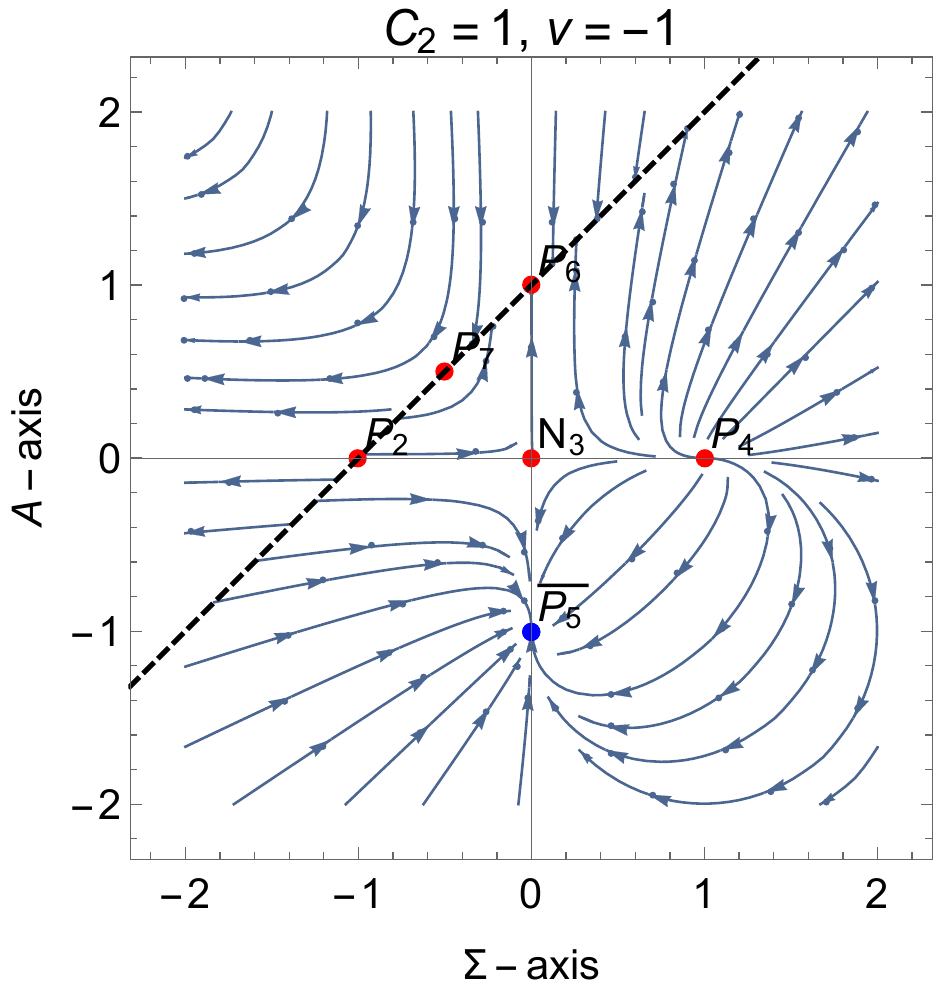}}
	\subfigure[]{\includegraphics[scale=0.6]{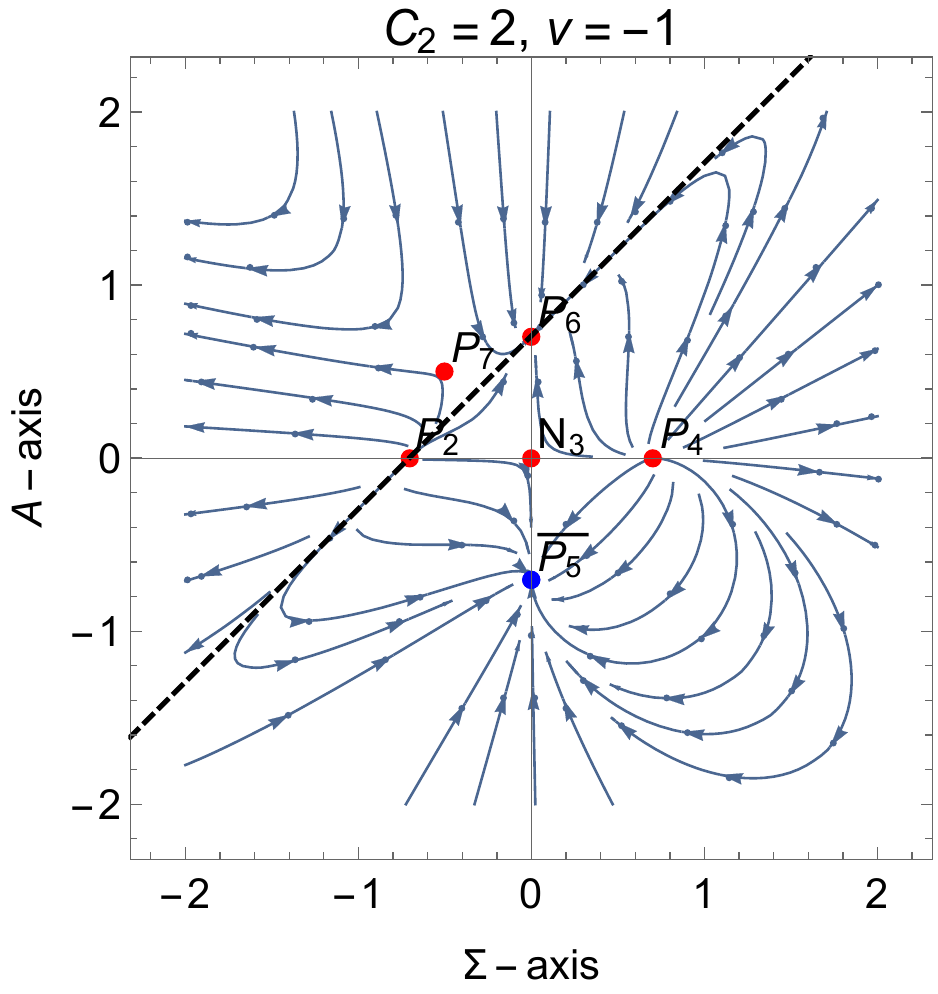}}
	\caption[Some orbits of the systems \eqref{extreme_tilt_0} with $v = 1, -1 $, for different choices of parameters.] {\label{FIG1a-d} Some orbits of the systems \eqref{extreme_tilt_0} with $v = 1, -1 $, for different parameter choices. The points $\overline {P_5}, \overline{P_6}, \overline{P_7} $ are included, where $ \overline{P_i}$ denotes the symmetric points of the point ${P_i}$. The dotted line represents $ H _{-}$ (resp. $ \overline{H_{-}} = H_{+}$) for $ C_2 = 1 $.}
\end{figure*}

\begin{figure*}
\centering
	\subfigure[]{\includegraphics[scale=0.6]{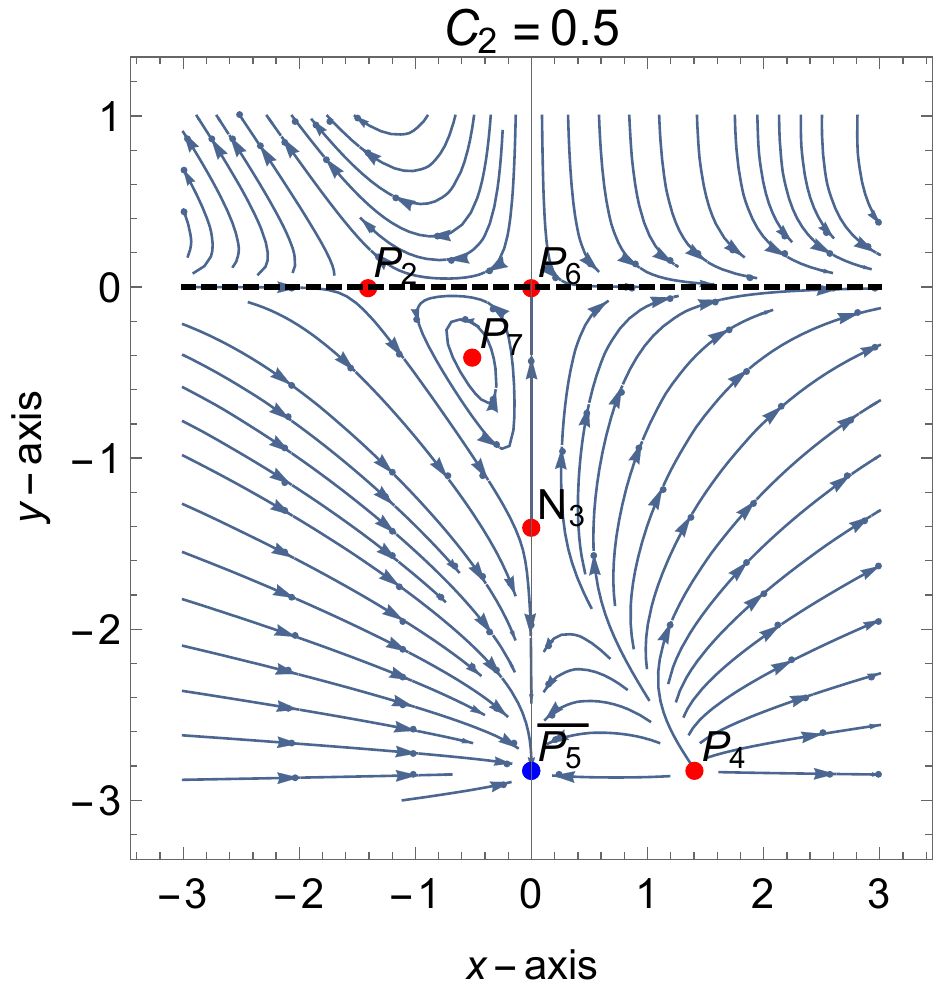}}
	\subfigure[]{\includegraphics[scale=0.6]{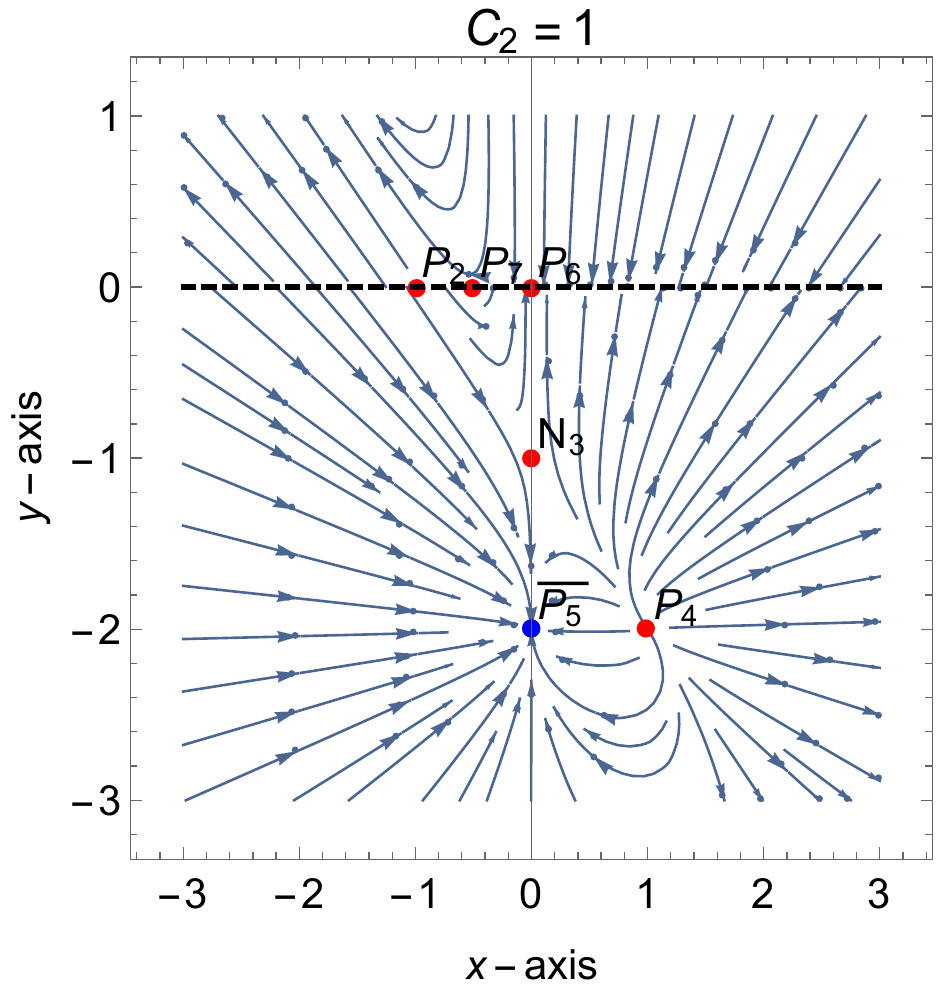}}
	\subfigure[]{\includegraphics[scale=0.6]{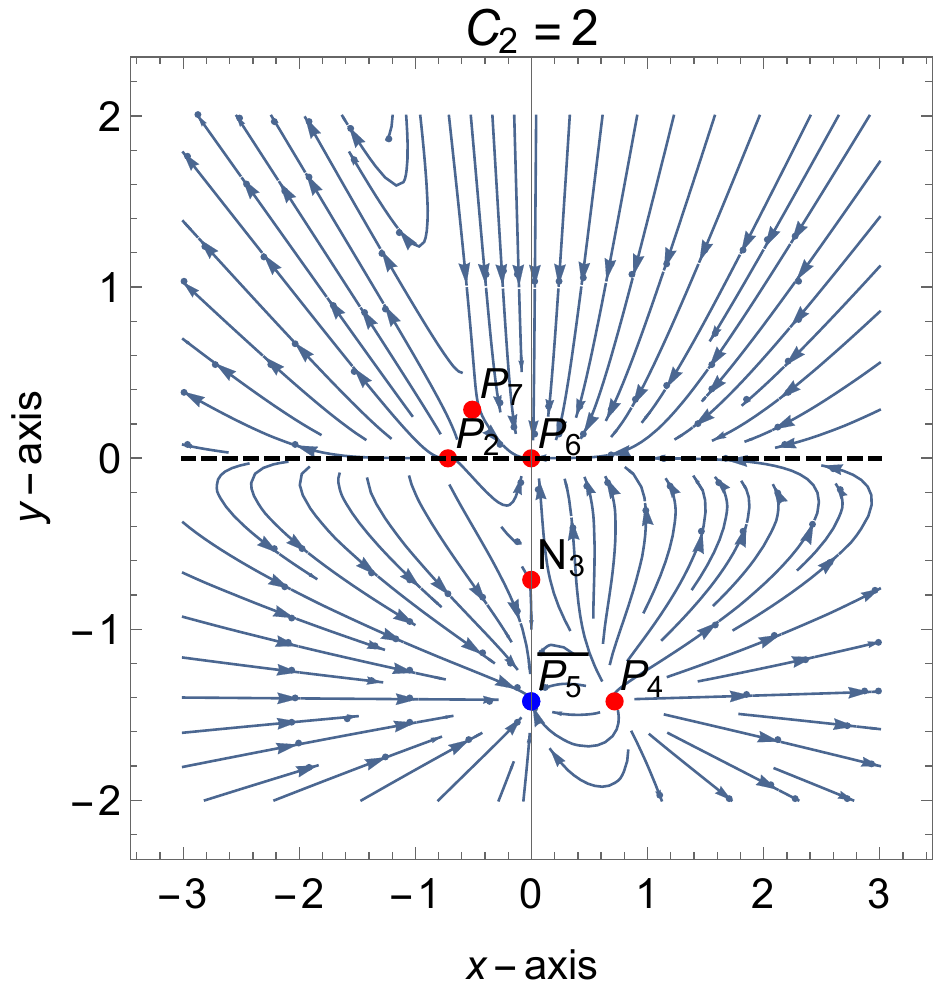}}
	\caption{\label{FIG2a-c} Some orbits of system   \eqref{variedad_extreme_tilt} for different choices of the parameter $C_2$.}
\end{figure*}

The systems \eqref{extreme_tilt_0} are related through the simultaneous change of $ A \rightarrow -A $, and the sign `` $ + $ '' by the sign `` $ - $ ''. Therefore, without loss of generality, the positive sign `` $ + $ '' is studied, with $ A \geq 0$.
\begin{table*}[t!]
\caption{\label{Tab1} Qualitative analysis of the equilibrium points of the systems \eqref{extreme_tilt_0} corresponding to extreme tilt:  $v=1,-1$.}
\begin{tabular*}{\textwidth}{@{\extracolsep{\fill}}lrrrl@{}}
\hline
Equil. & \multicolumn{1}{c}{$(\Sigma,A),(v)$} & \multicolumn{1}{c}{Eigenvalues} & \multicolumn{1}{c}{Stability} & \multicolumn{1}{c}{$(K,\Omega_t)$}  \\
Points \\
\hline
  $N_{2,3}$ & $(0,0),(\pm 1)$& $\left\{-1,1\right\}$  & hyperbolic saddle.  & $(1,0)$\\\hline
	$P_{1,2}$ & $(-\frac{1}{\sqrt{C_2}},0), (\pm 1)$  & $\left\{2-\frac{2}{\sqrt{C_2}},2\right\}$ & hyperbolic saddle for $0<C_2<1$. & \\
	&&& non-hyperbolic for  $C_2=1$. & \\
	&&& hyperbolic source for $C_2>1$.& (0,0)\\\hline
 $P_{3,4}$ & $(\frac{1}{\sqrt{C_2}},0),( \pm1)$ & $\left\{\frac{2}{\sqrt{C_2}}+2,2\right\}$ & hyperbolic source for $C_2>0$.  & $(0,0)$\\\hline
  $P_5$ & $\left(0, \frac{1}{\sqrt{C_2}}\right), (1) $& $\left\{-\frac{2}{\sqrt{C_2}}-2,-2\right\}$  & hyperbolic sink for $C_2>0$. & $(0,0)$\\\hline
	 $P_6$ & $\left(0, \frac{1}{\sqrt{C_2}}\right),(- 1) $& $\left\{\frac{2}{\sqrt{C_2}}-2,-2\right\}$  & hyperbolic saddle for $0<C_2<1$. & \\ &&& non-hyperbolic for  $C_2=1$. & \\	&&& hyperbolic sink for $C_2>1$. & $(0,0)$\\\hline
	$P_{7}$ & $\left(-\frac{1}{2}, \frac{1}{2}\right),(- 1)$ & $\left\{\pm\sqrt{C_2-1}\right\}$ & hyperbolic saddle for $C_2>1$. & \\
	&&& non-hyperbolic $C_2\leq 1$.  &$\scriptstyle{(1-C_2, \frac{C_2}{2})}$\\
 \hline
\end{tabular*}
\end{table*}
Table \ref{Tab1} presents the qualitative analysis of the equilibrium points of the systems \eqref{extreme_tilt_0} corresponding to the cases of extreme  tilt $ v = 1, -1 $, which are the following:
\begin{enumerate}
    \item[$N_{2,3}$:] $(\Sigma,A)=(0,0)$, $v=\pm 1$, with eigenvalues $\{-1,1\}$ is a hyperbolic saddle.
    \item[$P_{1,2}$:] $(\Sigma,A)=(-\frac{1}{\sqrt{C_2}},0)$ , $v=\pm 1$, with eigenvalues \newline $\left\{2-\frac{2}{\sqrt{C_2}},2\right\}$ is:
    \begin{enumerate}
        \item a hyperbolic saddle for $0<C_2<1$. 
        \item non-hyperbolic for $C_2=1$.
        \item a hyperbolic source for  $C_2>1$.
    \end{enumerate}
    \item[$P_{3,4}$:] $(\Sigma,A)=(\frac{1}{\sqrt{C_2}},0)$, $v=\pm 1$, with eigenvalues \newline $\left\{2+\frac{2}{\sqrt{C_2}},2\right\}$ is a hyperbolic source for $C_2>0$.
    \item[$P_5$:] $(\Sigma,A)=\left(0, \frac{1}{\sqrt{C_2}}\right)$, $v=1$, with eigenvalues \newline $\left\{-\frac{2}{\sqrt{C_2}}-2,-2\right\}$ is a hyperbolic sink for $C_2>0$.
    
    \item[$P_6$:] $(\Sigma,A)=\left(0, \frac{1}{\sqrt{C_2}}\right)$, $v=-1$, with eigenvalues \newline $\left\{\frac{2}{\sqrt{C_2}}-2,-2\right\}$ is:
    \begin{enumerate}
        \item a hyperbolic saddle for $0<C_2<1$. 
        \item non-hyperbolic for  $C_2=1$.
        \item a hyperbolic sink for  $C_2>1$.
    \end{enumerate}
    \item[$P_7$:] $(\Sigma,A)=\left(-\frac{1}{2}, \frac{1}{2}\right)$, $v=-1$, with eigenvalues \newline $\left\{-\sqrt{C_2-1},\sqrt{C_2-1}\right\}$ is  
    \begin{enumerate}
        \item a hyperbolic saddle for $C_2>1$.
        \item non-hyperbolic for $C_2\leq 1$.
    \end{enumerate}
\end{enumerate}
Figure \ref{FIG1a-d} shows some orbits of the system \eqref{extreme_tilt_0} with $ v = 1, -1 $, for different choices of parameters. The points $ \overline{P_5}, \overline{P_6}, \overline{P_7} $ are included, where $ \overline{P_i} $ denotes the symmetric points of the points ${P_i} $. The dotted invariant line represents $H_-$ (resp. $H_{+}$) for $C_2 = 1$. The analytical results discussed above are confirmed. Figure \ref{FIG2a-c} shows some orbits of the system \eqref{variedad_extreme_tilt} for different choices of parameters.

\subsection{Invariant manifolds of $P_6$}
\label{invP6}
Concerning the equilibrium point $P_6$ with $v=-1$, the invariant line  $A-\Sigma-\frac{1}{\sqrt{C_2}}$ corresponds to its unstable manifold for $C_2<1$. This line is stable for $C_2=1$, and for $C_2>1$ that line belongs to the stable 2-dimensional manifold of $P_6$. Introducing the change of variables: 
\begin{subequations}
\begin{align}
& x= \Sigma, \\
& y= A-\Sigma-\frac{1}{\sqrt{C_2}},   
\end{align}
the following equations are obtained:
\end{subequations}
\begin{subequations}
\label{variedad_extreme_tilt}
\begin{align}
&{x}'=x \left(-1+x^2 (1+\mu )^2-(1+x+y+(x+y) \mu )^2 \right. \nonumber \\ & \left. +2 \left(x+y+\frac{1}{1+\mu }\right)\right),\\
 & y'=-y (2+y+y \mu ) (1+y+y \mu +2 x(1+\mu )),
	\end{align}
\end{subequations}
where the following parameter is introduced: 
\begin{equation}
    \mu= \sqrt{C_2}-1, \quad C_2=(\mu +1)^2. 
\end{equation}
The eigenvalues of  $P_6$ are $\left\{-\frac{2 \mu }{\mu
   +1},-2\right\}$. For $\mu>0$,  i.e.,  $C_2>1$, the equilibrium point  $P_6$ is a hyperbolic sink in the invariant set  $v=-1$. For $\mu<0$,  i.e.,  $C_2<1$, the equilibrium point  $P_6$ has an unstable manifold tangent to the axis $x$. This unstable manifold of  $P_6$ can be expressed locally by the graph: 
\begin{equation}
    \left\{(x, y): y=h(x) , |x|<\delta\right\},
\end{equation}
where $h$ satisfies the initial value problem: 
\begin{align}
& h (2+(1+\mu ) h) (1+2 x (1+\mu )+(1+\mu ) h)\nonumber \\
& +x \left(-1+x^2 (1+\mu )^2+2 \left(x+\frac{1}{1+\mu }+h\right) \right. \nonumber \\ & \left. -(1+x+h+\mu
 (x+h))^2\right) h'=0, \\ & h(0)=0, \quad h'(0)=0.  
\end{align}
The previous differential equation admits the first integral 
\begin{align}
    & \frac{-x (1+\mu ) h^{p}(1+x+x \mu +(1+\mu ) h)}{ (2+(1+\mu ) h)^{q}} =C_1, \nonumber\\ 
    & p=-\frac{\mu }{1+\mu }>0, \quad  q=\frac{2+\mu }{1+\mu }<0, \quad -1< \mu<0. 
\end{align} 
where $C_1$ is an integration constant. 
\begin{figure*}
    \includegraphics[scale=0.35]{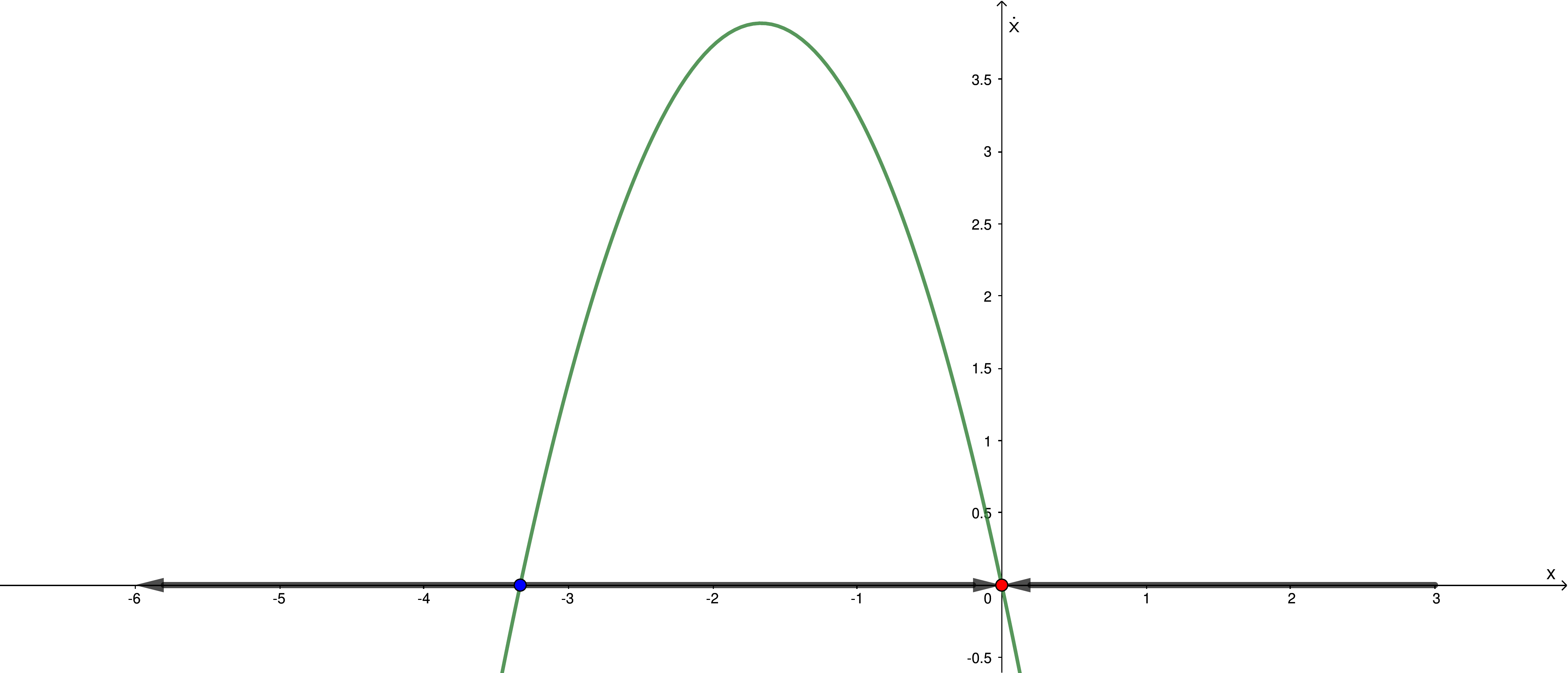}
    \caption{1-dimensional flow of the system \eqref{L3.49}, for $\mu =-0.7$. All the flows are topologically equivalent for $\mu$ on the interval $-1 < \mu < 0$.}
    \label{fig:x-mu}
\end{figure*}
Imposing the condition  $h(0)=0$, it is obtained $C_1=0$. Solving the resulting equation for $h$ are obtained two solutions:  $$h(x)=0,$$ and 
$$h(x)= -x-\frac{1}{1+\mu }.$$\
The last solution is discarded since $h'(x)=-1$, which implies that the tangential condition $h'(0)=0$ is not fulfilled. Then, the unstable solution is given locally by the trivial solution:
\begin{equation}
    \left\{(x, y): y=0 , |x|<\delta\right\}.
\end{equation}
The dynamics on the invariant manifold is given by:
\begin{equation}
\label{L3.49}
 {x}'=-\frac{2 \mu  x (\mu  x+x+1)}{\mu +1},   
\end{equation}
whose solution passing by $x(0)=x_0$, is:  
\begin{equation}
    x(\eta )= \frac{x_0}{e^{\frac{2 \eta  \mu }{\mu +1}} (\mu  x_0+x_0+1)-(\mu +1) x_0}, 
\end{equation}
with $-1 < \mu < 0$. Finally, $\lim_{\eta\rightarrow -\infty} x(\eta)=0$. That is, the solutions generically approaches the origin as $\eta\rightarrow -\infty$, tending towards the past to $P_6$.

Figure \ref{fig:x-mu} shows the 1-dimensional flow of \eqref{L3.49} for $ -1 <\mu <0 $, where it is illustrated that the origin of the system \eqref{L3.49} is stable. Then, applying the Unstable Manifold Theorem, it is confirmed that $P_6$ is a hyperbolic saddle, as it was previously commented in table \ref{Tab1}.
\subsection{Ideal gas $\gamma=1$}
\label{fluidosinpresion}
For an ideal gas with $\gamma=1$  (pressureless fluid; dust) the equations \eqref{YYYYY} and restrictions \eqref{YYYYY-rest} become: 
\begin{small}
\begin{subequations}
\label{dust-1}
\begin{align}
&{\Sigma}'=\frac{A^2  \left(v^2 (1-2 C_2 \Sigma )+1\right) }{2  v^2} \nonumber \\
& + \frac{\left(C_2 \Sigma ^2-1\right) \left(2 C_2 \Sigma  v^2+v^2+1\right)+K
   \left(v^2+1\right)}{2 C_2 v^2},\\
& A'=-C_2 A^3 +A \Sigma  (C_2 \Sigma +2)+A,\\
& K'=	2 C_2 K \left(\Sigma ^2-A^2\right),\\
& v'=\frac{\left(v^2-1\right) (A v-\Sigma )}{v},
	\end{align}
\end{subequations}
\end{small}

      \begin{figure*}[t!]
    \centering
    \includegraphics[scale=0.4]{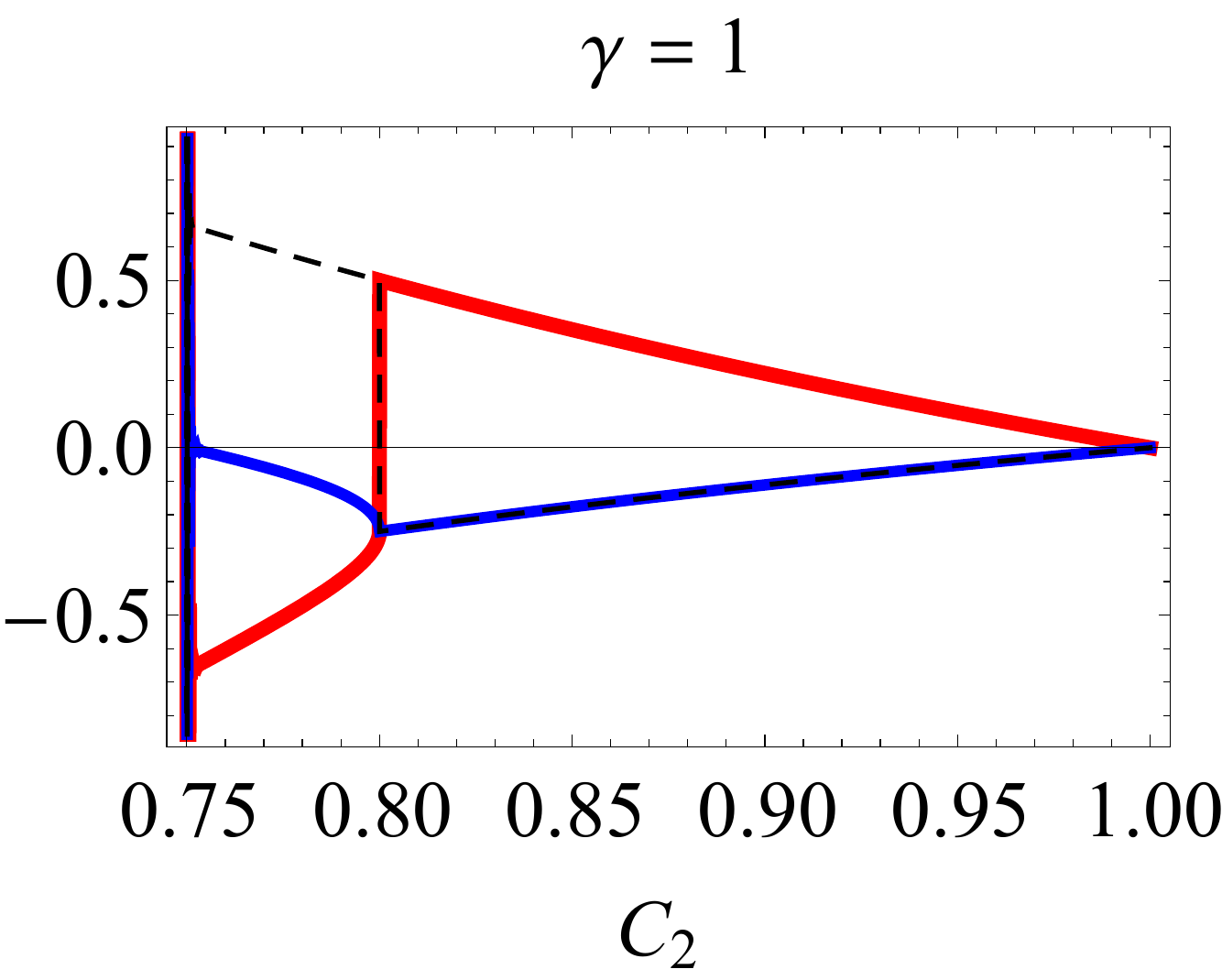} \hspace{2cm}
    \includegraphics[scale=0.4]{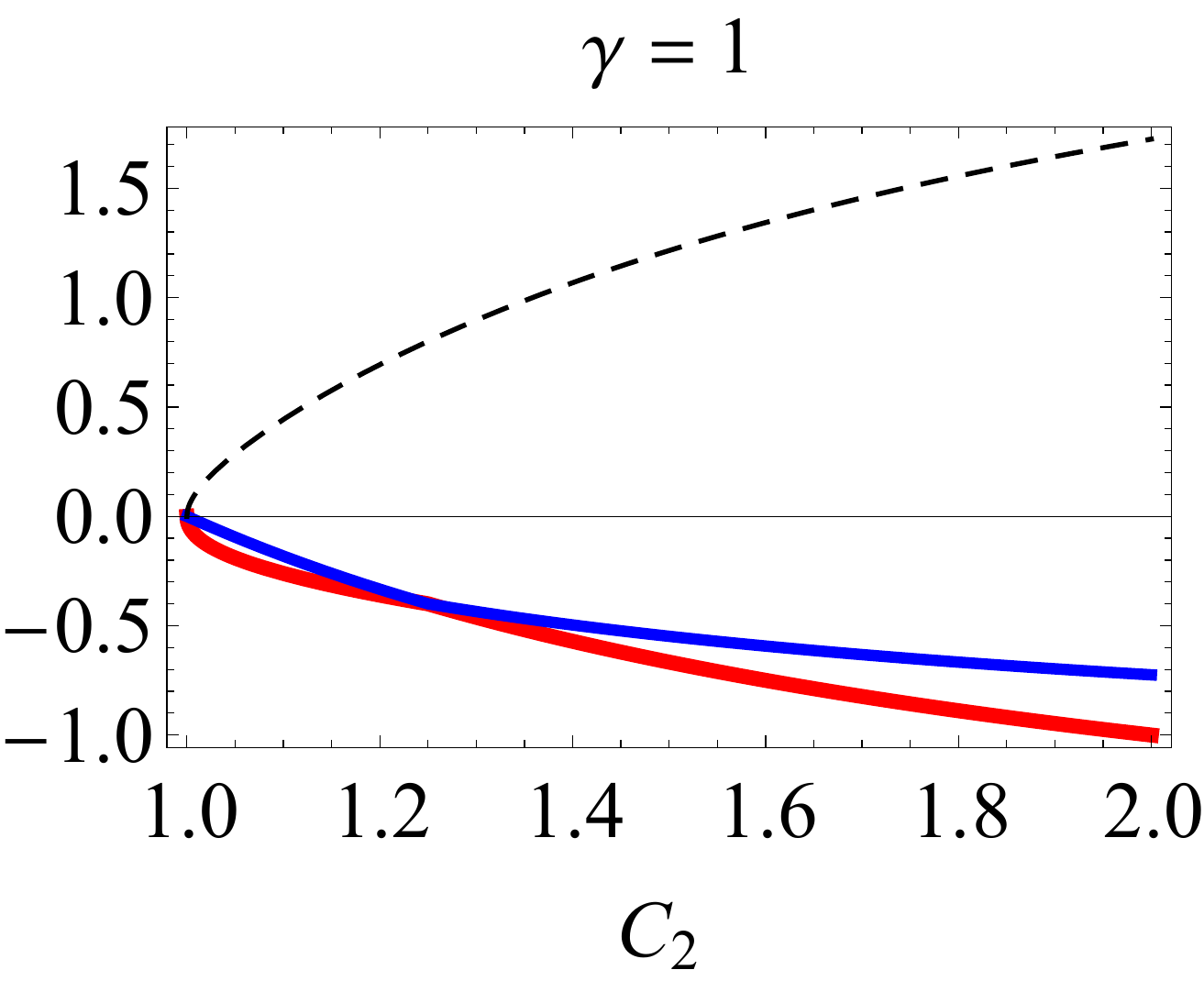}
    \caption[{Real part of the eigenvalues of the equilibrium point $P_8 $ for $ C_2 \geq \frac{3}{4} $ and $\gamma = 1 $.}] {\label{eigenP8gamma1} Real parts of $\lambda_i $ corresponding to the equilibrium point $P_8$ for $ C_2 \geq \frac{3}{4} $ and $\gamma = 1 $, showing that in general it has saddle behavior.}
   \end{figure*}

\noindent where the expressions $\Omega_{t}$ and $r$ given by: 
\begin{small}
\begin{align}
& \Omega_{t}=\frac{1-C_2 A^2 -C_2 \Sigma ^2-K}{v^2}, \quad  r=\frac{1-C_2 A^2 +C_2 \Sigma ^2+\Sigma}{\sqrt{3}},
	\end{align}
\end{small}
were used. 

The restriction  \eqref{constraintmod1} is reduced to: 
\begin{equation}
K=-C_2 A^2 -2 A C_2 \Sigma  v-C_2 \Sigma ^2+1.
\end{equation}
This allows us to study the reduced 3-dimensional system:
\begin{subequations}
\label{3.41-3.43}
\begin{align}
&{\Sigma}'=-\frac{\Sigma  \left(A \left(A C_2 v+v^2+1\right)+v\right)}{v},\\
& A'=-C_2 A^3 +A \Sigma  (C_2 \Sigma +2)+A,\\
& v'=\frac{\left(v^2-1\right) (A v-\Sigma )}{v},
	\end{align}
\end{subequations}
defined on the phase space: 
\begin{small}
\begin{align}
    & \Bigg\{(\Sigma, A, v)\in\mathbb{R}^3:  C_2(A^2 + \Sigma ^2)\leq 1, \nonumber \\
    & \quad C_2 \left(A^2 +2 A \Sigma  v+ \Sigma ^2\right) \leq 1,  \quad v \in [-1, 0)\cup (0, 1]\Bigg\}.
\end{align}
\end{small}
The equilibrium points of system \eqref{3.41-3.43} are the following:
\begin{enumerate}
    \item[$N$:] $(\Sigma,A,v)=(0,0,v)$  with eigenvalues  $\{0,-1,1\}$ is a non-hyperbolic saddle.
    \item[$N_{2,3}$:] $(\Sigma,A,v)=(0,0,\pm 1) $ with eigenvalues $\{-1,1,0\}$ are non-hyperbolic saddles.
   The points $N_1$, $N_2$ and $N_3$ are joined in a line called $N$.
    \item[$P_{1,2}$:] $ (\Sigma,A,v)=\left(-\frac{1}{\sqrt{C_2}},0,\pm 1\right)$ with eigenvalues \newline  $\left\{\frac{2}{\sqrt{C_2}},2-\frac{2}{\sqrt{C_2}},2\right\}$ are: \begin{enumerate}
        \item hyperbolic sources for  $C_2>1$.
        \item hyperbolic saddles for $C_2<1$. 
        \item non-hyperbolic for  $C_2=1$.
    \end{enumerate} 
    \item[$P_{3,4}$:] $(\Sigma,A,v)=\left(\frac{1}{\sqrt{C_2}},0,\pm 1\right)$ with eigenvalues \newline $\left\{-\frac{2}{\sqrt{C_2}},2+\frac{2}{\sqrt{C_2}},2\right\}$ are  hyperbolic saddles for  $C_2>0$.
    \item[$P_5$:] $(\Sigma,A,v)=\left(0, \frac{1}{\sqrt{C_2}}, 1\right)$ with eigenvalues  \newline $\left\{\frac{2}{\sqrt{C_2}},-\frac{2}{\sqrt{C_2}}-2,-2\right\}$  is a hyperbolic saddle for $C_2>0$. 
    \item[$P_6$:] $(\Sigma,A,v)=\left(0, \frac{1}{\sqrt{C_2}},- 1\right)$ with eigenvalues  \newline $\left\{-\frac{2}{\sqrt{C_2}},\frac{2}{\sqrt{C_2}}-2,-2\right\}$ is
    \begin{enumerate}
        \item a hyperbolic sink for  $C_2>1$.
        \item a hyperbolic saddle for  $C_2<1$.
        \item non-hyperbolic for  $C_2=1$.
    \end{enumerate}
    \item[$P_{7}$:] $(\Sigma,A,v)=\left(-\frac{1}{2},\frac{1}{2},- 1\right)$ with eigenvalues  $\left\{0,\pm \sqrt{C_2-1}\right\}$ is:	
    \begin{enumerate}
	\item a non-hyperbolic saddle for $C_2>1$.
	\item non-hyperbolic with three zero eigenvalues for $C_2=1$.
	\item non-hyperbolic with a zero eigenvalue and two purely imaginary eigenvalues for  $0\leq
   C_2<1$.
	\end{enumerate}
    \item[$P_8$:] $(\Sigma,A,v)={\left(-\frac{1}{2 C_2}, \frac{\sqrt{4 C_2-3}}{2 C_2}, - \frac{1}{\sqrt{4 C_2-3}}\right)}$ with eigenvalues  $\left\{k \lambda_1,k \lambda_2, k \lambda_3\right\}$, where $k=\frac{1}{2 C_2^2 (4 C_2-3)^{3/2}}$ and $\lambda_1, \lambda_2$ and  $\lambda_3$ are the roots of the polynomial in  $\lambda$:  $P(\lambda)=-16 (C_2-1)^2 C_2^3 (4   C_2-3)^{11/2}-4 (C_2-1)  C_2^2 (8 C_2-7) (4
   C_2-3)^3 \lambda +\lambda ^3$. Figure \ref{eigenP8gamma1} shows the real part $ \lambda_i $ (which differs from  the eigenvalues in an overall multiplicative factor $k$) corresponding to the equilibrium point $ P_8 $ for $ C_2 \geq \frac{3}{4} $ and $\gamma = 1 $, showing that in general it has saddle behavior.

   \item[$H_{\pm}$:] $(\Sigma,A,v)=\left(\Sigma_0,-\varepsilon(1+\Sigma_0), \varepsilon\right)$, $\varepsilon=\pm 1$, exists for $C_2=1$. The eigenvalues are $\left\{0,-4 \Sigma _c-2\right\}$. These lines of equilibrium points, which do not exist in Einstein-æther theory ($C_2 \neq 1$), are associated with a change of causality of the homothetic vector field.
\end{enumerate}

\begin{table*}[t!]
\caption{\label{Tab2} Qualitative analysis of the equilibrium points of system \eqref{3.41-3.43} for a pressureless fluid ($\gamma =1$) and $v\neq 0$. Line $N$ is included.}
\begin{tabular*}{\textwidth}{@{\extracolsep{\fill}}lrrrl@{}}
\hline
Equil. & \multicolumn{1}{c}{$(\Sigma,A, v)$} & \multicolumn{1}{c}{Eigenvalues} & \multicolumn{1}{c}{Stability} & \multicolumn{1}{c}{$(K,\Omega_t)$}  \\
Points \\
\hline
 $N$ & $(0,0,v)$  & $\{0,-1,1\}$    & non-hyperbolic saddle. & (1,0)\\\hline
$N_{2,3}$ & $(0,0,\pm 1) $& $\{-1,1,0\}$  & non-hyperbolic saddle. & $(1,0)$\\\hline
  $P_{1,2}$ & $\left(-\frac{1}{\sqrt{C_2}},0,\pm 1\right)$ & $\left\{\frac{2}{\sqrt{C_2}},2-\frac{2}{\sqrt{C_2}},2\right\}$ & hyperbolic source for $C_2>1$ & \\ &&& hyperbolic saddle for $C_2<1$. & \\ &&& non-hyperbolic for  $C_2=1$.  & $(0,0)$\\\hline
$P_{3,4}$ & $\left(\frac{1}{\sqrt{C_2}},0,\pm 1\right)$&$\left\{-\frac{2}{\sqrt{C_2}},2+\frac{2}{\sqrt{C_2}},2\right\}$ & hyperbolic saddle for $C_2>0$. & $(0,0)$\\\hline
$P_5$ & $\left(0, \frac{1}{\sqrt{C_2}}, 1\right)$& $\left\{\frac{2}{\sqrt{C_2}},-\frac{2}{\sqrt{C_2}}-2,-2\right\}$ &  hyperbolic saddle for $C_2>0$. &$(0,0)$\\\hline
$P_6$ & $\left(0, \frac{1}{\sqrt{C_2}},- 1\right)$& $\left\{-\frac{2}{\sqrt{C_2}},\frac{2}{\sqrt{C_2}}-2,-2\right\}$  & hyperbolic sink for $C_2>1$. & \\ &&& hyperbolic saddle for $C_2<1$ & \\ &&& non-hyperbolic for $C_2=1$. &$(0,0)$\\\hline
$P_{7}$ & $\left(-\frac{1}{2},\frac{1}{2},- 1\right)$& $\left\{0,\pm \sqrt{C_2-1}\right\}$ & non-hyperbolic saddle  for $C_2> 1$. & $\scriptstyle{(1-C_2,\newline \frac{C_2}{2})}$\\\hline
$P_8$ &${\left(-\frac{1}{2 C_2}, \frac{\sqrt{4 C_2-3}}{2 C_2}, - \frac{1}{\sqrt{4 C_2-3}}\right)}$ & see text. & see text.  & ${\left(0,\newline \frac{4 C_2-3}{2 C_2}\right)}$\\ \hline
$H_{-}$ &  ${\left(\Sigma_0,1+\Sigma_0, - 1\right), C_2=1}$ & $\left\{0,-2(1+ 2\Sigma _c)\right\}$ & non-hyperbolic. &  ${\left(0, -2 \Sigma_0 (1+\Sigma_0)\right)}$\\\hline
$H_{+}$ &  ${\left(\Sigma_0,-1-\Sigma_0,  1\right), C_2=1}$ & $\left\{0,-2(1+ 2\Sigma _c)\right\}$ & non-hyperbolic. &  ${\left(0, -2 \Sigma_0 (1+\Sigma_0)\right)}$\\
 \hline
\end{tabular*}
\end{table*}

The stability criteria of the equilibrium points of the system \eqref{3.41-3.43} for pressureless fluid  ($\gamma = 1 $) and $ v \neq 0 $ are summarized in table \ref{Tab2}.
\subsection{General case $v\neq 0$}
\label{general}
In the general case $v\neq 0$ is possible reduce the system's dimension when the restriction \eqref{constraintmod1} in non-degenerated. The restrictions \eqref{constraintmod1} and \eqref{defnOmegat} can be globally solved for ${\Omega}_{t}$ y $K$ (assuming  $\gamma v \neq 0$ and $\gamma  v^2-v^2+1 \neq 0$):
\begin{small}
\begin{subequations}
\begin{align}
 & {\Omega}_t = \frac{2 A C_2 {\Sigma} \left((\gamma -1) v^2+1\right)}{\gamma
 v}, \\ 
 & K=- C_2 A^2-\frac{2 A C_2 \Sigma  \left(\gamma
   +v^2-1\right)}{\gamma  v}-C_2 \Sigma ^2+1.
\end{align}
\end{subequations}
 \end{small}
Then, a 3-dimensional dynamical system is obtained:

\begin{small}
\begin{subequations}
\label{reducedsyst}
\begin{align}
&{\Sigma}'=\Sigma  \Big(-C_2 A^2+C_2 \Sigma^2-1 \nonumber \\
& +\frac{A\left(-3 \gamma -2 (\gamma -1) C_2 \Sigma +v^2(\gamma +2 (\gamma -1) C_2 \Sigma -2)+2\right)}{\gamma  v}\Big),
\\
& A'=A \left(-C_2 A^2+\frac{2 A (\gamma -1) C_2 \Sigma \left(v^2-1\right)}{\gamma v}+\Sigma (C_2 \Sigma +2)+1\right),\\
& v'=\frac{\left(v^2-1\right)\left(A \left(3 \gamma ^2-5\gamma +(\gamma -2) v^2+2\right)+\gamma  v (\gamma (\Sigma +2)-2)\right)}{\gamma \left(\gamma -v^2-1\right)}.
	\end{align}
\end{subequations}
\end{small}	
\noindent The dynamical system \eqref{reducedsyst} admits some invariant sets. These are: $ v = \pm 1 $, corresponding to extreme tilt, and the invariant sets $ A = 0 $ and $ \Sigma = 0 $. The equilibrium points of the system \eqref{reducedsyst} are the following.
\begin{enumerate}
\item[$P_{1,2}$:] $(\Sigma,A,v)=(-\frac{1}{\sqrt{C_2}},0,\pm 1)$, with eigenvalues \newline $\left\{2-\frac{2}{\sqrt{C_2}},2,\frac{2 \gamma \left(\frac{1}{\sqrt{C_2}}-2\right)+4}{2-\gamma }\right\}$. They are:
	\begin{enumerate}
	\item hyperbolic sources for  $1<\gamma <2, 
   1<C_2<\frac{\gamma
   ^2}{4 \gamma ^2-8 \gamma +4}$.
   	\item hyperbolic saddles for: 
	   \begin{enumerate}
	       \item $1<\gamma <2,0<C_2<1$, or   
	       \item $1<\gamma <2,C_2>\frac{\gamma ^2}{4 \gamma ^2-8 \gamma +4}$.
	   \end{enumerate}
	\item non-hyperbolic for:
	 \begin{enumerate}
	     \item $1<\gamma <2, C_2=\frac{4\gamma ^2}{(\gamma-1)^2}$, or  
         \item $1<\gamma <2, C_2=1$. 
	 \end{enumerate}
	\end{enumerate}
\item[$P_{3,4}$:] $(\Sigma,A,v)=(\frac{1}{\sqrt{C_2}},0,\pm 1)$, with eigenvalues \newline $\left\{\frac{2}{\sqrt{C_2}}+2,2,\frac{-4 \gamma -\frac{2 \gamma}{\sqrt{C_2}}+4}{2-\gamma }\right\}$. They are hyperbolic saddles for  $1<\gamma <2,  C_2>0$.
\item[$P_5$:] $(\Sigma,A,v)=(0,\frac{1}{\sqrt{C_2}},1)$, with eigenvalues \newline $\left\{-\frac{2}{\sqrt{C_2}}-2,-2,\frac{6 \gamma +4 (\gamma -1)
   \sqrt{C_2}-8}{(\gamma -2) \sqrt{C_2}}\right\}$. It is: 
	\begin{enumerate}
	\item a hyperbolic sink for:
	  \begin{enumerate}
	      \item $1<\gamma \leq \frac{4}{3}, C_2>\frac{9 \gamma^2-24 \gamma +16}{4 \gamma^2-8 \gamma +4}$, or  
	      \item $\frac{4}{3}<\gamma <2, C_2>0$. 
	  \end{enumerate} 
	\item a hyperbolic saddle for $1<\gamma <\frac{4}{3},  0<C_2<\frac{(4-3 \gamma )^2}{4 (\gamma -1)^2}$
	\item non-hyperbolic for $1<\gamma <\frac{4}{3},  C_2=\frac{9 \gamma^2-24 \gamma +16}{4 \gamma^2-8 \gamma +4}$.
	\end{enumerate}
	\item[$P_6$:] $(\Sigma,A,v)=(0,\frac{1}{\sqrt{C_2}},-1)$, with eigenvalues \newline $\left\{\frac{2}{\sqrt{C_2}}-2,-2,\frac{-6 \gamma +4 (\gamma -1)
   \sqrt{C_2}+8}{(\gamma -2) \sqrt{C_2}}\right\}$. It is: 
	\begin{enumerate}
	\item a hyperbolic sink for $1<\gamma <2,  C_2>1$.
	\item a hyperbolic saddle for: \begin{enumerate}
	    \item $1<\gamma <2, 0<C_2<1$, or  
	    \item $\frac{4}{3}<\gamma <2,  0<C_2<\frac{(4-3 \gamma )^2}{4 (\gamma -1)^2}$, or  
	    \item $1<\gamma \leq \frac{4}{3},  0<C_2<1$.
	\end{enumerate}
	\item non-hyperbolic for:
	  \begin{enumerate}
	      \item $1<\gamma \leq \frac{4}{3}, C_2=1$, or   
	      \item $\frac{4}{3}<\gamma <2,  C_2=\frac{9 \gamma^2-24 \gamma +16}{4 \gamma^2-8 \gamma +4}$, or  
	      \item $\frac{4}{3}<\gamma <2,  C_2=1$.
	  \end{enumerate}
	\end{enumerate}
 \item[$P_{7}$:] $(\Sigma,A,v)=(-\frac{1}{2},\frac{1}{2},-1)$, with eigenvalues \newline $\left\{0,-\sqrt{C_2-1},\sqrt{C_2-1}\right\}$ is: 
	\begin{enumerate}
	\item a non-hyperbolic saddle for $C_2>1$. \item non-hyperbolic with three zero eigenvalues for $C_2=1$.
	\item non-hyperbolic with a zero eigenvalue and two purely imaginary eigenvalues for $0\leq    C_2<1$.
	\end{enumerate}
  \item[$P_8$:] $(\Sigma,A,v)=\left(-\frac{1}{2 C_2},-\frac{1}{2 C_2 \Delta },\Delta \right)$, where $\Delta=-\sqrt{\frac{2-3 \gamma }{\gamma -4 \gamma  C_2+2}}$. For this point it is satisfied the equation $\Sigma =A v=-\frac{1}{2 C_2}$. $P_8$ exists for  $1\leq \gamma <2,  C_2>\frac{\gamma +2}{4 \gamma }$.
When $\gamma=1$, the eigenvalues are $\left\{k \lambda_1,k \lambda_2, k \lambda_3\right\}$, where  $k=\frac{1}{2 C_2^2 (4 C_2-3)^{3/2}}$, and $\lambda_1, \lambda_2$ and $\lambda_3$ are the roots of the polynomial in  $\lambda$:  
$P(\lambda)=-16 (C_2-1)^2 C_2^3 (4   C_2-3)^{11/2}-4 (C_2-1)  C_2^2 (8 C_2-7) (4
   C_2-3)^3 \lambda +\lambda ^3$. 
   For $C_2=1+\delta + \mathcal{O}(\delta^2)$,  $\lambda_1= -2 \delta +\mathcal{O}\left(\delta ^2\right)$,  $\lambda_2=-\frac{\left(3 \gamma ^2-8 \gamma +4\right) \sqrt{\delta }}{(\gamma
   -2) (3 \gamma -2)}+\frac{\left(-3 \gamma ^2+6 \gamma -4\right) \delta }{(\gamma -2) (3 \gamma
   -2)}+\mathcal{O}\left(\delta ^{3/2}\right)$, and 
   $\lambda_3=\frac{\left(3 \gamma ^2-8 \gamma +4\right) \sqrt{\delta }}{(\gamma -2) (3
   \gamma -2)}+\frac{\left(-3 \gamma ^2+6 \gamma -4\right) \delta }{(\gamma -2) (3 \gamma -2)}+\mathcal{O}\left(\delta
   ^{3/2}\right)$. This shows a hyperbolic saddle behavior for values of the parameters close to the values of General Relativity. For example, for a pressureless fluid($\gamma = 1 $) the $\lambda_i $ are approximately  $\left \{- 2 \delta , \delta + \sqrt{\delta}, \delta - \sqrt {\delta} \right\} $. For $ \delta<0 $ there are two complex imaginary eigenvalues  with  negative real parts and a positive real eigenvalue, while for $\delta>0$ there is a negative eigenvalue and the others have different signs. In the figure \ref{eigenP83D} the real part of $\lambda_i$ (which differ from the eigenvalues associated with the equilibrium point $P_8$ by the overall factor  $k$) are represented graphically for $C_2\geq\frac {2+ \gamma}{4\gamma} $ and $\gamma\in [1, 2] $. The figure illustrates that the equilibrium point is generally a hyperbolic saddle or is non-hyperbolic.
   
\begin{widetext} 
   \item[$P_9:$]   $(\Sigma,A,v)=(0, \lambda_{+} v_{-}, v_{-})$, where 
$v_{-}=\frac{\sqrt{(\gamma -1) \left(\gamma  \left(2 \gamma ^2 C_2-\gamma  (2 C_2+3)-2 \sqrt{\gamma
   -1} \sqrt{C_2} \sqrt{\gamma  (\gamma  ((\gamma -1) C_2-3)+8)-4}+8\right)-4\right)}}{\gamma
   -2}$,  
and $\lambda_{+}= \frac{\gamma +\frac{\sqrt{\gamma  (\gamma  ((\gamma -1) C_2-3)+8)-4}}{\sqrt{\gamma -1}
   \sqrt{C_2}}}{2-3 \gamma }$, such that $C_2=\frac{1}{\lambda_{+}^2 v_{-}^2}$.
The eigenvalues are $\mu_1=-2$, $\mu_2=-\gamma + \frac{\sqrt{8 \gamma +\gamma ^3 C_2-\gamma ^2 (C_2+3)-4}}{\sqrt{\gamma -1} \sqrt{C_2}}$, and
   $\mu_3=\frac{2 (\gamma -1) \gamma  \left(-\gamma -v_{-}^4+3 \gamma  v_{-}^2-4 v_{-}^2+1\right)-2 (\gamma -2) \lambda  v_{-}^2 \left(-3
   \gamma +v_{-}^2+3\right) \left(\gamma +v_{-}^2-1\right)}{\gamma  \left(-\gamma +v_{-}^2+1\right)^2}$. 
Noting that there is at least one change of sign in two eigenvalues, that is $ \mu_1 \mu_2 <0 $, for $ 1 <\gamma <2, C_2> 0 $, it is concluded that it is a hyperbolic saddle. 
\item[$P_{10}$:] $(\Sigma,A,v)=(0, \lambda_{-} v_{+}, v_{+})$, where 
  $v_{+}=\frac{\sqrt{(\gamma -1) \left(\gamma  \left(2 \gamma ^2 C_2-\gamma  (2 C_2+3)+2 \sqrt{\gamma -1} \sqrt{C_2} \sqrt{\gamma 
   (\gamma  ((\gamma -1) C_2-3)+8)-4}+8\right)-4\right)}}{2-\gamma }$,
and $\lambda_{-}=\frac{\gamma -\frac{\sqrt{\gamma  (\gamma  ((\gamma -1) C_2-3)+8)-4}}{\sqrt{\gamma -1} \sqrt{C_2}}}{2-3 \gamma }$, such that $C_2=\frac{1}{\lambda_{-}^2 v_{+}^2}$. The eigenvalues are $\nu_1=-2, \nu_2=-\gamma -\frac{\sqrt{\gamma  (\gamma  ((\gamma -1) C_2-3)+8)-4}}{\sqrt{\gamma -1} \sqrt{C_2}}$, and  $\nu_3= \frac{-2 (\gamma -2) \lambda  v_{+}^2 \left(-3 \gamma +v_{+}^2+3\right) \left(\gamma +v_{+}^2-1\right)-2 (\gamma -1) \gamma  \left(\gamma +v_{+}^4+(4-3 \gamma )
   v_{+}^2-1\right)}{\gamma  \left(-\gamma +v_{+}^2+1\right)^2}$.
   \end{widetext}
 
   \begin{figure*}[t!]
    \centering
    \includegraphics[scale=0.55]{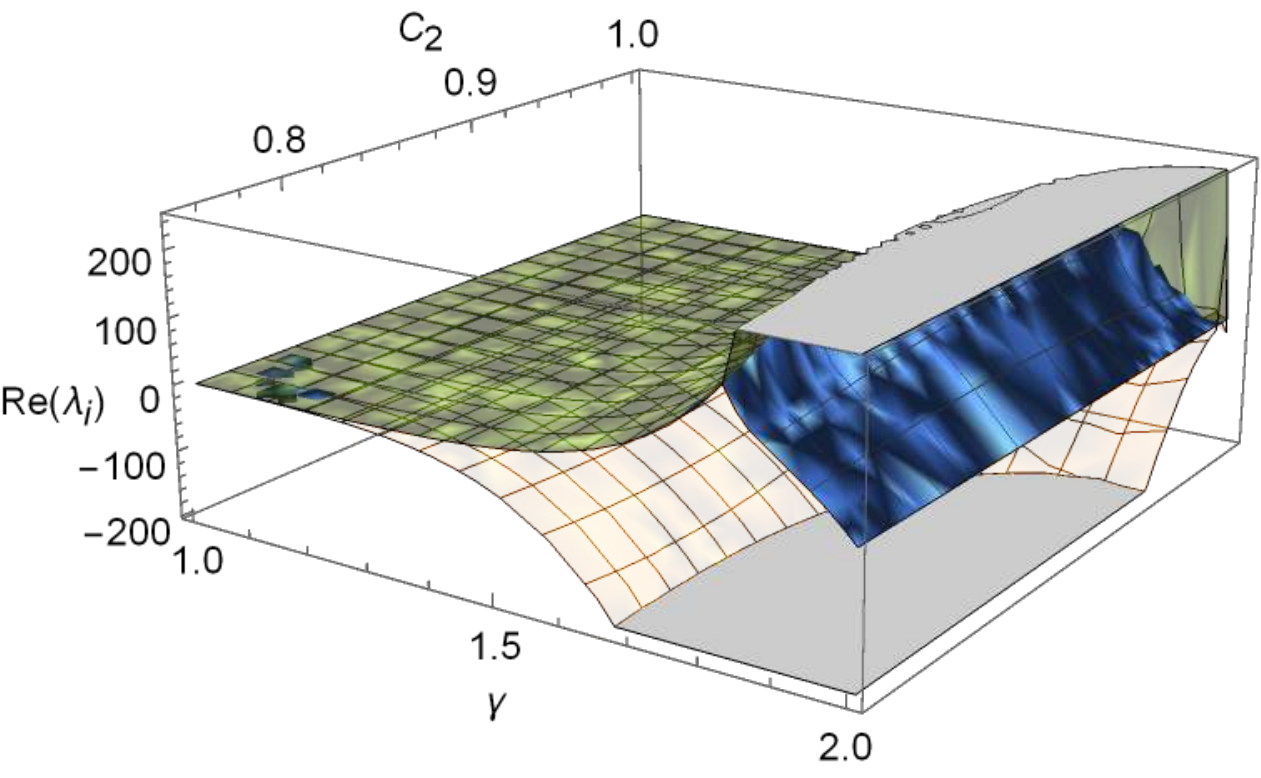} \hspace{2cm}
    \includegraphics[scale=0.55]{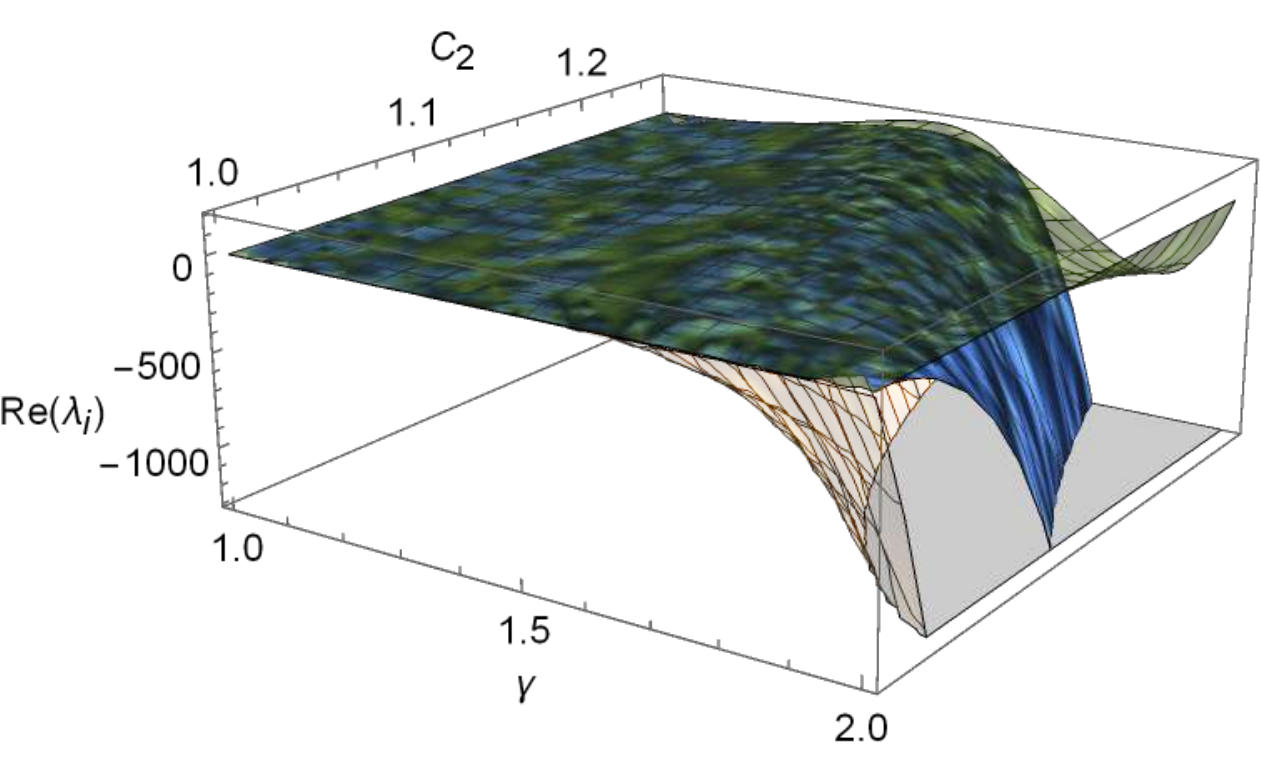}
    \caption[{Real parts of the eigenvalues of the equilibrium point $P_8 $ for $ C_2 \geq \frac {2+ \gamma} {4 \gamma} $ and $ \gamma \in [1, 2] $.}] {\label{eigenP83D} Real parts of $\lambda_i $ corresponding to the equilibrium point $P_8$ for $C_2 \geq \frac {2+ \gamma} {4 \gamma}$ and $\gamma \in [1, 2] $ .}
   \end{figure*}
It is a hyperbolic sink for: 
\begin{enumerate}
    \item $1<\gamma \leq \gamma_0, 0<C_2<\frac{4 \gamma -4}{\gamma ^2}$, or    
    \item $1<\gamma <\gamma_0, \frac{4 \gamma -4}{\gamma
   ^2}<C_2<\frac{(4-3 \gamma )^2}{4 (\gamma -1)^2}$, or    
 
   \item $\gamma_0<\gamma <\frac{4}{3}, 0<C_2<\frac{(4-3 \gamma )^2}{4 (\gamma -1)^2}$,
\end{enumerate}
where \newline $\gamma_0=-\frac{2}{27} \left(-11-\frac{14}{\sqrt[3]{27 \sqrt{57}-197}}+\sqrt[3]{27 \sqrt{57}-197}\right)$ \newline $\approx 1.22033$. \newline 
Is a hyperbolic saddle for: 
 \begin{enumerate}
     \item $1<\gamma \leq \gamma_0, C_2>\frac{(4-3 \gamma )^2}{4 (\gamma -1)^2}$, or   
     \item $\gamma_0<\gamma \leq \frac{4}{3}, 
   \frac{(4-3 \gamma )^2}{4 (\gamma -1)^2}<C_2<\frac{4 (\gamma -1)}{\gamma ^2}$, or   
   \item $\gamma_0<\gamma \leq \frac{4}{3}, 
   C_2>\frac{4 (\gamma -1)}{\gamma ^2}$, or    
   \item $\frac{4}{3}<\gamma <2,  0<C_2<\frac{4 (\gamma -1)}{\gamma ^2}$, or   
   \item $\frac{4}{3}<\gamma <2,  C_2>\frac{4 (\gamma -1)}{\gamma ^2}$.
 \end{enumerate}
\end{enumerate}
 \subsection{Invariant set $v= A=0$.}
\label{vcero} 
The focus of this section is the stability analysis of the equilibrium points of the system \eqref{Eq:22} in the invariant set $A = v = 0$.
In the following list the stability analysis is done with preserving the four eigenvalues.
\begin{enumerate}
    \item[$N_1$:] $(\Sigma,A,K,v)=(0,0,1,0)$ has eigenvalues $\{-2,-1,1,2\}$, then, is a hyperbolic saddle. 
        \item[$P_{11}$:] $(\Sigma,A,K,v)=\left(-\frac{1}{\sqrt{C_2}}, 0, 0, 0\right)$.  The eigenvalues are \\
    $\left\{2-\frac{2}{\sqrt{C_2}},2,\frac{\gamma }{(\gamma -1) \sqrt{C_2}}-2,\frac{-3 \gamma
   +4 (\gamma -1) \sqrt{C_2}+2}{(\gamma -1) \sqrt{C_2}}\right\}$.
    \begin{enumerate}
        \item It is a hyperbolic sink for $\gamma >1,  \frac{(2-3 \gamma )^2}{16 (\gamma -1)^2}<C_2<\frac{\gamma ^2}{4 (\gamma -1)^2}$. 
        \item It is a hyperbolic saddle for:
         \begin{enumerate}
             \item $1<\gamma <2,   C_2>\frac{\gamma ^2}{4(\gamma -1)^2}$, or  
             \item $1<\gamma <2, 
   0<C_2<\frac{(2-3 \gamma )^2}{16 (\gamma -1)^2}$
         \end{enumerate}
         \item It is non-hyperbolic if: 
         \begin{enumerate}
             \item $1<\gamma <2,  C_2=1$, or   
             \item $1<\gamma <2,  C_2=\frac{(2-3 \gamma )^2}{16 (\gamma -1)^2}$, or  
             \item $1<\gamma <2,  C_2=\frac{\gamma ^2}{4 (\gamma -1)^2}$.
         \end{enumerate}
    \end{enumerate}
    
    \item[$P_{12}$:] $(\Sigma,A,K,v)=\left(\frac{1}{\sqrt{C_2}}, 0, 0,  0\right)$, with eigenvalues \\$\left\{\frac{2}{\sqrt{C_2}}+2,2,-\frac{\gamma }{(\gamma -1) \sqrt{C_2}}-2,\frac{3 \gamma
   +4 (\gamma -1) \sqrt{C_2}-2}{(\gamma -1) \sqrt{C_2}}\right\}$. 
   It is a saddle for $
   1<\gamma <2,  C_2>0$. 
   
    \item[$P_{13}$:] $(\Sigma,A,K,v)=\left(-\frac{2-3 \gamma }{4
   C_2(1- \gamma)}, 0, 0, 0\right)$.  $\Omega_t\geq 0$ for ${ C_2}\geq \frac{(2-3 \gamma )^2}{16 (\gamma -1)^2}$.  The eigenvalues are 
   $\Big\{-\frac{(\gamma -2) (3 \gamma -2)}{8 (\gamma -1)^2 C_2},\frac{(2-3 \gamma )^2}{4 (\gamma
   -1)^2 C_2}-2$, \newline $\frac{\gamma  (3 \gamma -2)}{4 (\gamma -1)^2 C_2}-2,\frac{(2-3 \gamma
   )^2}{8 (\gamma -1)^2 C_2}-2\Big\}$. It is: 
    \begin{enumerate}
        \item a hyperbolic source for $1<\gamma <2,  0<C_2<\frac{(2-3 \gamma )^2}{16 (\gamma -1)^2}$. 
        \item a hyperbolic saddle for: 
         \begin{enumerate}
             \item $1<\gamma <2,  \frac{(2-3 \gamma )^2}{16 (\gamma -1)^2}<C_2<\frac{\gamma  (3 \gamma -2)}{8(\gamma -1)^2}$, or \item $1<\gamma <2,  \frac{\gamma  (3 \gamma -2)}{8 (\gamma -1)^2}<C_2<\frac{(2-3 \gamma )^2}{8(\gamma -1)^2}$
             \item $1<\gamma <2,  C_2>\frac{(2-3 \gamma )^2}{8 (\gamma -1)^2}$.
         \end{enumerate}
    \end{enumerate}
  \item[$P_{14}$:] $(\Sigma,A,K,v)=\left(-\frac{2 (\gamma -1)}{3 \gamma -2}, 0,  \frac{(2-3 \gamma )^2-8 (\gamma -1)^2 C_2}{(2-3 \gamma )^2}, 0 \right)$. \newline The eigenvalues are
   $\{\lambda_1,  \lambda_2, \lambda_3, \lambda_4\}=$ \newline $\Bigg\{\frac{2-\gamma }{3 \gamma -2},-\frac{4 (\gamma -1)}{3 \gamma -2},-\frac{1}{2}+\frac{\sqrt{64 (\gamma
   -1)^2 C_2-7 (2-3 \gamma )^2}}{4-6 \gamma }$, \newline $ -\frac{1}{2}-\frac{\sqrt{64 (\gamma -1)^2
   C_2-7 (2-3 \gamma )^2}}{6 \gamma -4}\Bigg\}$. It is:
   \begin{enumerate}
       \item  non-hyperbolic for: 
        \begin{enumerate}
            \item $C_2=\frac{(2-3 \gamma )^2}{8 (\gamma -1)^2}, 1<\gamma \leq 2$, or 
            \item $C_2\geq  0,  \gamma =1$, or 
            \item $C_2\geq  0, \gamma =2$.
        \end{enumerate}
        \item a saddle otherwise. 
   \end{enumerate}
 Figure \ref{fig:my_label} shows the real parts of the eigenvalues $\lambda_i$ for the equilibrium point \newline $(\Sigma, A,K,v)=\left(-\frac{2 (\gamma -1)}{3 \gamma -2}, 0,  \frac{(2-3 \gamma )^2-8 (\gamma -1)^2 C_2}{(2-3 \gamma )^2}, 0 \right)$. It represents static solutions for $1\leq \gamma\leq 2$ y $C_2\geq 0$. The figure shows that the equilibrium point is non-hyperbolic in the cases (a)-i,ii,iii previously described, or, it is a hyperbolic saddle. 
   \begin{figure*}
       \centering
       \includegraphics[scale=0.45]{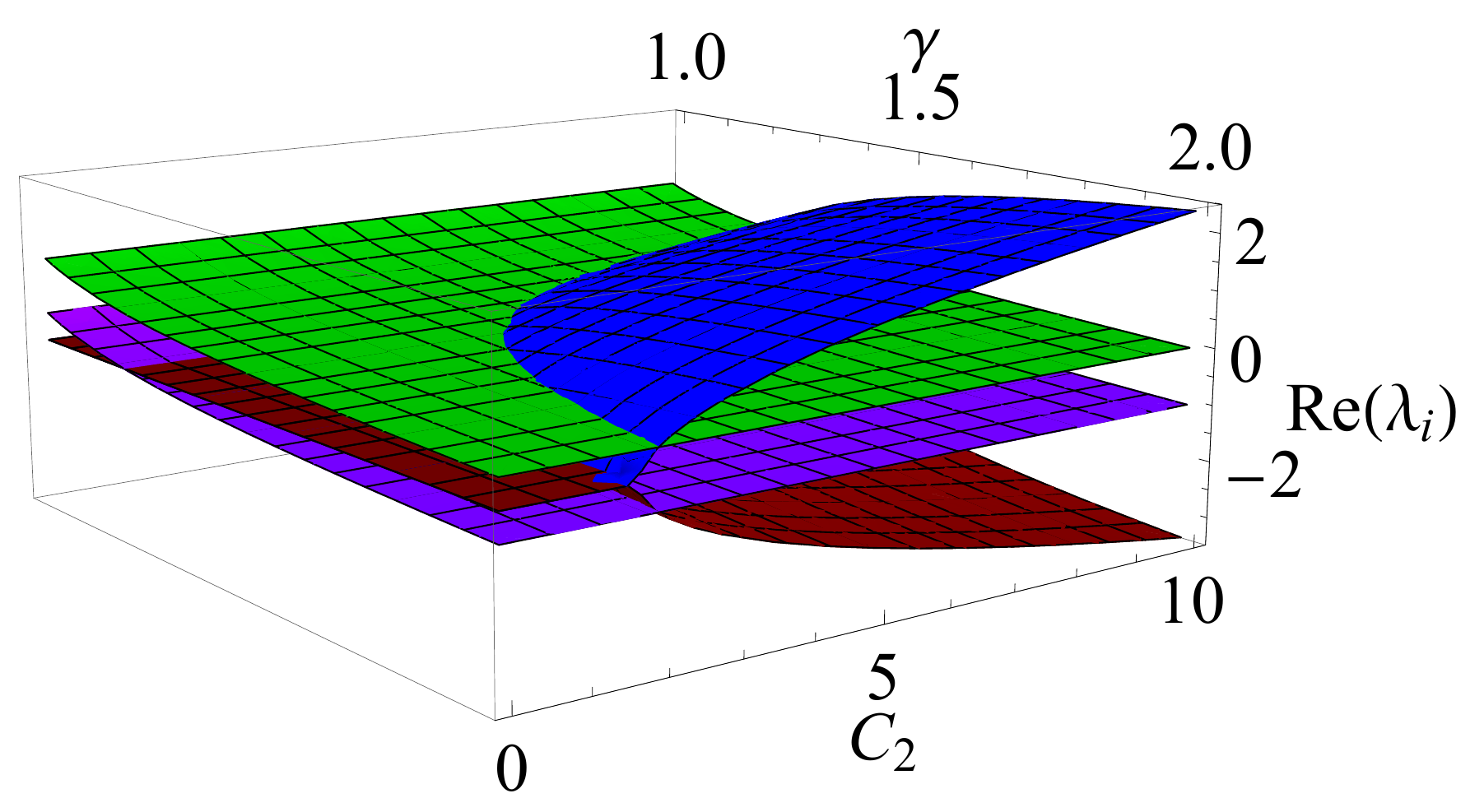}
       \caption{ \label{fig:my_label} Real part of the eigenvalues $\lambda_i$ for the equilibrium point $(\Sigma, A,K,v)=\left(-\frac{2 (\gamma -1)}{3 \gamma -2}, 0,  \frac{(2-3 \gamma )^2-8 (\gamma -1)^2 C_2}{(2-3 \gamma )^2}, 0 \right)$ for $1\leq \gamma\leq 2$ y $C_2\geq 0$.}
         \end{figure*}
\end{enumerate}

\subsubsection{Reduced system}
\label{sect:3.7.1}
When $A=v=0$, the restriction \eqref{constraintmod1} is trivially satisfied. On the other hand, from equation \eqref{defnOmegat} it follows 
\begin{equation}
(\gamma-1){\Omega}_{t}= \left(1-C_2{{\Sigma}}^2 -K\right).
\end{equation}
Imposing the energy condition  $\Omega_t\geq 0$ and choosing $\gamma\in(1,2]$, the following reduced dynamical system is obtained: 
\begin{subequations}
\label{eq:3.57}
\begin{align}
& \Sigma'= \frac{\left(C_2 \Sigma ^2-1\right) (3 \gamma +4 (\gamma -1) C_2 \Sigma -2)}{2 (\gamma -1) C_2} \nonumber \\
& + \frac{K (3
   \gamma +2 (\gamma -1) C_2 \Sigma -2)}{2 (\gamma -1) C_2},\\
& K'= 2 K \left(2 C_2 \Sigma^2+K-1\right), 
\end{align}
\end{subequations}
defined in the  phase space: 
\begin{equation}
    \left\{(\Sigma, K)\in \mathbb{R}^2 :  C_2{{\Sigma}}^2 +K \leq 1, K \geq 0\right\}.
\end{equation}
The qualitative analysis of system \eqref{eq:3.57} is given in Table \ref{Tab2Avcero}.
\begin{table*}
\caption{\label{Tab2Avcero} Qualitative analysis of the equilibrium points of the system \eqref{eq:3.57} with $v = A = 0$. Line $ N_1 $ is included. }
\begin{tabular*}{\textwidth}{@{\extracolsep{\fill}}lrrrrl@{}}
\hline
Equil. & \multicolumn{1}{c}{$(\Sigma, K)$} & \multicolumn{1}{c}{Eigenvalues (plane  $\Sigma$--$K$)} & \multicolumn{1}{c}{Stability (plane  $\Sigma$--$K$)} & \multicolumn{1}{c}{${ \Omega_t}$}  \\
Points \\
\hline
$N_1$ & $(0, 1)$ & $\{-1,2\}$  & saddle & $0$\\ \hline
$P_{11}$ & $\left(-\frac{1}{\sqrt{C_2}}, 0, \right)$ & $\left\{2,\frac{2-3 \gamma }{(\gamma -1) \sqrt{C_2}}+4\right\}$  &
saddle for 
$ 0<C_2<\frac{(2-3 \gamma )^2}{16 (\gamma -1)^2}.$ & \\ &&&
non-hyperbolic for  $ C_2=\frac{(2-3 \gamma )^2}{16 (\gamma -1)^2}$. & \\ &&&  local source for
$ C_2>\frac{(2-3 \gamma )^2}{16 (\gamma -1)^2}$. & $0$ \\\hline
$P_{12}$ & $\left(\frac{1}{\sqrt{C_2}}, 0 \right)$ & $\left\{2,\frac{3 \gamma -2}{(\gamma -1) \sqrt{C_2}}+4\right\}$ & local source for $C_2>0$. & $0$ \\\hline
$P_{13}$ & $\left(-\frac{2-3 \gamma }{4 C_2(1- \gamma)}, 0\right)$ & $\Bigg\{\frac{(2-3 \gamma )^2-16 (\gamma -1)^2 C_2}{8 (\gamma -1)^2 C_2},$ &    local source for $ 0<C_2<\frac{(2-3 \gamma )^2}{16 (\gamma -1)^2}$. & \\ && $ \frac{(2-3 \gamma
   )^2-8 (\gamma -1)^2 C_2}{4 (\gamma -1)^2 C_2}\Bigg\}$ &  saddle for
   $ \frac{(2-3 \gamma )^2}{16 (\gamma -1)^2}<C_2<\frac{(2-3 \gamma )^2}{8 (\gamma
   -1)^2}$.
   & \\ &&&
   non-hyperbolic for
   $   C_2=\frac{(2-3 \gamma )^2}{16 (\gamma -1)^2}$, or  & \\ &&&  $   C_2=\frac{(2-3 \gamma )^2}{8 (\gamma
   -1)^2}$. 
   & \\ &&&
     local attractor for
    $ C_2>\frac{(2-3 \gamma )^2}{8 (\gamma -1)^2}$ & $\frac{16 (\gamma -1)^2 C_2-(2-3 \gamma )^2}{16 (\gamma -1)^3 C_2}$\\\hline
   $P_{14}$ & $\Bigg(-\frac{2 (\gamma -1)}{3 \gamma -2},   \frac{(2-3 \gamma )^2-8 (\gamma -1)^2 C_2}{(2-3 \gamma )^2}\Bigg)$. & $ \Bigg\{-\frac{1}{2}+\frac{\sqrt{64 (\gamma
   -1)^2 C_2-7 (2-3 \gamma )^2}}{4-6 \gamma },  $ & local attractor for $ 0<C_2\leq \frac{7 (2-3 \gamma )^2}{64 (\gamma -1)^2}$, or   & \\ &&$ -\frac{1}{2}-\frac{\sqrt{64 (\gamma -1)^2
   C_2-7 (2-3 \gamma )^2}}{6 \gamma -4}\Bigg\}$ & $ \frac{7 (2-3 \gamma )^2}{64 (\gamma -1)^2}<C_2<\frac{(2-3 \gamma
   )^2}{8 (\gamma -1)^2}$.
\\ &&& non-hyperbolic if
   $ C_2=\frac{(2-3 \gamma )^2}{8 (\gamma -1)^2}$.
 & \\ &&& saddle if   $ C_2>\frac{(2-3 \gamma )^2}{8 (\gamma -1)^2}$ & $ \frac{4 (\gamma -1) C_2}{(2-3 \gamma )^2}$. \\\hline
\end{tabular*}
\end{table*}

\begin{figure*}
    \centering
    \includegraphics[scale=0.55]{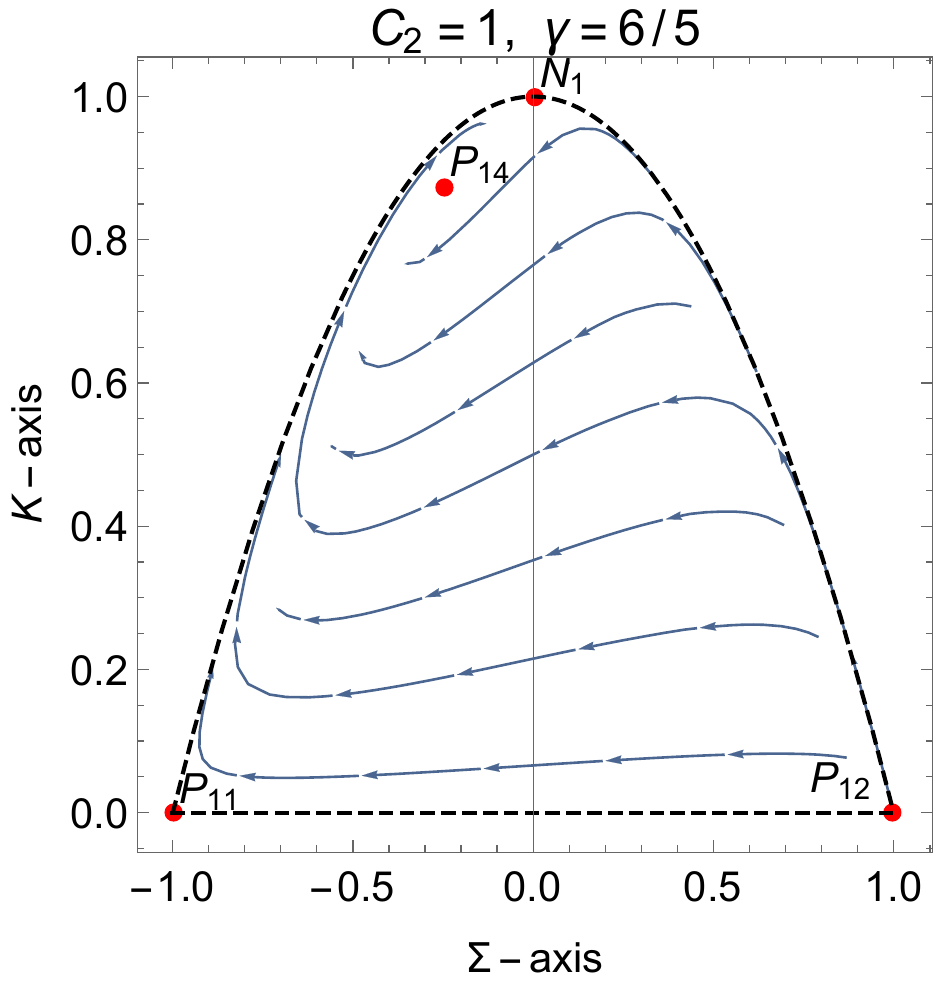} 
    \includegraphics[scale=0.55]{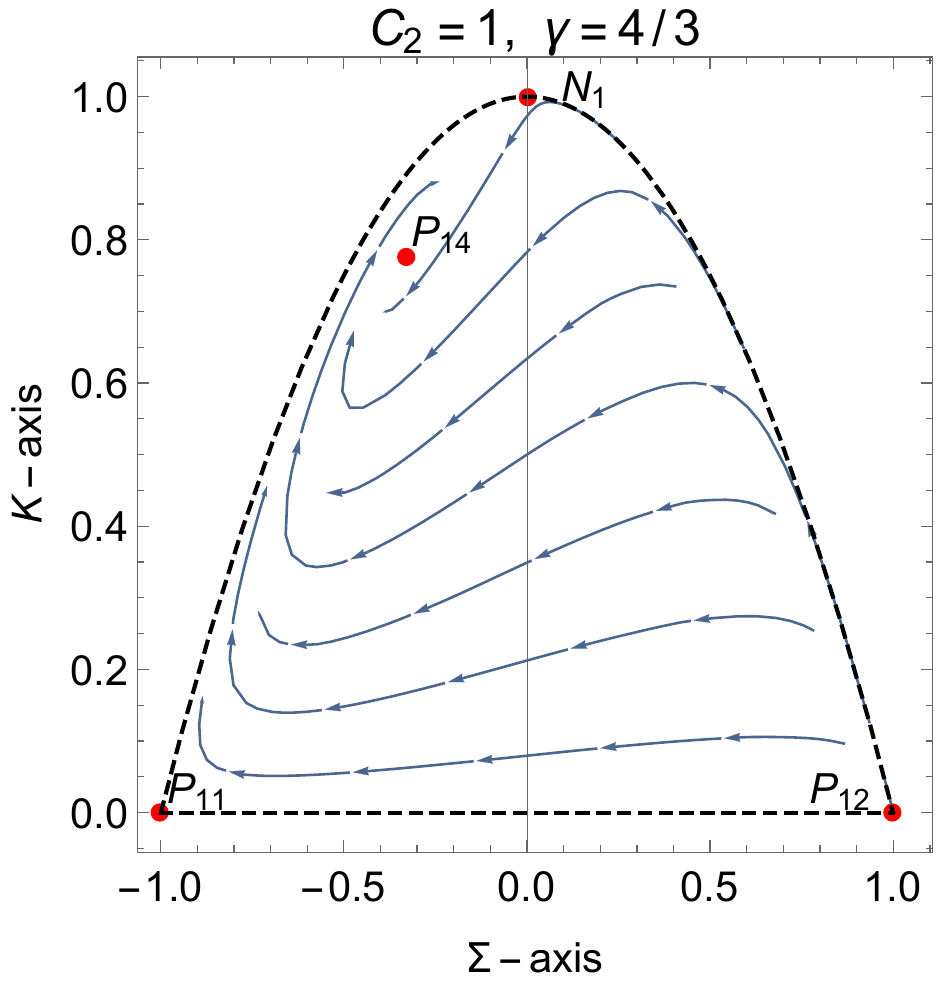} 
    \includegraphics[scale=0.55]{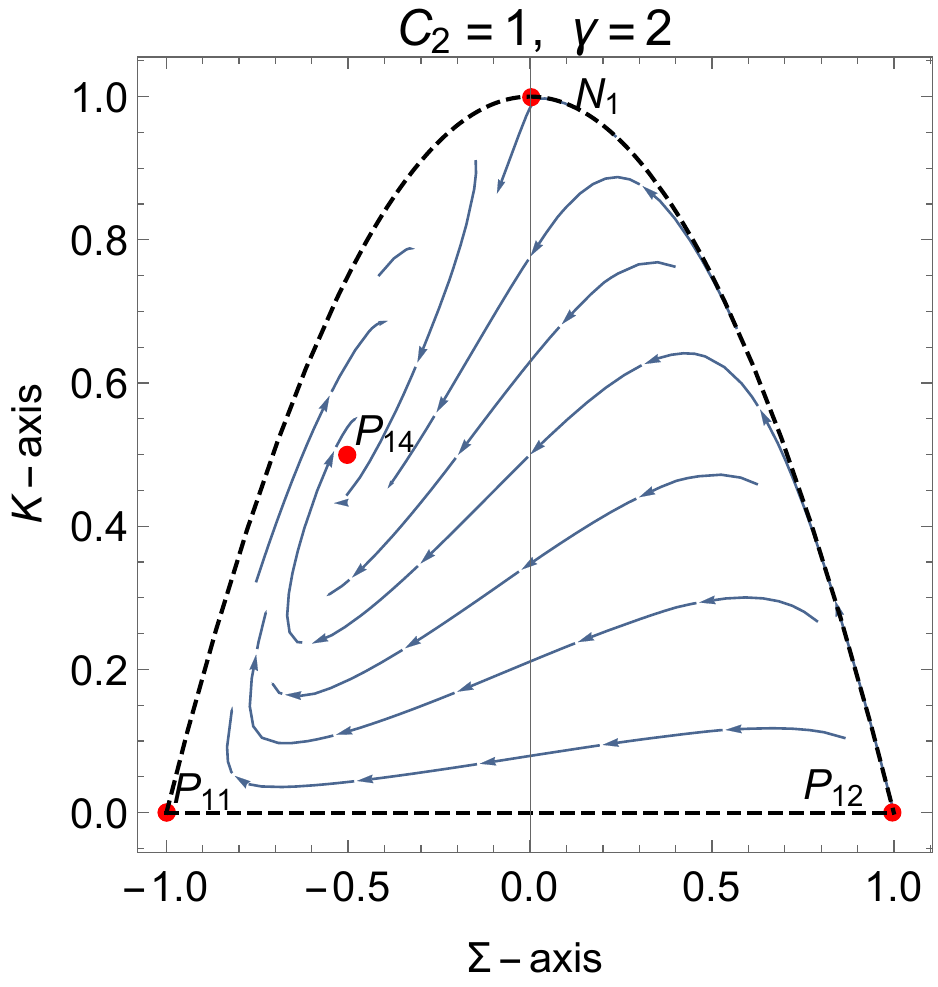} 
    \includegraphics[scale=0.55]{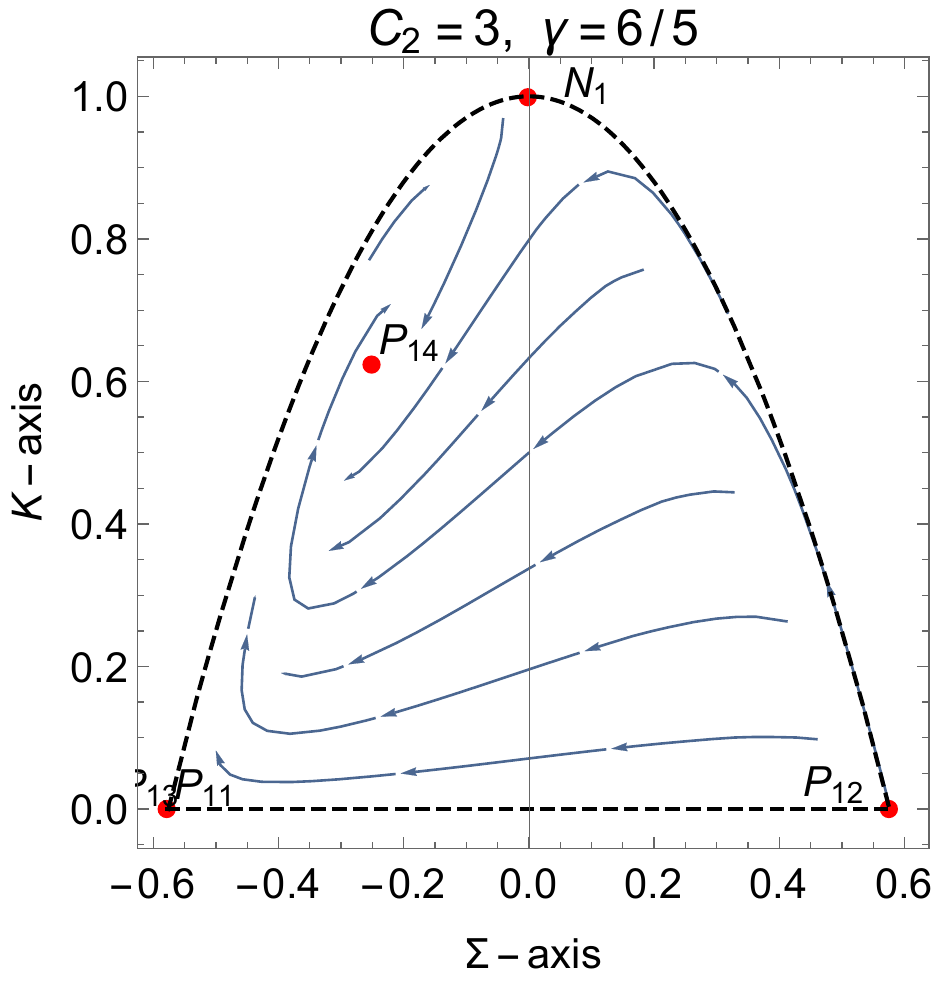} 
    \includegraphics[scale=0.55]{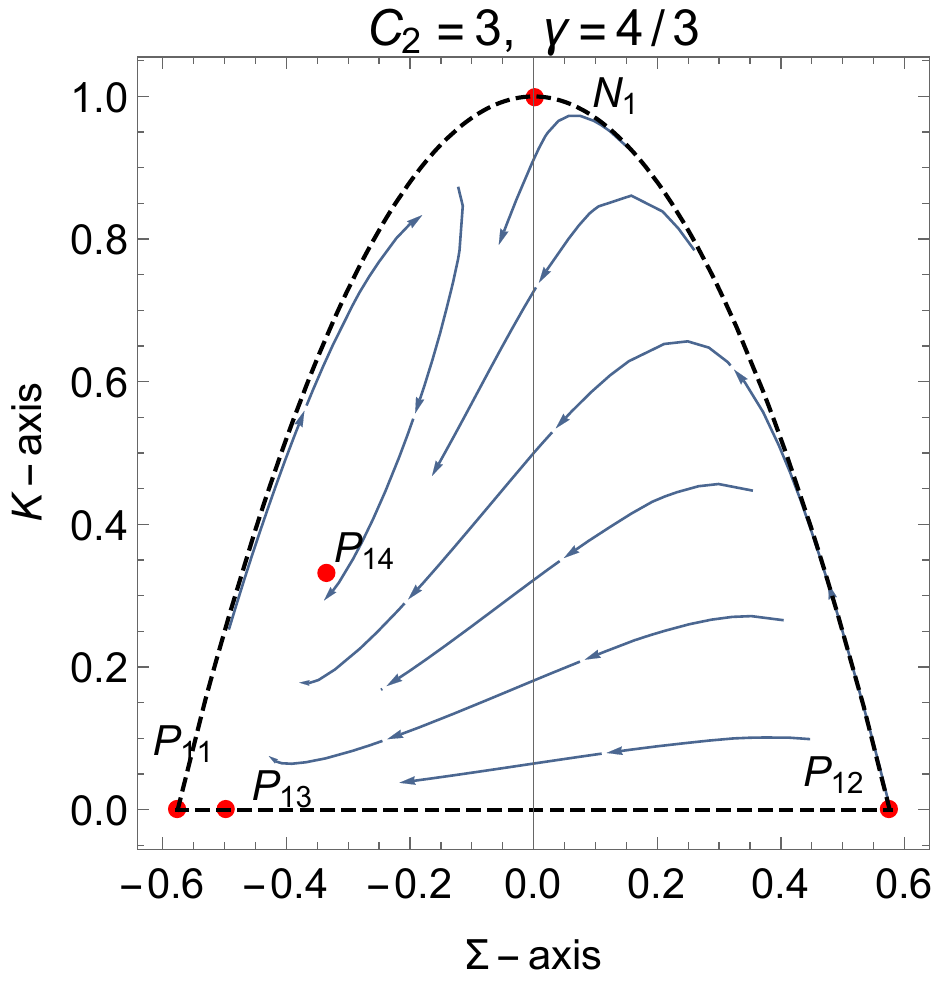} 
    \includegraphics[scale=0.41]{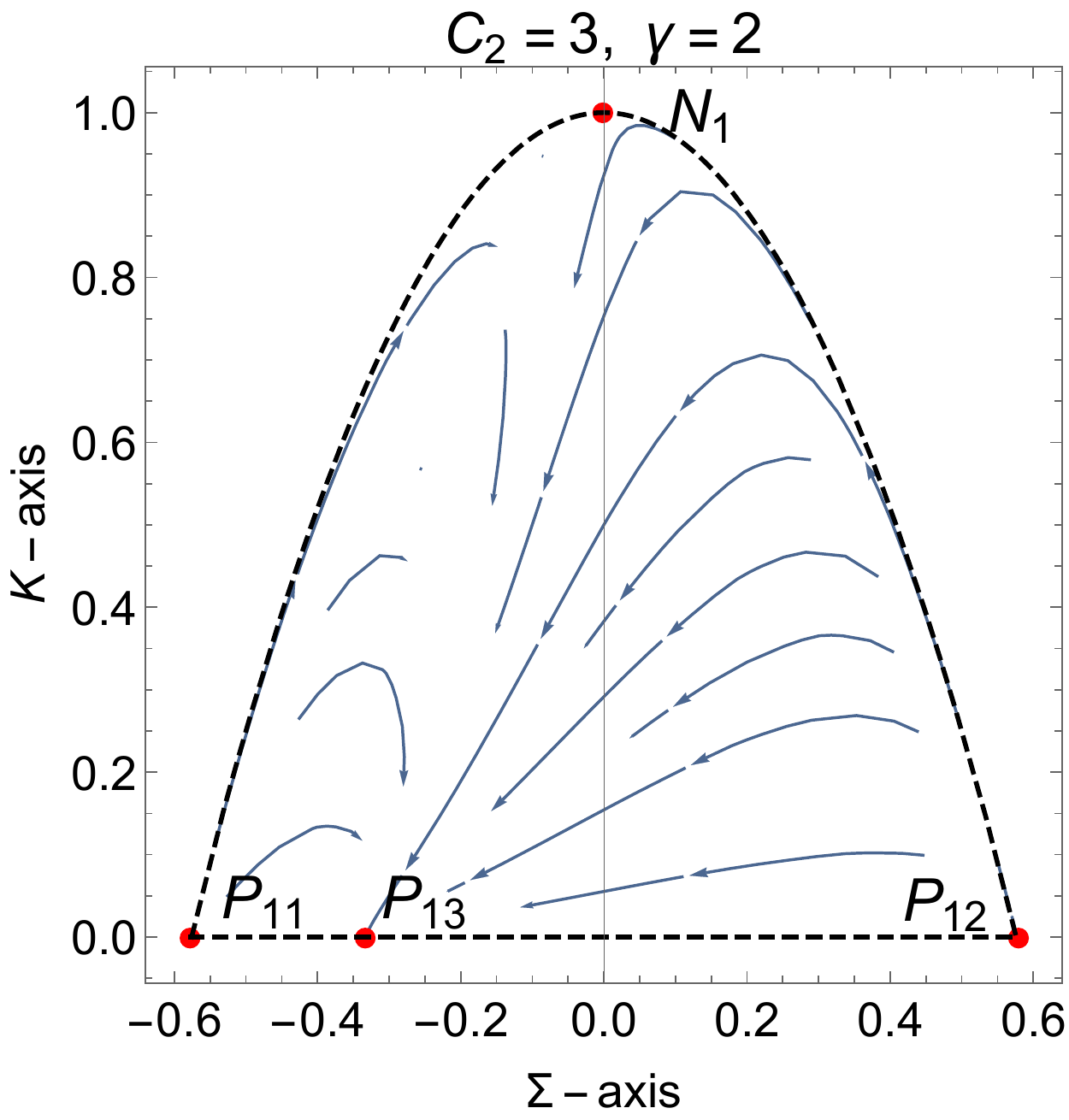}
    \includegraphics[scale=0.55]{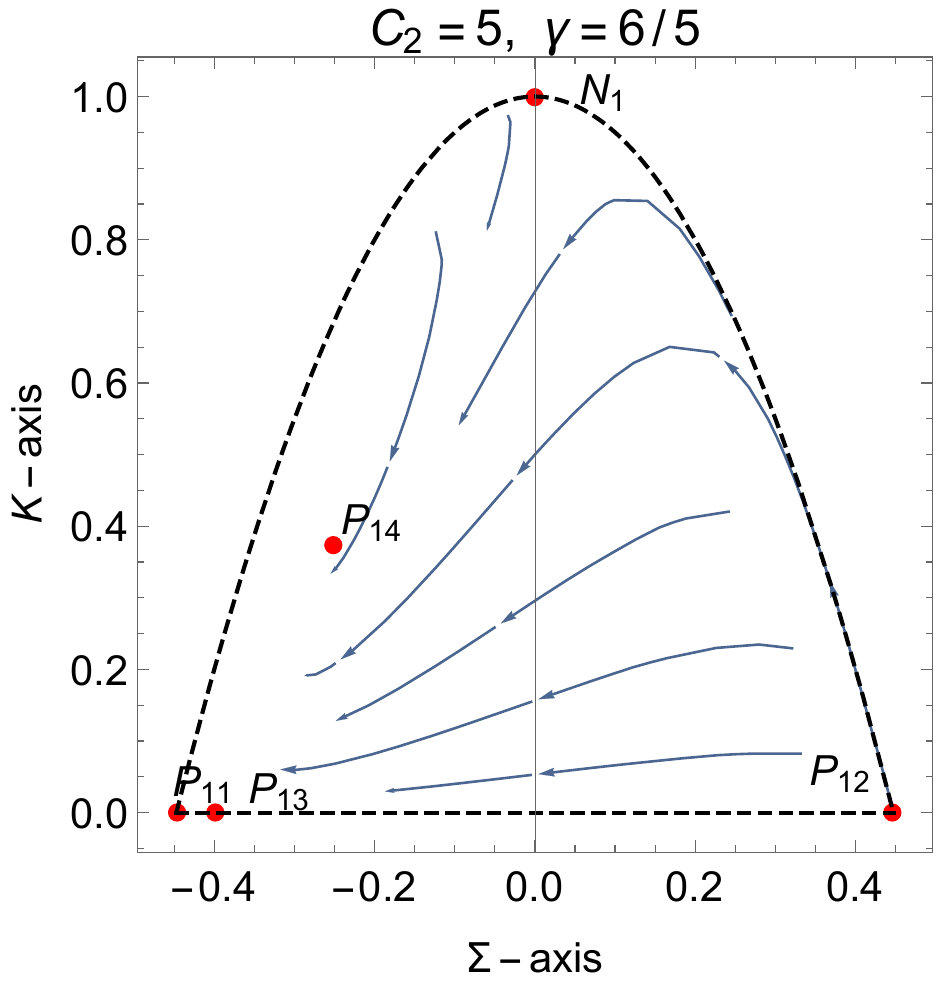} 
    \includegraphics[scale=0.55]{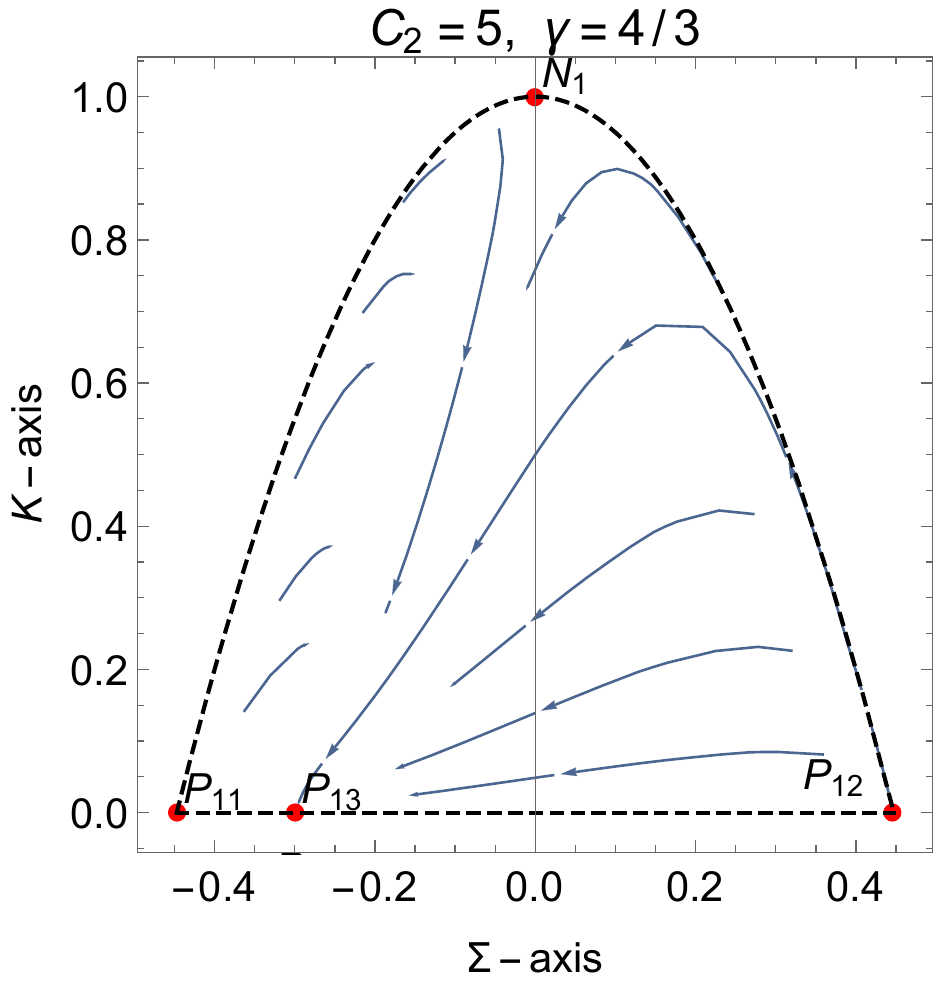} 
    \includegraphics[scale=0.55]{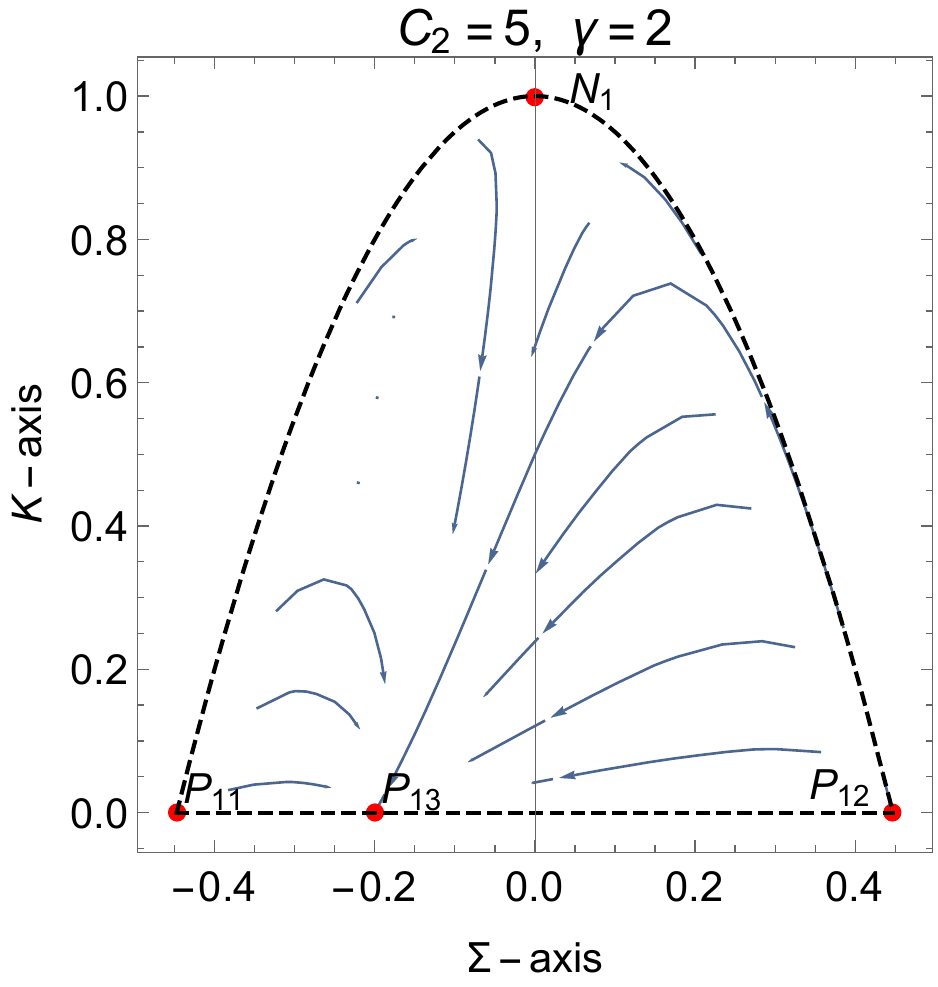} 
    \caption[{Orbits of system \eqref{eq:3.57} with $v=0, \quad A=0$ with $1<\gamma \leq 2$}]{\label{Bla} Orbits of system \eqref{eq:3.57} with $v=0, \quad A=0$ with $1<\gamma \leq 2$ for different choices of $\gamma$ and $C_2$.}
   \end{figure*}

  \begin{table*}[t!]
\caption{\label{Tab0} Qualitative analysis of the equilibrium points of  \eqref{system29} with $v\neq 0$. The notations
   $v_{\pm}=\frac{\sqrt{(\gamma -1) \left(\gamma  \left(2 \gamma ^2 C_2-\gamma  (2 C_2+3)\pm 2 \sqrt{\gamma
   -1} \sqrt{C_2} \sqrt{\gamma  (\gamma  ((\gamma -1) C_2-3)+8)-4}+8\right)-4\right)}}{\gamma
   -2}$,  
 $\lambda_{\pm}= \frac{\gamma \pm \frac{\sqrt{\gamma  (\gamma  ((\gamma -1) C_2-3)+8)-4}}{\sqrt{\gamma -1}
   \sqrt{C_2}}}{2-3 \gamma }$, $\gamma_0=-\frac{2}{27} \left(-11-\frac{14}{\sqrt[3]{27 \sqrt{57}-197}}+\sqrt[3]{27 \sqrt{57}-197}\right)\approx 1.22033$, $\Delta=-\sqrt{\frac{2-3 \gamma }{\gamma -4 \gamma  C_2+2}}$,  $k=\frac{1}{2 C_2^2 (4 C_2-3)^{3/2}}$, y $\lambda_1, \lambda_2$ and $\lambda_3$ satisfy $P(\lambda_i)=0$, where
$P(\lambda)=-16 (C_2-1)^2 C_2^3 (4   C_2-3)^{11/2}-4 (C_2-1)  C_2^2 (8 C_2-7) (4
   C_2-3)^3 \lambda +\lambda ^3$ are used.}
\begin{tabular*}{\textwidth}{@{\extracolsep{\fill}}lrrrl@{}}
\hline
Equil. & \multicolumn{1}{c}{$(\Sigma,A, v, K)$} & \multicolumn{1}{c}{Eigenvalues} & \multicolumn{1}{c}{Stability}  \\
Points \\
\hline
 $N$ & $(0,0,v,1)$, $\gamma=1$ & $\{0,-1,1,2\}$ & non-hyperbolic saddle. \\\hline
 $N_{1}$ & $(0,0,0,1)$& $\{-2,-1,1,2\}$  & hyperbolic saddle. \\\hline
$N_{2,3}$ & $(0,0,\pm 1,1) $ & $\left\{0,-1,1,\frac{4}{\gamma -2}+4\right\}$  & non-hyperbolic saddle. \\\hline
  $P_{1,2}$ & $(-\frac{1}{\sqrt{C_2}},0,\pm 1,0)$  & $\left\{2-\frac{2}{\sqrt{C_2}},2,\frac{2 \gamma \left(\frac{1}{\sqrt{C_2}}-2\right)+4}{2-\gamma }\right\}$  &   hyperbolic source for $1<\gamma <2,  1<C_2<\frac{\gamma
   ^2}{4 \gamma ^2-8 \gamma +4}$. \\
   &&& 
   	  hyperbolic saddle for 
	   $ 1<\gamma <2,0<C_2<1$,  $1<\gamma <2, C_2>\frac{\gamma ^2}{4 \gamma ^2-8 \gamma +4}$. \\ &&&
	  non-hyperbolic for $1<\gamma <2, C_2=\frac{4\gamma ^2}{(\gamma-1)^2}$, or  \\
	  &&& $1<\gamma <2, C_2=1$. \\\hline
$P_{3,4}$ & $\left(\frac{1}{\sqrt{C_2}},0,\pm 1,0\right)$ & $\left\{\frac{2}{\sqrt{C_2}}+2,2,\frac{-4 \gamma -\frac{2 \gamma}{\sqrt{C_2}}+4}{2-\gamma }\right\}$  & hyperbolic saddle  $1<\gamma <2,  C_2>0$. \\\hline
$P_5$ & $\left(0, \frac{1}{\sqrt{C_2}}, 1, 0\right)$& $ { \left\{-\frac{2}{\sqrt{C_2}}-2,-2,\frac{6 \gamma +4 (\gamma -1)
   \sqrt{C_2}-8}{(\gamma -2) \sqrt{C_2}}\right\}}$  &    hyperbolic sink   (see text). \\
   &&&
	  hyperbolic saddle (see text).
	   \\
	   &&&
      non-hyperbolic  (see text).\\
	 \hline 
$P_6$ & $ \left(0, \frac{1}{\sqrt{C_2}}, - 1, 0\right)$& ${\left\{\frac{2}{\sqrt{C_2}}-2,-2,\frac{-6 \gamma +4 (\gamma -1)
   \sqrt{C_2}+8}{(\gamma -2) \sqrt{C_2}}\right\}}$ &    hyperbolic sink for $1<\gamma <2,  C_2>1$.
	\\
	&&&
	  hyperbolic saddle for
	 $1<\gamma <2, 0<C_2<1 $, or  	\\
	&&& 
	      $\frac{4}{3}<\gamma <2,  0<C_2<\frac{(4-3\gamma )^2}{4 (\gamma -1)^2}$, or   	\\
	&&& $\frac{4}{3}<\gamma <2, C_2>1$.
	\\
	&&&
	  non-hyperbolic for 
	  $1<\gamma \leq \frac{4}{3}, C_2=1$, or  	\\
	&&&  $\frac{4}{3}<\gamma <2,  C_2=\frac{9 \gamma^2-24 \gamma +16}{4 \gamma^2-8 \gamma +4}$, or   	\\
	&&&
	    $\frac{4}{3}<\gamma <2,  C_2=1$. \\\hline
$P_{7}$ & $\left(-\frac{1}{2},\frac{1}{2},- 1, 1-C_2\right)$& $\left\{0,-\sqrt{C_2-1},\sqrt{C_2-1}\right\}$ & 
  non-hyperbolic saddle for $C_2>1, 1<\gamma <2$. \\ &&&
  non-hyperbolic with three zero eigenvalues for $C_2=1,  \gamma \neq 0$. \\&&&
  non-hyperbolic with a zero and two purely imaginary eigenvalues\\ &&& for $1<\gamma <2,  0\leq
   C_2\leq 1$. \\\hline
$P_8$ & $\left(-\frac{1}{2 C_2},-\frac{1}{2 C_2 \Delta },\Delta \right)$ & 
$\left\{k \lambda_1,k \lambda_2, k \lambda_3\right\}$ if $\gamma=1$ & saddle.  \\\hline
$P_9$ &$(0, \lambda_{+} v_{-}, v_{-})$  & $\{\mu_1, \mu_2, \mu_3\}$  & saddle.  \\
 \hline
 $P_{10}$ &$(0, \lambda_{-} v_{+}, v_{+})$  & $\{\nu_1, \nu_2, \nu_3\}$ &   hyperbolic sink  (see text).
\\ &&&
  hyperbolic saddle 
 (see text). \\
 \hline
\end{tabular*}
\end{table*}

\subsection{Summary of sources and sinks for a perfect fluid}
\label{summary6}
\noindent 
A summary of the equilibrium points classified as sinks or sources of the model with perfect fluid is presented.
\subsubsection{Non-extensibility surface of solutions}
$SL_{-}$ is stable for the regions described in Figure \ref{SL-stability}.
\subsubsection{Invariant sets $v=\pm 1$}
\begin{enumerate}
    \item[$P_{1,2}$:] $(\Sigma,A)=(-\frac{1}{\sqrt{C_2}},0)$, $v=\pm 1$,  are hyperbolic sources for  $C_2>1$.
    \item[$P_{3,4}$:] $(\Sigma,A)=(\frac{1}{\sqrt{C_2}},0)$, $v=\pm 1$, are hyperbolic sources for $C_2>0$.
    \item[$P_5$:] $(\Sigma,A)=\left(0, \frac{1}{\sqrt{C_2}}\right)$,  $v=1$, is a hyperbolic sink for $C_2>0$.
    \item[$P_6$:] $(\Sigma,A)=\left(0, \frac{1}{\sqrt{C_2}}\right)$, $v=-1$, is a hyperbolic sink for $C_2>1$.
\end{enumerate}
\subsubsection{Ideal gas $\gamma=1$}
\begin{enumerate}
    \item[$P_{1,2}$:] $(\Sigma,A,v)=\left(-\frac{1}{\sqrt{C_2}},0,\pm 1\right)$  are hyperbolic sources for  \newline $C_2>1$.
    \item[$P_6$:] $(\Sigma,A,v)=\left(0, \frac{1}{\sqrt{C_2}},- 1\right)$ is hyperbolic sink for $C_2>1$.
\end{enumerate}
\subsubsection{General case $v\neq 0$}
\begin{enumerate}
\item[$P_{1,2}$:] $(\Sigma,A,v)=(-\frac{1}{\sqrt{C_2}},0,\pm 1)$,  are hyperbolic sources for  $1<\gamma <2, 
   1<C_2<\frac{\gamma
   ^2}{4 \gamma ^2-8 \gamma +4}$.
\item[$P_5$:] $(\Sigma,A,v)=(0,\frac{1}{\sqrt{C_2}},1)$,  is a hyperbolic sink for:
	  \begin{enumerate}
	      \item $1<\gamma \leq \frac{4}{3}, C_2>\frac{9 \gamma^2-24 \gamma +16}{4 \gamma^2-8 \gamma +4}$, or  
	      \item $\frac{4}{3}<\gamma <2, C_2>0$. 
	  \end{enumerate}
	\item[$P_6$:] $(\Sigma,A,v)=(0,\frac{1}{\sqrt{C_2}},-1)$, is a hyperbolic sink for $1<\gamma <2,  C_2>1$.
\item[$P_{10}$:] $(\Sigma,A,v)=(0, \lambda_{-} v_{+}, v_{+})$, is a hyperbolic sink for: 
\begin{enumerate}
    \item $1<\gamma \leq \gamma_0, 0<C_2<\frac{4 \gamma -4}{\gamma ^2}$, or    
    \item $1<\gamma <\gamma_0, \frac{4 \gamma -4}{\gamma
   ^2}<C_2<\frac{(4-3 \gamma )^2}{4 (\gamma -1)^2}$, or 
   \item $\gamma_0<\gamma <\frac{4}{3}, 0<C_2<\frac{(4-3 \gamma )^2}{4 (\gamma -1)^2}$,
\end{enumerate}
where \newline $\gamma_0=-\frac{2}{27} \left(-11-\frac{14}{\sqrt[3]{27 \sqrt{57}-197}}+\sqrt[3]{27 \sqrt{57}-197}\right)$ \newline $\approx 1.22033$.
\end{enumerate}
\subsubsection{Invariant set $v=A=0$}
\begin{enumerate}
    \item[$P_{11}$:] $(\Sigma,A,K,v)=\left(-\frac{1}{\sqrt{C_2}}, 0, 0, 0\right)$, is a hyperbolic source for $\gamma >1$,   $\frac{(2-3 \gamma )^2}{16 (\gamma -1)^2}<C_2<\frac{\gamma ^2}{4 (\gamma -1)^2}$.
    \item[$P_{13}$:] $(\Sigma,A,K,v)=\left(-\frac{2-3 \gamma }{4
   C_2(1- \gamma)}, 0, 0, 0\right)$, is a hyperbolic source for   $1<\gamma <2,  0<C_2<\frac{(2-3 \gamma )^2}{16 (\gamma -1)^2}$. 
\end{enumerate}
\paragraph{Reduced system:}
\begin{enumerate}
    \item[$P_{11}$:] $(\Sigma,K)=\left(-\frac{1}{\sqrt{C_2}}, 0 \right)$, is a local source for
$C_2>\frac{(2-3 \gamma )^2}{16 (\gamma -1)^2}$.
    \item[$P_{12}$:] $(\Sigma,K)=\left(\frac{1}{\sqrt{C_2}}, 0 \right)$ is a local source for $C_2>0$.
    \item[$P_{13}$:] $(\Sigma,K)=\left(-\frac{2-3 \gamma }{4 C_2(1- \gamma)}, 0\right)$ is:
    \begin{enumerate}
        \item a local source for  $0<C_2<\frac{(2-3 \gamma )^2}{16 (\gamma -1)^2}$.
        \item a local attractor for $C_2>\frac{(2-3 \gamma )^2}{8 (\gamma -1)^2}$.
    \end{enumerate}
    \item[$P_{14}$:] $(\Sigma,K)=\Bigg(-\frac{2 (\gamma -1)}{3 \gamma -2}, \frac{(2-3 \gamma )^2-8 (\gamma -1)^2 C_2}{(2-3 \gamma )^2}\Bigg)$ is a local sink for: 
    \begin{enumerate}
        \item  $0<C_2\leq \frac{7 (2-3 \gamma )^2}{64 (\gamma -1)^2}$, or  
        \item $\frac{7 (2-3 \gamma )^2}{64 (\gamma -1)^2}<C_2<\frac{(2-3 \gamma
   )^2}{8 (\gamma -1)^2}$.
    \end{enumerate}
\end{enumerate}

\subsection{Discussion}
\label{DiscusionC1}
In this section, timelike self-similar spherically symmetric  metrics in Einstein-æther theory with perfect fluid as matter content were studied using the  homothetic diagonal formulation, which gives the propagation equations \eqref{modelouno} plus algebraic restrictions. The homothetic diagonal formalism has some disadvantages. Symmetry surfaces generally change causality. Then, in the homothetic diagonal formulation spacetime must be covered with two coordinate systems (two charts); one when  homothetic Killing vector is timelike, and another when  homothetic Killing vector is spacelike. These two regions have to be matched in the region where the Killing vector is null \cite{Goliath:1998mx}. However, the formulation has more advantages than disadvantages. The main one is that it allows the field equations, which are a well-defined system of first-order partial derivative equations (PDE), in two variables (from the $ 1 + 3 $ formalism), to be written as a system of ordinary differential equations using the symmetries that come from the Killing vectors. The resulting equations are very similar to those of the models with homogeneous hypersurfaces. In turn, it is possible to write these equations as a dynamical system, which makes it possible to study the model using the techniques of the qualitative theory of dynamical systems. This makes it possible to obtain a complete description in a  phase space, which leads to a better understanding of the dynamics of the model. In this regard, the $\theta$--normalized equations were presented. 

Four specific models were studied; these are: extreme tilt \eqref{extreme_tilt_0}; pressureless perfect fluid \eqref{3.41-3.43}; the reduced system \eqref{eq:3.57} in the invariant set $A = v = 0$; and the general system \eqref{reducedsyst}. Hyperbolic points were classified according to their stability conditions using the Hartman-Grobman theorem; while non-hyperbolic points were classified as saddles. Furthermore, it was possible to retrieve the results obtained in \cite{Goliath:1998mx}. The following list shows the points obtained in \cite{Goliath:1998mx} and their correspondence with the points discussed in this section: 
\begin{enumerate}
\item[$SL_{\pm}$:] sonic lines given by ${A}=-\frac{\gamma 
   \varepsilon  (\gamma  (\Sigma
   +2)-2)}{4 (\gamma
   -1)^{3/2}}$,
$v=\varepsilon\sqrt{\gamma -1}$, were analyzed in section \ref{SL}. Unlike General Relativity, for $1<\gamma<2$ and $C_2= \frac{\gamma
   ^2}{4 (\gamma -1)^2}$ the system \eqref{reducedsyst} admits the following equilibrium points:  
\begin{enumerate}
    \item[$SL_1$:] $\Sigma =
   \frac{2 (\gamma -1)}{\gamma
   },v=\sqrt{\gamma -1}, A=
  - \frac{\gamma  (\gamma  (\Sigma
   +2)-2)}{4 (\gamma
   -1)^{3/2}}$, 
    \item[$SL_2$:] $\Sigma =
   -\frac{2 (\gamma -1)}{\gamma
   },v= -\sqrt{\gamma -1}, A=
   \frac{\gamma  (\gamma  (\Sigma
   +2)-2)}{4 (\gamma
   -1)^{3/2}}$,
\end{enumerate}
which lie on the sonic line. If $ \gamma = 2, C_2 = 1 $ these points exist, and since $ \gamma = 2 $ the fluid behaves like stiff matter. Additionally, if $ \gamma = 2, C_2 = 1 $, these points correspond to models with extreme tilt ($ v = \varepsilon $), $SL_1: \Sigma = 1, A = -2, v = 1 $, and $SL_2: \Sigma = -1, A = 0, v = -1 $. $ SL_ {\pm}$ corresponds to a flat FLRW space and static orbits depending on the parameter $\gamma$.

    \item[$C^0$:] $({\Sigma}, {A}, v)=(0,0,0),\quad ({K},\Omega_t)=(1,0)$, corresponds to $N_1$. 
    
    \item[$C^{\pm }$:] $({\Sigma}, {A}, v)=(0,0,\pm 1),\quad ({K},\Omega_t)=(1,0)$, correspond to $N_{2,3}$. 
    
    \item[$K^{\pm}_-$:] $({\Sigma}, {A}, v)=(-1,0,\pm 1),\quad ({K},\Omega_t)=(0,0)$, correspond to $P_{1,2}$ for $C_2=1$.
    
    \item[$K^{\pm}_+$:] $({\Sigma}, {A}, v)=(1,0,\pm 1),\quad ({K},\Omega_t)=(0,0)$, correspond to $P_{3,4}$ for $C_2=1$.
    
    \item[$M^+$:] $({\Sigma}, {A}, v)=(0,1,1),\quad ({K},\Omega_t)=(0,0)$, corresponds to  $P_5$ for $C_2=1$.
    
    \item[$M^-$:] $({\Sigma}, {A}, v)=(0,1,-1),\quad ({K},\Omega_t)=(0,0)$, corresponds to $P_6$ for  $C_2=1$.
    
    \item[$H^-$:] The line $ {A}( {\Sigma})= {\Sigma}  +1$, $v( {\Sigma})=-1$, $(0,-2 {\Sigma} {A})$. 
    
    \item[$K^0_-$:] $({\Sigma}, {A}, v)=(-1,0,0),\quad ({K},\Omega_t)=(0,0)$, corresponds to $P_{11}$ for  $C_2=1$.
    
    \item[$K^0_+$:] $({\Sigma}, {A}, v)=(1,0,0),\quad ({K},\Omega_t)=(0,0)$,  corresponds to $P_{12}$ for  $C_2=1$. 
    
    \item[$T$:] $({\Sigma}, {A}, v)=\left(-2\frac{\gamma-1}{3\gamma-2},0,0\right)$, 
    $ ({K},\Omega_t)=\left(\frac{\gamma^2+4(\gamma-1)}{(3\gamma -2},\frac{4(\gamma-1)}{(3\gamma -2}\right)$,  corresponds to $P_{13}$ for  $C_2=1$. 
    
    \item[$\widetilde{M}^{\pm}$:] $({\Sigma}, {A}, v)=\left(0, 1, \frac{(\gamma -1) \gamma  \pm \left(\sqrt{(\gamma -1) \left((\gamma
   -1) \gamma ^2+(2-\gamma ) (3 \gamma -2)\right)}\right)}{2-\gamma }\right)$,\newline $ ({K},\Omega_t)=(0,0)$,  exists for $C_2=1$.
    
   \end{enumerate}
The authors in \cite{Goliath:1998mx} used the notation $ \text{Kernel}^ {\text{sgn} (v)} _ {\text{sgn} ({\Sigma})} $, when there is no confusion  $ \text{sgn} (v) $ or $ \text{sgn} ({\Sigma}) $ is omitted. The kernel indicates the interpretation of the point: $M, C$ represent Minkowski spacetime; $K$ represents a Kasner solution; $T$ corresponds to static solutions; $SL_ {\pm}$ corresponds to a flat FLRW space and static orbits depending on the parameter $\gamma $; The equilibrium point line $ H^- $ is associated with a change of causality of the homothetic vector field. The points $\widetilde{M}^{\pm} $ are equilibrium points of \eqref{reducedsyst}, only if $C_2 = 1$.

Table \ref{Tab2Avcero} summarizes the results of the qualitative analysis of the equilibrium points of the system \eqref{Eq:22} with $v = A = 0$. Line $ N_1 $ is included.
The figure \ref{Bla} shows some orbits in the phase space of the system \eqref{eq:3.57} with $v = A = 0 $ and $1 <\gamma \leq 2$.

\section{Timelike self-similar spherically symmetric   models with scalar field}
\label{SECT:4.1}
In order for a non-homogeneous scalar field $\phi (t, x)$ with potential $V(\phi(t, x))$ to fulfill the static conformal symmetry they have to have the form \cite{Coley:2002je}:
\begin{small}
\begin{align*}
& \phi(t,x)=\psi (x)-\lambda t, \quad  V(\phi(t,x))= e^{-2 t} U(\psi(x)),  \quad  U(\psi)=U_0 e^{-\frac{2 \psi}{\lambda}}.    
\end{align*}
\end{small}
where it is assumed, for convenience, $\lambda> 0 $, such that for $\psi> 0 $, $ U \rightarrow 0$ as $ \lambda \rightarrow 0 $.
\newline Then, the energy-momentum tensor of the scalar field is given by: 
\begin{small}
\begin{equation}
{T^{\psi}}_{a}^{b}=\left(
\begin{array}{cccc}
 \mu_\phi & q_\phi  & 0 & 0 \\
 q_\phi & p_\phi-2\pi_\phi & 0 & 0 \\
 0 & 0 & p_\phi+\pi_\phi & 0 \\
 0 & 0 & 0 & p_\phi+\pi_\phi \\
\end{array}
\right),
\end{equation}
\end{small}
\noindent where $\widehat{...}$ denotes the derivative with respect the spatial variable $x$, and:  
\begin{small}
\begin{subequations}
\begin{align}
& \mu_\phi:= \frac{1}{2} \e_0(\phi)^2 + \frac{1}{2}\e_1(\phi)^2 +V(\phi) = \nonumber \\
& \frac{1}{2} \lambda ^2 e^{-2 t} {b_1}^2+U_0 e^{-2 t-\frac{2 \psi (x)}{\lambda }}+\frac{1}{2} e^{-2 t} {\widehat{\psi}}^2,
\\
& p_\phi:= \frac{1}{2} \e_0(\phi)^2 - \frac{1}{6}\e_1(\phi)^2 -V(\phi) = \nonumber \\
& \frac{1}{2} \lambda ^2 e^{-2 t} {b_1}^2-U_0 e^{-2 t-\frac{2 \psi (x)}{\lambda
   }}-\frac{1}{6} e^{-2 t} {\widehat{\psi}}^2,\\
& q_\phi:=  -\e_0(\phi)  \e_1(\phi)=\lambda  e^{-2 t} {b_1} {\widehat{\psi}},\\
&\pi_\phi:= - \frac{1}{3}\e_1(\phi)^2=-\frac{1}{3} e^{-2 t} {\widehat{\psi}}^2.   
\end{align}
\end{subequations}
\end{small}
The field equations are the following. \newline
\noindent{\bf Propagation equations:} 
\begin{subequations}
\label{modelodos}
\begin{align}
&\widehat{ {\theta}}= -\sqrt{3} {b_2}^2-\frac{\sigma \left(2 C_2 \sigma+\theta\right)}{\sqrt{3}}- \sqrt{3} \Psi^2-\frac{\sqrt{3} \gamma \mu_{t} v^2}{(\gamma -1) v^2+1},
\\
&\widehat{ {\sigma}}=-\frac{\sqrt{3} \lambda ^2 {b_1}^2}{C_2}+\frac{\sqrt{3} U_0 e^{-\frac{2 \psi}{\lambda
   }}}{C_2}-\frac{\sigma  (2 \theta +\sigma)}{\sqrt{3}}  \nonumber \\
   & +\frac{\sqrt{3} \mu _{t} \left(-3 \gamma +(\gamma -2) v^2+2\right)}{2 C_2 \left((\gamma -1) v^2+1\right)}, \\
&\widehat{b_1}= \frac{b_1  {\sigma}}{\sqrt{3}}, \\
&\widehat{b_2}= -\frac{b_2 ( {\theta}+ {\sigma})}{\sqrt{3}},\\
&\widehat{v}=  \frac{\left(v^2-1\right)}{\sqrt{3} \gamma  \left(\gamma -v^2-1\right)} \Big\{\gamma 
   v (2 (\gamma -1) \theta +\gamma  \sigma ) \nonumber \\
   & + \sqrt{3} b_1 \left(3 \gamma ^2-5 \gamma +(\gamma -2) v^2+2\right)\Big\},
\\
&\widehat{\Psi}= 2 \lambda  {b_1}^2-\frac{(2 \theta+\sigma) \Psi}{\sqrt{3}}-\frac{2 U_0 e^{-\frac{2 \psi}{\lambda }}}{\lambda },
\\
& \widehat{\psi}=\Psi.
\end{align}
\end{subequations}
\noindent{\bf	Auxiliary equation:}
\begin{subequations}
\begin{align}
& \widehat{\mu _{t}}=\frac{\mu_{t}}{\sqrt{3} \left(\gamma -v^2-1\right) \left((\gamma -1) v^2+1\right)} \times \nonumber \\
& \Big\{\gamma  \left(\sigma+(\gamma -1) v^4 (2 \theta +\sigma )-v ^2 ((4 \gamma -6) \theta+\gamma  \sigma )\right) \nonumber \\
& +2 \sqrt{3} b_1 v \left((7-3 \gamma ) \gamma +(\gamma  (2 \gamma -5)+4) v^2-4\right)\Big\}.
\end{align}
\end{subequations}

\noindent{\bf Restriction:}
\begin{equation}
3 \gamma  \mu_{t}  v- b_1\left((\gamma -1) v^2+1\right) \left(2 \sqrt{3} C_2 \sigma+3 \lambda  \Psi\right)=0.\end{equation}
\noindent{\bf Equation for  $\mu_t$:}
	\begin{align}
	&{b_1}^2 \left(C_2+\frac{\lambda ^2}{2}\right)+{b_2}^2+\frac{1}{3} C_2 \sigma^2+\frac{1}{2} \Psi^2= \nonumber \\
	& \frac{\theta^2}{3}+U_0 e^{-\frac{2 \psi}{\lambda }}-\frac{\mu_{t} \left(\gamma +v^2-1\right)}{(\gamma -1) v^2+1}   \leq \theta^2+U_0 e^{-\frac{2 \psi}{\lambda }}.
	\end{align}
In general, for $\mu_t \geq 0, 0 <\gamma \leq 1, 1- \gamma \leq v^ 2 \leq 1$, or $ \mu_t \geq 0, 1 <\gamma <2, -1 \leq v \leq 1 $ assuming $ C_2 \geq 0 $, the term $ \theta^ 2$ is dominant, which suggests using $\theta $ -normalized variables.
 
\section{$\theta$-Normalized equations}
\label{Sect:4.2}
The following $\theta$-normalized variables are introduced:
\begin{align}
&{\Sigma} =\frac{\sigma}{\theta},\quad  A=\frac{\sqrt{3} b_1}{\theta},\quad K=\frac{3 b_2^2}{\theta^2},\quad {\Omega}_t =\frac{3\mu_t}{\theta^2}, \nonumber \\
& u= \sqrt{\frac{3}{2}} \frac{\Psi}{\theta}, \quad w=\frac{ e^{-\frac{ \psi}{\lambda }}\sqrt{3 U_0}}{\theta},
\end{align}
\newline along with the radial coordinate: 
$$\frac{df}{d \eta} := \frac{\sqrt{3}\widehat{f}}{\theta}.$$
The parameter $ {r} $ is defined, analogously to the ``Hubble gradient parameter'' $ {r} $, by  $$\widehat{\theta}=-r {\theta}^2,$$ where 
\begin{equation}
  \sqrt{3} r=  2 C_2 \Sigma ^2+K+\frac{\gamma  {\Omega_t}  v^2}{(\gamma -1) v^2+1}+\Sigma +2 u^2.
\end{equation}
In these variables, the dynamical system is given by:
\begin{subequations}
\begin{small}
\begin{align}
& \Sigma'= -\frac{A^2 \lambda ^2}{C_2}+2 C_2 \Sigma ^3+\frac{w^2}{C_2}+\Sigma  \left(K+2 u^2-2\right) \nonumber \\
& +\frac{{\Omega_t}  \left(-3 \gamma +2 \gamma  C_2 \Sigma  v^2+(\gamma -2) v^2+2\right)}{2 C_2 \left((\gamma -1) v^2+1\right)},\\
& A'=A \left(2 C_2 \Sigma ^2+K+2 \left(\Sigma +u^2\right)+\frac{\gamma  v^2 {\Omega_t} }{(\gamma -1) v^2+1}\right),\\
& K'=2 K \left(2 C_2 \Sigma ^2+K+2 u^2+\frac{\gamma  v^2 {\Omega_t} }{(\gamma -1) v^2+1}-1\right),\\
& u'= \sqrt{2} A^2 \lambda +2 C_2 \Sigma ^2 u+u \left(K+\frac{\gamma  v^2 {\Omega_t} }{(\gamma -1) v^2+1}-2\right) +2 u^3-\frac{\sqrt{2} w^2}{\lambda },\end{align}
\begin{align}
& w'= w\left(2 C_2 \Sigma ^2+K+\Sigma +2 u^2-\frac{\sqrt{2} u}{\lambda }+\frac{\gamma  v^2 {\Omega_t} }{(\gamma -1) v^2+1}\right),\\
& v'= \frac{\left(v^2-1\right) \left(A \left(3 \gamma ^2-5 \gamma +(\gamma -2) v^2+2\right)+\gamma  v (\gamma  (\Sigma +2)-2)\right)}{\gamma  \left(\gamma -v^2-1\right)},\\
& {\Omega_t}'={\Omega_t}  \Bigg(4 C_2 \Sigma
   ^2+2 K+2 \Sigma +4 u^2+\frac{2 \gamma  v^2 {\Omega_t} }{(\gamma -1) v^2+1} \Bigg)  \\ \nonumber &  +{\Omega_t}\Bigg(\frac{2 A v \left((7-3
   \gamma ) \gamma +(\gamma  (2 \gamma -5)+4) v^2-4\right)}{\left((\gamma -1) v^2+1\right) \left(\gamma -v^2-1\right)}\Bigg)\\ \nonumber & +{\Omega_t}\Bigg(\frac{\gamma  \left(\Sigma +v^2 \left(-\gamma  (\Sigma +4)+(\gamma -1) (\Sigma +2) v^2+6\right)\right)}{\left((\gamma -1) v^2+1\right) \left(\gamma -v^2-1\right)}\Bigg).
\end{align}
\end{small}
\end{subequations}
The restrictions are: 
\begin{small}
\begin{subequations}
\begin{align}
& C_2 \left(A^2+\Sigma ^2\right)+\frac{A^2 \lambda ^2}{2}+K+u^2-w^2  +\frac{{\Omega_t} \left(\gamma +v^2-1\right)}{(\gamma -1) v^2+1} =1,\\
&\gamma  {\Omega_t}  v -A \left((\gamma -1) v^2+1\right) \left(2 C_2 \Sigma +\sqrt{2} \lambda  u\right)=0. 
\end{align}
\end{subequations}
\end{small}

These can be globally resolved for $\Omega_t $ and $K$ to get: 
\begin{subequations}
\begin{small}
\begin{align}
&\Omega_t= \frac{A \left(\gamma  v^2-v^2+1\right) \left(2 C_2 \Sigma +\sqrt{2} \lambda  u\right)}{\gamma  v},\\
& K= \frac{1}{2} \left(-2 A^2 C_2 -\lambda ^2A^2 -2 C_2 \Sigma ^2-2 u^2+2 w^2+2\right) \nonumber\\ & -\frac{A \left(\gamma +v^2-1\right) \left(\gamma  v^2-v^2+1\right) \left(2 C_2 \Sigma +\sqrt{2} \lambda 
   u\right)}{\gamma  v \left((\gamma -1) v^2+1\right)}.
\end{align}
\end{small}
\end{subequations}
Finally, the following reduced 5-dimensional system is obtained:
\begin{widetext}
\begin{subequations}
\label{reducedsystSF}
\begin{align}
& \Sigma'=-\frac{A^2 \lambda ^2 \left(C_2 \Sigma +2\right)}{2 C_2}+C_2 \left(\Sigma ^3-A^2 \Sigma \right)+\frac{w^2}{C_2}+\Sigma  \left(u^2+w^2-1\right) \nonumber \\
   & +\frac{A \left(2 C_2 \Sigma +\sqrt{2} \lambda  u\right) \left(-3 \gamma +2 (\gamma -1) C_2 \Sigma  \left(v^2-1\right)+(\gamma -2) v^2+2\right)}{2 \gamma  C_2 v},\\
& A'= -\frac{1}{2} A^3 \lambda ^2+A \left(C_2 \left(\Sigma ^2-A^2\right)+2 \Sigma +u^2+w^2+1\right) +\frac{A^2 (\gamma -1) \left(v^2-1\right) \left(2
   C_2 \Sigma +\sqrt{2} \lambda  u\right)}{\gamma  v},\\
& v'=\frac{\left(v^2-1\right) \left(A \left(3 \gamma ^2-5 \gamma +(\gamma -2) v^2+2\right)+\gamma  v (\gamma  (\Sigma +2)-2)\right)}{\gamma  \left(\gamma -v^2-1\right)},
\\
& u'= u \left(C_2 \left(\Sigma ^2-A^2\right)+u^2+w^2-1\right)+\sqrt{2} A^2 \lambda -\frac{1}{2} A^2 \lambda ^2 u  +\frac{A (\gamma -1) u \left(v^2-1\right) \left(2 C_2 \Sigma +\sqrt{2} \lambda  u\right)}{\gamma  v}-\frac{\sqrt{2} w^2}{\lambda },\end{align}
\begin{align}
& w'=-\frac{1}{2} A^2 \lambda ^2
   w+\frac{\sqrt{2} A (\gamma -1) \lambda  u \left(v^2-1\right) w}{\gamma  v}-\frac{\sqrt{2} u w}{\lambda }\nonumber \\
 & + \frac{w \left(C_2 \left(A^2 \gamma  (-v)+2 A (\gamma -1) \Sigma  \left(v^2-1\right)+\gamma  \Sigma ^2 v\right)+\gamma  v \left(\Sigma +u^2+w^2+1\right)\right)}{\gamma  v}.
\end{align}
\end{subequations}
\end{widetext}
If $ u = w = 0 $ and the limit $\lambda \rightarrow 0$ is taken, system \eqref{reducedsyst} is recovered.
Given the computational difficulty of obtaining (analytically) the stability conditions for  all the equilibrium points of the system \eqref{reducedsystSF}, in the following sections some subcases of \eqref{reducedsystSF} of special interest will be studied: perfect fluid in the form of ideal gas  \eqref{scalar-field-A} , the invariant set $\Sigma=0$ \eqref{scalar-field-B}, the extreme tilt case \eqref{scalar-field-C} and the invariant set  $A=v=0$ \eqref{campo-escalar-D}.
An exhaustive analysis (analytical or numerical) of the stability conditions is provided for these particular cases. 
Relaxing the condition $1\leq \gamma \leq 2$, interestingly, a cosmological fluid in the form of an ideal gas with equation of state $p_m = (\gamma-1) \mu_m $, with $ \gamma = 2/3 $ describes a FLRW spacetime
with non-zero curvature. 
\subsection{Special case $\Omega_t=v=0$ and $\gamma=2/3$}
\label{gamma=2/3}
The equations are:
\begin{subequations}
\begin{align}
& \Sigma'=-\frac{\left(A^2 \lambda ^2-w^2\right)}{C_2}+2 C_2 \Sigma ^3+\Sigma  \left(K+2 u^2-2\right),\\ 
& A'=A \left(2 C_2 \Sigma ^2+K+2 \left(\Sigma +u^2\right)\right), \\
& K'=2 K \left(2 C_2 \Sigma ^2+K+2 u^2-1\right),
\\
& u'= \frac{\sqrt{2} \left(A^2 \lambda ^2-w^2\right)}{\lambda }+u \left(2 C_2 \Sigma ^2+K-2\right)+2 u^3,\\
& w'= w \left(2 C_2 \Sigma ^2+K+\Sigma +2 u^2-\frac{\sqrt{2} u}{\lambda }\right),  \end{align}
\end{subequations}
with restrictions
\begin{subequations}
\begin{align}
& -\frac{1}{3} A \left(2 C_2 \Sigma +\sqrt{2} \lambda  u\right)=0,\\
& -2 C_2 \left(A^2+\Sigma ^2\right)-A^2 \lambda ^2-2 K-2 u^2+2 w^2+2=0.
\end{align}
\end{subequations}
Afterwards, the restrictions are solved to find: 
\begin{subequations}
\begin{align}
   & u= -\frac{\sqrt{2} C_2 \Sigma }{\lambda }, \\
   & K=  -C_2 \left(A^2+\Sigma ^2\right)-\frac{A^2 \lambda ^2}{2}-\frac{2 C_2^2 \Sigma ^2}{\lambda ^2}+w^2+1.
\end{align}
\end{subequations}
Therefore, the following reduced  3-dimensional dynamical system is obtained :
\begin{subequations}
\label{scalar-field-A}
\begin{align}
&\Sigma'=C_2 \left(\Sigma ^3-A^2 \Sigma \right)+\frac{1}{2} \Sigma  \left(-A^2 \lambda ^2+2 w^2-2\right)\nonumber \\
& +\frac{(w-A \lambda ) (A \lambda +w)}{C_2}+\frac{2 C_2^2 \Sigma ^3}{\lambda ^2},\\
& A'= -\frac{1}{2} A^3 \lambda ^2+A C_2 \left(\Sigma
   ^2-A^2\right)+\frac{2 A C_2^2 \Sigma ^2}{\lambda ^2} \nonumber \\
   & +A \left(2 \Sigma +w^2+1\right),\\
& w'=w \left(\frac{C_2 \left(\lambda ^2 \left(\Sigma ^2-A^2\right)+2 C_2 \Sigma ^2+2 \Sigma \right)}{\lambda ^2}\right. \nonumber \\ & \left. -\frac{A^2 \lambda ^2}{2}  +\Sigma
   +w^2+1\right).
\end{align}
\end{subequations}

 \begin{figure*}[!t]
    \centering
    \includegraphics[scale=0.6]{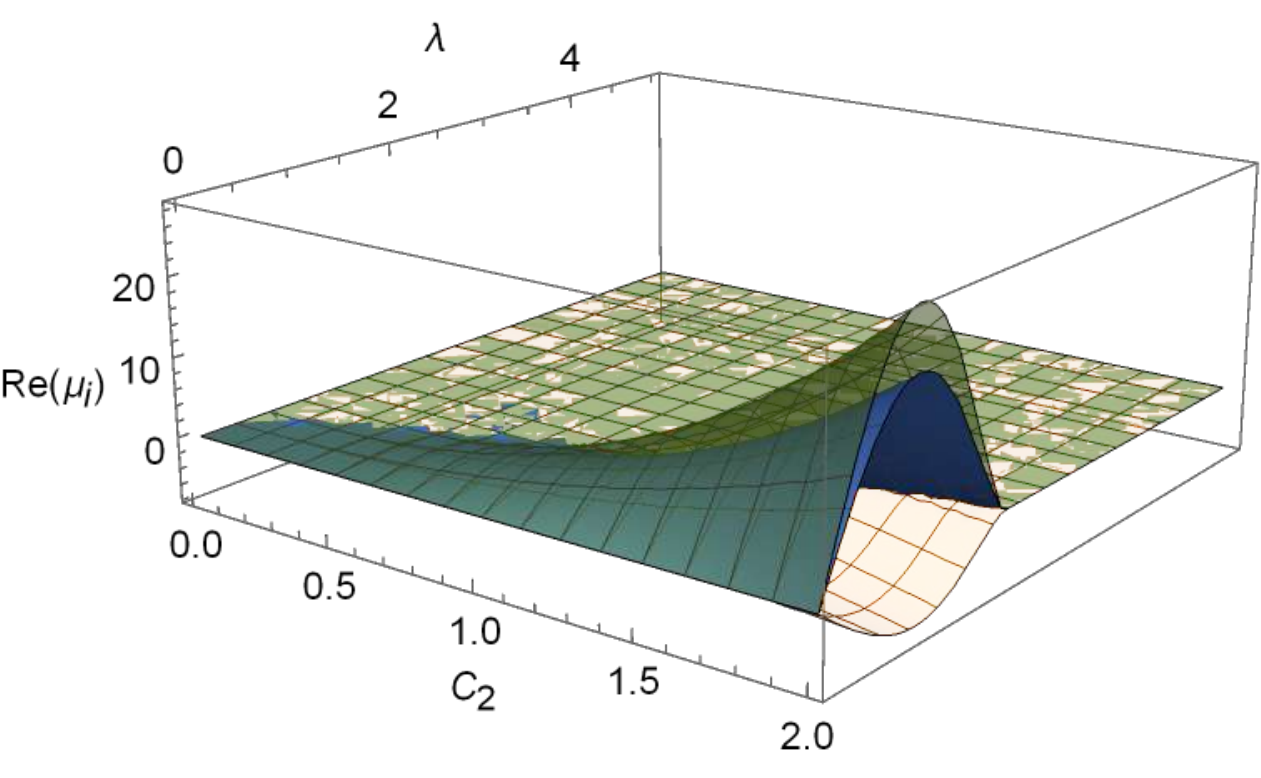}  \caption[{Real parts of the  $\mu_i$'s associated to the equilibrium points $(\Sigma,A,w)=\left(0, \frac{1}{\sqrt{C_2-\frac{\lambda ^2}{2}}},\pm \frac{\lambda }{\sqrt{C_2-\frac{\lambda
   ^2}{2}}}\right)$.}]{\label{parte real}{Real parts of the  $\mu_i$'s associated to the equilibrium points $Q_{7,8}: (\Sigma,A,w)=\left(0, \frac{1}{\sqrt{C_2-\frac{\lambda ^2}{2}}},\pm \frac{\lambda }{\sqrt{C_2-\frac{\lambda
   ^2}{2}}}\right)$. The equilibrium points are typically saddles or non-hyperbolic with three zero eigenvalues, depending on the choice of parameters.}}
   \end{figure*}
   
The equilibrium points of the system  \eqref{scalar-field-A} are the following. 
\begin{enumerate}
   \item[$N_1$:] $(\Sigma,A,w)=(0,0,0)$ with eigenvalues $\{-1,1,1\}$ is a hyperbolic saddle.
    \item[$Q_1$:] $(\Sigma,A,w)=\left(-\frac{1}{2}, \frac{\sqrt{C_2}}{\sqrt{2} \lambda },0\right)$, with eigenvalues \\$\Big\{\frac{1}{2}-\frac{{C_2}}{\lambda ^2},\frac{1}{2} \left(-\frac{\sqrt{8
   {C_2}^2+(4 {C_2}-7) \lambda ^2}}{\lambda }-1\right)$, \newline $\frac{1}{2}
   \left(\frac{\sqrt{8 {C_2}^2+(4 {C_2}-7) \lambda ^2}}{\lambda
   }-1\right)\Big\}$. It is:
   \begin{enumerate}
       \item a hyperbolic sink for:
       \begin{enumerate}
           \item $0<\lambda \leq \frac{\sqrt{7}}{2}, \; \frac{\lambda}{4} \sqrt{\lambda ^2+14}-\frac{\lambda^2}{4}<{C_2}<
           \frac{\lambda}{4}\sqrt{\lambda ^2+16}-\frac{\lambda ^2}{4}$, or    
            \item $\frac{\sqrt{7}}{2}<\lambda <\sqrt{2}, \; \frac{\lambda ^2}{2}<{C_2}<\frac{\lambda}{4} \sqrt{\lambda ^2+16}-\frac{\lambda ^2}{4}$, or    
             \item $0<\lambda <\frac{\sqrt{7}}{2}, \; \frac{\lambda ^2}{2}<{C_2}\leq \frac{1}{4} \sqrt{\lambda ^4+14
   \lambda ^2}-\frac{\lambda ^2}{4}$.
       \end{enumerate}
       \item and a hyperbolic saddle for:
       \begin{enumerate}
           \item  $0<\lambda \leq \sqrt{2}, \; {C_2}>\frac{\lambda}{4} \sqrt{\lambda ^2+16 }-\frac{\lambda ^2}{4}$, or    
           \item $\lambda >\sqrt{2}, \; {C_2}>\frac{\lambda ^2}{2}$, or    
           \item $\lambda >\sqrt{2}, \; \frac{\lambda}{4} \sqrt{\lambda ^2+16}-\frac{\lambda
   ^2}{4}<{C_2}<\frac{\lambda ^2}{2}$, or    
            \item $\frac{\sqrt{7}}{2}<\lambda \leq \sqrt{2}, \; \frac{\lambda}{4} \sqrt{\lambda ^2+14}-\frac{\lambda
   ^2}{4}<{C_2}<\frac{\lambda ^2}{2}$, or    
            \item $\lambda >\sqrt{2}, \; \frac{\lambda}{4} \sqrt{\lambda ^2+14}-\frac{\lambda
   ^2}{4}<{C_2}<\frac{\lambda}{4} \sqrt{\lambda ^2+16}-\frac{\lambda ^2}{4}$, or    
            \item $0<\lambda \leq \frac{\sqrt{7}}{2}, \; 0<{C_2}<\frac{\lambda ^2}{2}$, or    
            \item $\lambda >\frac{\sqrt{7}}{2}, \; 0<{C_2}\leq \frac{\lambda}{4} \sqrt{\lambda ^2+14}-\frac{\lambda ^2}{4}$.
       \end{enumerate}
       \item non-hyperbolic for:
       \begin{enumerate}
           \item $\lambda >0, \; {C_2}=\frac{\lambda ^2}{2}$, or    
           \item $\lambda >0, \; {C_2}=\frac{\lambda}{4} \sqrt{\lambda ^2+16}-\frac{\lambda ^2}{4}$.
       \end{enumerate}
   \end{enumerate}
   \item[$Q_2$:] $(\Sigma,A,w)=\left(-\frac{\lambda ^2}{C_2 \left(2 C_2+\lambda ^2\right)}, \frac{\sqrt{4 C_2^2+2
   (C_2-1) \lambda ^2}}{\sqrt{C_2} \left(2 C_2+\lambda ^2\right)},0\right)$ with eigenvalues \\ $\left\{\frac{1}{C_2}-\frac{4}{2 C_2+\lambda ^2},-\frac{8}{2 C_2+\lambda
   ^2}+\frac{4}{C_2}-2,-\frac{4}{2 C_2+\lambda ^2}+\frac{2}{C_2}-2\right\}$. It exist for  $\lambda >0,\\   {C_2}\geq \frac{1}{4} \lambda 
   \sqrt{\lambda ^2+8}-\frac{\lambda ^2}{4}$, and it is: 
    \begin{enumerate}
        \item a hyperbolic sink for: 
        \begin{enumerate}
            \item $0<\lambda \leq \sqrt{2}, C_2>\frac{\lambda}{4} \sqrt{\lambda ^2+16}-\frac{\lambda ^2}{4}$, or    
            \item $\lambda >\sqrt{2}, C_2>\frac{\lambda ^2}{2}$.
        \end{enumerate}
        \item a hyperbolic saddle for:
        \begin{enumerate}
            \item $0<\lambda \leq 1, \frac{\lambda}{4} \sqrt{\lambda ^2+8}-\frac{\lambda
   ^2}{4}<C_2<\frac{\lambda}{4} \sqrt{\lambda ^2+16}-\frac{\lambda ^2}{4}$, or    
            \item $1<\lambda <\sqrt{2}, \frac{\lambda ^2}{2}<C_2<\frac{\lambda}{4} \sqrt{\lambda ^2+16}-\frac{\lambda ^2}{4}$, or    
            \item $1<\lambda \leq \sqrt{2}, \frac{\lambda}{4} \sqrt{\lambda ^2+8}-\frac{\lambda
   ^2}{4}<C_2<\frac{\lambda ^2}{2}$, or    
            \item $\lambda >\sqrt{2}, \frac{\lambda}{4} \sqrt{\lambda ^2+8}-\frac{\lambda
   ^2}{4}<C_2<\frac{\lambda}{4} \sqrt{\lambda ^2+16}-\frac{\lambda ^2}{4}$, or    
            \item $\lambda >\sqrt{2}, \frac{\lambda}{4} \sqrt{\lambda ^2+16}-\frac{\lambda
   ^2}{4}<C_2<\frac{\lambda ^2}{2}$.
        \end{enumerate}
        \item non-hyperbolic for:
        \begin{enumerate}
            \item $\lambda \geq 1, C_2=\frac{\lambda ^2}{2}$, or    
            \item $C_2=\frac{1}{4} \left(-\lambda ^2-\sqrt{\lambda ^2+16} \lambda \right)$, or    
            \item $C_2=\frac{1}{4} \left(\lambda  \sqrt{\lambda ^2+16}-\lambda ^2\right)$, or    
            \item $C_2=\frac{1}{4} \left(-\lambda ^2-\sqrt{\lambda ^2+8} \lambda \right)$, or    
            \item $C_2=\frac{1}{4} \left(\lambda  \sqrt{\lambda ^2+8}-\lambda ^2\right)$.
        \end{enumerate}
    \end{enumerate}
    \item[$Q_3$:] $(\Sigma,A,w)=\left(-\frac{\lambda }{\sqrt{C_2 \left(2 C_2+\lambda ^2\right)}},0,0\right)$ with eigenvalues\\ $\left\{2,2-\frac{2 \lambda }{\sqrt{C_2 \left(2 C_2+\lambda
   ^2\right)}},2-\frac{\sqrt{C_2 \left(2 C_2+\lambda ^2\right)}}{C_2 \lambda }\right\}$. It is: 
   \begin{enumerate}
       \item a hyperbolic source for: 
       \begin{enumerate}
           \item $\frac{1}{\sqrt{2}}<\lambda \leq 1, C_2>\frac{\lambda ^2}{4 \lambda ^2-2}$, or    
           \item $\lambda >1, C_2>\frac{\lambda}{4} \sqrt{\lambda ^2+8}-\frac{\lambda ^2}{4}$.
       \end{enumerate}
       \item a hyperbolic saddle for:
       \begin{enumerate}
           \item $\lambda >0, 0<C_2<\frac{\lambda}{4} \sqrt{\lambda ^2+8}-\frac{\lambda ^2}{4}$, or    
           \item $0<\lambda \leq \frac{1}{\sqrt{2}},  {C_2}>0$, or   
           \item $\lambda >1, \frac{\lambda ^2}{4 \lambda ^2-2}<C_2<\frac{1}{4}
   \sqrt{\lambda ^4+8 \lambda ^2}-\frac{\lambda ^2}{4}$, or   
           \item $0<\lambda \leq \frac{1}{\sqrt{2}}, C_2>\frac{1}{4} \sqrt{\lambda ^4+8
   \lambda ^2}-\frac{\lambda ^2}{4}$, or   
           \item $\frac{1}{\sqrt{2}}<\lambda <1, \frac{1}{4} \sqrt{\lambda ^4+8 \lambda
   ^2}-\frac{\lambda ^2}{4}<C_2<\frac{\lambda ^2}{4 \lambda ^2-2}$, or   
           \item $\lambda >\frac{1}{\sqrt{2}}, 
   0<{C_2}<\frac{\lambda ^2}{4 \lambda ^2-2}$.
       \end{enumerate}
       \item non-hyperbolic for:
       \begin{enumerate}
           \item $0<\lambda \leq \frac{1}{\sqrt{2}}, C_2=\frac{1}{4} \sqrt{\lambda ^4+8
   \lambda ^2}-\frac{\lambda ^2}{4}$, or   
           \item $\frac{1}{\sqrt{2}}<\lambda <1, C_2=\frac{1}{4} \sqrt{\lambda ^4+8
   \lambda ^2}-\frac{\lambda ^2}{4}$, or   
           \item $\frac{1}{\sqrt{2}}<\lambda <1, C_2=\frac{\lambda ^2}{4 \lambda ^2-2}$, or   
           \item $\lambda =1, C_2=\frac{1}{2}$, or   
           \item $\lambda >1, C_2=\frac{\lambda ^2}{4 \lambda ^2-2}$, or   
           \item $\lambda >1, C_2=\frac{1}{4} \sqrt{\lambda ^4+8 \lambda
   ^2}-\frac{\lambda ^2}{4}$.
       \end{enumerate}
   \end{enumerate}

   \item[$Q_4$:] $(\Sigma,A,w)=\left(\frac{\lambda }{\sqrt{C_2 \left(2 C_2+\lambda ^2\right)}},0,0\right)$ with eigenvalues \\ $\left\{2,\frac{2 \lambda }{\sqrt{C_2 \left(2 C_2+\lambda ^2\right)}}+2,\frac{\sqrt{C_2
   \left(2 C_2+\lambda ^2\right)}}{C_2 \lambda }+2\right\}$. It is a hyperbolic source for $\lambda >0, C_2>0$.

   \item[$Q_{5,6}$:] $(\Sigma,A,w)=\left(-\frac{1}{2 {C_2}},0,\pm \frac{\sqrt{{C_2} \left(2-4 \lambda ^2\right)+\lambda ^2}}{2
   \sqrt{{C_2}} \lambda }\right)$ with eigenvalues  $\left\{\frac{1}{\lambda ^2}-\frac{1}{2 {C_2}},\frac{1}{{C_2}}+\frac{2}{\lambda ^2}-2,\frac{1}{2
   {C_2}}+\frac{1}{\lambda ^2}-2\right\}$. Exists for $0<\lambda \leq \frac{1}{\sqrt{2}},  {C_2}>0$ o $\lambda >\frac{1}{\sqrt{2}},  0<{C_2}\leq \frac{\lambda ^2}{4
   \lambda ^2-2}$. It is:
   \begin{enumerate}
       \item a hyperbolic source for: 
       \begin{enumerate}
           \item $0<\lambda \leq \frac{1}{\sqrt{2}}, \; {C_2}>\frac{\lambda ^2}{2}$, or    
           \item $\frac{1}{\sqrt{2}}<\lambda <1, \; \frac{\lambda ^2}{2}<{C_2}<\frac{\lambda ^2}{4 \lambda ^2-2}$.
       \end{enumerate}
       \item a hyperbolic saddle for: 
       \begin{enumerate}
           \item $0<\lambda \leq 1, 0<C_2<\frac{\lambda ^2}{2}$, or    
           \item $\lambda >1, 0<C_2<\frac{\lambda ^2}{4 \lambda ^2-2}$. 
       \end{enumerate}
       \item non-hyperbolic for: 
       \begin{enumerate}
           \item $C_2=\frac{1}{2}, \lambda =1$, or    
           \item $C_2=\frac{\lambda ^2}{2}, 0<\lambda <1$, or    
           \item $C_2=\frac{\lambda ^2}{4 \lambda ^2-2}, \lambda >1$, or    
           \item $C_2=\frac{\lambda ^2}{4 \lambda ^2-2}, \frac{1}{\sqrt{2}}<\lambda <1$.
       \end{enumerate}
   \end{enumerate}

   \item[$Q_{7,8}$:] $(\Sigma,A,w)=\left(0, \frac{1}{\sqrt{C_2-\frac{\lambda ^2}{2}}},\pm \frac{\lambda }{\sqrt{C_2-\frac{\lambda
   ^2}{2}}}\right)$ with eigenvalues\\  $\left\{-\frac{\mu_1(\lambda,C_2)}{C_2 \lambda  \left(2 C_2-\lambda ^2\right)^{3/2}}  ,-\frac{\mu_2(\lambda,C_2)}{C_2 \lambda  \left(2 C_2-\lambda ^2\right)^{3/2}}  ,-\frac{\mu_3(\lambda,C_2)}{C_2 \lambda  \left(2 C_2-\lambda ^2\right)^{3/2}}  \right\}$ \newline where the expressions for $\mu_i$ depend on $\lambda, C_2$. They exist for $C_2\geq \frac{\lambda ^2}{2}$. In Figure  \ref{parte real},  the real parts of the $\mu_i$'s are depicted, so,   the equilibrium points are typically saddles or non-hyperbolic with three zero eigenvalues. 
\end{enumerate}
  
  \bigskip 
   
\subsection{Invariant set $\Sigma=0$}
\label{Isotrop}
Imposing $\Sigma=0$ the following restrictions are deduced: 
\begin{subequations}
\begin{align}
& \Omega = -\frac{2 \left(\gamma  v^2-v^2+1\right) \left(w^2-A^2 \lambda ^2\right)}{-3 \gamma +\gamma  v^2-2 v^2+2}, \\
& K=\frac{1}{2} \left(A^2 \left(-\left(2 C_2+\lambda ^2\right)\right) -2 u^2+2 w^2+2 \right. \nonumber \\
& \left. +\frac{4 \left(\gamma +v^2-1\right) (w-A \lambda ) (A \lambda +w)}{-3 \gamma +(\gamma -2) v^2+2}\right), \\
& 2 A^2 \gamma  \lambda ^2 v+\sqrt{2} A \lambda  u \left(3 \gamma -(\gamma -2) v^2-2\right)-2 \gamma  v w^2=0.     
\end{align}
\end{subequations}
Finally a reduced dynamical system for the invariant set $\Sigma=0$ is obtained
\begin{small}
\begin{subequations}
\label{scalar-field-B}
\begin{align}
& A'= \frac{1}{2} A \left(-2 A^2 C_2+A^2 \lambda ^2+\frac{\sqrt{2} A \lambda  u \left(v^2+1\right)}{v}+2 u^2+2\right), \\
& v'= \frac{\left(v^2-1\right) \left(A \left(3 \gamma ^2-5 \gamma +(\gamma -2) v^2+2\right)+2 (\gamma -1) \gamma  v\right)}{\gamma  \left(\gamma -v^2-1\right)}, 
\end{align}
\begin{align}
& u'=\frac{ u \left(-2 A^2 C_2 v+A \left(A \lambda ^2 v+\sqrt{2} \lambda  u \left(v^2+1\right)+2 v^2-6\right)+2 \left(u^2-1\right) v\right)}{2  v} \nonumber \\
& -\frac{2 A u \left(v^2-1\right)}{\gamma v}.   
\end{align}
\end{subequations}
\end{small}

The equilibrium points of system  \eqref{scalar-field-B} are the following. 
\begin{enumerate}
\item[$N_{2,3}$:] $(A,v,u)=(0,\pm 1,0)$ has eigenvalues $\left\{-1,1,\frac{4}{\gamma -2}+4\right\}$. They are non-hyperbolic saddles for $\lambda >0,  \gamma =1,  C_2>0$, or, generically,  they are saddles because two eigenvalues have opposite signs.
\item[$M_{1,2}$:] $(A,v,u)=(0,1,\pm 1)$ and 
\item[$M_{3,4}$:] $(A,v,u)=(0,-1,\pm 1)$ 
with eigenvalues $\left\{2,2,\frac{4}{\gamma -2}+4\right\}$. They are:
	\begin{enumerate}
	\item hyperbolic sources for  $\lambda >0,  0<\gamma <1,  C_2>0$, or \item hyperbolic saddles for  $\lambda >0,  1<\gamma <2,  C_2>0$, or \item non-hyperbolic for $\lambda >0,  \gamma =1,  C_2>0$.
	\end{enumerate}
\item[$Q_9$:] $(A,v,u)=\left(\frac{1}{\sqrt{C_2-\frac{\lambda ^2}{2}}},1,0\right)$  exists for  $\lambda >0,  C_2\geq \frac{\lambda ^2}{2}$. The eigenvalues are \newline $\left\{-2,-\frac{2}{\sqrt{C_2-\frac{\lambda ^2}{2}}}-2,\frac{-4 \gamma
   -\frac{6 \gamma }{\sqrt{C_2-\frac{\lambda
   ^2}{2}}}+\frac{8}{\sqrt{C_2-\frac{\lambda ^2}{2}}}+4}{2-\gamma
   }\right\}$. $Q_9$ is:
	\begin{enumerate}
	\item a hyperbolic sink for 
	\begin{enumerate}
	\item $\lambda >0,  1<\gamma \leq \frac{4}{3},  C_2>\frac{1}{4}
   \left(\frac{(4-3 \gamma )^2}{(\gamma -1)^2}+2 \lambda ^2\right)$, or    
	\item $\lambda >0,  \frac{4}{3}<\gamma <2,  2 C_2>\lambda ^2$.
	\end{enumerate}
	\item a hyperbolic saddle for 	\begin{enumerate}
	    \item  $1<\gamma <\frac{4}{3},  \lambda >0,  C_2=\frac{1}{4}
   \left(\frac{(4-3 \gamma )^2}{(\gamma -1)^2}+2 \lambda ^2\right)$, or      \item $\lambda >0,  0<\gamma \leq 1,  2 C_2>\lambda ^2$.
	\end{enumerate}
	\item non-hyperbolic for $1<\gamma <\frac{4}{3},  \lambda >0$, \newline $ C_2=\frac{1}{4} \left(\frac{(4-3
   \gamma )^2}{(\gamma -1)^2}+2 \lambda ^2\right)$.
	\end{enumerate}

\item[$Q_{10}$:] $(A,v,u)=\left(\frac{1}{\sqrt{C_2-\frac{\lambda ^2}{2}}},-1,0\right)$ exists for $C_2\geq \frac{\lambda ^2}{2}$. The eigenvalues are \newline $\left\{-2,\frac{2}{\sqrt{C_2-\frac{\lambda ^2}{2}}}-2,\frac{-4 \gamma
   +\frac{6 \gamma }{\sqrt{C_2-\frac{\lambda
   ^2}{2}}}-\frac{8}{\sqrt{C_2-\frac{\lambda ^2}{2}}}+4}{2-\gamma
   }\right\}$.  $Q_{10}$ is: 
	\begin{enumerate}
	\item a hyperbolic sink for:
	\begin{enumerate}
	\item $\lambda >0,  2 C_2>\lambda ^2+2,  1\leq \gamma <2$, or    
	\item $\lambda >0,  0<\gamma <1$, \newline $\frac{1}{2} \left(\lambda
   ^2+2\right)<C_2<\frac{1}{4} \left(\frac{(4-3 \gamma )^2}{(\gamma
   -1)^2}+2 \lambda ^2\right)$.
	\end{enumerate}
	\item a hyperbolic saddle for: 
	\begin{enumerate}
	\item $\lambda >0,  0<\gamma <2,  0<2 C_2-\lambda ^2<2$, or    
	\item $\lambda >0,  0<\gamma <1,  C_2>\frac{1}{4} \left(\frac{(4-3 \gamma
   )^2}{(\gamma -1)^2}+2 \lambda ^2\right)$, or    
	\item $\lambda >0,  \frac{4}{3}<\gamma <2$,  $\frac{\lambda
   ^2}{2}<C_2<\frac{1}{4} \left(\frac{(4-3 \gamma )^2}{(\gamma -1)^2}+2
   \lambda ^2\right)$, or   
   \item $\lambda >0,  0<\gamma \leq \frac{4}{3},  0<2 C_2-\lambda ^2<2$, or    
   \item $\lambda >0,  3 \gamma >4$, \newline $\frac{1}{4} \left(\frac{(4-3 \gamma )^2}{(\gamma
   -1)^2}+2 \lambda ^2\right)<C_2<\frac{1}{2} \left(\lambda ^2+2\right)$. 
	\end{enumerate}
	\item non-hyperbolic for: 
	\begin{enumerate}
	\item $\lambda >0,  0<\gamma <2,  2 C_2=\lambda ^2+2$, or     
	\item $C_2=\frac{1}{4} \left(\frac{(4-3 \gamma )^2}{(\gamma -1)^2}+2 \lambda
   ^2\right),  0<\gamma <1, 
   \lambda >0$, or    
   \item $C_2=\frac{1}{4} \left(\frac{(4-3 \gamma )^2}{(\gamma -1)^2}+2 \lambda
   ^2\right),  \frac{4}{3}<\gamma <2, 
   \lambda >0$.
	\end{enumerate}
	\end{enumerate}

\item[$Q_{11}$:] $(A,v,u)=\left(1,-1,\frac{\lambda }{\sqrt{2}}-\sqrt{C_2-1}\right)$ exists for $C_2\geq 1$. The eigenvalues are \newline $\Big\{-2,-\sqrt{-2 \sqrt{2} \sqrt{{C_2}-1} \lambda +4 {C_2}-3}-1$,\newline $\sqrt{-2 \sqrt{2} \sqrt{{C_2}-1} \lambda +4 {C_2}-3}-1\Big\}$. $Q_{11}$ is:
	\begin{enumerate}
	\item a hyperbolic sink for: 
	\begin{enumerate}
	\item $1<C_2<2,  \lambda =\sqrt{2},  0<\gamma <2$, or    
	\item $C_2>1,  0<\lambda <\sqrt{2},  0<\gamma <2,  2 C_2<\lambda
   ^2+2$, or    	
	\item $C_2>1,  \lambda >\sqrt{2},  0<\gamma <2,  4 C_2+\lambda 
   \sqrt{\lambda ^2-2}<\lambda ^2+3$, or    
	\item $\lambda >\sqrt{2},  0<\gamma <2,  2 C_2<\lambda ^2+2,  \lambda 
   \left(\sqrt{\lambda ^2-2}+\lambda \right)+3<4 C_2$, or    
	\item $\lambda >\sqrt{2},  0<\gamma <2,  \lambda  \left(\sqrt{\lambda
   ^2-2}+\lambda \right)+3\geq 4 C_2,  4 C_2+\lambda 
   \sqrt{\lambda ^2-2}\geq \lambda ^2+3$.
	\end{enumerate}
	\item a hyperbolic saddle for $\lambda >0,  0<\gamma <2,  C_2>\frac{1}{2} \left(\lambda ^2+2\right)$.
	\item non-hyperbolic for: 
	\begin{enumerate}
	\item $C_2=1 ,  0<\gamma <2,  \lambda >0$, or    
	\item $2 C_2=\lambda ^2+2 ,  0<\gamma <2,   \lambda >0$.
	\end{enumerate}
	\end{enumerate}
	\item[$Q_{12}$:] $(A,v,u)=\left(1,-1,\frac{\lambda }{\sqrt{2}}+\sqrt{C_2-1}\right)$ exists for $C_2\geq 1$. The eigenvalues are $\Big\{-2,-\sqrt{2 \sqrt{2} \sqrt{{C_2}-1} \lambda +4 {C_2}-3}-1$, \newline $\sqrt{2 \sqrt{2} \sqrt{{C_2}-1} \lambda +4 {C_2}-3}-1\Big\}$.  $Q_{12}$ is: 
	\begin{enumerate}
	\item a hyperbolic saddle for $\lambda >0,  0<\gamma <2,  {C_2}>1$.
	\item non-hyperbolic for $\lambda >0,  0<\gamma <2,  {C_2}=1$.
	\end{enumerate}
 \begin{widetext}	
 \item[$Q_{13}$:] $(A,v,u)=\left\{\frac{\Delta  \left(\Gamma +(\gamma -1)^2 \gamma ^2 \left(2
   {C_2}-\lambda ^2\right)\right)}{(\gamma -1)^2 \gamma  (3 \gamma -2)
   \left(2 {C_2}-\lambda ^2\right)},-\Delta ,0\right\}$ where $\Delta=\sqrt{\frac{-\Gamma +\gamma  \left(\gamma  \left((\gamma -1)^2 \left(-\lambda
   ^2\right)-3 \gamma +2 (\gamma -1)^2 {C_2}+11\right)-12\right)+4}{(\gamma
   -2)^2}}$ and \newline
     $\Gamma=\sqrt{(\gamma -1)^3 \gamma ^2 \left(2 {C_2}-\lambda ^2\right) \left(\gamma 
   \left(\gamma  \left(-(\gamma -1) \lambda ^2+2 (\gamma -1)
   {C_2}-6\right)+16\right)-8\right)}$, 
   \end{widetext}  
   exists for:
	\begin{enumerate}
		\item $\lambda >0, \gamma =1, C_2>0$, or    
		\item $\lambda >0, 2 C_2\geq \lambda ^2, 0<\gamma \leq \frac{2}{3}$, or    
		\item $\lambda >0, 2 C_2>\lambda ^2, 1<\gamma <2$ o
        \item $\lambda >0, \frac{2}{3}<\gamma <1$, \newline $\gamma  \left(\gamma  \left(-(\gamma
   -1) \lambda ^2+2 (\gamma -1) C_2-6\right)+16\right)<8$.
	\end{enumerate}
The eigenvalues are 
 $\Big\{-2,\frac{\Gamma }{(\gamma -1)^2 \gamma  \left(2 {C_2}-\lambda
   ^2\right)}-\gamma$, \newline $\gamma +\frac{(\gamma -2) \gamma  \left(4 \gamma  (\gamma
   -1)^2+\Gamma \right)}{2 (\gamma -1)^3 \left(\gamma  \left(-\gamma  \lambda
   ^2+2 \gamma  {C_2}-8\right)+8\right)}+\frac{(3 \gamma -4) \Gamma }{2
   (\gamma -1)^3 \gamma  \left(\lambda ^2-2 {C_2}\right)}-2\Big\}$. 
$Q_{13}$ is: 
	\begin{enumerate}
		\item a hyperbolic sink for $\lambda >0,  \frac{2}{3}<\gamma <1$, \newline ${C_2}>\frac{\gamma ^3 \lambda
   ^2-\gamma ^2 \lambda ^2+6 \gamma ^2-16 \gamma +8}{2 \gamma ^3-2 \gamma ^2}$.
		\item a hyperbolic saddle for: 
		\begin{enumerate}
		\item $\lambda >0,  1<\gamma <2,  2 {C_2}>\lambda ^2$, or    
		\item $\lambda >0,  0<\gamma <\frac{2}{3},  2 {C_2}>\lambda ^2$, or    
		\item $\lambda >0,  1<\gamma \leq \frac{4}{3},  \frac{\lambda
   ^2}{2}<{C_2}<\frac{\gamma ^2 \lambda ^2+8 \gamma -8}{2 \gamma ^2}$, or    
		\item $\lambda >0,  \frac{4}{3}<\gamma <2,  \frac{2 \gamma ^2 \lambda ^2+9 \gamma
   ^2-4 \gamma  \lambda ^2-24 \gamma +2 \lambda ^2+16}{4 \gamma ^2-8 \gamma
   +4}<{C_2}<\frac{\gamma ^2 \lambda ^2+8 \gamma -8}{2 \gamma ^2}$.
		\end{enumerate}
		\item non-hyperbolic for: 
		\begin{enumerate}
		    \item $\lambda >0,  \frac{4}{3}<\gamma <2,  {C_2}=\frac{2 \gamma ^2 \lambda
   ^2+9 \gamma ^2-4 \gamma  \lambda ^2-24 \gamma +2 \lambda ^2+16}{4 \gamma ^2-8
   \gamma +4}$
             \item $\lambda >0, \gamma =\frac{2}{3}, C_2>\frac{\lambda ^2}{2}$.
		\end{enumerate}
	\end{enumerate}
	
  \begin{widetext}	
    \item[$Q_{14}$:] $(A,v,u)=\left\{\frac{\Delta  \left(-\Gamma +(\gamma -1)^2 \gamma ^2 \left(2
   {C_2}-\lambda ^2\right)\right)}{(\gamma -1)^2 \gamma  (3 \gamma -2)
   \left(2 {C_2}-\lambda ^2\right)},-\Delta ,0\right\}$ where 
    $\Delta=\sqrt{\frac{-\Gamma +\gamma  \left(\gamma  \left((\gamma -1)^2 \left(-\lambda
   ^2\right)-3 \gamma +2 (\gamma -1)^2 {C_2}+11\right)-12\right)+4}{(\gamma
   -2)^2}}$ and \newline $\Gamma=\sqrt{(\gamma -1)^3 \gamma ^2 \left(2 {C_2}-\lambda ^2\right) \left(\gamma 
   \left(\gamma  \left(-(\gamma -1) \lambda ^2+2 (\gamma -1)
   {C_2}-6\right)+16\right)-8\right)}$,  
   \end{widetext}
   exists for: 
	\begin{enumerate}
		\item $\lambda >0, C_2>0, \gamma =1$, or 
		\item $\lambda >0, C_2>0, 3 \gamma =2$, or    
		\item $\lambda >0, 2 C_2\geq \lambda ^2, 0<\gamma <\frac{2}{3}$, or  
		\item $\lambda >0, 2 C_2\geq \lambda ^2, 1<\gamma <2$, or   
		\item $\lambda >0, \frac{2}{3}<\gamma <1$, \newline $\gamma  \left(\gamma  \left(-(\gamma
   -1) \lambda ^2+2 (\gamma -1) C_2-6\right)+16\right)<8$.
	\end{enumerate}
The eigenvalues are  $\Big\{-2,\frac{-\Gamma }{(\gamma -1)^2 \gamma  \left(2 {C_2}-\lambda
   ^2\right)}-\gamma$, \newline $\gamma +\frac{(\gamma -2) \gamma  \left(4 (\gamma -1)^2 \gamma -\Gamma
   \right)}{2 (\gamma -1)^3 \left(\gamma  \left(-\gamma  \lambda ^2+2 \gamma 
   {C_2}-8\right)+8\right)}-\frac{(3 \gamma -4) \Gamma }{2 (\gamma -1)^3
   \gamma  \left(\lambda ^2-2 {C_2}\right)}-2\Big\}$. 
$Q_{14}$ is: 
	\begin{enumerate}
	    \item a hyperbolic sink for: 
	    \begin{enumerate}
	        \item $\lambda >0, 0<\gamma <\frac{2}{3}, C_2>\frac{1}{4} \left(\frac{(4-3
   \gamma )^2}{(\gamma -1)^2}+2 \lambda ^2\right)$, or  
	        \item $\lambda >0, 3 \gamma =2, 2 C_2>\lambda ^2+18$, or  
	        \item $\lambda >0, \frac{2}{3}<\gamma <1, C_2>\frac{1}{4} \left(\frac{(4-3
   \gamma )^2}{(\gamma -1)^2}+2 \lambda ^2\right)$, or  
	        \item $\lambda >0, 3 \gamma +\sqrt{13}=7$, \newline $ 2 C_2<\lambda
   ^2+\sqrt{13}+7, \left(7 \sqrt{13}-31\right) \left(2 C_2-\lambda
   ^2\right)<12 \left(\sqrt{13}-4\right)$, or  
	        \item $\lambda >0, 3 \gamma +\sqrt{13}=7, 18 C_2+\sqrt{13}<9 \lambda
   ^2+11, \lambda ^2<2 C_2$, or  
	        \item $\lambda >0, C_2<\frac{1}{4} \left(\frac{(4-3 \gamma )^2}{(\gamma
   -1)^2}+2 \lambda ^2\right), \frac{4 (\gamma -1)}{\gamma ^2}+\frac{\lambda
   ^2}{2}<C_2, \gamma >1, 3 \gamma +\sqrt{13}<7$, or  
	        \item $\lambda >0, C_2<\frac{1}{4} \left(\frac{(4-3 \gamma )^2}{(\gamma
   -1)^2}+2 \lambda ^2\right), 3 \gamma +\sqrt{13}>7, \gamma +\mu _1<0$ where $\mu _1\approx -1.22033$ is the real root of $P(\mu)=9 \mu ^3+22 \mu ^2+20 \mu +8$, or  
	        \item $\lambda >0, C_2<\frac{1}{4} \left(\frac{(4-3 \gamma )^2}{(\gamma
   -1)^2}+2 \lambda ^2\right), 3 \gamma <4$,  $\lambda ^2<2 C_2,
   \gamma +\mu _1>0$ where $\mu _1\approx -1.22033$ is the real root of $P(\mu)=9 \mu ^3+22 \mu ^2+20 \mu +8$, or  
    \item $\lambda >0, 2 C_2>\lambda ^2, \frac{8}{\gamma ^2}+2
   C_2<\frac{8}{\gamma }+\lambda ^2, 3 \gamma +\sqrt{13}>7, \gamma
   +\mu _1\leq 0$ where $\mu _1\approx -1.22033$ is the real root of $P(\mu)=9 \mu ^3+22 \mu ^2+20 \mu +8$, or  
        \item $\lambda >0, 2 C_2>\lambda ^2, \frac{8}{\gamma ^2}+2
   C_2<\frac{8}{\gamma }+\lambda ^2, 3 \gamma +\sqrt{13}<7$.
	    \end{enumerate}
	    \item a hyperbolic saddle for: 
	    \begin{enumerate}
	        \item $\lambda >0, \frac{2}{3}<\gamma <1$, \newline   $\frac{(\gamma -2) (3 \gamma
   -2)}{(\gamma -1) \gamma ^2}+\frac{\lambda ^2}{2}<C_2<\frac{1}{4}
   \left(\frac{(4-3 \gamma )^2}{(\gamma -1)^2}+2 \lambda ^2\right)$, or  
	        \item $\lambda >0, 3 \gamma +\sqrt{13}=7, 2 C_2>\lambda ^2+\sqrt{13}+7$, or  
	        \item $\lambda >0, C_2>\frac{1}{4} \left(\frac{(4-3 \gamma )^2}{(\gamma
   -1)^2}+2 \lambda ^2\right), \gamma >1, 3 \gamma +\sqrt{13}<7$, or  
	        \item $\lambda >0, C_2>\frac{1}{4} \left(\frac{(4-3 \gamma )^2}{(\gamma
   -1)^2}+2 \lambda ^2\right), 3 \gamma +\sqrt{13}>7, \gamma +\mu _1<0$ where $\mu _1\approx -1.22033$ is the real root of $P(\mu)=9 \mu ^3+22 \mu ^2+20 \mu +8$, or   
	        \item $\lambda >0, C_2<\frac{4 (\gamma -1)}{\gamma ^2}+\frac{\lambda
   ^2}{2}, 3 \gamma <4$, \newline $\frac{1}{4} \left(\frac{(4-3 \gamma )^2}{(\gamma
   -1)^2}+2 \lambda ^2\right)<C_2$, or   
	        \item $\lambda >0, \gamma <2, 2 C_2>\lambda ^2, 3 \gamma \geq 4,
   \frac{8}{\gamma ^2}+2 C_2<\frac{8}{\gamma }+\lambda ^2$, or   
	        \item $\lambda >0, \gamma <2, \frac{8}{\gamma ^2}+2 C_2>\frac{8}{\gamma
   }+\lambda ^2, \gamma +\mu _1\geq 0$, or   
	        \item $\lambda >0, 0<\gamma <\frac{2}{3}, \frac{\lambda
   ^2}{2}<C_2<\frac{1}{4} \left(\frac{(4-3 \gamma )^2}{(\gamma -1)^2}+2
   \lambda ^2\right)$, or  
	        \item $\lambda >0, 3 \gamma =2, 0<2 C_2-\lambda ^2<18$.
	    \end{enumerate}
	    \item non-hyperbolic for: 
	    \begin{enumerate}
	    \item $C_2=\frac{1}{4} \left(\frac{(4-3 \gamma )^2}{(\gamma -1)^2}+2 \lambda
   ^2\right), 0<\gamma <1$, or
	        \item $C_2=\frac{1}{4} \left(\frac{(4-3 \gamma )^2}{(\gamma -1)^2}+2 \lambda
   ^2\right), 1<\gamma <\chi _1$  (where  $\chi _1\approx 1.22033$ is the real root of $P(\chi)=9 \chi ^3-22 \chi ^2+20 \chi -8$), or  
	        \item $\lambda >0, \chi _1 <\gamma <\frac{4}{3}$ where  $\chi _1\approx 1.22033$ is the real root of $P(\chi)=9 \chi ^3-22 \chi ^2+20 \chi -8$.
	    \end{enumerate}
	\end{enumerate}
\end{enumerate}

\subsection{Invariant sets $v=\pm 1$}
\label{tilt2}
Assuming $v=\varepsilon=\pm 1$, the system is reduced to the following 4-dimensional dynamical system:
\begin{subequations}
\label{scalar-field-C}
\begin{small}
\begin{align}
&\Sigma'= \frac{C_2 \Sigma  \left(2 C_2 \left(\Sigma ^2-A^2\right)-A \left(A \lambda ^2+4 \varepsilon \right)+2 u^2+2 w^2-2\right)}{2 C_2} \nonumber \\
& -\frac{A \lambda  \left(A \lambda +\sqrt{2} u \varepsilon \right)-2 w^2}{C_2},\\
& A'=-\frac{1}{2} A^3 \lambda ^2+A
   C_2 \left(\Sigma ^2-A^2\right)+A \left(2 \Sigma +u^2+w^2+1\right),\\
& u'= C_2 u \left(\Sigma ^2-A^2\right)+u \left(-\frac{A^2 \lambda ^2}{2}+w^2-1\right)+\frac{\sqrt{2} \left(A^2 \lambda ^2-w^2\right)}{\lambda }+u^3,\\
& w'=w \left(C_2\left(\Sigma ^2-A^2\right)-\frac{A^2 \lambda ^2}{2}+\Sigma +u^2-\frac{\sqrt{2} u}{\lambda }+w^2+1\right).
\end{align}
\end{small}
\end{subequations}

The equilibrium points of the systems   \eqref{scalar-field-C} for $\varepsilon=\pm 1$ are the following:
\begin{enumerate}
    \item[$N_1$:] $(\Sigma,A,u,w)=(0,0,0,0)$ with eigenvalues $\{-1,-1,1,1\}$ is a hyperbolic saddle.
   \item[$Q_{15,16}$:] $(\Sigma,A,u,w)=\left(\Sigma_0 ,0,\varepsilon \sqrt{1-C_2 \Sigma_0 ^2},0\right)$ with eigenvalues\\ $\left\{0,2,2 (\Sigma_0 +1),-\varepsilon \frac{\sqrt{2-2 C_2 \Sigma_0 ^2}}{\lambda }+\Sigma_0
   +2\right\}$.  This line of equilibrium points contains the equilibrium points $P_1,P_2,P_3,P_4$ studied in section \ref{model3}. It exists for  $\lambda >0, 0<C_2\leq \frac{1}{\Sigma ^2}$ y $\Sigma \in \mathbb{R}$. This line is a normally-hyperbolic invariant set. Indeed, the parametric curve can be expressed as: 
    \begin{align*}
  \textbf{r}(\Sigma_0)= \left(\Sigma_0, 0,  \varepsilon\sqrt{1-C_2 \Sigma_0 ^2}, 0\right),
 \end{align*}
Its tangent vector evaluated at $\Sigma_0$: $$ \textbf{r}'(\Sigma_0)=\left(1,0,-\frac{\varepsilon C_2 \Sigma_0}{\sqrt{1-C_2 \Sigma_0 ^2}},0\right),$$  is parallel to the eigenvector corresponding to the zero eigevalue:  $$\textbf{v}(\Sigma_0)=\left(-\frac{\sqrt{1-C_2 \Sigma_0 ^2}}{\varepsilon C_2 \Sigma_0},0,1,0\right).$$ In this particular case, the stability can be studied considering only the signs of the real parts of the non-zero eigenvalues. In this way it is concluded that:
   \begin{enumerate}
       \item $Q_{15}$  is a hyperbolic source for: 
       \begin{enumerate}
           \item $C_2>0,  \Sigma_0=\frac{1}{\sqrt{C_2}},  \lambda >0,  1<\gamma <2$, or 
           \item $C_2>1,  \Sigma_0=-\frac{1}{\sqrt{C_2}},  \lambda >0,  1<\gamma <2$, or 
           \item $0<C_2\leq 1,  -1<\Sigma_0<\sqrt{\frac{1}{C_2}}$, \newline $ \lambda >\sqrt{2} \sqrt{\frac{1-C_2 \Sigma_0^2}{(\Sigma_0+2)^2}},  1<\gamma <2$, or 
           \item $C_2>1,    -\sqrt{\frac{1}{C_2}}<\Sigma_0<\sqrt{\frac{1}{C_2}}$, \newline $  \lambda >\sqrt{2} \sqrt{\frac{1-C_2 \Sigma_0^2}{(\Sigma_0+2)^2}},  1<\gamma <2$, or 
           \item $C_2\leq 0, 
   \Sigma_0>-1$, \newline $  \lambda >\sqrt{2} \sqrt{\frac{1-C_2 \Sigma_0^2}{(\Sigma_0+2)^2}},  1<\gamma <2$
       \end{enumerate}
       \item $Q_{15}$  is a hyperbolic saddle for:  
       \begin{enumerate}
           \item $\frac{1}{4}<C_2<1, \Sigma_0=-\frac{1}{\sqrt{C_2}}, \lambda >0, 1<\gamma <2$, or  
           \item $0<C_2\leq \frac{1}{4}, -2<\Sigma_0<-1$, \newline $ \lambda >\sqrt{2}
   \sqrt{\frac{1-C_2 \Sigma_0^2}{(\Sigma_0+2)^2}}, 1<\gamma <2$, or  
   \item $\frac{1}{4}<C_2<1, -\sqrt{\frac{1}{C_2}}<\Sigma_0<-1$, \newline $ \lambda >\sqrt{2}
   \sqrt{\frac{1-C_2 \Sigma_0^2}{(\Sigma_0+2)^2}}, 1<\gamma <2$, or  
   \item $0<C_2\leq 1, -1<\Sigma_0<\sqrt{\frac{1}{C_2}}$, \newline $ 0<\lambda <\sqrt{2} \sqrt{\frac{1-C_2
   \Sigma_0^2}{(\Sigma_0+2)^2}}, 1<\gamma <2$, or  
   \item $C_2>1, -\sqrt{\frac{1}{C_2}}<\Sigma_0<\sqrt{\frac{1}{C_2}}$, \newline $ 0<\lambda <\sqrt{2} \sqrt{\frac{1-C_2
   \Sigma_0^2}{(\Sigma_0+2)^2}}, 1<\gamma <2$, or  
   \item $0<C_2<\frac{1}{4}, -\sqrt{\frac{1}{C_2}}\leq \Sigma_0\leq -2$, \newline $ \lambda >0, 1<\gamma <2$, or  
   \item $\frac{1}{4}\leq C_2<1, -\sqrt{\frac{1}{C_2}}<\Sigma_0<-1$, \newline $ 0<\lambda <\sqrt{2} \sqrt{\frac{1-C_2 \Sigma_0^2}{(\Sigma_0+2)^2}}, 1<\gamma <2$, or  
   \item $0<C_2<\frac{1}{4}, -2<\Sigma_0<-1$, \newline $ 0<\lambda <\sqrt{2} \sqrt{\frac{1-C_2 \Sigma_0^2}{(\Sigma_0+2)^2}}, 1<\gamma <2$.
       \end{enumerate}
       \item $Q_{15}$ is non-hyperbolic for: 
       \begin{enumerate}
           \item $C_2=\frac{1}{4},  \Sigma_0=-2$, \newline $  \lambda >0,  1<\gamma <2$, or  
           \item $C_2=\frac{1}{4},  \Sigma_0=-1$, \newline $  \lambda >0,  1<\gamma <2$, or  
           \item $0<C_2<\frac{1}{4},  \Sigma_0=-1$, \newline $  \lambda >0,  1<\gamma <2$, or  
   \item $\frac{1}{4}<C_2\leq 1,  \Sigma_0=-1$, \newline $  \lambda >0,  1<\gamma <2$, or  
   \item $C_2>1,  -\sqrt{\frac{1}{C_2}}<\Sigma_0<\sqrt{\frac{1}{C_2}}$, \newline $\lambda =\sqrt{\frac{2(1-C_2 \Sigma_0^2)}{(\Sigma_0+2)^2}},  1<\gamma <2$, or  
   \item $C_2=\frac{1}{4},  -2<\Sigma_0<-1$, \newline $  \lambda =\frac{1}{2} \sqrt{\frac{8}{\Sigma_0+2}-2},  1<\gamma <2$, or  
   \item $C_2=\frac{1}{4},  -1<\Sigma_0<2$, \newline $  \lambda
   =\frac{1}{2} \sqrt{\frac{8}{\Sigma_0+2}-2},  1<\gamma <2$, or  
   \item $0<C_2<\frac{1}{4},  -1<\Sigma_0<\sqrt{\frac{1}{C_2}}$, \newline $  \lambda =\sqrt{\frac{2(1-C_2 \Sigma_0^2)}{(\Sigma_0+2)^2}},  1<\gamma <2$, or  
   \item $0<C_2<\frac{1}{4},  -2<\Sigma_0<-1 $, \newline $  \lambda =\sqrt{\frac{2(1-C_2 \Sigma_0^2)}{(\Sigma_0+2)^2}},  1<\gamma
   <2$, or  
   \item $\frac{1}{4}<C_2\leq 1,  -1<\Sigma_0<\sqrt{\frac{1}{C_2}}$, \newline $ \lambda =\sqrt{\frac{2(1-C_2 \Sigma_0^2)}{(\Sigma_0+2)^2}},  1<\gamma <2$, or  
   \item $\frac{1}{4}<C_2<1,  -\sqrt{\frac{1}{C_2}}<\Sigma_0<-1$, \newline $  \lambda =\sqrt{\frac{2(1-C_2 \Sigma_0^2)}{(\Sigma_0+2)^2}},  1<\gamma <2$.
       \end{enumerate}
   \end{enumerate}
Furthermore,
   \begin{enumerate}
       \item $Q_{16}$ is a hyperbolic source for: 
       \begin{enumerate}
           \item $C_2\leq 0, \Sigma_0>-1, \lambda >0, 1<\gamma <2$, or  
           \item $0<C_2\leq 1, -1<\Sigma_0\leq \sqrt{\frac{1}{C_2}}$, \newline $ \lambda >0, 1<\gamma <2$, or  
           \item $C_2>1,
   -\sqrt{\frac{1}{C_2}}\leq \Sigma_0 \leq \sqrt{\frac{1}{C_2}}$, \newline $ \lambda >0, 1<\gamma <2$.
       \end{enumerate}
       \item $Q_{16}$ is a hyperbolic saddle for: 
       \begin{enumerate}
           \item $C_2=\frac{1}{4}, -2<\Sigma_0<-1$, \newline $ \lambda >0, 1<\gamma <2$, or  
           \item $0<C_2<\frac{1}{4}, -\sqrt{\frac{1}{C_2}}<\Sigma_0<-2$, \newline $ 0<\lambda <\sqrt{2}
   \sqrt{\frac{1-C_2 \Sigma_0^2}{(\Sigma_0+2)^2}}, 1<\gamma <2$, or  \item $\frac{1}{4}<C_2<1, -\sqrt{\frac{1}{C_2}}\leq \Sigma_0<-1$, \newline $ \lambda >0, 1<\gamma
   <2$, or  
   \item $0<C_2<\frac{1}{4}, -2\leq \Sigma_0<-1$, \newline $ \lambda >0, 1<\gamma <2$, or  
   \item $0<C_2<\frac{1}{4}, \Sigma_0=-\frac{1}{\sqrt{C_2}}$, \newline $ \lambda >0,
   1<\gamma <2$, or  
   \item $0<C_2<\frac{1}{4}, -\sqrt{\frac{1}{C_2}}<\Sigma_0<-2$, \newline $ \lambda >\sqrt{2} \sqrt{\frac{1-C_2 \Sigma_0^2}{(\Sigma_0+2)^2}}, 1<\gamma <2$.
       \end{enumerate}
       \item $Q_{16}$ is non-hyperbolic for 
       \begin{enumerate}
           \item $C_2=\frac{1}{4},  \Sigma_0=-2,  \lambda >0,  1<\gamma <2$, or  
           \item $0<C_2\leq 1,  \Sigma_0=-1,  \lambda >0,  1<\gamma <2$, or  
           \item $0<C_2<\frac{1}{4}, 
   -\sqrt{\frac{1}{C_2}}<\Sigma_0<-2$, \newline $  \lambda =\sqrt{\frac{2-2 C_2 \Sigma_0^2}{(\Sigma_0+2)^2}},  1<\gamma <2$.
       \end{enumerate}
   \end{enumerate} 
   \item[$Q_{17,18}$:] $(\Sigma,A,u,w)=\left(-\frac{1}{2}, -\varepsilon \frac{1}{2},\frac{\lambda }{2 \sqrt{2}},0\right)$ where one eigenvalue is zero and the other three eigenvalues are the roots of the polynomial in $\mu$: \newline $P(\mu)=\mu ^3+\mu ^2+\mu  \left(\frac{\lambda ^2}{2 C_2}-C_2-\frac{\lambda
   ^2}{2}+1\right)-\frac{\lambda ^2}{2 C_2}-C_2+\frac{\lambda
   ^2}{2}+1$. It is either a non-hyperbolic saddle or a center depending on the choice of parameters. Figure \ref{P4eigen} graphically represents the real part of the $\mu_i $ corresponding to the equilibrium point $Q_{17,18}$,  illustrating that the equilibrium points have saddle behavior or they are non-hyperbolic.

\begin{figure*}[!htb]
    \centering
    \includegraphics[scale=0.4]{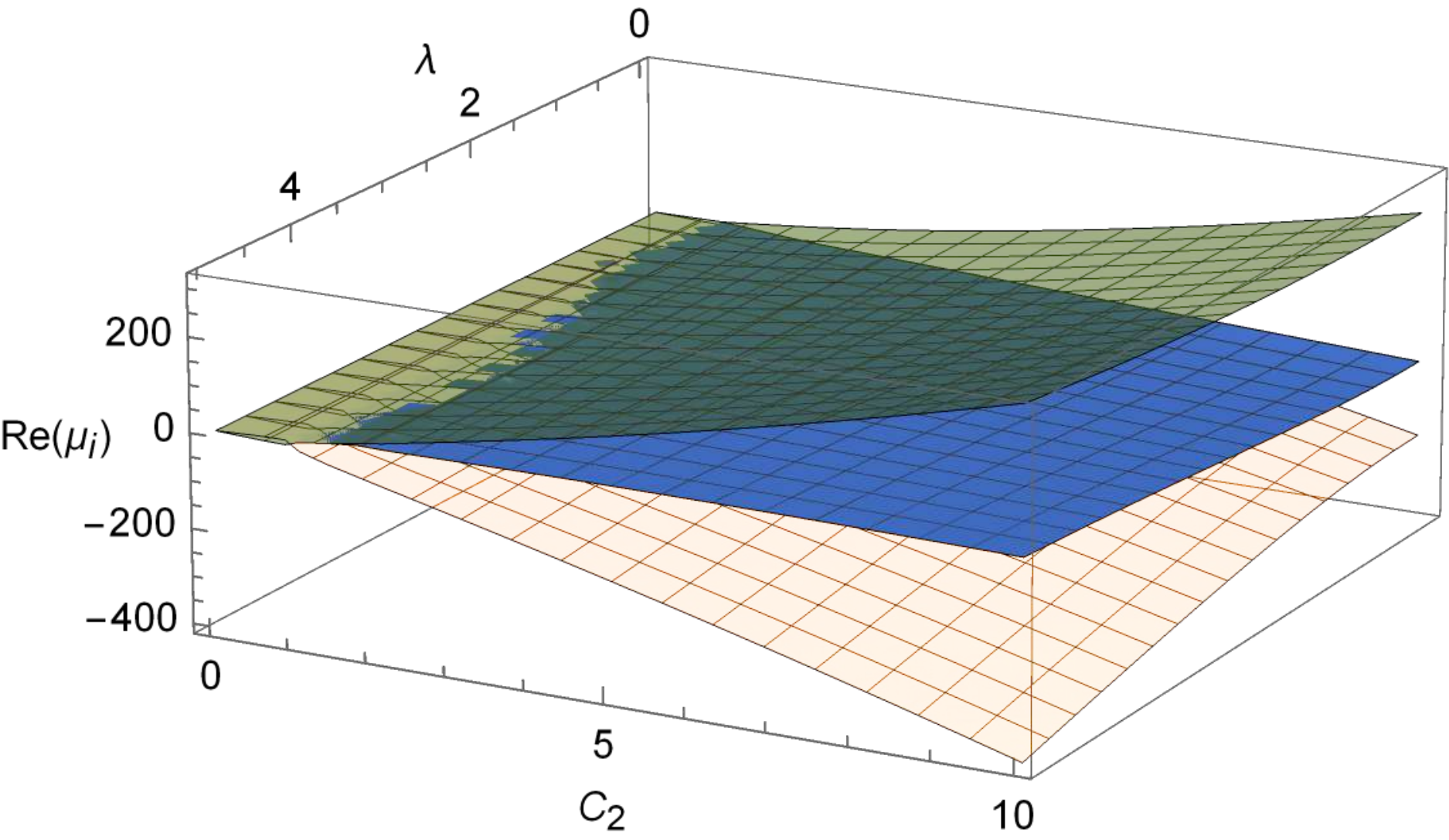}  \caption[Real parts of the $\mu_i$'s corresponding to  $Q_{17,18}:(\Sigma,A,u,w)=\left(-\frac{1}{2},-\frac{1}{2} \varepsilon ,\frac{\lambda }{2 \sqrt{2}},0\right)$.]{\label{P4eigen} Real parts of the $\mu_i$'s corresponding to  $Q_{17,18}:(\Sigma,A,u,w)=\left(-\frac{1}{2},-\frac{1}{2} \varepsilon ,\frac{\lambda }{2 \sqrt{2}},0\right)$.}
   \end{figure*}
   
   \item[$Q_{19}$:] $(\Sigma,A,u,w)=\left(-\frac{1}{2 C_2},0,\frac{1}{\sqrt{2} \lambda },  \frac{1}{2} 
   \sqrt{\frac{1}{C_2}+\frac{2}{\lambda ^2}-4}\right)$ with eigenvalues $\Big\{\frac{1}{\lambda ^2}-\frac{1}{2 C_2}, \frac{1}{2} \left(\frac{1}{C_2}+\frac{2}{\lambda
   ^2}-4\right)$, \newline $\frac{1}{2} \left(\frac{1}{C_2}+\frac{2}{\lambda
   ^2}-4\right), \frac{1}{C_2}+\frac{2}{\lambda ^2}-2\Big\}$. \\ It exists for $0<\lambda \leq \frac{1}{\sqrt{2}}, C_2>0$, or $\lambda
   >\frac{1}{\sqrt{2}}, 0<C_2\leq \frac{\lambda ^2}{4 \lambda ^2-2}$.
   \begin{enumerate}
   \item is a hyperbolic source for:
   \begin{enumerate}
       \item $0<\lambda \leq \frac{1}{\sqrt{2}}, C_2>\frac{\lambda ^2}{2}$, or   
       \item $\frac{1}{\sqrt{2}}<\lambda <1, \frac{\lambda ^2}{2}<C_2<\frac{\lambda
   ^2}{4 \lambda ^2-2}$.
   \end{enumerate}
   \item is a hyperbolic saddle for: 
   \begin{enumerate}
       \item $\lambda >1, 0<C_2<\frac{\lambda ^2}{4 \lambda ^2-2}$, or   
       \item $0<\lambda \leq 1,  0<C_2<\frac{\lambda ^2}{2}$.
   \end{enumerate}
   \item is non-hyperbolic for: 
   \begin{enumerate}
       \item $C_2=\frac{1}{2},  \lambda =1$, or   
       \item $C_2=\frac{\lambda ^2}{2},  0<\lambda <1$, or   
       \item $C_2=\frac{\lambda ^2}{4 \lambda ^2-2},  \frac{1}{\sqrt{2}}<\lambda <1$, or    
       \item $C_2=\frac{\lambda ^2}{4 \lambda ^2-2},  \lambda >1$.
   \end{enumerate}
   \end{enumerate}
   \item[$Q_{20,21}$:] $(\Sigma,A,u,w)=\left(-\frac{1}{2}, \varepsilon \frac{\sqrt{C_2}}{\sqrt{2} \lambda },\frac{C_2}{\sqrt{2}
   \lambda },0\right)$ with eigenvalues \\$\Big\{\frac{1}{2}-\frac{C_2}{\lambda ^2}, \frac{\sqrt{2} \sqrt{C_2}}{\lambda
   }-1, -\frac{1}{2} \left(1+\frac{\sqrt{8 C_2^2+4 C_2 \lambda ^2-7 \lambda
   ^2}}{\lambda }\right)$, \newline $-\frac{1}{2} \left(1-\frac{\sqrt{8 C_2^2+4 C_2 \lambda ^2-7 \lambda
   ^2}}{\lambda }\right)\Big\}$. They are:
   \begin{enumerate}
       \item hyperbolic saddles for: 
       \begin{enumerate}
       \item $0<\lambda \leq \frac{\sqrt{7}}{2}$, \newline $ \frac{1}{4} \sqrt{\lambda ^4+14 \lambda
   ^2}-\frac{\lambda ^2}{4}<C_2<\frac{1}{4} \sqrt{\lambda ^4+16 \lambda
   ^2}-\frac{\lambda ^2}{4}$, or   
       \item $\frac{\sqrt{7}}{2}<\lambda <\sqrt{2}, \frac{\lambda
   ^2}{2}<C_2<\frac{1}{4} \sqrt{\lambda ^4+16 \lambda ^2}-\frac{\lambda
   ^2}{4}$, or   
       \item $0<\lambda <\frac{\sqrt{7}}{2}$, \newline $ \frac{\lambda ^2}{2}<C_2\leq
   \frac{1}{4} \sqrt{\lambda ^4+14 \lambda ^2}-\frac{\lambda ^2}{4}$, or    
        \item $\lambda >\sqrt{2}, C_2>\frac{\lambda ^2}{2}$
        \item $0<\lambda \leq \sqrt{2}, C_2>\frac{1}{4} \sqrt{\lambda ^4+16 \lambda
   ^2}-\frac{\lambda ^2}{4}$, or    
        \item $\lambda >\sqrt{2}, \frac{1}{4} \sqrt{\lambda ^4+16 \lambda ^2}-\frac{\lambda
   ^2}{4}<C_2<\frac{\lambda ^2}{2}$, or   
        \item $\frac{\sqrt{7}}{2}<\lambda \leq \sqrt{2}$, \newline $ \frac{1}{4} \sqrt{\lambda ^4+14
   \lambda ^2}-\frac{\lambda ^2}{4}<C_2<\frac{\lambda ^2}{2}$, or    
        \item $\lambda >\sqrt{2}, \frac{1}{4} \sqrt{\lambda ^4+14 \lambda ^2}-\frac{\lambda
   ^2}{4}<C_2<\frac{1}{4} \sqrt{\lambda ^4+16 \lambda ^2}-\frac{\lambda
   ^2}{4}$, or  
         \item $0<\lambda \leq \frac{\sqrt{7}}{2}$,  $ 0<C_2<\frac{\lambda ^2}{2}$
        \item $\lambda >\frac{\sqrt{7}}{2}$,   $ 0<C_2\leq \frac{1}{4} \sqrt{\lambda ^4+14
   \lambda ^2}-\frac{\lambda ^2}{4}$.
       \end{enumerate}
       \item non-hyperbolic for: 
       \begin{enumerate}
           \item $\lambda >0, C_2=\frac{\lambda ^2}{2}$, or   
           \item $\lambda >0, C_2=\frac{1}{4} \sqrt{\lambda ^4+16 \lambda
   ^2}-\frac{\lambda ^2}{4}$.
       \end{enumerate}
   \end{enumerate}
    \item[$Q_{22,23}$:] $(\Sigma,A,u,w)=\left(-\frac{1}{2}, -\varepsilon \frac{\sqrt{C_2}}{\sqrt{2} \lambda },\frac{C_2}{\sqrt{2}
   \lambda },0\right)$ with eigenvalues \newline $\Big\{\frac{1}{2}-\frac{C_2}{\lambda ^2}, -\frac{\sqrt{2} \sqrt{C_2}}{\lambda
   }-1, -\frac{1}{2} \left(1+\frac{\sqrt{8 C_2^2+4 C_2 \lambda ^2-7 \lambda
   ^2}}{\lambda }\right)$, \newline $ -\frac{1}{2} \left(1-\frac{\sqrt{8 C_2^2+4 C_2 \lambda ^2-7 \lambda
   ^2}}{\lambda }\right)\Big\}$. They are: 
   \begin{enumerate}
   \item hyperbolic sinks for: 
   \begin{enumerate}
       \item $0<\lambda \leq \frac{\sqrt{7}}{2}$, \newline $ \frac{1}{4} \sqrt{\lambda ^4+14 \lambda
   ^2}-\frac{\lambda ^2}{4}<C_2<\frac{1}{4} \sqrt{\frac{\lambda ^6+14
   \lambda ^4+2}{\lambda ^2}}-\frac{\lambda ^2}{4}$, or   
       \item $\frac{\sqrt{7}}{2}<\lambda <1.35095$, \newline $ \frac{\lambda
   ^2}{2}<C_2<\frac{1}{4} \sqrt{\frac{\lambda ^6+14 \lambda ^4+2}{\lambda
   ^2}}-\frac{\lambda ^2}{4}$, or   
       \item $0<\lambda <\frac{\sqrt{7}}{2}$,   $ \frac{\lambda ^2}{2}<C_2\leq \frac{1}{4}
   \sqrt{\lambda ^4+14 \lambda ^2}-\frac{\lambda ^2}{4}$.
   \end{enumerate}
       \item hyperbolic saddles for:
       \begin{enumerate}
           \item $0<\lambda \leq \frac{\sqrt{7}}{2}$, \newline $ \frac{1}{4} \sqrt{\lambda ^4+14 \lambda
   ^2}-\frac{\lambda ^2}{4}<C_2<\frac{1}{4} \sqrt{\lambda ^4+16 \lambda
   ^2}-\frac{\lambda ^2}{4}$, or   
           \item $\frac{\sqrt{7}}{2}<\lambda <\sqrt{2}, \frac{\lambda
   ^2}{2}<C_2<\frac{1}{4} \sqrt{\lambda ^4+16 \lambda ^2}-\frac{\lambda
   ^2}{4}$, or   
           \item $0<\lambda <\frac{\sqrt{7}}{2}$, \newline $ \frac{\lambda ^2}{2}<C_2\leq
   \frac{1}{4} \sqrt{\lambda ^4+14 \lambda ^2}-\frac{\lambda ^2}{4}$, or   
           \item $0<\lambda \leq \sqrt{2}, C_2>\frac{1}{4} \sqrt{\lambda ^4+16 \lambda
   ^2}-\frac{\lambda ^2}{4}$, or   
            \item $\lambda >\sqrt{2}, C_2>\frac{\lambda ^2}{2}$, or   
            \item $\lambda >\sqrt{2}, \frac{1}{4} \sqrt{\lambda ^4+16 \lambda ^2}-\frac{\lambda
^2}{4}<C_2<\frac{\lambda ^2}{2}$, or   
            \item $\frac{\sqrt{7}}{2}<\lambda \leq \sqrt{2}, \frac{1}{4} \sqrt{\lambda ^4+14
   \lambda ^2}-\frac{\lambda ^2}{4}<C_2<\frac{\lambda ^2}{2}$, or   
            \item $\lambda >\sqrt{2}$, \newline $\frac{1}{4} \sqrt{\lambda ^4+14 \lambda ^2}-\frac{\lambda
   ^2}{4}<C_2<\frac{1}{4} \sqrt{\lambda ^4+16 \lambda ^2}-\frac{\lambda
   ^2}{4}$, or   
             \item $0<\lambda \leq \frac{\sqrt{7}}{2}$, \newline $ 0<C_2<\frac{\lambda ^2}{2}$, or   
             \item $\lambda >\frac{\sqrt{7}}{2}, 0<C_2\leq \frac{1}{4} \sqrt{\lambda ^4+14
   \lambda ^2}-\frac{\lambda ^2}{4}$.
             \end{enumerate}
             \item non-hyperbolic for: 
                \begin{enumerate}
                    \item $\lambda >0, C_2=\frac{\lambda ^2}{2}$, or   
                    \item $\lambda >0, C_2=\frac{1}{4} \sqrt{\lambda ^4+16 \lambda
   ^2}-\frac{\lambda ^2}{4}$.
                \end{enumerate}
   \end{enumerate}
   \item[$Q_{24,25}$:] $(\Sigma,A,u,w)=\left(-\frac{\lambda ^2}{2 C_2^2+C_2 \lambda ^2},-\varepsilon \delta,\frac{\sqrt{2} \lambda }{2 C_2+\lambda ^2},0\right)$ with eigenvalues $\Big\{\frac{1}{C_2}-\frac{4}{2
   C_2+\lambda ^2},-\frac{8}{2
   C_2+\lambda
   ^2}+\frac{4}{C_2}-2$, \newline $ -\frac{4}{2
   C_2+\lambda ^2}+\frac{2}{C_2}-2,2
  \delta+\frac{2 \lambda ^2}{C_2
   \left(2 C_2+\lambda ^2\right)}-2\Big\}$,\\ where $\delta=\sqrt{\frac{4 C_2^2+2 (C_2-1)
   \lambda ^2}{C_2 \left(2 C_2+\lambda
   ^2\right)^2}}$. They exist for: \newline $\lambda >0, C_2\geq \frac{1}{4} \sqrt{\lambda ^4+8 \lambda ^2}-\frac{\lambda
   ^2}{4}$. They are:  
   \begin{enumerate}
       \item hyperbolic sinks for: 
       \begin{enumerate}
           \item $0<\lambda \leq \sqrt{2},  C_2>1$, or  
           \item $\lambda >\sqrt{2},  C_2>\frac{\lambda ^2}{2}$.
       \end{enumerate}
       \item hyperbolic saddles for:
       \begin{enumerate}
           \item $0<\lambda <\sqrt{2},  \frac{1}{4} \sqrt{\lambda ^4+16 \lambda
   ^2}-\frac{\lambda ^2}{4}<C_2<1$, or   
   \item $0<\lambda \leq 1,  \frac{1}{4} \sqrt{\lambda ^4+8 \lambda ^2}-\frac{\lambda
   ^2}{4}<C_2<\frac{1}{4} \sqrt{\lambda ^4+16 \lambda ^2}-\frac{\lambda
   ^2}{4}$, or   
   \item $1<\lambda <\sqrt{2},  \frac{\lambda ^2}{2}<C_2<\frac{1}{4} \sqrt{\lambda
   ^4+16 \lambda ^2}-\frac{\lambda ^2}{4}$o, 
   \item $\lambda >\sqrt{2},  \frac{1}{4} \sqrt{\lambda ^4+16 \lambda ^2}-\frac{\lambda
   ^2}{4}<C_2<\frac{\lambda ^2}{2}$, or   
   \item $\lambda >\sqrt{2},  1<C_2<\frac{1}{4} \sqrt{\lambda ^4+16 \lambda
   ^2}-\frac{\lambda ^2}{4}$.
       \end{enumerate}
       \item non-hyperbolic for
       \begin{enumerate}
           \item $\lambda \geq 1,  C_2=\frac{\lambda ^2}{2}$, or   
           \item $\lambda >0,  C_2=\frac{1}{4} \sqrt{\lambda ^4+8 \lambda
   ^2}-\frac{\lambda ^2}{4}$, or   
   \item $\lambda >0, C_2=\frac{1}{4} \sqrt{\lambda ^4+16 \lambda
   ^2}-\frac{\lambda ^2}{4}$, or   
   \item $\lambda >0,  C_2=1$.
       \end{enumerate}
   \end{enumerate}
   \item[$Q_{26,27}$:] $(\Sigma,A,u,w)=\left(-\frac{\lambda ^2}{2 C_2^2+C_2 \lambda ^2},\varepsilon \delta,\frac{\sqrt{2} \lambda }{2 C_2+\lambda ^2},0\right)$  with eigenvalues $\Big\{\frac{1}{C_2}-\frac{4}{2
   C_2+\lambda ^2},-\frac{8}{2
   C_2+\lambda
   ^2}+\frac{4}{C_2}-2$, \newline $ -\frac{4}{2
   C_2+\lambda ^2}+\frac{2}{C_2}-2,-2
   \delta+\frac{2 \lambda ^2}{C_2
   \left(2 C_2+\lambda ^2\right)}-2\Big\}$,\\where $\delta=\sqrt{\frac{4 C_2^2+2 (C_2-1)
   \lambda ^2}{C_2 \left(2 C_2+\lambda
   ^2\right)^2}}$. They exist for: \newline $\lambda >0, C_2\geq \frac{1}{4} \sqrt{\lambda ^4+8 \lambda ^2}-\frac{\lambda
   ^2}{4}$. They are: 
      \begin{enumerate}
       \item hyperbolic sinks for:  
       \begin{enumerate}
           \item $0<\lambda \leq \sqrt{2},  C_2>\frac{1}{4} \sqrt{\lambda ^4+16 \lambda
   ^2}-\frac{\lambda ^2}{4}$, or   
           \item $\lambda >\sqrt{2},  C_2>\frac{\lambda ^2}{2}$. \end{enumerate}
       \item hyperbolic saddles for:
       \begin{enumerate}
           \item $0<\lambda \leq 1$, \newline $ \frac{1}{4} \sqrt{\lambda ^4+8 \lambda ^2}-\frac{\lambda
   ^2}{4}<C_2<\frac{1}{4} \sqrt{\lambda ^4+16 \lambda ^2}-\frac{\lambda
   ^2}{4}$, or    
           \item $1<\lambda <\sqrt{2},  \frac{\lambda ^2}{2}<C_2<\frac{1}{4} \sqrt{\lambda
   ^4+16 \lambda ^2}-\frac{\lambda ^2}{4}$, or   
           \item $\lambda >\sqrt{2},  \frac{1}{4} \sqrt{\lambda ^4+16 \lambda ^2}-\frac{\lambda
   ^2}{4}<C_2<\frac{\lambda ^2}{2}$, or   
           \item $1<\lambda \leq \sqrt{2},  \frac{1}{4} \sqrt{\lambda ^4+8 \lambda
   ^2}-\frac{\lambda ^2}{4}<C_2<\frac{\lambda ^2}{2}$, or   
           \item $\lambda >\sqrt{2}$, \newline $ \frac{1}{4} \sqrt{\lambda ^4+8 \lambda ^2}-\frac{\lambda
   ^2}{4}<C_2<\frac{1}{4} \sqrt{\lambda ^4+16 \lambda ^2}-\frac{\lambda
   ^2}{4}$.
       \end{enumerate}
       \item non-hyperbolic for:
       \begin{enumerate}
           \item $\lambda \geq 1,  C_2=\frac{\lambda ^2}{2}$, or   
           \item $\lambda >0, C_2=\frac{1}{4} \sqrt{\lambda ^4+16 \lambda
   ^2}-\frac{\lambda ^2}{4}$, or   
           \item $\lambda >0,  C_2=\frac{1}{4} \sqrt{\lambda ^4+8 \lambda
   ^2}-\frac{\lambda ^2}{4}$.
       \end{enumerate}
   \end{enumerate}
   \item[$Q_{28}$:]  $(\Sigma,A,u,w)=$
 \newline
  $\left(\frac{C_2-1}{\lambda ^2}-\frac{1}{2},\frac{1-C_2}{\lambda
   ^2}-\frac{1}{2},\frac{-2 C_2+\lambda ^2+2}{2 \sqrt{2} \lambda },\frac{ 
   \sqrt{(C_2-1) \left(2 C_2+\lambda ^2-2\right)}}{\lambda }\right)
   $.  Exists for $\lambda >0, C_2>0, 2 C_2+\lambda ^2\leq 2$ or $\lambda >0, C_2\geq 1$ with eigenvalues $\left\{\frac{4-4 C_2}{\lambda ^2},\frac{e_1(\lambda,C_2)}{32 C_2 \lambda ^{12}},\frac{e_2(\lambda,C_2)}{32 C_2 \lambda ^{12}},\frac{e_3(\lambda,C_2)}{32 C_2 \lambda ^{12}}\right\}$. In Figure \ref{P11eigen} the real parts of $ e_i $ are shown, where it can be seen that the point is a hyperbolic saddle or a non-hyperbolic behavior.
   \begin{figure*}
    \centering
    \includegraphics[scale=0.5]{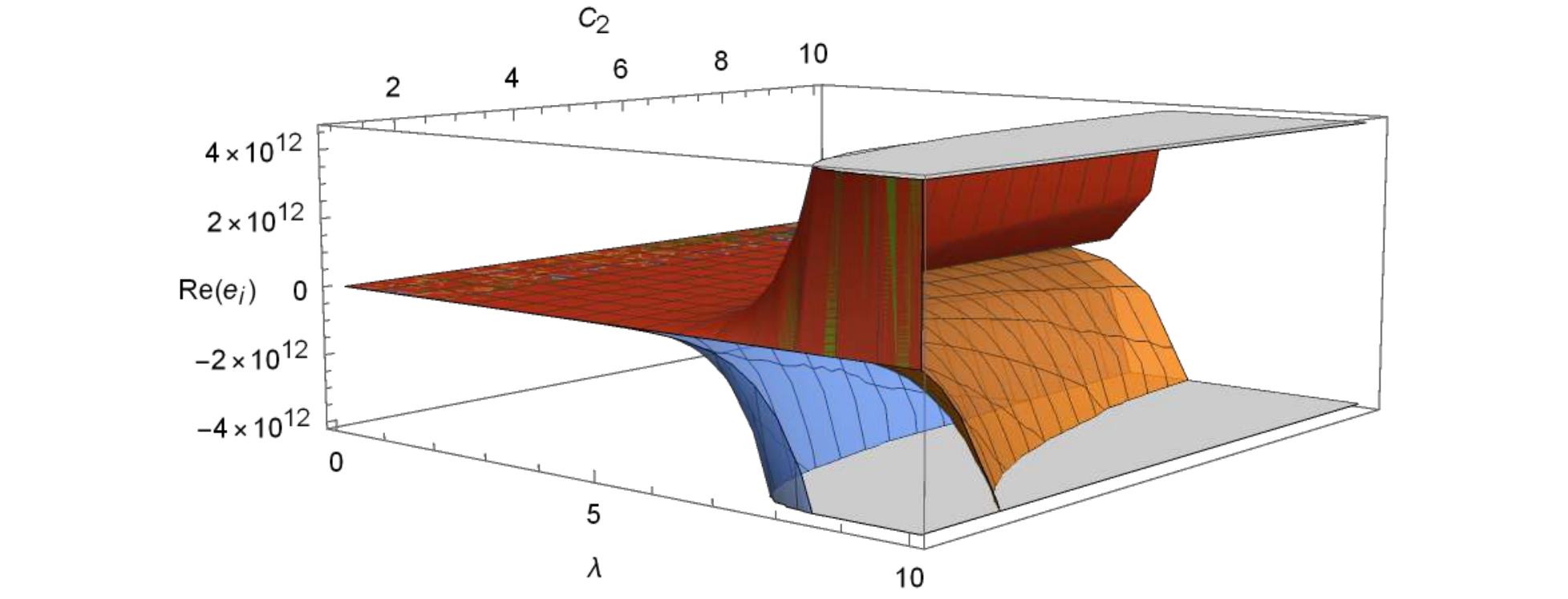}  \caption[Real parts of $e_i$ corresponding to the equilibrium point $Q_{28}: (\Sigma,A,u,w)=\left(\frac{C_2-1}{\lambda ^2}-\frac{1}{2},\frac{1-C_2}{\lambda
   ^2}-\frac{1}{2},\frac{-2 C_2+\lambda ^2+2}{2 \sqrt{2} \lambda },\frac{ 
   \sqrt{(C_2-1) \left(2 C_2+\lambda ^2-2\right)}}{\lambda }\right)$.]{\label{P11eigen}Reals parts of $e_i$ corresponding to the equilibrium point $Q_{28}: (\Sigma,A,u,w)=\left(\frac{C_2-1}{\lambda ^2}-\frac{1}{2},\frac{1-C_2}{\lambda
   ^2}-\frac{1}{2},\frac{-2 C_2+\lambda ^2+2}{2 \sqrt{2} \lambda },\frac{ 
   \sqrt{(C_2-1) \left(2 C_2+\lambda ^2-2\right)}}{\lambda }\right)$.}
   \end{figure*}
   
   \item[$Q_{29}$:] $(\Sigma,A,u,w)= \Big(-\frac{\lambda ^2}{2 C_2+\lambda ^2},0,\frac{\sqrt{2} C_2 \lambda }{2
   C_2+\lambda ^2}$, \newline $\sqrt{-\frac{(1-4 C_2)^2 C_2 \lambda
   ^2 \left(2 (4 C_2-1) \lambda ^2-4
   C_2+\lambda ^4\right)^2}{2
   C_2+\lambda ^2}}\Big)$. It exists for: 
   \begin{enumerate}
       \item $C_2=\frac{1}{4}, \lambda >0$, or 
       \item $C_2=\frac{\lambda ^2 \left(\lambda ^2-2\right)}{4-8 \lambda ^2}, 1<\lambda<\sqrt{2}$, or
      \item  $C_2=\frac{\lambda ^2 \left(\lambda ^2-2\right)}{4-8 \lambda ^2},
   \frac{1}{\sqrt{2}}<\lambda <1$.
   \end{enumerate}  Figure \ref{P12eigen} graphically represents the real part of the eigenvalues of the equilibrium point $Q_{29}: (\Sigma,A,u,w)=$ \newline $\left(-\frac{\lambda ^2}{2 C_2+\lambda ^2},0,\frac{\sqrt{2} C_2 \lambda }{2
   C_2+\lambda ^2}, 0\right)$,  $C_2 \in \left\{\frac{1}{4}, \frac{\lambda ^2 \left(\lambda ^2-2\right)}{4-8 \lambda ^2}\right\}$, for different choices of the parameter $\lambda$.
This figure illustrates that the point has a general saddle behavior.
     \begin{figure*}[!htb]
    \centering
    \includegraphics[scale=0.45]{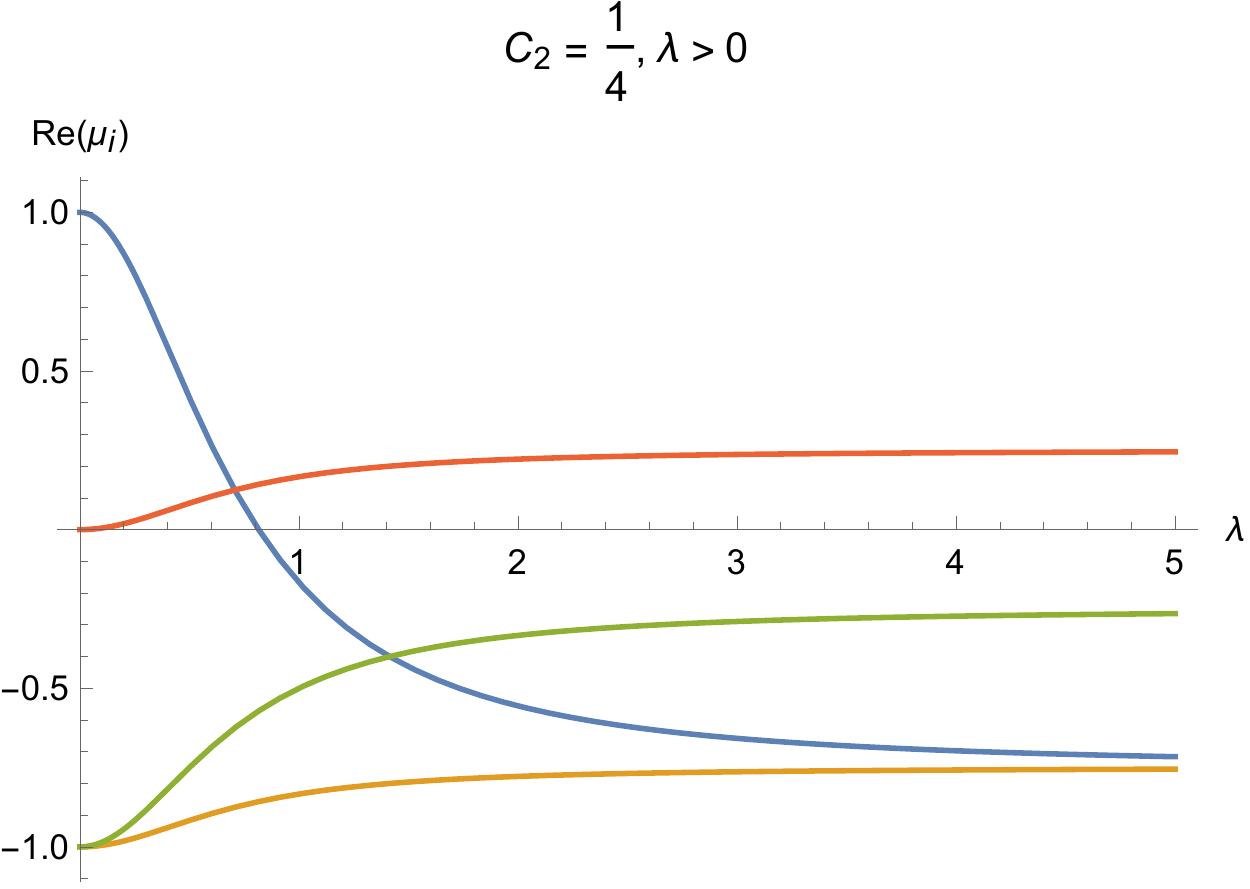}  
    \includegraphics[scale=0.45]{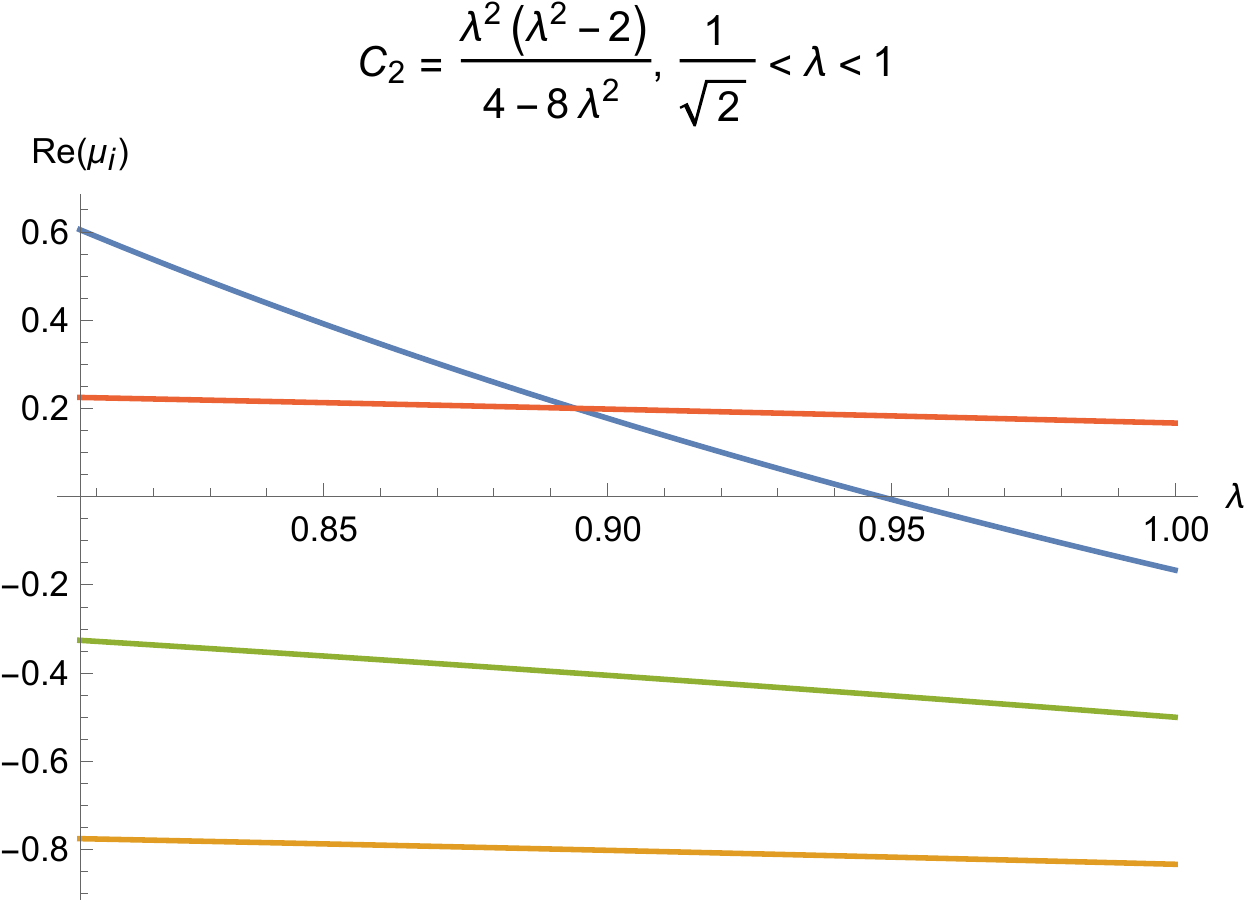}  
    \includegraphics[scale=0.45]{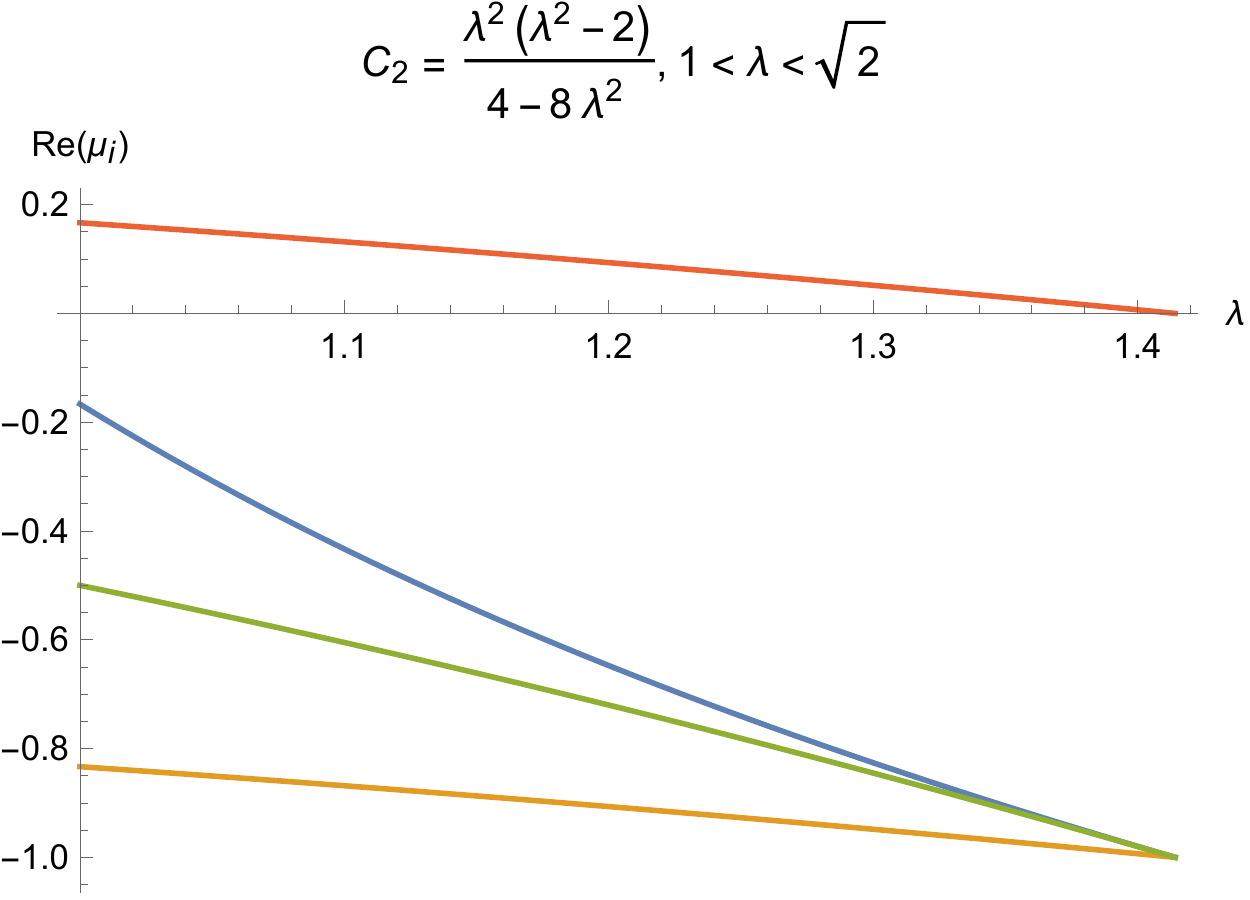}
    \caption[{Real parts of the eigenvalues of the equilibrium point  $(\Sigma,A,u,w)=\left(-\frac{\lambda ^2}{2 C_2+\lambda ^2},0,\frac{\sqrt{2} C_2 \lambda }{2
   C_2+\lambda ^2}, 0\right)$, $C_2 \in \left\{\frac{1}{4}, \frac{\lambda ^2 \left(\lambda ^2-2\right)}{4-8 \lambda ^2}\right\}$.}]{\label{P12eigen} Real parts of the eigenvalues of the equilibrium point  $(\Sigma,A,u,w)=\left(-\frac{\lambda ^2}{2 C_2+\lambda ^2},0,\frac{\sqrt{2} C_2 \lambda }{2
   C_2+\lambda ^2}, 0\right)$, $C_2 \in \left\{\frac{1}{4}, \frac{\lambda ^2 \left(\lambda ^2-2\right)}{4-8 \lambda ^2}\right\}$, for different choices $\lambda$.}
   \end{figure*} 
   \item[$Q_{30}$:] $(\Sigma,A,u,w)=\Big(0,  \frac{1}{\sqrt{C_2-\frac{\lambda ^2}{2}}},0$, \newline $\sqrt{2} \sqrt{-\frac{\lambda ^2
   \left(-2 C_2+\lambda ^2+8\right)^2 \left(2 (2 C_2+1) \lambda ^2+4
   (C_2-1) C_2+\lambda ^4\right)^2}{\lambda ^2-2 C_2}}\Big)$ with $\varepsilon=1$ exists for $\lambda >0, C_2\geq \frac{\lambda ^2}{2}$,  with eigenvalues \\ $\Big\{-\frac{f_1(\lambda,C_2)}{2 C_2 \lambda  \left(\lambda ^2-2 C_2\right)^2}, -\frac{f_2(\lambda,C_2)}{2 C_2 \lambda  \left(\lambda ^2-2 C_2\right)^2}, -\frac{f_3(\lambda,C_2)}{2 C_2 \lambda  \left(\lambda ^2-2 C_2\right)^2}$, \newline $ -\frac{f_4(\lambda,C_2)}{2 C_2 \lambda  \left(\lambda ^2-2 C_2\right)^2}\Big\}$. 
   The real parts of $f_i$'s  are represented in figure  \ref{P13P14eigen} (left panel).
   
   \item[$Q_{31}$:] $(\Sigma,A,u,w)=\Big(0, \frac{1}{\sqrt{C_2-\frac{\lambda ^2}{2}}},0$, \newline $\sqrt{2} \sqrt{-\frac{\lambda ^2
   \left(-2 C_2+\lambda ^2+8\right)^2 \left(2 (2 C_2+1) \lambda ^2+4
   (C_2-1) C_2+\lambda ^4\right)^2}{\lambda ^2-2 C_2}}\Big)$ with $\varepsilon=-1$ exists for $\lambda >0, C_2\geq \frac{\lambda ^2}{2}$ with eigenvalues  \\
   $\Big\{ \frac{g_1(\lambda,C_2)}{C_2 \lambda  \left(2 C_2-\lambda ^2\right)^{3/2}},  \frac{g_2(\lambda,C_2)}{C_2 \lambda  \left(2 C_2-\lambda ^2\right)^{3/2}},  \frac{g_3(\lambda,C_2)}{C_2 \lambda  \left(2 C_2-\lambda ^2\right)^{3/2}}$, \newline $  \frac{g_4(\lambda,C_2)}{C_2 \lambda  \left(2 C_2-\lambda ^2\right)^{3/2}}\Big\}$. The real parts of the   $g_i$'s  are represented in Figure \ref{P13P14eigen} (right panel).
  Summarizing, according to Figure \ref{P13P14eigen} these equilibrium points,  $Q_{30}$ and $Q_{31}$, are either sources or they are non-hyperbolic (with four pure imaginary eigenvalues).

   \begin{figure*}
    \centering
    \includegraphics[scale=0.42]{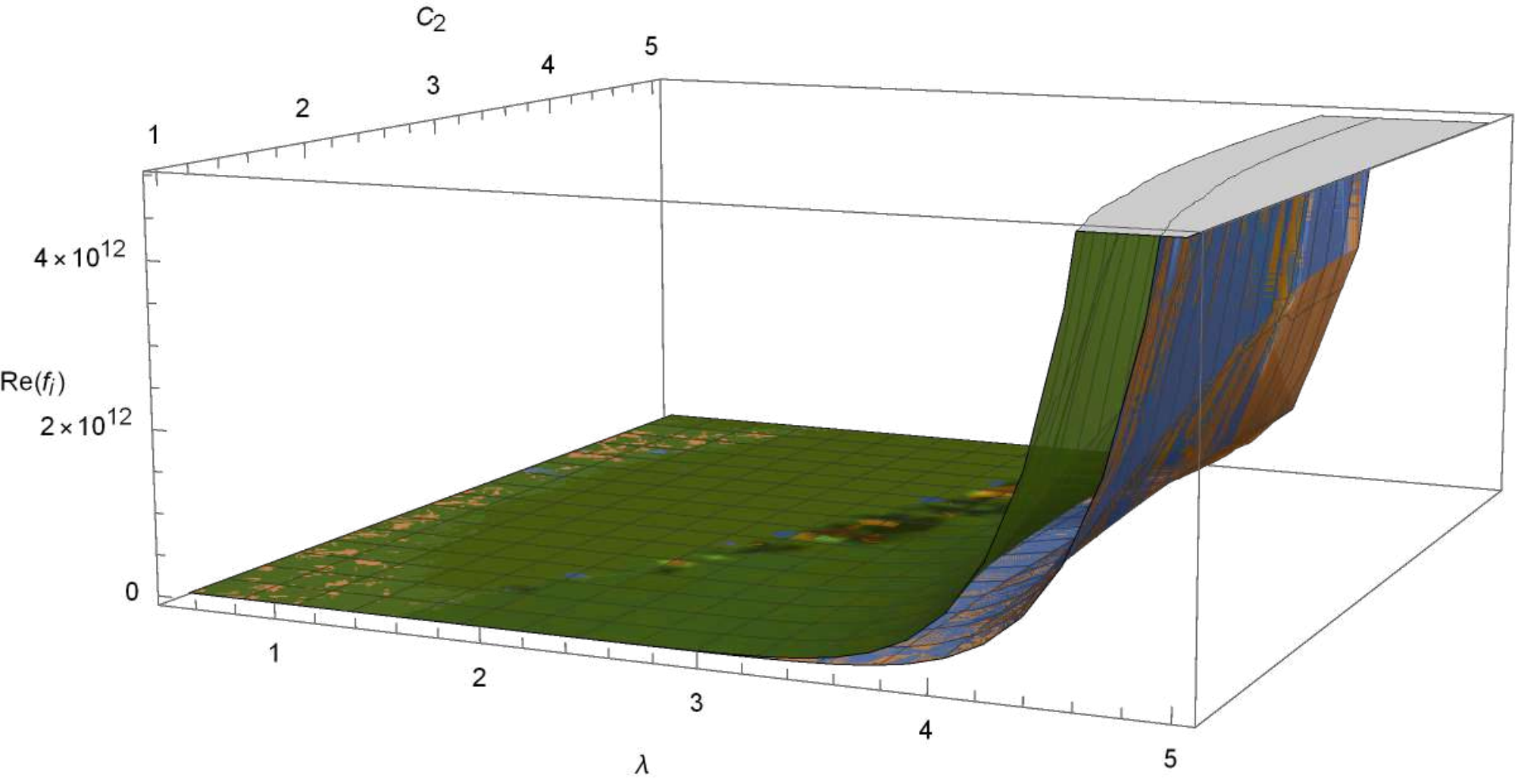} 
    \includegraphics[scale=0.42]{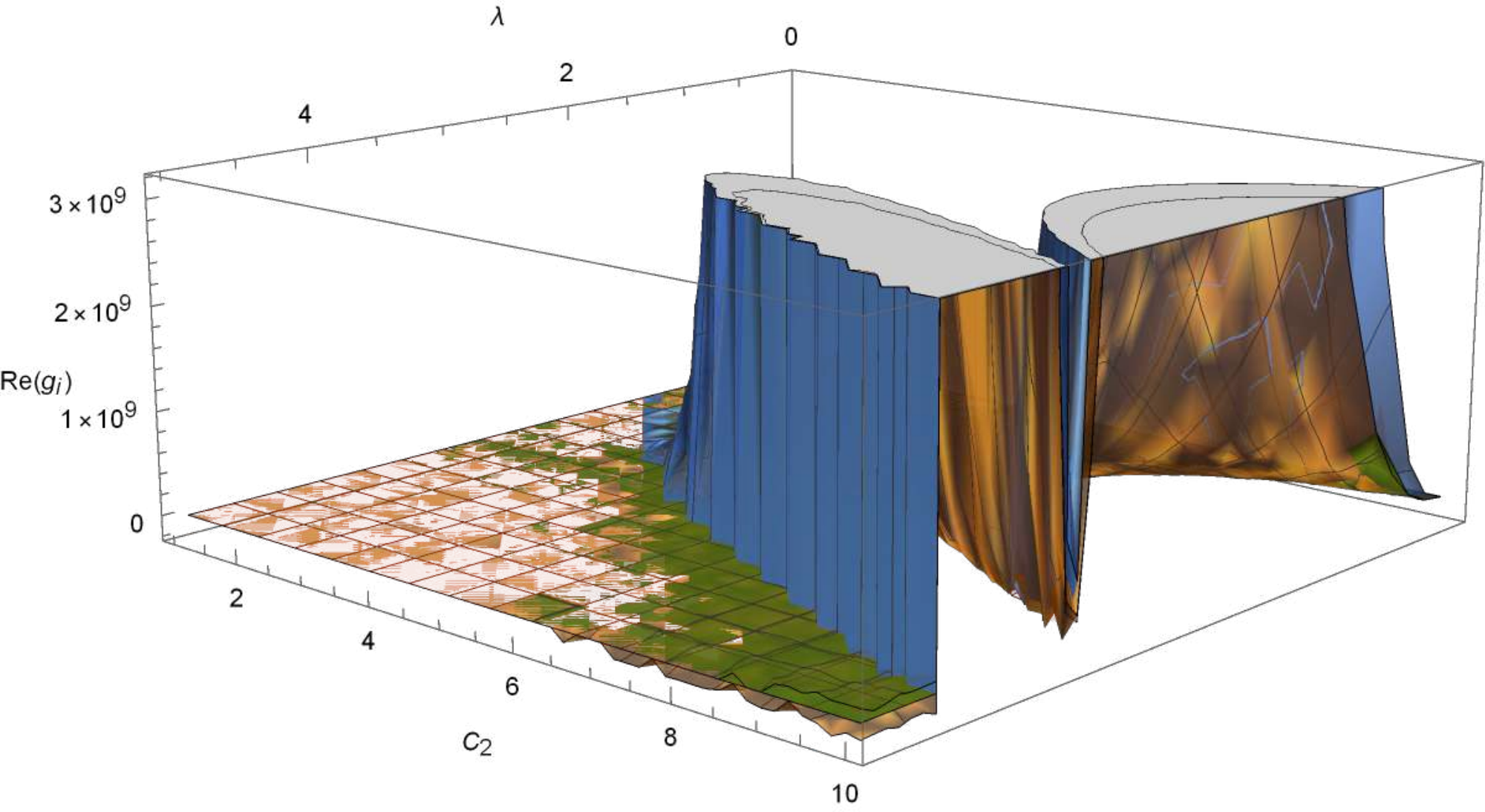}
    \caption[{\small{Real parts of $f_i$ (left panel) and  $g_i$ (right panel) corresponding to the equilibrium points  $Q_{30}: (\Sigma,A,u,w)=\left(0,  \frac{1}{\sqrt{C_2-\frac{\lambda ^2}{2}}},0, \sqrt{2} \sqrt{-\frac{\lambda ^2
   \left(-2 C_2+\lambda ^2+8\right)^2 \left(2 (2 C_2+1) \lambda ^2+4
   (C_2-1) C_2+\lambda ^4\right)^2}{\lambda ^2-2 C_2}}\right), \varepsilon=1$ and  $Q_{31}: (\Sigma,A,u,w)=\left(0,  \frac{1}{\sqrt{C_2-\frac{\lambda ^2}{2}}},0, \sqrt{2} \sqrt{-\frac{\lambda ^2
   \left(-2 C_2+\lambda ^2+8\right)^2 \left(2 (2 C_2+1) \lambda ^2+4
   (C_2-1) C_2+\lambda ^4\right)^2}{\lambda ^2-2 C_2}}\right), \varepsilon=-1$, respectively.}}]{\label{P13P14eigen} {{Real parts of $f_i$ (left panel) and  $g_i$ (right panel) corresponding to the equilibrium points  $Q_{30}: (\Sigma,A,u,w)=\left(0,  \frac{1}{\sqrt{C_2-\frac{\lambda ^2}{2}}},0, \sqrt{2} \sqrt{-\frac{\lambda ^2
   \left(-2 C_2+\lambda ^2+8\right)^2 \left(2 (2 C_2+1) \lambda ^2+4
   (C_2-1) C_2+\lambda ^4\right)^2}{\lambda ^2-2 C_2}}\right), \varepsilon=1$ and  $Q_{31}: (\Sigma,A,u,w)=\left(0,  \frac{1}{\sqrt{C_2-\frac{\lambda ^2}{2}}},0, \sqrt{2} \sqrt{-\frac{\lambda ^2
   \left(-2 C_2+\lambda ^2+8\right)^2 \left(2 (2 C_2+1) \lambda ^2+4
   (C_2-1) C_2+\lambda ^4\right)^2}{\lambda ^2-2 C_2}}\right), \varepsilon=-1$, respectively. Then,  $Q_{30}$ and $Q_{31}$, are either sources or they are non-hyperbolic (with four pure imaginary eigenvalues).}}}
   \end{figure*}
\end{enumerate}
\subsection{Invariant set $A=v=0$}
\label{section4.7}
At the invariant set $A=v=0$, the equations are reduced to: 
\begin{subequations}
\label{campo-escalar-D}
\begin{align}
    & \Sigma'= 2 C_2 \Sigma ^3+\frac{-\frac{3 \gamma  \Omega }{2}+w^2+\Omega }{C_2}+\Sigma  \left(K+2 u^2-2\right), \\
    & K'= 2 K \left(2 C_2 \Sigma ^2+K+2 u^2-1\right), \\
    & u'=u \left(2 C_2 \Sigma ^2+K-2\right)+2 u^3-\frac{\sqrt{2} w^2}{\lambda }, \\
    & w'= w \left(2 C_2 \Sigma ^2+K+\Sigma +2 u^2-\frac{\sqrt{2} u}{\lambda }\right),
\end{align}
with restriction 
\begin{equation}
\label{eq4:72}
  C_2 \Sigma^2+K+u^2 -(1-\gamma)\Omega_t-w^2 =1.  
\end{equation}
For $1<\gamma\leq 2$, we have the auxiliary equation: 
\begin{equation}
 \Omega_t'= \Omega_t  \left(\left(\frac{1}{\gamma -1}+3\right) \Sigma +4 C_2 \Sigma ^2+2 K+4 u^2\right).
\end{equation}
\end{subequations}
\subsubsection{Reduced system}
For $1<\gamma\leq 2$, the restriction \eqref{eq4:72} can be globally solved for $\Omega_t$, leading to the reduced system: 
\begin{subequations}
\label{eq:4.74}
\begin{align}
& \Sigma'=\frac{(3 \gamma -2) \left(C_2 \Sigma ^2+K+u^2-w^2-1\right)}{2 (\gamma -1) C_2} \nonumber \\
& +2 C_2 \Sigma ^3+\frac{w^2}{C_2}+\Sigma  \left(K+2 u^2-2\right), \\
& K'= 2 K \left(2 C_2 \Sigma ^2+K+2 u^2-1\right) 
\end{align}
\begin{align}
& u'= u \left(2 C_2 \Sigma ^2+K+2 u^2-2\right)-\frac{\sqrt{2} w^2}{\lambda }, \\
& w'= w \left(2 C_2 \Sigma ^2+K+\Sigma +2 u^2-\frac{\sqrt{2} u}{\lambda }\right). 
\end{align}
\end{subequations}
For $u=w=0$ and $\lambda \rightarrow 0$ the system  
\eqref{eq:3.57} is recovered. Therefore, the equilibrium points, and their stability conditions studied in section \ref{sect:3.7.1} are retrieved. 
 By definition $ K \geq 0 $, and $ w \geq 0 $ (if $ \theta> 0 $). Given that the system \eqref{eq:4.74} is invariant to the simultaneous change $ (u, \lambda) \rightarrow (-u, - \lambda) $, it can be assumed $ \lambda> 0 $. In the following discussion the analysis is restricted to $\lambda>0, u \geq 0 $ (the sign of $u$ corresponds to the sign of $\Psi$, if $\theta> 0$). The following lists contains the equilibrium points of the reduced system \eqref{eq:4.74}.
\begin{enumerate}
\item[$N_1$:] $(\Sigma, K, u, w)= \left(0 ,  1 ,  0 ,  0\right)$, $\Omega_t= 0$. The eigenvalues are $\{-1,-1,1,2\}$, then it is a hyperbolic saddle. 
\item[$L$:] $(\Sigma, K, u, w)=\left(\Sigma_0, 0 , \sqrt{1-\Sigma_0 ^2 C_2}, 0 \right)$,  $\Omega_t=0$. This line of equilibrium points contains the equilibrium points  $P_{11}$ and $P_{12}$ studied in section  \ref{sect:3.7.1} for $u=0$. The eigenvalues are  $\left\{0,\frac{(3 \gamma -2) \Sigma_0 }{\gamma -1}+4,-\frac{\sqrt{2-2 C_2
   \Sigma_0 ^2}}{\lambda }+\Sigma_0 +2,2\right\}$.    This line is normally-hyperbolic, i. e., given the curve parametrization: 
$$
  \textbf{r}(\Sigma_0)= \left(\Sigma_0, 0,  \sqrt{1-C_2 \Sigma_0 ^2}, 0\right),
$$ the tangent vector at $\Sigma_0$: $$ \textbf{r}'(\Sigma_0)=\left(1,0,-\frac{ C_2 \Sigma_0}{\sqrt{1-C_2 \Sigma_0 ^2}},0\right),$$ is parallel to the eigenvector associated to the zero eigenvalue, say, $\textbf{v}(\Sigma_0)=\left(-\frac{\sqrt{1-C_2 \Sigma _0^2}}{C_2 \Sigma _0},0,1,0\right).$ In this particular case, the stability of the curve of equilibrium point can be studied considering only the signs of the real part of the non-zero eigenvalues, concluding that:
   \begin{enumerate}
       \item $L$ is a hyperbolic source for: 
       \begin{enumerate}
           \item $C_2>0,  \Sigma_0=\frac{1}{\sqrt{C_2}},  \lambda >0,  1<\gamma <2$, or   
           \item $C_2>1,  \Sigma_0=-\sqrt{\frac{1}{C_2}},  \lambda >0,  \frac{2 \Sigma_{0}+4}{3 \Sigma_0+4}<\gamma <2$, or   
           \item $C_2>1,  0<\Sigma_0<\sqrt{\frac{1}{C_2}},  \lambda >\sqrt{2} \sqrt{\frac{1-C_2 \Sigma_0^2}{(\Sigma_0+2)^2}}$, \newline $ 1<\gamma <2$, or   
           \item $C_2>0,  0<\Sigma_0<\sqrt{\frac{1}{C_2}},  \lambda >\sqrt{2} \sqrt{\frac{1-C_2 \Sigma_0^2}{(\Sigma_0+2)^2}}$, \newline $  1<\gamma <2$, or   
           \item $C_2>1,  -\sqrt{\frac{1}{C_2}}<\Sigma_0\leq 0,  \lambda >\sqrt{2} \sqrt{\frac{1-C_2 \Sigma_0^2}{(\Sigma_0+2)^2}}$, \newline $  \frac{2 \Sigma_0+4}{3 \Sigma_0+4}<\gamma <2$, or   
           \item $0<C_2\leq 1,  -1<\Sigma_0\leq 0,  \lambda >\sqrt{2} \sqrt{\frac{1-C_2 \Sigma_0^2}{(\Sigma_0+2)^2}}$, \newline $  \frac{2 \Sigma_0+4}{3
   \Sigma_0+4}<\gamma <2$, or   
   \item $C_2>1,  -\sqrt{\frac{1}{C_2}}<\Sigma_0\leq 0,  \lambda >\sqrt{2} \sqrt{\frac{1-C_2 \Sigma_0^2}{(\Sigma_0+2)^2}}$, \newline $  \frac{2
   \Sigma_0+4}{3 \Sigma_0+4}<\gamma <2$, or   
   \item $0<C_2\leq 1,  -1<\Sigma_0\leq 0,  \lambda >\sqrt{2} \sqrt{\frac{1-C_2 \Sigma_0^2}{(\Sigma_0+2)^2}}$, \newline $ 
   \frac{2 \Sigma_0+4}{3 \Sigma_0+4}<\gamma <2$, or   
   \item $C_2>1,  -\sqrt{\frac{1}{C_2}}<\Sigma_0\leq 0,  \lambda >\sqrt{2} \sqrt{\frac{1-C_2 \Sigma_0^2}{(\Sigma_0+2)^2}}$, \newline $  \frac{2 \Sigma_0+4}{3 \Sigma_0+4}<\gamma <2$, or   
   \item $C_2>1,  0<\Sigma_0<\sqrt{\frac{1}{C_2}},  \lambda >\sqrt{2}
   \sqrt{\frac{1-C_2 \Sigma_0^2}{(\Sigma_0+2)^2}}$, \newline $  1<\gamma <2$, or   
   \item $0<C_2\leq 1,  -1<\Sigma_0\leq 0,  \lambda >\sqrt{2} \sqrt{\frac{1-C_2 \Sigma_0^2}{(\Sigma_0+2)^2}}$, \newline $  \frac{2 \Sigma_0+4}{3 \Sigma_0+4}<\gamma <2$, or   
   \item $0<C_2\leq 1,  0<\Sigma_0<\sqrt{\frac{1}{C_2}}$, \newline $  \lambda >\sqrt{2}
   \sqrt{\frac{1-C_2 \Sigma_0^2}{(\Sigma_0+2)^2}},  1<\gamma <2$, or   
   \item $C_2>1,  -\sqrt{\frac{1}{C_2}}<\Sigma_0\leq 0$, \newline $\lambda >\sqrt{2} \sqrt{\frac{1-C_2
   \Sigma_0^2}{(\Sigma_0+2)^2}},  \frac{2 \Sigma_0+4}{3 \Sigma_0+4}<\gamma <2$, or   
   \item $C_2>1,  0<\Sigma_0<\sqrt{\frac{1}{C_2}}$, \newline $  \lambda >\sqrt{2}
   \sqrt{\frac{1-C_2 \Sigma_0^2}{(\Sigma_0+2)^2}},  1<\gamma <2$, or   
   \item $0<C_2\leq 1,  -1<\Sigma_0\leq 0$, \newline $  \lambda >\sqrt{2} \sqrt{\frac{1-C_2 \Sigma_0^2}{(\Sigma_0+2)^2}},  \frac{2 \Sigma_0+4}{3 \Sigma_0+4}<\gamma <2$, or   
   \item $0<C_2\leq 1,  0<\Sigma_0<\sqrt{\frac{1}{C_2}}$, \newline $  \lambda >\sqrt{2}
   \sqrt{\frac{1-C_2 \Sigma_0^2}{(\Sigma_0+2)^2}},  1<\gamma <2$.
       \end{enumerate}
       \item $L$ is a hyperbolic saddle for:
       \begin{enumerate}
        \item $0<C_2\leq \frac{1}{4},  -2<\Sigma_0\leq -1$, \newline $  \lambda >\sqrt{2} \sqrt{\frac{1-C_2 \Sigma_0^2}{(\Sigma_0+2)^2}},  1<\gamma <2$, or  
           \item $0<C_2\leq \frac{1}{4}, 
   -1<\Sigma_0<0$, \newline $  \lambda >\sqrt{2} \sqrt{\frac{1-C_2 \Sigma_0^2}{(\Sigma_0+2)^2}},  1<\gamma <\frac{2 \Sigma_0+4}{3 \Sigma_0+4}$, or  
   \item $\frac{1}{4}<C_2\leq 1,  \Sigma_0=-\frac{1}{\sqrt{C_2}},  \lambda >0,  1<\gamma <2$, or  
   \item $\frac{1}{4}<C_2<1,  -\sqrt{\frac{1}{C_2}}<\Sigma_0\leq -1$, \newline $ 
   \lambda >\sqrt{2} \sqrt{\frac{1-C_2 \Sigma_0^2}{(\Sigma_0+2)^2}},  1<\gamma <2$, or  
   \item $\frac{1}{4}<C_2\leq 1,  -1<\Sigma_0<0$, \newline $ \lambda >\sqrt{2} \sqrt{\frac{1-C_2
   \Sigma_0^2}{(\Sigma_0+2)^2}},  1<\gamma <\frac{2 \Sigma_0+4}{3 \Sigma_0+4}$, or  
   \item $C_2>1,  \Sigma_0=-\sqrt{\frac{1}{C_2}},  \lambda >0,  1<\gamma
   <\frac{2 \Sigma_0+4}{3 \Sigma_0+4}$, or  
   \item $C_2>1,  -\sqrt{\frac{1}{C_2}}<\Sigma_0<0$, \newline $  \lambda >\sqrt{2} \sqrt{\frac{1-C_2 \Sigma_0^2}{(\Sigma_0+2)^2}},  1<\gamma <\frac{2 \Sigma_0+4}{3 \Sigma_0+4}$, or  
   \item $0<C_2\leq 1,  0<\Sigma_0<\sqrt{\frac{1}{C_2}}$, \newline $  0<\lambda <\sqrt{2} \sqrt{\frac{1-C_2
   \Sigma_0^2}{(\Sigma_0+2)^2}},  1<\gamma <2$, or  
   \item $0<C_2\leq 1,  -1<\Sigma_0\leq 0$, \newline $  0<\lambda <\sqrt{2} \sqrt{\frac{1-C_2 \Sigma_0^2}{(\Sigma_0+2)^2}},  \frac{2 \Sigma_0+4}{3 \Sigma_0+4}<\gamma <2$, or  
   \item $C_2>1,  0<\Sigma_0<\sqrt{\frac{1}{C_2}}$, \newline $  0<\lambda <\sqrt{2} \sqrt{\frac{1-C_2 \Sigma_0^2}{(\Sigma_0+2)^2}},  1<\gamma <2$, or  
   \item $C_2>1,  -\sqrt{\frac{1}{C_2}}<\Sigma_0\leq 0$, \newline $  0<\lambda <\sqrt{2} \sqrt{\frac{1-C_2 \Sigma_0^2}{(\Sigma_0+2)^2}},  \frac{2 \Sigma_0+4}{3 \Sigma_0+4}<\gamma <2$, or  
   \item $0<C_2<\frac{1}{4},  -\sqrt{\frac{1}{C_2}}\leq \Sigma_0\leq -2,  \lambda >0$, \newline $  1<\gamma <2$, or  
   \item $0<C_2<\frac{1}{4},  -1<\Sigma_0<0$, \newline $  0<\lambda <\sqrt{2} \sqrt{\frac{1-C_2 \Sigma_0^2}{(\Sigma_0+2)^2}},  1<\gamma <\frac{2 \Sigma_0+4}{3 \Sigma_0+4}$, or  
   \item $0<C_2<\frac{1}{4},  -2<\Sigma_0\leq -1$, \newline $  0<\lambda <\sqrt{2} \sqrt{\frac{1-C_2 \Sigma_0^2}{(\Sigma_0+2)^2}},  1<\gamma <2$, or  
   \item $\frac{1}{4}\leq C_2<1,  -\sqrt{\frac{1}{C_2}}<\Sigma_0\leq -1$, \newline $  0<\lambda <\sqrt{2} \sqrt{\frac{1-C_2 \Sigma_0^2}{(\Sigma_0+2)^2}},  1<\gamma <2$, or  
   \item $\frac{1}{4}\leq C_2\leq 1,  -1<\Sigma_0<0$, \newline $  0<\lambda <\sqrt{2} \sqrt{\frac{1-C_2 \Sigma_0^2}{(\Sigma_0+2)^2}},  1<\gamma <\frac{2 \Sigma_0+4}{3 \Sigma_0+4}$, or  
   \item $C_2>1,  -\sqrt{\frac{1}{C_2}}<\Sigma_0<0$, \newline $  0<\lambda <\sqrt{2} \sqrt{\frac{1-C_2 \Sigma_0^2}{(\Sigma_0+2)^2}},  1<\gamma <\frac{2 \Sigma_0+4}{3 \Sigma_0+4}$.
       \end{enumerate}
    \item $L$ is non-hyperbolic for: 
     \begin{enumerate}
         \item $0<C_2<\frac{1}{4}, -1<\Sigma_0<0, \gamma =\frac{2 (\Sigma_0+2)}{3 \Sigma_0+4}$, \newline $ 0<\lambda <\sqrt{2} \sqrt{\frac{1-C_2 \Sigma_0^2}{(\Sigma_0+2)^2}}$, or  
         \item $0<C_2<\frac{1}{4}, -1<\Sigma_0<0, \gamma =\frac{2 (\Sigma_0+2)}{3 \Sigma_0+4}$, \newline $ \lambda >\sqrt{2} \sqrt{\frac{1-C_2 \Sigma_0^2}{(\Sigma_0+2)^2}}$, or  
         \item $0<C_2<\frac{1}{4}, -1<\Sigma_0<0, \sqrt{2} \sqrt{\frac{1-C_2 \Sigma_0^2}{(\Sigma_0+2)^2}}=\lambda , 1<\gamma <2$, or  
         \item $0<C_2<\frac{1}{4}, \sqrt{2} \sqrt{\frac{1-C_2 \Sigma_0^2}{(\Sigma_0+2)^2}}=\lambda , 1<\gamma <2, -2<\Sigma_0\leq -1$, or  
         \item $0<C_2<\frac{1}{4}, 
   \sqrt{2} \sqrt{\frac{1-C_2 \Sigma_0^2}{(\Sigma_0+2)^2}}=\lambda , 1<\gamma <2, 0\leq \Sigma_0<\frac{1}{\sqrt{C_2}}$, or  
   \item $C_2=\frac{1}{4}, -1<\Sigma_0<0, 
   \gamma =\frac{2 (\Sigma_0+2)}{3 \Sigma_0+4}$, \newline $ 0<\lambda <\sqrt{\frac{2}{\Sigma_0+2}-\frac{1}{2}}$, or  
   \item $C_2=\frac{1}{4}, -1<\Sigma_0<0, \gamma =\frac{2 (\Sigma_0+2)}{3 \Sigma_0+4}$, \newline $ 2 \lambda >\sqrt{\frac{8}{\Sigma_0+2}-2}$, or  
   \item $C_2=\frac{1}{4}, -1<\Sigma_0<0, \gamma =\frac{2 (\Sigma_0+2)}{3 \Sigma_0+4}$, \newline $ 
   \sqrt{\frac{8}{\Sigma_0+2}-2}=2 \lambda , 1<\gamma <2$, or  
   \item $C_2=\frac{1}{4}, 1<\gamma <2$, \newline $ \sqrt{\frac{8}{\Sigma_0+2}-2}=2 \lambda , -2<\Sigma_0\leq -1$, or  
   \item $4
   C_2=1, 1<\gamma <2, \sqrt{\frac{8}{\Sigma_0+2}-2}=2 \lambda , 0\leq \Sigma_0<2$, or  
   \item $C_2=\frac{1}{4}, 1<\gamma <2, \Sigma_0=-2, \lambda >0$, or  
   \item $\frac{1}{4}<C_2\leq 1, -1<\Sigma_0<0, \gamma =\frac{2 (\Sigma_0+2)}{3 \Sigma_0+4}$, \newline $ 0<\lambda <\sqrt{2} \sqrt{\frac{1-C_2 \Sigma_0^2}{(\Sigma_0+2)^2}}$, or  
   \item $\frac{1}{4}<C_2\leq 1, -1<\Sigma_0<0$, \newline $ \gamma =\frac{2 (\Sigma_0+2)}{3 \Sigma_0+4}, \lambda >\sqrt{2} \sqrt{\frac{1-C_2 \Sigma_0^2}{(\Sigma_0+2)^2}}$, or  
   \item $\frac{1}{4}<C_2\leq 1, -1<\Sigma_0<0, \sqrt{2} \sqrt{\frac{1-C_2 \Sigma_0^2}{(\Sigma_0+2)^2}}=\lambda$, \newline $  1<\gamma
   <2$, or  \item $\frac{1}{4}<C_2\leq 1, \sqrt{2} \sqrt{\frac{1-C_2 \Sigma_0^2}{(\Sigma_0+2)^2}}=\lambda , 1<\gamma <2$, \newline $  0\leq \Sigma_0<\frac{1}{\sqrt{C_2}}$, or  
   \item $\frac{1}{4}<C_2\leq 1, \sqrt{2} \sqrt{\frac{1-C_2 \Sigma_0^2}{(\Sigma_0+2)^2}}=\lambda , 1<\gamma <2$, \newline $  -\frac{1}{\sqrt{C_2}}<\Sigma_0\leq -1$, or  
   \item $C_2>1, \gamma =\frac{2 (\Sigma_0+2)}{3 \Sigma_0+4}, \frac{1}{\sqrt{C_2}}+\Sigma_0=0, \lambda >0$, or  
   \item $C_2>1, \gamma =\frac{2 (\Sigma_0+2)}{3 \Sigma_0+4}, -\frac{1}{\sqrt{C_2}}<\Sigma_0<0$, \newline $  0<\lambda <\sqrt{2} \sqrt{\frac{1-C_2 \Sigma_0^2}{(\Sigma_0+2)^2}}$, or  
   \item $C_2>1, \gamma
   =\frac{2 (\Sigma_0+2)}{3 \Sigma_0+4}, -\frac{1}{\sqrt{C_2}}<\Sigma_0<0$, \newline $  \lambda >\sqrt{2} \sqrt{\frac{1-C_2 \Sigma_0^2}{(\Sigma_0+2)^2}}$, or  
   \item $C_2>1, \sqrt{2} \sqrt{\frac{1-C_2 \Sigma_0^2}{(\Sigma_0+2)^2}}=\lambda , 1<\gamma <2, 0\leq \Sigma_0<\frac{1}{\sqrt{C_2}}$, or  
   \item $C_2>1, \sqrt{2} \sqrt{\frac{1-C_2 \Sigma_0^2}{(\Sigma_0+2)^2}}=\lambda , 1<\gamma <2, -\frac{1}{\sqrt{C_2}}<\Sigma_0<0$.
     \end{enumerate}
   \end{enumerate}

The following list contains new points  that were not studied in section \ref{sect:3.7.1}. 
 \item $P_{13}(\lambda): (\Sigma, K, u, w)= \Bigg(-\frac{2 (\gamma -1) (3 \gamma -2)}{8 C_2 (\gamma -1)^2+\gamma ^2
   \lambda ^2} ,  0 ,  \frac{\gamma  (3 \gamma -2) \lambda }{\sqrt{2}
   \left(8 C_2 (\gamma -1)^2+\gamma ^2 \lambda ^2\right)},$ \newline $\frac{
   \sqrt{\gamma } \sqrt{3 \gamma -2} \lambda  \sqrt{4- 16 C_2 (\gamma -1)^2-\gamma  \left(\gamma 
   \left(2 \lambda ^2-9\right)+12\right)}}{\sqrt{2} \left(8 C_2 (\gamma -1)^2+\gamma
   ^2 \lambda ^2\right)}\Bigg)$. \newline $\Omega_t= \frac{\left(\gamma  \lambda ^2-4 (\gamma
   -1) C_2\right) \left(4- 16 C_2 (\gamma -1)^2-\gamma  \left(\gamma 
   \left(2 \lambda ^2-9\right)+12\right)\right)}{\left(8 C_2 (\gamma
   -1)^2+\gamma ^2 \lambda ^2\right){}^2}$. The eigenvalues are 
   \noindent\( {\left\{-2+\frac{(2-3 \gamma )^2}{\gamma ^2 \lambda ^2+8 (-1+\gamma )^2 C_2},-2+\frac{2 (2-3 \gamma )^2}{\gamma ^2 \lambda ^2+8 (-1+\gamma
)^2 C_2},\right.} \) 
\begin{widetext}
\( {-\left((-1+\gamma ) C_2 \left(\gamma ^2 \lambda ^2+8 (-1+\gamma )^2 C_2\right) \left(-4+\gamma  \left(12+\gamma  \left(-9+2 \lambda ^2\right)\right)+16
(-1+\gamma )^2 C_2\right)+\right.}\\
 {\surd \left((-1+\gamma ) C_2 \left(\gamma ^2 \lambda ^2+8 (-1+\gamma )^2 C_2\right){}^2 \left(-4+\gamma  \left(12+\gamma  \left(-9+2 \lambda
^2\right)\right)+16 (-1+\gamma )^2 C_2\right) \right.}\\
 {\left.\left.\left.\left(2 \gamma ^2 (-2+3 \gamma ) \lambda ^2+(-1+\gamma ) C_2 \left(-4+\gamma  \left(28+\gamma  \left(-33+2 \lambda ^2\right)\right)+16
(-1+\gamma )^2 C_2\right)\right)\right)\right)\right/} 
 {\left(2 (-1+\gamma ) C_2 \left(\gamma ^2 \lambda ^2+8 (-1+\gamma )^2 C_2\right){}^2\right) ,}\) \({\left(-(-1+\gamma ) C_2 \left(\gamma ^2 \lambda ^2+8
(-1+\gamma )^2 C_2\right) \right.} 
 {\left(-4+\gamma  \left(12+\gamma  \left(-9+2 \lambda ^2\right)\right)+16 (-1+\gamma )^2 C_2\right)+}\\
 {\surd \left((-1+\gamma ) C_2 \left(\gamma ^2 \lambda ^2+8 (-1+\gamma )^2 C_2\right){}^2 \left(-4+\gamma  \left(12+\gamma  \left(-9+2 \lambda
^2\right)\right)+16 (-1+\gamma )^2 C_2\right) \right.}\\
 {\left.\left.\left.\left(2 \gamma ^2 (-2+3 \gamma ) \lambda ^2+(-1+\gamma ) C_2 \left(-4+\gamma  \left(28+\gamma  \left(-33+2 \lambda ^2\right)\right)+16
(-1+\gamma )^2 C_2\right)\right)\right)\right)\right/} \\
 {\left.\left(2 (-1+\gamma ) C_2 \left(\gamma ^2 \lambda ^2+8 (-1+\gamma )^2 C_2\right){}^2\right)\right\}}\).
\end{widetext}
\begin{enumerate}
    \item $P_{13}(\lambda)$ is a hyperbolic sink for:
    \begin{enumerate}
        \item $1<C_2\leq 2,  \frac{4 C_2-2}{4 C_2-3}<\gamma <2$, \newline $   \sqrt{\frac{(2-3 \gamma )^2-8 (\gamma -1)^2 C_2}{\gamma ^2}}<\lambda <2 \sqrt{\frac{(\gamma -1) C_2}{\gamma }}$, or   
        \item $C_2>2,  \frac{4 C_2-2}{4 C_2-3}<\gamma <\frac{2 \left(4 C_2-3\right)}{8 C_2-9}+2 \sqrt{2} \sqrt{\frac{C_2}{\left(8 C_2-9\right){}^2}}$, \newline $   \sqrt{\frac{(2-3 \gamma )^2-8 (\gamma -1)^2 C_2}{\gamma ^2}}<\lambda <2 \sqrt{\frac{(\gamma-1)C_2}{\gamma }}$, or 
   \item $C_2>2,  \frac{2 \left(4 C_2-3\right)}{8 C_2-9}+2 \sqrt{2} \sqrt{\frac{C_2}{\left(8 C_2-9\right){}^2}}\leq \gamma <2,  0<\lambda <2
   \sqrt{\frac{(\gamma -1)C_2}{\gamma }}$.
    \end{enumerate}
    
    \item $P_{13}(\lambda)$ is a hyperbolic source for:
     \begin{enumerate}
         \item $0<C_2\leq \frac{1}{2},  1<\gamma <2$, \newline $   2 \sqrt{\frac{(\gamma-1)C_2}{\gamma }}<\lambda <\frac{1}{2} \sqrt{\frac{2 (2-3 \gamma )^2-32 (\gamma -1)^2 C_2}{\gamma ^2}}$, or 
         \item $C_2>\frac{1}{2},  1<\gamma
   <\frac{8 C_2-2}{8 C_2-3}$, \newline $   2 \sqrt{\frac{(\gamma-1)C_2}{\gamma }}<\lambda <\frac{1}{2} \sqrt{\frac{2 (2-3 \gamma )^2-32 (\gamma -1)^2 C_2}{\gamma ^2}}$.
     \end{enumerate}
     \item $P_{13}(\lambda)$ is non-hyperbolic for: 
     \begin{enumerate}
         \item $C_2>\frac{1}{2},  1<\gamma <\frac{8 C_2-2}{8 C_2-3},  \lambda =2 \sqrt{\frac{(\gamma-1)C_2}{\gamma }}$, or 
         \item $C_2>\frac{1}{2},  \frac{8 C_2-2}{8 C_2-3}<\gamma <2,  \lambda =2 \sqrt{\frac{\gamma 
   C_2-C_2}{\gamma }}$, or 
   \item $C_2>1,  1<\gamma <\frac{2 \left(8 C_2-3\right)}{16 C_2-9}+4 \sqrt{\frac{C_2}{\left(16 C_2-9\right){}^2}}$, \newline $  \lambda =\sqrt{\frac{(2-3 \gamma )^2-8 (\gamma -1)^2 C_2}{\gamma
   ^2}}$, or 
   \item $C_2>2,  1<\gamma <\frac{2 \left(4 C_2-3\right)}{8 C_2-9}+2 \sqrt{2} \sqrt{\frac{C_2}{\left(8 C_2-9\right){}^2}}$, \newline $   \lambda =\sqrt{\frac{(2-3 \gamma )^2-8 (\gamma -1)^2 C_2}{\gamma ^2}}$, or 
   \item $0<C_2\leq \frac{1}{2},  1<\gamma <2,  \lambda =2 \sqrt{\frac{(\gamma-1)C_2}{\gamma }}$, or 
   \item $0<C_2\leq 1,  1<\gamma <2,  \lambda =\sqrt{\frac{(2-3 \gamma )^2-8 (\gamma -1)^2 C_2}{\gamma ^2}}$, or 
   \item $0<C_2\leq 2,  1<\gamma <2,  \lambda =\sqrt{\frac{(2-3 \gamma )^2-8 (\gamma -1)^2 C_2}{\gamma ^2}}$.
     \end{enumerate}
    \item $P_{13}(\lambda)$ is a hyperbolic saddle for: 
    \begin{enumerate}
        \item $0<C_2\leq \frac{1}{2},  1<\gamma <2,  0<\lambda <2 \sqrt{\frac{(\gamma-1)C_2}{\gamma }}$, or 
        \item $0<C_2\leq \frac{1}{2},  1<\gamma <2$, \newline $   2 \sqrt{\frac{(\gamma-1)C_2}{\gamma }}<\lambda <\sqrt{\frac{(2-3
   \gamma )^2-8 (\gamma -1)^2 C_2}{\gamma ^2}}$, or 
   \item $0<C_2\leq \frac{1}{2},  1<\gamma <2,  \lambda >\sqrt{\frac{(2-3 \gamma )^2-8 (\gamma -1)^2 C_2}{\gamma ^2}}$, or  
   \item $\frac{1}{2}<C_2\leq 1,  1<\gamma \leq \frac{8 C_2-2}{8 C_2-3},  0<\lambda <2 \sqrt{\frac{(\gamma-1)C_2}{\gamma }}$, or  
   \item $\frac{1}{2}<C_2\leq 1,  1<\gamma <\frac{8 C_2-2}{8 C_2-3}$, \newline $  2
   \sqrt{\frac{(\gamma-1)C_2}{\gamma }}<\lambda <\sqrt{\frac{(2-3    \gamma )^2-8 (\gamma -1)^2 C_2}{\gamma ^2}}$, or 
   \item $\frac{1}{2}<C_2\leq 1,  1<\gamma <\frac{8 C_2-2}{8
   C_2-3},  \lambda >\sqrt{\frac{(2-3    \gamma )^2-8 (\gamma -1)^2 C_2}{\gamma ^2}}$, or 
   \item $\frac{1}{2}<C_2\leq 1,  \frac{8 C_2-2}{8 C_2-3}<\gamma <2$, \newline $ 0<\lambda
   <\sqrt{\frac{(2-3    \gamma )^2-8 (\gamma -1)^2 C_2}{\gamma ^2}}$, or 
   \item $\frac{1}{2}<C_2\leq 1,  \frac{8 C_2-2}{8 C_2-3}\leq \gamma <2,  \lambda >2 \sqrt{\frac{(\gamma-1)C_2}{\gamma }}$, or  
   \item $C_2>1,  1<\gamma \leq \frac{8 C_2-2}{8 C_2-3},  0<\lambda <2 \sqrt{\frac{(\gamma-1)C_2}{\gamma }}$, or  
   \item $C_2>1,  1<\gamma <\frac{8 C_2-2}{8 C_2-3}$, \newline $   2 \sqrt{\frac{(\gamma-1)C_2}{\gamma }}<\lambda
   <\sqrt{\frac{(2-3    \gamma )^2-8 (\gamma -1)^2 C_2}{\gamma ^2}}$, or  
   \item $C_2>1,  1<\gamma <\frac{8 C_2-2}{8 C_2-3},  \lambda >\sqrt{\frac{(2-3    \gamma )^2-8 (\gamma -1)^2 C_2}{\gamma ^2}}$, or  
   \item $C_2>1,  \frac{8 C_2-2}{8 C_2-3}<\gamma <\frac{2 \left(8 C_2-3\right)}{16 C_2-9}+4 \sqrt{\frac{C_2}{\left(16 C_2-9\right){}^2}}$, \newline $ 0<\lambda
   <\sqrt{\frac{(2-3    \gamma )^2-8 (\gamma -1)^2 C_2}{\gamma ^2}}$, or  
   \item $C_2>1,  \frac{8 C_2-2}{8 C_2-3}\leq \gamma <\frac{2 \left(8 C_2-3\right)}{16 C_2-9}+4
   \sqrt{\frac{C_2}{\left(16 C_2-9\right){}^2}}$, \newline $   \lambda >2 \sqrt{\frac{(\gamma-1)C_2}{\gamma }}$, or  
   \item $C_2>1,  \frac{2 \left(8 C_2-3\right)}{16 C_2-9}+4 \sqrt{\frac{C_2}{\left(16 C_2-9\right){}^2}}\leq \gamma
   <2$, \newline $   \lambda >2 \sqrt{\frac{(\gamma-1)C_2}{\gamma }}$.
    \end{enumerate}
\end{enumerate}

 \item $P_{14}(\lambda): (\Sigma, K, u, w)=$ \newline $\left(-\frac{2 (\gamma -1)}{3 \gamma -2} ,  \frac{4-8 C_2 (\gamma -1)^2-\gamma
    \left(\gamma  \left(\lambda ^2-9\right)+12\right)}{(2-3 \gamma
   )^2} ,  \frac{\gamma  \lambda }{\sqrt{2} (3 \gamma -2)} , 
   \frac{\lambda }{\sqrt{\frac{4}{\gamma }-6}}\right)$. $ \Omega_t=\frac{4 (\gamma -1)
   C_2-\gamma  \lambda ^2}{(2-3 \gamma )^2}$.
   The real parts of the $\mu_i $'s are represented in the figure \ref{A} for some values of $ C_2 $, where it is shown that $P_{14}(\lambda) $ is typically a hyperbolic saddle for the given values of the parameter $ C_2 $ (or is it non-hyperbolic).
    \begin{figure*}
       \centering
       \includegraphics[scale=0.45]{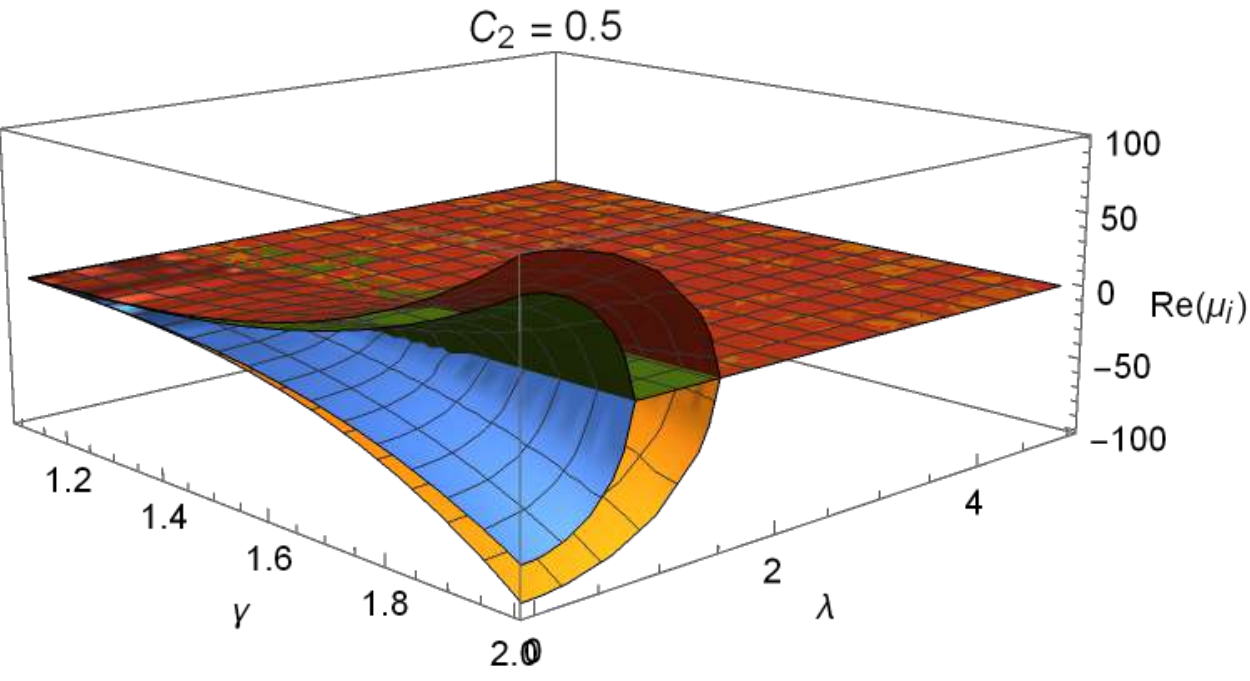} 
       \includegraphics[scale=0.45]{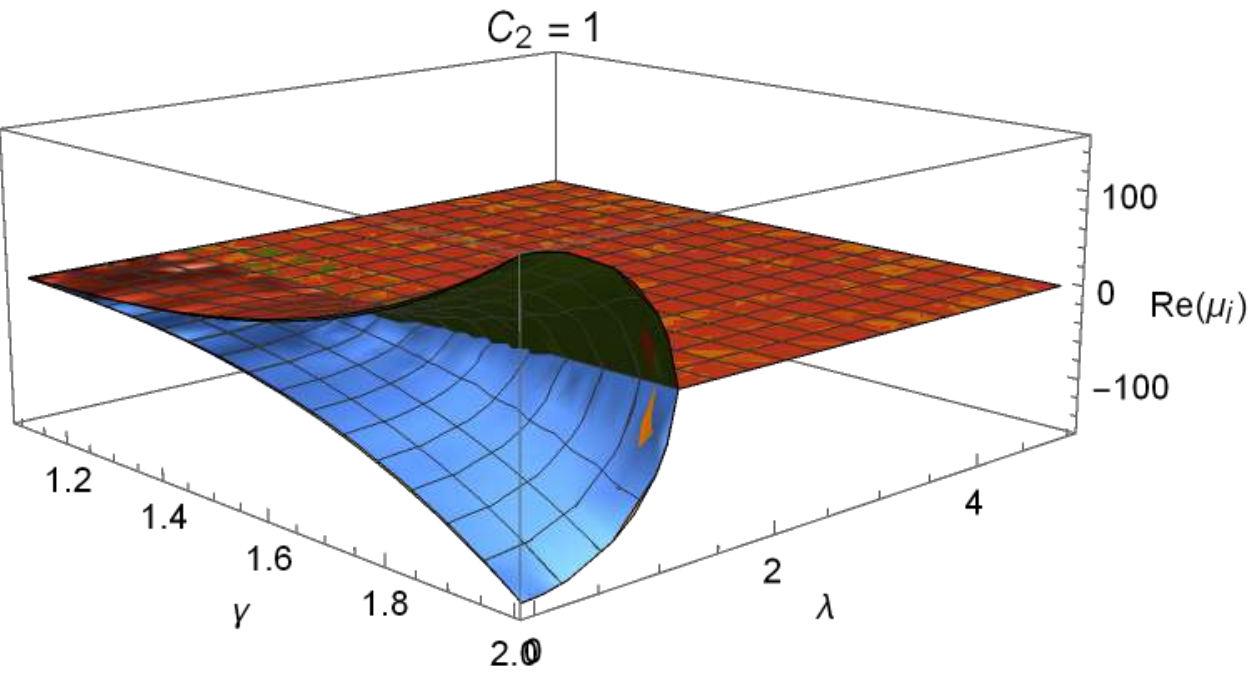} 
       \includegraphics[scale=0.45]{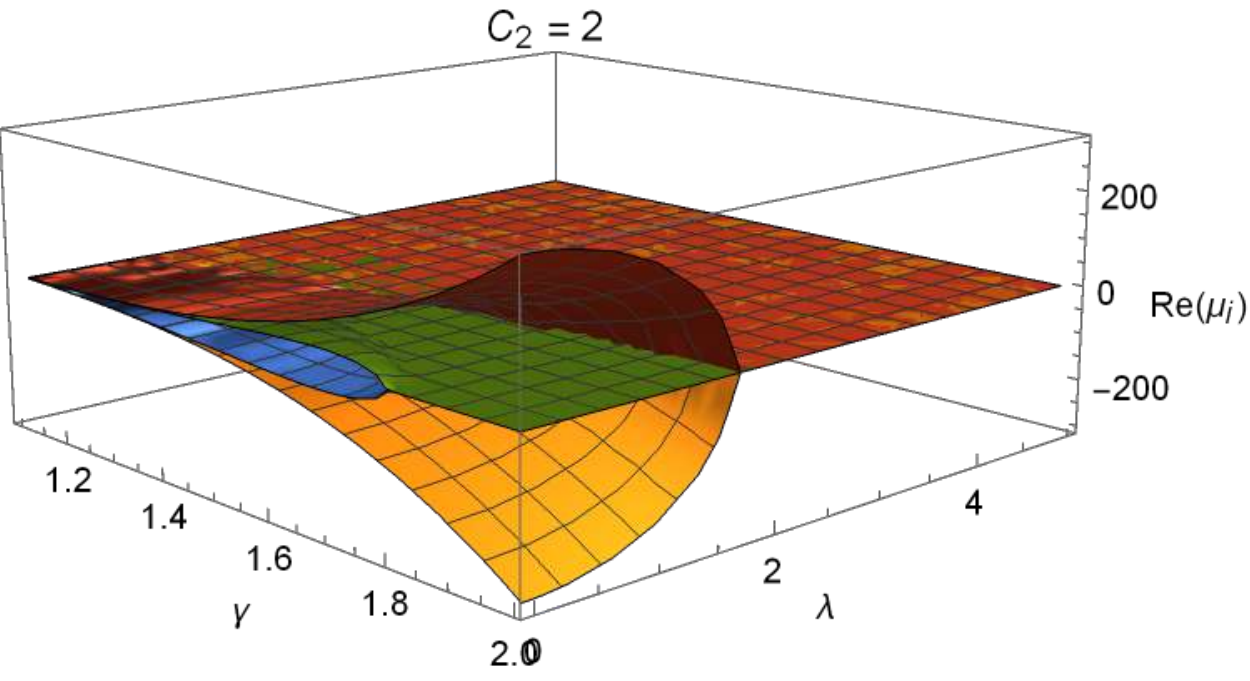}
       \caption{ \label{A} Real parts of the $\mu_i$'s corresponding to $P_{14}(\lambda)$ for some values of $C_2$. }
         \end{figure*}
         \newline 
   The following points are recovered:  $\lim_{\lambda\rightarrow 0} P_{13}(\lambda)=P_{13}$ and $\lim_{\lambda\rightarrow 0} P_{14}(\lambda)=P_{14}$.
         \item $P_{13}: (\Sigma, K, u, w)= \left(\frac{2-3 \gamma }{4 (\gamma -1) C_2} ,  0 ,  0 ,  0\right)$,  $\Omega_t=\frac{1}{\gamma -1}-\frac{(2-3 \gamma )^2}{16 (\gamma -1)^3 C_2}$. The eigenvalues are: \newline $\left\{\frac{\gamma  (3 \gamma -2)}{8 (\gamma -1)^2 C_2},\frac{(2-3
   \gamma )^2}{8 (\gamma -1)^2 C_2}-2,\frac{(2-3 \gamma )^2}{8 (\gamma
   -1)^2 C_2}-2,\frac{(2-3 \gamma )^2}{4 (\gamma -1)^2 C_2}-2\right\}$.
   
   \item $P_{14}: (\Sigma, K, u, w)=  \left( -\frac{2 (\gamma -1)}{3 \gamma -2} ,  1-\frac{8 (\gamma -1)^2 C_2}{(2-3
   \gamma )^2} ,  0 ,  0\right)$,  $\Omega_t= \frac{4 (\gamma -1) C_2}{(2-3 \gamma )^2}$. The eigenvalues are: \newline
   $\Big\{-1,\frac{\gamma }{3 \gamma -2},-\frac{1}{2}- \frac{\sqrt{ \left(64 (\gamma -1)^2 C_2-7 (2-3
   \gamma )^2\right)}}{2 (3 \gamma -2)}$, \newline $ -\frac{1}{2}+ \frac{\sqrt{ \left(64 (\gamma -1)^2 C_2-7 (2-3
   \gamma )^2\right)}}{2 (3 \gamma -2)}\Big\}$.

\item $P_{15}(\lambda): (\Sigma, K, u, w)= \left(-\frac{1}{2 C_2} ,  0 ,  \frac{1}{\sqrt{2} \lambda } ,  \frac{\sqrt{\lambda ^2+\left(2-4 \lambda ^2\right) C_2}}{2 \lambda  \sqrt{C_2}}\right)$,  $\Omega_t= 0$, with eigenvalues \newline $\left\{\frac{1}{C_2}+\frac{2}{\lambda ^2}-2,\frac{\gamma }{2 C_2-2
   \gamma  C_2}+\frac{2}{\lambda ^2},\frac{1}{2 C_2}+\frac{1}{\lambda
   ^2}-2,\frac{1}{2 C_2}+\frac{1}{\lambda ^2}-2\right\}$. 
   
   \begin{enumerate}
       \item $P_{15}(\lambda)$ is a hyperbolic sink for: 
       \begin{enumerate}
           \item $\frac{1}{2}<C_2\leq 1, 1<\gamma <2, \lambda >\sqrt{2} \sqrt{\frac{C_2}{2 C_2-1}}$, or 
        \item $C_2>1, 1<\gamma \leq \frac{4 C_2-2}{4 C_2-3}, \lambda >\sqrt{2} \sqrt{\frac{C_2}{2 C_2-1}}$, or 
        \item $C_2>1, \frac{4 C_2-2}{4 C_2-3}<\gamma <2, \lambda >2 \sqrt{\frac{(\gamma-1)C_2}{\gamma }}$
       \end{enumerate}
       \item $P_{15}(\lambda)$ is a hyperbolic source for:
       \begin{enumerate}
        \item $0<C_2\leq \frac{1}{2}, 1<\gamma <2, 0<\lambda <2 \sqrt{\frac{(\gamma-1)C_2}{\gamma }}$, or 
        \item $C_2>\frac{1}{2}, 1<\gamma \leq \frac{8 C_2-2}{8 C_2-3}, 0<\lambda <2 \sqrt{\frac{(\gamma-1)C_2}{\gamma }}$, or  
        \item $C_2>\frac{1}{2}, \frac{8 C_2-2}{8 C_2-3}<\gamma <2, 0<\lambda <\sqrt{2} \sqrt{\frac{C_2}{4 C_2-1}}$.
       \end{enumerate}
       \item $P_{15}(\lambda)$ is non-hyperbolic for:
       \begin{enumerate}
           \item $0<C_2\leq \frac{1}{4},  1<\gamma <2,  \lambda =2 \sqrt{\frac{(\gamma -1) C_2}{\gamma }}$, or  \item $\frac{1}{4}<C_2\leq \frac{1}{2},  1<\gamma <2,  \lambda =2 \sqrt{\frac{(\gamma -1) C_2}{\gamma
   }}$, or  
           \item $\frac{1}{4}<C_2\leq \frac{1}{2},  1<\gamma <2,  \lambda =\sqrt{2} \sqrt{\frac{C_2}{4 C_2-1}}$, or  
           \item $\frac{1}{2}<C_2\leq 1,  \gamma =\frac{8 C_2-2}{8 C_2-3},  \lambda =\sqrt{2}
   \sqrt{\frac{C_2}{4 C_2-1}}$, or  
           \item $\frac{1}{2}<C_2\leq 1,  \gamma =\frac{8 C_2-2}{8 C_2-3},  \lambda =\sqrt{2} \sqrt{\frac{C_2}{2 C_2-1}}$, or \item $\frac{1}{2}<C_2\leq 1,  \frac{8 C_2-2}{8
   C_2-3}<\gamma <2,  \lambda =2 \sqrt{\frac{(\gamma -1) C_2}{\gamma }}$, or  
           \item $\frac{1}{2}<C_2\leq 1,  1<\gamma <\frac{8 C_2-2}{8 C_2-3},  \lambda =2 \sqrt{\frac{(\gamma -1) C_2}{\gamma }}$, or  
           \item $\frac{1}{2}<C_2\leq 1,  \frac{8 C_2-2}{8 C_2-3}<\gamma <2,  \lambda =\sqrt{2} \sqrt{\frac{C_2}{4 C_2-1}}$, or  
           \item $\frac{1}{2}<C_2\leq 1,  1<\gamma <\frac{8 C_2-2}{8 C_2-3},  \lambda =\sqrt{2}
   \sqrt{\frac{C_2}{4 C_2-1}}$, or  
           \item $\frac{1}{2}<C_2\leq 1,  \frac{8 C_2-2}{8 C_2-3}<\gamma <2,  \lambda =\sqrt{2} \sqrt{\frac{C_2}{2 C_2-1}}$, or  
           \item $\frac{1}{2}<C_2\leq 1,  1<\gamma <\frac{8 C_2-2}{8
   C_2-3},  \lambda =\sqrt{2} \sqrt{\frac{C_2}{2 C_2-1}}$, or           \item $C_2>1,  \frac{4 C_2-2}{4 C_2-3}<\gamma <2,  \lambda =2 \sqrt{\frac{(\gamma -1) C_2}{\gamma }}$, or  
            \item $C_2>1,  \frac{8 C_2-2}{8
   C_2-3}<\gamma <\frac{4 C_2-2}{4 C_2-3},  \lambda =2 \sqrt{\frac{(\gamma -1) C_2}{\gamma }}$, or  
            \item $C_2>1,  1<\gamma <\frac{8 C_2-2}{8 C_2-3},  \lambda =2 \sqrt{\frac{(\gamma -1) C_2}{\gamma }}$, or 
           \item $C_2>1,  \gamma =\frac{4 C_2-2}{4 C_2-3},  \lambda =\sqrt{2} \sqrt{\frac{C_2}{4 C_2-1}}$, or  
           \item $C_2>1,  \gamma =\frac{8 C_2-2}{8 C_2-3},  \lambda =\sqrt{2} \sqrt{\frac{C_2}{4 C_2-1}}$, or  
           \item $C_2>1,  \frac{4 C_2-2}{4 C_2-3}<\gamma <2,  \lambda =\sqrt{2} \sqrt{\frac{C_2}{4 C_2-1}}$, or  
           \item $C_2>1,  \frac{8 C_2-2}{8 C_2-3}<\gamma <\frac{4 C_2-2}{4 C_2-3},  \lambda =\sqrt{2} \sqrt{\frac{C_2}{4
   C_2-1}}$, or  
           \item $C_2>1,  1<\gamma <\frac{8 C_2-2}{8 C_2-3},  \lambda =\sqrt{2} \sqrt{\frac{C_2}{4 C_2-1}}$, or  
           \item $C_2>1,  \gamma =\frac{4 C_2-2}{4 C_2-3},  \lambda =\sqrt{2} \sqrt{\frac{C_2}{2   C_2-1}}$, or  
           \item $C_2>1,  \gamma =\frac{8 C_2-2}{8 C_2-3},  \lambda =\sqrt{2} \sqrt{\frac{C_2}{2 C_2-1}}$, or  
           \item $C_2>1,  \frac{4 C_2-2}{4 C_2-3}<\gamma <2,  \lambda =\sqrt{2} \sqrt{\frac{C_2}{2   C_2-1}}$, or  
           \item $C_2>1,  \frac{8 C_2-2}{8 C_2-3}<\gamma <\frac{4 C_2-2}{4 C_2-3},  \lambda =\sqrt{2} \sqrt{\frac{C_2}{2 C_2-1}}$, or 
           \item $C_2>1,  1<\gamma <\frac{8 C_2-2}{8 C_2-3},  \lambda =\sqrt{2} \sqrt{\frac{C_2}{2 C_2-1}}$.
       \end{enumerate}
      \item $P_{15}(\lambda)$ is a hyperbolic saddle for: 
       \begin{enumerate}
           \item $\frac{1}{2}<C_2\leq 1,  \frac{8 C_2-2}{8 C_2-3}<\gamma <2$, \newline $  \sqrt{2} \sqrt{\frac{C_2}{4 C_2-1}}<\lambda <2 \sqrt{\frac{(\gamma-1)C_2}{\gamma }}$, or  
           \item $C_2>1,  \frac{4 C_2-2}{4 C_2-3}<\gamma <2$, \newline $  \sqrt{2} \sqrt{\frac{C_2}{4 C_2-1}}<\lambda <\sqrt{2} \sqrt{\frac{C_2}{2 C_2-1}}$, or 
           \item $C_2>1,  \frac{8 C_2-2}{8 C_2-3}<\gamma \leq \frac{4 C_2-2}{4 C_2-3}$, \newline $   \sqrt{2} \sqrt{\frac{C_2}{4 C_2-1}}<\lambda <2\sqrt{\frac{(\gamma-1)C_2}{\gamma }}$, or 
           \item $0<C_2\leq \frac{1}{4},  1<\gamma <2,  \lambda >2 \sqrt{\frac{(\gamma-1)C_2}{\gamma }}$, or 
           \item $\frac{1}{4}<C_2\leq \frac{1}{2},  1<\gamma <2$, \newline $   2 \sqrt{\frac{(\gamma-1)C_2}{\gamma }}<\lambda <\sqrt{2} \sqrt{\frac{C_2}{4 C_2-1}}$, or 
           \item $C_2>\frac{1}{2},  1<\gamma <\frac{8 C_2-2}{8 C_2-3}$, \newline $   2 \sqrt{\frac{(\gamma-1)C_2}{\gamma }}<\lambda <\sqrt{2}    \sqrt{\frac{C_2}{4 C_2-1}}$, or 
           \item $\frac{1}{4}<C_2\leq \frac{1}{2},  1<\gamma <2,  \lambda >\sqrt{2} \sqrt{\frac{C_2}{4 C_2-1}}$, or 
           \item $\frac{1}{2}<C_2\leq 1,  \frac{8 C_2-2}{8 C_2-3}<\gamma <2$, \newline $   2 \sqrt{\frac{(\gamma-1)C_2}{\gamma }}<\lambda <\sqrt{2} \sqrt{\frac{C_2}{2 C_2-1}}$, or 
           \item $\frac{1}{2}<C_2\leq 1,  1<\gamma \leq \frac{8 C_2-2}{8 C_2-3}$, \newline $   \sqrt{2} \sqrt{\frac{C_2}{4 C_2-1}}<\lambda <\sqrt{2} \sqrt{\frac{C_2}{2 C_2-1}}$, or 
           \item $C_2>1,  \frac{8 C_2-2}{8 C_2-3}<\gamma <\frac{4 C_2-2}{4 C_2-3}$, \newline $   2 \sqrt{\frac{(\gamma-1)C_2}{\gamma }}<\lambda <\sqrt{2} \sqrt{\frac{C_2}{2 C_2-1}}$, or 
           \item $C_2>1,  1<\gamma \leq \frac{8 C_2-2}{8 C_2-3}$, \newline $   \sqrt{2} \sqrt{\frac{C_2}{4 C_2-1}}<\lambda <\sqrt{2} \sqrt{\frac{C_2}{2 C_2-1}}$, or 
           \item $C_2>1,  \frac{4 C_2-2}{4 C_2-3}<\gamma <2$, \newline $   \sqrt{2} \sqrt{\frac{C_2}{2 C_2-1}}<\lambda <2 \sqrt{\frac{(\gamma-1)C_2}{\gamma }}$.
       \end{enumerate}
   \end{enumerate}

\end{enumerate}

\subsection{Discussion}
\label{SECT:4-3}
In this section,  timelike self-similar spherically symmetric models with scalar field \eqref{modelodos}, were  qualitatively analyzed using dynamical systems tools. The first notable feature of the present model is that for non- homogeneous scalar field $\phi(t, x)$ and its potential $ V(\phi (t, x)) $ to satisfy the homothetic symmetry imposed by the metric,  it is required \cite{Coley:2002je}: 
\begin{small}
\begin{align*}
& \phi(t,x)=\psi (x)-\lambda t, \quad  V(\phi(t,x))= e^{-2 t} U(\psi(x)),  \quad  U(\psi)=U_0 e^{-\frac{2 \psi}{\lambda}}.    
\end{align*}
\end{small}
It is assumed that $\lambda>0$, such that for $\psi>0$, $U\rightarrow 0$ as $\lambda \rightarrow 0$.
The equations were normalized with the variable $\theta$. 

Due to the computational complexity of the resulting problem, it was not possible to obtain and analytically treat all the equilibrium points of the system \eqref{reducedsystSF}. Hence, only some special cases of physical interest were considered, being \eqref{scalar-field-A} corresponding to a perfect fluid in the form of an ideal gas, \eqref{scalar-field-B} corresponding to the solutions in the invariant set $\Sigma = 0 $, \eqref{scalar-field-C} corresponding to the case of extreme inclination. Finally, the invariant set $A = v = 0 $ sets of system \eqref{campo-escalar-D} was studied. The hyperbolic points were completely classified according to their stability conditions.

\section{Conclusions}\label{ch_5}
\noindent 

In this paper the space of the solutions of the differential equations that result from considering perfect fluid and/ or scalar field as the matter content  in the Einstein- æther theory was studied. Einstein- æther  theory of gravity consists of General Relativity coupled to a vector field of unit time type, called the æther. In this effective theory, the Lorentz invariance is violated, but the locality and the covariance are preserved in the presence of the vector field. 

In section \ref{section2.2} the $ 1 + 3 $ formalism was discussed. This formalism is useful to write the field equations as a system of partial differential equations in two variables for spherically symmetric metrics. Furthermore, using the homothetic diagonal formulation, it was possible to write the partial differential equations as ordinary differential equations using the fact that the metric adapts to homothetic symmetry. The resulting equations (with algebraic restrictions) are very similar to those of the models with homogeneous spatial hypersurfaces \cite{Goliath:1998mw}. It was then possible to use the techniques of the qualitative theory of dynamical systems for the stability analysis of the solutions of the models. The analytical results were verified by numerical integration.

In the section \ref{aetheory} the Einstein-æther theory of gravity was presented, which contains the theory of General Relativity as a limit.  Conformally static metrics were studied in Einstein-æther theory for models of physical interest, such as pressure-free perfect fluids, perfect fluids, and models with extreme tilt. The stability criteria of the equilibrium points of the dynamical systems were obtained and discussed, imposing restrictions on the parameter space. Phase portraits were also presented to illustrate the qualitative behavior of the solutions.
The equilibrium points obtained by \cite{Goliath:1998mx} are recovered as particular cases of the present model.
In the notation $\text{Kernel} ^ {\text{sgn} (v)}_ {\text{sgn} ({\Sigma})} $ the kernel indicates the interpretation of the point: $ M, C $ represent the Minkowski spacetime; $ K $ represents a Kasner solution; $ T $ corresponds to static solutions; $ SL _ {\pm} $ corresponds to a flat FLRW space and static orbits depending on the parameter $\gamma $. $ H $ is associated with a change of causality of the homothetic vector field.  
The following results were retrieved: 
\begin{enumerate}
\item[$SL_{\pm}$]: Sonic lines defined by  ${A}=-\frac{\gamma 
   \varepsilon  (\gamma  (\Sigma
   +2)-2)}{4 (\gamma
   -1)^{3/2}}$,
$v=\varepsilon\sqrt{\gamma -1}$, were analyzed in section \ref{SL}. As a difference with General Relativity, for $1<\gamma<2$ and $C_2= \frac{\gamma
   ^2}{4 (\gamma -1)^2}$ the system \eqref{reducedsyst} admits the following equilibrium points:  
\begin{enumerate}
    \item[$SL_1$:] $\Sigma =
   \frac{2 (\gamma -1)}{\gamma
   },v=\sqrt{\gamma -1}, A=
  - \frac{\gamma  (\gamma  (\Sigma
   +2)-2)}{4 (\gamma
   -1)^{3/2}}$, 
    \item[$SL_2$:] $\Sigma =
   -\frac{2 (\gamma -1)}{\gamma
   },v= -\sqrt{\gamma -1}, A=
   \frac{\gamma  (\gamma  (\Sigma
   +2)-2)}{4 (\gamma
   -1)^{3/2}}$,
\end{enumerate}
which lie on the sonic line. If $ \gamma = 2, C_2 = 1 $ these points exist, and since $ \gamma = 2 $ the fluid behaves like stiff matter. Additionally, if $ \gamma = 2, C_2 = 1 $, these points correspond to models with extreme tilt ($ v = \varepsilon $), $SL_1: \Sigma = 1, A = -2, v = 1 $, and $SL_2: \Sigma = -1, A = 0, v = -1 $. $ SL_ {\pm}$ corresponds to a flat FLRW space and static orbits depending on the parameter $\gamma$.

\item[$\widetilde{M}^{\pm}:$] $({\Sigma}, {A}, v)=\left(0,1,\frac{(\gamma -1) \gamma  \pm \left(\sqrt{(\gamma -1) \left((\gamma
   -1) \gamma ^2+(2-\gamma ) (3 \gamma -2)\right)}\right)}{2-\gamma }\right)$, $({K},\Omega_t)=(0,0)$, exist for $C_2=1$. They represent the Minkowski space-time.
   
\item[$M^+:$] $({\Sigma}, {A}, v)=(0,1,1),\quad ({K},\Omega_t)=(0,0)$, corresponds to  $P_5$ for $C_2=1$.   It represents the Minkowski space-time.
    
\item[$M^-:$] $({\Sigma}, {A}, v)=(0,1,-1),\quad ({K},\Omega_t)=(0,0)$, corresponds to $P_6$ for $C_2=1$.  It represents the Minkowski space-time.
    
\item[$C^0:$] $({\Sigma}, {A}, v)=(0,0,0),\quad ({K},\Omega_t)=(1,0)$, corresponds to $N_1$. It represents the Minkowski space-time.
    
\item[$C^{\pm }:$] $({\Sigma}, {A}, v)=(0,0,\pm 1),\quad ({K},\Omega_t)=(1,0)$, correspond to $N_{2,3}$. They represent the Minkowski space-time. 
    
\item[$K^0_-:$] $({\Sigma}, {A}, v)=(-1,0,0),\quad ({K},\Omega_t)=(0,0)$, corresponds to $P_{11}$ for $C_2=1$.  It represents the Kasner solution.
    
\item[$K^0_+:$] $({\Sigma}, {A}, v)=(1,0,0),\quad ({K},\Omega_t)=(0,0)$, corresponds to $P_{12}$ for $C_2=1$.   It represents the Kasner solution.
    
\item$K^{\pm}_-:$] $({\Sigma}, {A}, v)=(-1,0,\pm 1),\quad ({K},\Omega_t)=(0,0)$, correspond to $P_{1,2}$ for $C_2=1$.    They represent the Kasner solution.
    
\item[$K^{\pm}_+:$] $({\Sigma}, {A}, v)=(1,0,\pm 1),\quad ({K},\Omega_t)=(0,0)$, correspond to $P_{3,4}$ for $C_2=1$.   They represent the Kasner solution.
    
\item[$T:$] $({\Sigma}, {A}, v)=\left(-2\frac{\gamma-1}{3\gamma-2},0,0\right)$, \newline $ ({K},\Omega_t)=\left(\frac{\gamma^2+4(\gamma-1)}{(3\gamma -2},\frac{4(\gamma-1)}{(3\gamma -2}\right)$,  corresponds to $P_{13}$ for $C_2=1$. 
    
\item[$H^-$:] The curve of equilibrium points $ {A}( {\Sigma})= {\Sigma}  +1$, $v( {\Sigma})=-1$, $\quad (0,-2 {\Sigma} {A})$. 
This line of equilibrium points is associated with a change of causality of the homothetic vector field. 
   \end{enumerate}
These results are of interest in Cosmology and Astrophysics.
   
In section \ref{SECT:4.1} conformally static metrics were studied in Einstein-æther theory for models with tilted perfect fluid and inhomogeneous scalar field with exponential potential, so the model contains the model studied in section \ref{model3} and thus  contains the model studied in \cite{Goliath:1998mw}. Particular cases of interest in Physics were studied, such as the perfect fluid in the form of an ideal gas, solutions with $\Sigma =  0$, models with extreme tilt and the invariant set $A = v = 0$. It was possible to study a more general model than the one studied in \cite{Goliath:1998mw}, and the results obtained by the authors were reproduced through the use of techniques from the qualitative theory of dynamical systems. A qualitative analysis of some invariant points was also made for models with timelike self-similar spherically symmetric metrics with a perfect fluid and a scalar field.  New equilibrium points were obtained, and their stability conditions were found either numerically or analytically,  by imposing restrictions on the parameter space.

\textbf{Acknowledgments}
G. L. and A. M. acknowledges to Agencia Nacional de Investigaci\'on y Desarrollo - ANID for financial support through the program FONDECYT Iniciaci\'on grant no.
11180126. Additionally, this research was funded by Vicerrector\'ia de Investigaci\'on y Desarrollo Tecnol\'ogico at Universidad Católica del Norte.

\bibliographystyle{unsrt}
\bibliography{librero}
\end{document}